%% file: Emulator.tex
\documentclass[9pt]{article}
\usepackage{sp_template}
\usepackage{tikz}
\usepackage{capt-of}
\usepackage{needspace}
\input{tikz_macros}

\input{atlas_macros}

\newcommand{\latimg}[1]{\includegraphics[width=1.55cm]{#1}}
\newcommand{\latimgsmall}[1]{\includegraphics[width=1.38cm]{#1}}
\newcommand{\rme}{{\rm{e}}}
\newcommand{\rmi}{{\rm{i}}}

\hypersetup{
  pdftitle  = {A universal emulator for planar Ising lattices},
  pdfauthor = {Riccardo Ben Al\`i Zinati, Alessandro Codello},
}

\title{A universal emulator \\ for planar Ising lattices}

\author{%
  Riccardo Ben Al\`i Zinati\affilmark{1}%
  \and
  Alessandro Codello\affilmark{2,3}%
}
\affiliation[1]{Universit\`a della Svizzera italiana, 6900 Lugano, Switzerland.}
\affiliation[2]{DSMN, Ca'\ Foscari University of Venice, Via Torino 155, 30172 Venice, Italy}
\affiliation[3]{IFFI, Universidad de la Rep\'ublica, J.H.y Reissig 565, 11300 Montevideo, Uruguay}

\date{\today}

\begin{document}

\maketitle

\begin{abstract}
We introduce the notion of an Ising emulator for two-dimensional Ising models: a broad class of flat, unit-edge-length lattices can be represented as site- or bond-diluted supercells of a single host lattice, for which the Feynman--Vdovichenko/Kac--Ward solution is fixed once and for all.
We construct explicit square and triangular emulators and show that a single transition matrix, supplemented by lattice-specific binary masks, gives all the thermodynamic quantities of interest for both ferro- and antiferromagnetic couplings.
We apply the framework to all eleven Archimedean lattices, to all twenty $2$-uniform lattices -- whose thermodynamics is obtained here for the first time -- and to several pentagonal lattices, and show that the same construction extends directly to fractal and disordered Ising models, with no modification to the underlying machinery.
\end{abstract}

\section{Introduction}

Exact studies of planar Ising models on Archimedean, Laves, and more general $k$-uniform lattices have seen renewed activity in recent years, motivated both by the intrinsic richness of their critical behaviour and by their use as a clean testing ground for theoretical and analytical approaches \cite{Codello_2010,Laurent_2025,Portillo_2025,Joseph_2026,Joseph_2026b}.
At the same time, a recent generalization of the underlying combinatorial Feynman--Vdovichenko / Kac--Ward (FV/KW) method \cite{Kac_Ward_1952,Potts_Ward_1952,Sherman_1960,Vodvicenko_1965,Feynman_1972} has made it possible to treat broad families of non-homogeneous lattices, ranging from fractals \cite{Codello:2015bia,Zinati:2025npw} to randomly diluted quenched disorder \cite{Zinati_Gori_Codello_2026}.

Here we introduce a common formalism to treat all these lattice Ising models within a single framework.
It rests on two facts:
(1) lattices that can be continuously deformed into one another without altering their topology share identical thermodynamic properties, since the loop counting intrinsic to the FV/KW method is invariant under such deformations;
and (2) a broad class of flat, unit-edge-length lattices --- including all the Archimedean, $2$-uniform, and pentagonal lattices studied here --- can be represented by a site or bond dilution of the triangular lattice.
Together, these two facts imply that the Ising model on any lattice of this class has the same thermodynamics as one defined on a finite, periodically repeated supercell of the triangular host lattice.

The conceptual step we emphasize in this work is that such a framework should be viewed as an {\textbf{Ising emulator}}: a reusable computational primitive that encodes an entire family of lattices through simple binary masks specifying the bond (or site) embedding pattern.
Once the local compatibility rules of the host lattice are fixed, all model-dependent information is confined to a small set of site- or bond-occupation matrices, while the thermodynamic computation itself remains unchanged.
A single FV/KW transition matrix can, in this sense, emulate a remarkably large class of periodic, decorated, fractal, and disordered lattices.
The goal of this paper is to make this statement concrete, to spell out the construction rules, and to demonstrate how they organize a broad atlas of exact critical temperatures and thermodynamic observables.

A complementary route to unified matrix formulations is the direct fermionic and Pfaffian representation of the planar Ising model \cite{Hurst_Green_1960,Kasteleyn_1963,Fisher_1966, Schultz_Mattis_Lieb_1964,Samuel_1980,Itzykson_1982,Plechko_1985,Sedrakyan_1984,Khachatryan_Sedrakyan_2009}. This approach has recently been shown to yield simple finite-dimensional momentum-space matrices for several lattices, including the inhomogeneous kagome and dice lattices \cite{Khachatryan_Zhang_Sedrakyan_2026,Khachatryan_Sedrakyan_2026}. There the matrix is constructed lattice by lattice, with general couplings; here a single host matrix is fixed once and for all, and the lattices are selected by binary masks.

The paper is structured as follows. 
After stating the emulator framework in Section \ref{sec:emulator-principle}, we first introduce the square host as the simplest one in which the supercell rules are completely transparent;
we then turn to the triangular host, which subsumes the square construction and provides the universal platform used throughout the rest of the work.
In Section \ref{sec:thermodynamics} we show how the corresponding FV/KW transition matrices give direct access to the
free energy, internal energy, specific heat, entropy and other thermodynamic observables for the lattice families studied here.
In Section \ref{examples} we apply the emulator framework to $k$-uniform lattices; first we reproduce the known thermodynamics of the eleven Archimedean lattices and then we extend the analysis to cover all twenty $k=2$ lattices, the thermodynamics of which is analysed here for the first time.
We also cover some non-uniform pentagonal lattices to provide further examples of the general use of the emulator framework.
All results are summarized in the atlas reported in the Appendix.
Finally, Section \ref{sec:targets} outlines applications to non-homogeneous -- fractals and quenched-disordered -- Ising models and sketches a road-map for future work.

\section{The emulator framework}
\label{sec:emulator-principle}

The combinatorial solution of the Ising model on a planar lattice $\Lambda$ relies on expressing the partition function in terms of the determinant of the  FV/KW transition matrix of a related random walk on $\Lambda$ \cite{Kac_Ward_1952,Potts_Ward_1952,Vodvicenko_1965,Feynman_1972}.
Conventionally, this matrix must be constructed anew for each lattice, since the specific arrangement of sites and bonds of $\Lambda$ determines the allowed transitions and therefore the matrix entries.
Yet this dependence on the detailed geometry of $\Lambda$ is stronger than the thermodynamics itself requires.
In fact, this determinant encodes the high-temperature expansion of the partition function, written as a signed sum over the closed loops of the lattice -- a topological count, depending on which bonds each loop occupies and how it winds, not on how the lattice is drawn.
Two planar realizations of the same graph therefore have identical thermodynamics, however differently they are embedded in the plane.
It is precisely this redundancy that makes an Ising emulator possible.

We exploit this freedom by fixing a single regular {\it host} lattice $\mathcal H$ --- for us the square or triangular lattice --- and by constructing the FV/KW matrix $\mathbb W_{\mathcal H}$ of the Ising model with unspecified binary weights as couplings on a finite $L\times L$  supercell of $\mathcal{H}$.
The supercell is the smallest block of $L^2$ host sites containing the bond- or site-occupation pattern needed to reproduce the original target lattice $\Lambda$.\footnote{We take square supercells of side $L$ for convenience; rectangular supercells are equally admissible.} This pattern is specified by binary masks, which fix the weights to $0$ or $1$ and thus record which host bonds or sites are retained.
Inserting these masks into $\mathbb W_{\mathcal H}$ yields the matrix $\mathbb W_{\mathcal H}^{\Lambda}$ of the Ising model on $\Lambda\subset\mathcal{H}$, namely $\Lambda$ viewed as a sublattice of $\mathcal H$. 
Changing the target lattice $\Lambda$ therefore amounts to changing the masks alone, leaving the host construction fixed.
For a periodic target lattice, $L$ is the size of its unit cell of $\Lambda \subset \mathcal H$; for a non-homogeneous target --- such as a fractal generation or a disordered sample --- it is the size of a finite approximant, systematically refined by increasing $L$.

This separation has the character of the division between hardware and software in a computer. The host lattice and its FV/KW matrix $\mathbb W_{\mathcal H}$ are fixed hardware; the binary masks are lattice-specific programs. The same device solves every $\Lambda$ in its representable class, once supplied with the corresponding masks.

On a host with $\mu$ independent bond directions, we define a \emph{bond program} as a collection of binary masks
\begin{equation}
\bigl(w^{(1)},\ldots,w^{(\mu)}\bigr),
\qquad
w^{(a)}\in\{ 0,1 \}^{L\times L}.
\label{eq:bond-program}
\end{equation}
The entry $w^{(a)}_{ij}=1$ indicates that the bond starting from site $(i,j)$ in direction $a$ is retained, while $w^{(a)}_{ij}=0$ deletes it. A \emph{site program} is the special case in which a single binary matrix
\begin{equation}
w\in\{ 0,1\}^{L\times L}\,,
\label{eq:site-program}
\end{equation}
specifies which host sites are retained; the associated bond program is then obtained by removing all bonds incident on absent sites.
Throughout, the first index $i$ of a mask labels its rows and the second index $j$ its columns.
Inserting a program into the parent matrix $\mathbb{W}_{\mathcal{H}}$
gives the programmed matrix $\mathbb W^\Lambda_{\mathcal H}$. 
The corresponding FV/KW determinant of $\Lambda \subset \mathcal H$ is
\begin{equation}\label{eq:KW-determinant}
P^\Lambda_{\mathcal H}(v,k_1,k_2) = \det\!\big[\mathbb I-v\,\mathbb W^\Lambda_{\mathcal H}(k_1,k_2)\big]\,, \quad\quad v=\tanh(\xi/T)
\end{equation}
where $(k_1,k_2)$ are the lattice momenta, and $\xi=+1$ for ferromagnetic interactions while $\xi=-1$ for antiferromagnetic interactions.
Throughout this work we consider the nearest-neighbour Ising Hamiltonian $H=-\xi J\sum_{\langle ij\rangle}\sigma_i\sigma_j$ with a uniform coupling $J>0$ on all retained bonds, and we measure temperatures and energies in units of $J$ (with $k_B=1$), i.e.\ we set $J=1$. All bonds of a unit-edge-length lattice are thus treated on the same footing and carry the same coupling.
In Section \ref{sec:thermodynamics} all thermodynamic quantities are obtained from equation \eqref{eq:KW-determinant}.

We note that the representation $\Lambda \subset \mathcal H$ provided by a program is not unique. We refer to the equivalence class $[\Lambda]$ as the set of programs, on any admissible supercell size
$L\times L$, realizing lattices topologically equivalent to $\Lambda$.
Non-uniqueness within $[\Lambda]$ has several sources: the same lattice can be represented on supercells of different size; by a site program rather than a bond program, or vice versa; by embeddings related by a symmetry of the discrete $L\times L$ supercell (translations, rotations, reflections); or by an altogether different embedding into the host (a systematic account of this redundancy will be reported elsewhere). In practice one simply works with whichever representative of $[\Lambda]$ is most convenient.

We call \emph{representable class} $\mathcal R(\mathcal H)$ the set of periodic planar lattices that are topologically equivalent to a site- or bond-diluted periodic subgraph of the host $\mathcal H$. This is the precise sense in which we use the word universal: a single host and a single FV/KW matrix $\mathbb W_{\mathcal H}$ serve every lattice in $\mathcal R(\mathcal H)$. Membership is established constructively, by exhibiting a program, and all the lattices treated in this work belong to $\mathcal R(\triangle)$. A necessary condition is that $\Lambda$ have maximal coordination at most six (at most four for $\mathcal R(\square)$); flat unit-edge-length tilings with higher coordination therefore lie outside $\mathcal R(\triangle)$. We conjecture that, conversely, every periodic flat unit-edge-length lattice with maximal coordination at most six belongs to $\mathcal R(\triangle)$.

Finally, we remark that in the following we consider the square and triangular host lattices, but other host lattices might be useful in specific applications, for instance to reduce the computational overhead. 

\section{Square and triangular emulators}

The purpose of this section is to make the programming rule explicit for the two hosts used in this work.
We first describe the square emulator, which is the simplest case and suffices for lattices that can be embedded using two independent bond directions.
We then describe the triangular emulator, which contains the square emulator as a special sector and is the host used for the broader class of examples below.

\subsection{The square host}
\label{sec:square-emulator}

The square host lattice $\mathcal{H} = \square$ has vertices labelled by $(i,j)$ and two independent bond directions.
We denote them by \{1, 2\}, while the opposite directed bonds are labelled \{4, 5\}:
\begin{center}
\centering
\begin{tikzpicture}[scale=1.3, every node/.style={font=\small}]
\draw[gray!30, thin] (-1.75,0) -- (1.75,0);
\draw[gray!30, thin] (0,-1.75) -- (0,1.75);
\draw[gray!30, thin] (+1,-1.75) -- (+1,1.75);
\draw[gray!30, thin] (-1,-1.75) -- (-1,1.75);
\draw[gray!30, thin] (-1.75,+1) -- (1.75,+1);
\draw[gray!30, thin] (-1.75,-1) -- (1.75,-1);
\filldraw[gray!30] (1,1) circle (2pt);
\filldraw[gray!30] (-1,-1) circle (2pt);
\filldraw[gray!30] (-1,1) circle (2pt);
\filldraw[gray!30] (1,-1) circle (2pt);
\filldraw[black] (1,0) circle (2pt) node[right = 1pt, font=\scriptsize] {$(i+1,j)$};
\filldraw[black] (0,1) circle (2pt) node[above =1pt, font=\scriptsize] {$(i,j+1)$};
\filldraw[black] (-1,0) circle (2pt) node[left=1pt, font=\scriptsize] {$(i-1,j)$};
\filldraw[black] (0,-1) circle (2pt) node[below=2pt, font=\scriptsize] {$(i,j-1)$};
\draw[->, very thick, color=blue] (0,0) -- (1,0);
\draw[->, very thick, color=red] (0,0) -- (0,1);
\draw[->, very thick, color=gray] (0,0) -- (-1,0);
\draw[->, very thick, color=gray] (0,0) -- (0,-1);
\node[black, right=4pt] at (2,0) {${\color{blue}\bf{1}}$};
\node[black, above=4pt] at (0,1.5) {${\color{red}\bf{2}}$};
\node[black, left=1pt] at (-2,0) {$\bf{4}$};
\node[black, below=4pt] at (0,-1.5) {$\bf{5}$};
\filldraw[black] (0,0) circle (2pt) node[below left=1pt, font=\scriptsize] {$(i,j)$};
\end{tikzpicture}
\captionof{figure}{Bond-direction convention on the square host $\square$. Directions $\{1,2\}$ (blue, red) are the independent directions; directions $\{4,5\}$ (grey) are their opposites.}
\label{fig:square-directions}
\end{center}
The labels are chosen so that the square host embeds naturally into the triangular notation introduced below, where labels $1$--$6$ enumerate the six directions of the triangular host. The square emulator uses only the four labels $\{1, 2, 4, 5\}$; labels $3$ and $6$ are absent.
A square-lattice bond program is specified by two binary matrices
\begin{equation}
w^{(1)},\; w^{(2)} \;\in\; \{0,1\}^{L\times L}\,,
\end{equation}
encoding for each site in the supercell which bonds are present or absent along direction 1 or 2 respectively.
The remaining two masks, $w^{(4)}$ and $w^{(5)}$, are fixed by the requirement that the two orientations of the same undirected bond agree:
\begin{equation}
w^{(4)}_{ij} \;=\; w^{(1)}_{i-1,j}\, ,
\qquad\qquad
w^{(5)}_{ij} \;=\; w^{(2)}_{i,j-1}\, .
\label{eq:square-consistency}
\end{equation}
The square program therefore has $4L^2$ binary entries, of which half are independent. When site dilution suffices, a single binary matrix $w\in\{0,1\}^{L\times L}$ determines all four directed masks via $w^{(a)}_{ij} = w_{ij}$. Cut bonds incident on absent sites contribute nothing to the determinant.

We recall that the FV/KW combinatorial solution of the Ising model first maps the high temperature expansion to a loop counting problem on the lattice, which is then solved by constructing an auxiliary non-backtracking random walk with complex transition amplitudes, especially engineered so that the walk counting correctly matches the original loop counting.
This auxiliary random walk is analysed using a master equation for which the FV/KW matrix is the transition matrix.
Therefore, generalising the standard construction, $\mathbb W_\square$ is conveniently defined as the weighted adjacency matrix of the following graph:
\begin{equation}
\scalebox{0.9}{$
\begin{array}{cccccccccccccccccc}
\underset{(i,j)}{1} & \overset{1\,w_{ij}^{(1)}\rme^{\rmi k_1}}{\longleftrightarrow} & \underset{(i+1,j)}{1} &  &  &
\underset{(i,j)}{2} & \overset{\beta\,w_{ij}^{(2)}\rme^{\rmi k_2}}{\longleftrightarrow} & \underset{(i,j+1)}{1} &  &  &
\underset{(i,j)}{4} & \overset{0\,w_{ij}^{(4)}\rme^{-\rmi k_1}}{\longleftrightarrow} & \underset{(i-1,j)}{1} &  &  &
\underset{(i,j)}{5} & \overset{\alpha\,w_{ij}^{(5)}\rme^{-\rmi k_2}}{\longleftrightarrow} & \underset{(i,j-1)}{1} \\[6pt]
\underset{(i,j)}{1} & \overset{\alpha\,w_{ij}^{(1)}\rme^{\rmi k_1}}{\longleftrightarrow} & \underset{(i+1,j)}{2} &  &  &
\underset{(i,j)}{2} & \overset{1\,w_{ij}^{(2)}\rme^{\rmi k_2}}{\longleftrightarrow} & \underset{(i,j+1)}{2} &  &  &
\underset{(i,j)}{4} & \overset{\beta\,w_{ij}^{(4)}\rme^{-\rmi k_1}}{\longleftrightarrow} & \underset{(i-1,j)}{2} &  &  &
\underset{(i,j)}{5} & \overset{0\,w_{ij}^{(5)}\rme^{-\rmi k_2}}{\longleftrightarrow} & \underset{(i,j-1)}{2} \\[6pt]
\underset{(i,j)}{1} & \overset{0\,w_{ij}^{(1)}\rme^{\rmi k_1}}{\longleftrightarrow} & \underset{(i+1,j)}{4} &  &  &
\underset{(i,j)}{2} & \overset{\alpha\,w_{ij}^{(2)}\rme^{\rmi k_2}}{\longleftrightarrow} & \underset{(i,j+1)}{4} &  &  &
\underset{(i,j)}{4} & \overset{1\,w_{ij}^{(4)}\rme^{-\rmi k_1}}{\longleftrightarrow} & \underset{(i-1,j)}{4} &  &  &
\underset{(i,j)}{5} & \overset{\beta\,w_{ij}^{(5)}\rme^{-\rmi k_2}}{\longleftrightarrow} & \underset{(i,j-1)}{4} \\[6pt]
\underset{(i,j)}{1} & \overset{\beta\,w_{ij}^{(1)}\rme^{\rmi k_1}}{\longleftrightarrow} & \underset{(i+1,j)}{5} &  &  &
\underset{(i,j)}{2} & \overset{0\,w_{ij}^{(2)}\rme^{\rmi k_2}}{\longleftrightarrow} & \underset{(i,j+1)}{5} &  &  &
\underset{(i,j)}{4} & \overset{\alpha\,w_{ij}^{(4)}\rme^{-\rmi k_1}}{\longleftrightarrow} & \underset{(i-1,j)}{5} &  &  &
\underset{(i,j)}{5} & \overset{1\,w_{ij}^{(5)}\rme^{-\rmi k_2}}{\longleftrightarrow} & \underset{(i,j-1)}{5}
\end{array}$}
\label{eq:square-graph}
\end{equation}
where $\alpha = \rme^{\rmi \frac{\pi}{4}}$ and $\beta = 1/\alpha$.
The rationale behind the construction is the following:
direction $a$ at site $(i,j)$ is connected to the four directions at the neighbouring site, determined according to Fig.~\ref{fig:square-directions} above; this follows from the fact that a walker leaving site $(i,j)$ in direction $a$ reaches the neighbouring site, from which it can continue in any direction except the one leading back to $(i,j)$ (backtracking is excluded by the zero weight).
The graph weights on each link are the product of: (1) the FV phase factor to ensure the correct loop counting, (2) the binary mask weight $w^{(a)}_{i,j}$ and (3) the Fourier phase factor associated with that direction.    
The matrix $\mathbb W_\square$ has dimension $4L^2$ (as the number of binary entries) since there are four directions for each one of the $L^2$ sites of the square lattice supercell.

\paragraph{Worked examples: honeycomb and bathroom tile.}
The honeycomb lattice $(6,6,6)$ is the simplest target accommodated by the square emulator.
As is well known, the honeycomb lattice admits the \emph{brick-wall} realization, in which each hexagonal face is squished into a $1\times 2$ rectangle.
The result is a sublattice of the square host: every vertex of the brick wall sits on a vertex of $\square$, and every brick-wall bond is a unit step along directions $1$ or $2$. The bonds absent from the brick wall are the vertical bonds removed in a $1\times 2$ staggered pattern, which is captured by an $L=2$ supercell.
\begin{center}
\begin{tikzpicture}[
    paneltitle/.style={
        font=\normalsize,
        anchor=south
    },
    panellabel/.style={
        font=\bfseries,
        anchor=north west
    },
    arrow/.style={
       ->,
        black!60,
        line width=0.8pt
    },
    matrixnode/.style={
    anchor=center,
    inner sep=0pt,
    font=\normalsize
    },
    vertexnumber/.style={
        font=\tiny,
        inner sep=1pt,
        fill=white,
        fill opacity=0.85,
        text opacity=1
    },
    coordarrow/.style={
    ->,
    >=stealth,
    thin,
    black!70
}
]
 
\def\PanelW{4.2cm}
\def\PanelGap{0.75cm}
\def\ArrowGap{0.18cm}
\def\PanelTop{4.25cm}
 
\coordinate (A0) at (0,0);
\coordinate (B0) at ($(A0)+(\PanelW+\PanelGap,0)$);
\coordinate (C0) at ($(B0)+(\PanelW+\PanelGap,0)$);
 
\node[
    anchor=south west,
    inner sep=0pt
] (panelAimg) at (A0) {
    \includegraphics[width=\PanelW]{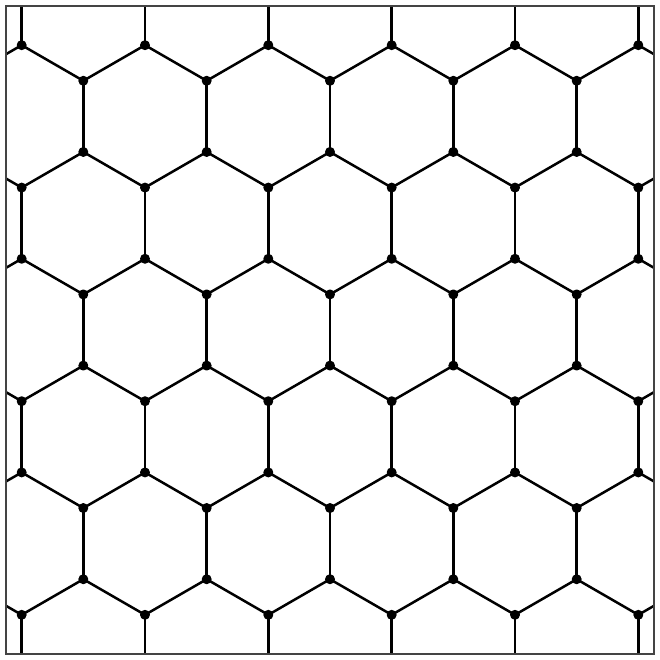}
};
 
\node[paneltitle] at ($(panelAimg.north)+(0,0.1cm)$)
    {$\Lambda = (6,6,6)$};
 
\node[
    anchor=south west,
    inner sep=0pt
] (panelBimg) at (B0) {
    \includegraphics[width=\PanelW]{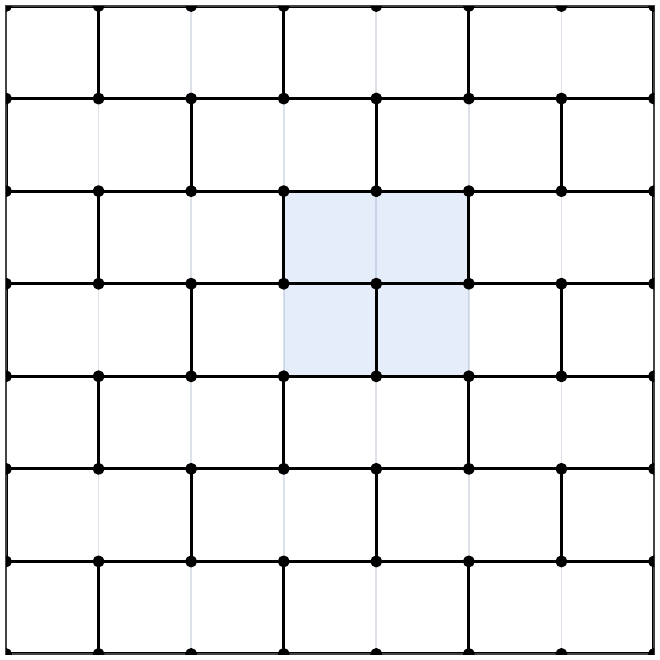}
};
 
\node[paneltitle] at ($(panelBimg.north)+(0,0.1cm)$)
    {$\Lambda \subset \square$};
 
\begin{scope}[
    shift={($(panelBimg.south west)$)},
    x={($(panelBimg.south east)-(panelBimg.south west)$)},
    y={($(panelBimg.north west)-(panelBimg.south west)$)}
]
 
\node[vertexnumber] at (0.43,0.50) {${\color{red}0}$};
\node[vertexnumber] at (0.43,0.64) {${\color{red}1}$};
\node[vertexnumber] at (0.57,0.50) {${\color{red}1}$};
\node[vertexnumber] at (0.57,0.64) {${\color{red}0}$};
\node[vertexnumber] at (0.50,0.43) {${\color{blue}1}$};
\node[vertexnumber] at (0.50,0.57) {${\color{blue}1}$};
\node[vertexnumber] at (0.64,0.43) {${\color{blue}1}$};
\node[vertexnumber] at (0.64,0.57) {${\color{blue}1}$};
 
\draw[coordarrow] (0.36,0.36) -- (0.36,0.5) node[pos=1.65, anchor=north,  font=\tiny] {$j$};
\draw[coordarrow] (0.36,0.36) -- (0.5,0.36) node[pos=1.45, anchor=east,  font=\tiny] {$i$};
 
\draw[arrow, color=red] (0.06,0.06) -- (0.06,0.2) node[pos=1.65, anchor=north,  font=\tiny] {${\color{red}(2)}$};
\draw[arrow, color=blue] (0.06,0.06) -- (0.2,0.06) node[pos=1.75, anchor=east,  font=\tiny] {${\color{blue}(1)}$};
 
\foreach \x/\y in {0.43/0.43,0.57/0.43,0.43/0.57,0.57/0.57}
  \fill[black] (\x,\y) circle (2.3pt);
\end{scope}
 
\coordinate (Csw) at (C0);
\coordinate (Cse) at ($(Csw)+(\PanelW,0)$);
\coordinate (Cnw) at ($(Csw)+(0,\PanelW)$);
\coordinate (Cne) at ($(Csw)+(\PanelW,\PanelW)$);
\coordinate (Ccenter) at ($(Csw)!0.5!(Cne)$);
 
 
\node[
    anchor=center,
    inner sep=0pt,
    text width=\PanelW,
    align=center,
    font=\normalsize
] at (Ccenter) {%
{\color{blue}
\[
\begin{array}{ccc}
w^{(1)} & \!\!=\!\! & \begin{array}{cc}
 & {\footnotesize j}\;\rightarrow\\
\!\!\!\! \begin{array}{c}
{\footnotesize i}\\
\downarrow
\end{array}\!\!\!\!\!\!\!\!\!& \left(\begin{array}{cc}
1 & 1 \\
1 & 1\\
\end{array}\right)\\
\\
\end{array}
\end{array}
\]}
\vspace{0.0em}
{\color{red}
\[
\begin{array}{ccc}
w^{(2)} & \!\!=\!\! & \begin{array}{cc}
 & {\footnotesize j}\;\rightarrow\\
\!\!\!\! \begin{array}{c}
{\footnotesize i}\\
\downarrow
\end{array}\!\!\!\!\!\!\!\!\! & \left(\begin{array}{cc}
0 & 1 \\
1 & 0\\
\end{array}\right)\\
\\
\end{array}
\end{array}
\]
}
};

\draw[arrow]
    ($(panelAimg.east)+(\ArrowGap,0)$)
    --
    ($(panelBimg.west)-(\ArrowGap,0)$);
 
\draw[arrow]
    ($(panelBimg.east)+(\ArrowGap,0)$)
    --
    ($(Csw)+(0,\PanelW/2)-(\ArrowGap,0)$);

\end{tikzpicture}
\captionof{figure}{Honeycomb lattice $\Lambda=(6,6,6)$ emulated on the square host. Left: the target honeycomb graph. Middle: its brick-wall realization $\Lambda \subset \square$, with the $L=2$ supercell (shaded, its four sites marked by enlarged dots), the supercell coordinates $(i,j)$, and the independent bond directions $\{1,2\}$; the number next to each bond leaving a supercell site is the corresponding mask entry. Right: the resulting binary bond masks $w^{(1)}$, $w^{(2)}$. The entry $w^{(a)}_{ij}$ is the label of the bond leaving the site at horizontal position $i$ and vertical position $j$ in the middle panel; since $i$ labels the rows and $j$ the columns of the masks, each $w^{(a)}$ is the transpose of the pattern of 0s and 1s drawn in the figure (with the vertical axis pointing upwards).}
\label{fig:honeycomb-embedding}
\end{center}
\vspace{5pt}
To read off the program, we walk over the four sites of the supercell (the enlarged dots in Fig.~\ref{fig:honeycomb-embedding}) and ask, at each one, which outgoing bonds are present.
Every horizontal bond of the host is retained, so the right-mask $w^{(1)}$ is the all-ones matrix.
The up-bonds, by contrast, follow the staggered brick-wall pattern: at sites $(1,1)$ and $(2,2)$ the up-bond is absent, while at $(1,2)$ and $(2,1)$ it is present. This gives
\begin{equation}
w^{(1)} \;=\; \begin{pmatrix} 1 & 1 \\ 1 & 1 \end{pmatrix}\,,
\qquad
w^{(2)} \;=\; \begin{pmatrix} 0 & 1 \\ 1 & 0 \end{pmatrix}\,,
\label{eq:honeycomb-program}
\end{equation}
where, following the convention stated after Eq.~\eqref{eq:site-program}, $i$ labels rows and $j$ columns; since in Fig.~\ref{fig:honeycomb-embedding} $i$ runs horizontally and $j$ vertically (upwards), each mask is the transpose of the pattern of 0s and 1s drawn in the figure (the distinction is immaterial here, the matrices being symmetric, but it matters for the larger programs below). The masks $w^{(4)}, w^{(5)}$ are determined by the consistency conditions~\eqref{eq:square-consistency}
\begin{equation}
w^{(4)} \;=\; \begin{pmatrix} 1 & 1 \\ 1 & 1 \end{pmatrix}\,,
\qquad
w^{(5)} \;=\; \begin{pmatrix} 1 & 0 \\ 0 & 1 \end{pmatrix}\,.
\end{equation}
It is easy to check that these last two matrices also agree with what can be read off from Fig.~\ref{fig:honeycomb-embedding}. 
The parent matrix at $L=2$ has dimension $\big|\mathbb{W}^{(6,6,6)}_\square\big| = 16$, and has the following form, which we report for illustrative purposes:
\begin{equation}
\scalebox{0.65}{$
\mathbb{W}^{(6,6,6)}_\square = \left(
\begin{array}{cccccccccccccccc}
 0 & \rme^{\rmi k_1} & \alpha\,\rme^{\rmi k_1} & 0 & \beta\,\rme^{\rmi k_1} & 0 & 0 & 0 & 0 & 0 & 0 & 0 &
   0 & 0 & 0 & 0 \\
 \rme^{\rmi k_1} & 0 & 0 & 0 & 0 & \alpha \,\rme^{\rmi k_1} & 0 & 0 & 0 & 0 & 0 & \beta \,\rme^{\rmi k_1} &
   0 & 0 & 0 & 0 \\
 0 & 0 & 0 & 0 & 0 & 0 & 0 & 0 & 0 & 0 & 0 & 0 & \beta \, \rme^{\rmi k_2} & \rme^{\rmi k_2} & \alpha \, \rme^{\rmi
   k_2} & 0 \\
 0 & 0 & 0 & 0 & 0 & \beta \, \rme^{-\rmi k_1} & 0 & 0 & 0 & 0 & \rme^{-\rmi k_1} & \alpha \, \rme^{-\rmi
   k_1} & 0 & 0 & 0 & 0 \\
 0 & 0 & 0 & 0 & 0 & 0 & 0 & 0 & 0 & 0 & 0 & 0 & 0 & 0 & 0 & 0 \\
 0 & 0 & 0 & 0 & 0 & 0 & 0 & 0 & 0 & 0 & 0 & 0 & 0 & 0 & 0 & 0 \\
 0 & 0 & 0 & 0 & 0 & 0 & 0 & 0 & 0 & 0 & 0 & 0 & \rme^{\rmi k_1} & \alpha \, \rme^{\rmi k_1} & 0 & \beta \, 
   \rme^{\rmi k_1} \\
 \beta \,  \rme^{\rmi k_2} & 0 & 0 & 0 & 0 & \rme^{\rmi k_2} & 0 & 0 & 0 & 0 & \alpha \, \rme^{\rmi k_2} & 0 &
   0 & 0 & 0 & 0 \\
 0 & 0 & 0 & 0 & 0 & 0 & 0 & 0 & 0 & 0 & 0 & 0 & 0 & \beta \,  \rme^{-\rmi k_1} & \rme^{-\rmi k_1} & \alpha \,
   \rme^{-\rmi k_1} \\
 0 & 0 & 0 & 0 & 0 & 0 & 0 & 0 & 0 & 0 & 0 & 0 & 0 & 0 & 0 & 0 \\
 0 & 0 & \beta \, \rme^{-\rmi k_1} & \rme^{-\rmi k_1} & \alpha  \, \rme^{-\rmi k_1} & 0 & 0 & 0 & 0 & 0 & 0 &
   0 & 0 & 0 & 0 & 0 \\
 0 & 0 & 0 & 0 & 0 & 0 & \alpha \,  \rme^{-\rmi k_2} & 0 & \beta \, \rme^{-\rmi k_2} & \rme^{-\rmi k_2} & 0 &
   0 & 0 & 0 & 0 & 0 \\
 0 & 0 & 0 & 0 & 0 & 0 & \rme^{\rmi k_1} & \alpha \, \rme^{\rmi k_1} & 0 & \beta \, \rme^{\rmi k_1} & 0 & 0 &
   0 & 0 & 0 & 0 \\
 0 & 0 & 0 & 0 & 0 & 0 & 0 & 0 & 0 & 0 & 0 & 0 & 0 & 0 & 0 & 0 \\
 0 & 0 & 0 & 0 & 0 & 0 & 0 & \beta \, \rme^{-\rmi k_1} & \rme^{-\rmi k_1} & \alpha \, \rme^{-\rmi k_1} & 0 &
   0 & 0 & 0 & 0 & 0 \\
 0 & \alpha \, \rme^{-\rmi k_2} & 0 & \beta \, \rme^{-\rmi k_2} & \rme^{-\rmi k_2} & 0 & 0 & 0 & 0 & 0 & 0 &
   0 & 0 & 0 & 0 & 0 \\
\end{array}
\right)
$}
\label{eq:honeycomb-W}
\end{equation}
This explicit form has no intrinsic meaning, since it depends on the chosen program and is not an observable;
to obtain a first validation of our emulator construction we therefore compute the simplest thermodynamic quantity: the critical temperature. This is directly related to the unique real eigenvalue $\lambda > 1$ of $\mathbb{W}^{(6,6,6)}_\square$ (see Section \ref{sec:thermodynamics} below).
A simple computation correctly reproduces the Wannier--Onsager critical temperature \cite{Onsager_1944}
\begin{equation}
T_c \;=\; \frac{2}{\log(2 + \sqrt{3})} \;=\; 1.518651...\,.
\end{equation}
%

The same emulator handles the bathroom-tile lattice $(4,8,8)$ at $L=4$ with parent matrix dimension  $\big| \mathbb{W}^{(4,8,8)}_\square\big| = 64$. The supercell program is constructed as follows (Fig.~\ref{fig:bathroom-tile-embedding}):
\begin{center}
\begin{tikzpicture}[
    paneltitle/.style={
        font=\normalsize,
        anchor=south
    },
    panellabel/.style={
        font=\bfseries,
        anchor=north west
    },
    arrow/.style={
       ->,
        black!60,
        line width=0.8pt
    },
    matrixnode/.style={
    anchor=center,
    inner sep=0pt,
    font=\normalsize
    },
    vertexnumber/.style={
        font=\tiny,
        inner sep=1pt,
        fill=white,
        fill opacity=0.85,
        text opacity=1
    },
    coordarrow/.style={
    ->,
    >=stealth,
    thin,
    black!70
}
]
 
\def\PanelW{4.2cm}
\def\PanelGap{0.75cm}
\def\ArrowGap{0.18cm}
\def\PanelTop{4.25cm}
 
\coordinate (A0) at (0,0);
\coordinate (B0) at ($(A0)+(\PanelW+\PanelGap,0)$);
\coordinate (C0) at ($(B0)+(\PanelW+\PanelGap,0)$);
 
\node[
    anchor=south west,
    inner sep=0pt
] (panelAimg) at (A0) {
    \includegraphics[width=\PanelW]{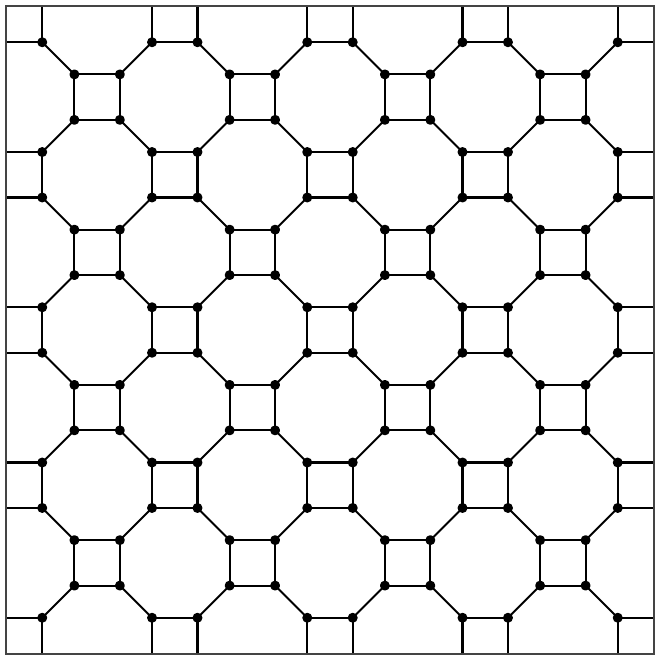}
};
 
\node[paneltitle] at ($(panelAimg.north)+(0,0.2cm)$)
    {$\Lambda = (4,8,8)$};
 
\node[
    anchor=south west,
    inner sep=0pt
] (panelBimg) at (B0) {
    \includegraphics[width=\PanelW]{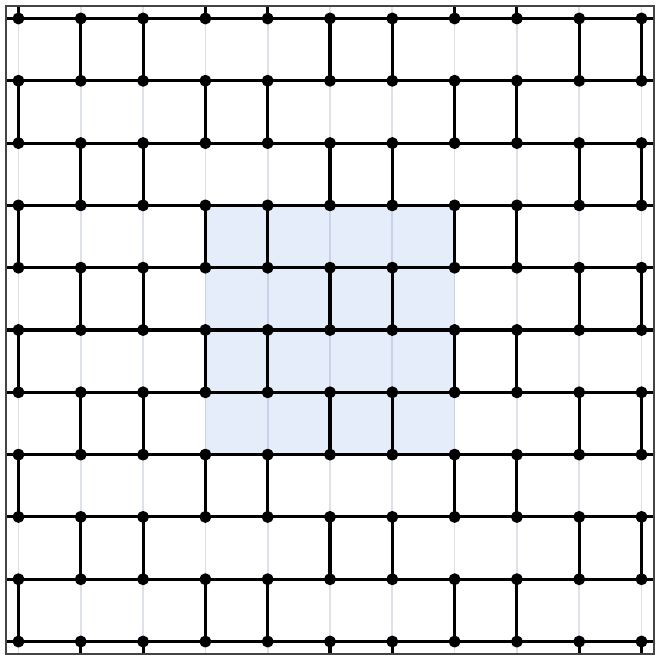}
};
 
\node[paneltitle] at ($(panelBimg.north)+(0,0.2cm)$)
    {$\Lambda \subset \square$};
 
\begin{scope}[
    shift={($(panelBimg.south west)$)},
    x={($(panelBimg.south east)-(panelBimg.south west)$)},
    y={($(panelBimg.north west)-(panelBimg.south west)$)}
]

\node[vertexnumber] at (0.312,0.64) {${\color{red}1}$};
\node[vertexnumber] at (0.312,0.545) {${\color{red}0}$};
\node[vertexnumber] at (0.312,0.45) {${\color{red}1}$};
\node[vertexnumber] at (0.312,0.355) {${\color{red}0}$};
\node[vertexnumber] at (0.406,0.64) {${\color{red}1}$};
\node[vertexnumber] at (0.406,0.545) {${\color{red}0}$};
\node[vertexnumber] at (0.406,0.45) {${\color{red}1}$};
\node[vertexnumber] at (0.406,0.355) {${\color{red}0}$};
\node[vertexnumber] at (0.5,0.64) {${\color{red}0}$};
\node[vertexnumber] at (0.5,0.545) {${\color{red}1}$};
\node[vertexnumber] at (0.5,0.45) {${\color{red}0}$};
\node[vertexnumber] at (0.5,0.355) {${\color{red}1}$};
\node[vertexnumber] at (0.594,0.64) {${\color{red}0}$};
\node[vertexnumber] at (0.594,0.545) {${\color{red}1}$};
\node[vertexnumber] at (0.594,0.45) {${\color{red}0}$};
\node[vertexnumber] at (0.594,0.355) {${\color{red}1}$};
\node[vertexnumber] at (0.359,0.31) {${\color{blue}1}$};
\node[vertexnumber] at (0.453,0.31) {${\color{blue}1}$};
\node[vertexnumber] at (0.547,0.31) {${\color{blue}1}$};
\node[vertexnumber] at (0.641,0.31) {${\color{blue}1}$};
\node[vertexnumber] at (0.359,0.405) {${\color{blue}1}$};
\node[vertexnumber] at (0.453,0.405) {${\color{blue}1}$};
\node[vertexnumber] at (0.547,0.405) {${\color{blue}1}$};
\node[vertexnumber] at (0.641,0.405) {${\color{blue}1}$};
\node[vertexnumber] at (0.359,0.5) {${\color{blue}1}$};
\node[vertexnumber] at (0.453,0.5) {${\color{blue}1}$};
\node[vertexnumber] at (0.547,0.5) {${\color{blue}1}$};
\node[vertexnumber] at (0.641,0.5) {${\color{blue}1}$};
\node[vertexnumber] at (0.359,0.595) {${\color{blue}1}$};
\node[vertexnumber] at (0.453,0.595) {${\color{blue}1}$};
\node[vertexnumber] at (0.547,0.595) {${\color{blue}1}$};
\node[vertexnumber] at (0.641,0.595) {${\color{blue}1}$};
 
\draw[coordarrow] (0.26,0.26) -- (0.26,0.4) node[pos=1.65, anchor=north,  font=\tiny] {$j$};
\draw[coordarrow] (0.26,0.26) -- (0.4,0.26) node[pos=1.45, anchor=east,  font=\tiny] {$i$};
 
\draw[arrow, color=red] (0.08,0.08) -- (0.08,0.22) node[pos=1.65, anchor=north,  font=\tiny] {${\color{red}(2)}$};
\draw[arrow, color=blue] (0.08,0.08) -- (0.22,0.08) node[pos=1.75, anchor=east,  font=\tiny] {${\color{blue}(1)}$};
 
\foreach \x in {0.312,0.406,0.5,0.594}
  \foreach \y in {0.31,0.405,0.5,0.595}
    \fill[black] (\x,\y) circle (1.5pt);
\end{scope}
 
\coordinate (Csw) at (C0);
\coordinate (Cse) at ($(Csw)+(\PanelW,0)$);
\coordinate (Cnw) at ($(Csw)+(0,\PanelW)$);
\coordinate (Cne) at ($(Csw)+(\PanelW,\PanelW)$);
\coordinate (Ccenter) at ($(Csw)!0.5!(Cne)$);
 
\node[
    anchor=center,
    inner sep=0pt,
    text width=\PanelW,
    align=center,
    font=\normalsize
] at (Ccenter) {%
{\color{blue}
\[
w^{(1)} \;=\; \begin{pmatrix} 1 & 1 & 1 &1 \\ 1 & 1 & 1 &1 \\ 1 & 1 & 1 &1 \\ 1 & 1 & 1 &1 \end{pmatrix}
\]
}
\vspace{0.35em}
{\color{red}
\[
w^{(2)} \;=\; \begin{pmatrix} 0 & 1 & 0 & 1 \\ 0 & 1 & 0 & 1 \\ 1 & 0 & 1 & 0 \\ 1 & 0 & 1 & 0 \end{pmatrix}
\]
}
};
 
\draw[arrow]
    ($(panelAimg.east)+(\ArrowGap,0)$)
    --
    ($(panelBimg.west)-(\ArrowGap,0)$);
 
\draw[arrow]
    ($(panelBimg.east)+(\ArrowGap,0)$)
    --
    ($(Csw)+(0,\PanelW/2)-(\ArrowGap,0)$);
 
\end{tikzpicture}
\captionof{figure}{Bathroom-tile lattice $\Lambda=(4,8,8)$ emulated on the square host. Left: the target graph. Middle: its embedding $\Lambda \subset \square$ at supercell size $L=4$ (the sixteen supercell sites are marked by enlarged dots), with the independent bond directions $\{1,2\}$ marked. Right: the resulting binary bond masks $w^{(1)}$, $w^{(2)}$, which, as in Fig.~\ref{fig:honeycomb-embedding}, are the transpose of the patterns drawn in the middle panel.}
\label{fig:bathroom-tile-embedding}
\end{center}
\vspace{5pt}
from which we immediately reproduce the Utiyama critical temperature \cite{Utiyama_1951}
\begin{equation}
T_c \;=\; \frac{2}{\log \frac{\sqrt{10+8 \sqrt{2}}-\sqrt{2}}{4-\sqrt{10+8 \sqrt{2}}+\sqrt{2}}} \;=\; 1.438695...\,.
\end{equation}
Lattices containing odd-sided polygons fall outside the reach of the square emulator and require the triangular host.

\subsection{The triangular host}\label{sec:triangular-emulator}


The triangular host $\mathcal{H} = \triangle$ carries three independent bond directions and is the universal version of the construction: it accommodates every flat unit-edge-length lattice we have examined. 
Vertices are labelled by $(i,j)$ at position $\mathbf{v} = i\,\mathbf{e}_1 + j\,\mathbf{e}_2$, where $\mathbf{e}_1 = (1, 0)$ and $\mathbf{e}_2 = (\tfrac{1}{2}, \tfrac{\sqrt{3}}{2})$.
Each vertex has six directions: \{1, 2, 3\} are the independent ones, while \{4, 5, 6\} are their opposites. They are represented as follows (Fig.~\ref{fig:triangular-directions}):
\begin{center}
\centering
\begin{tikzpicture}[scale=1.3, every node/.style={font=\small}]
\draw[gray!30, thin] (-1.25,0) -- (1.25,0);
\draw[gray!20, thin] (-1.75,0.866) -- (1.75,0.866);
\draw[gray!20, thin] (-1.75,-0.866) -- (1.75,-0.866);
\draw[gray!20, thin] (-1.7,1.22) -- (0,-1.732);
\draw[gray!20, thin] (-1.7,-1.22) -- (0,1.732);
\draw[gray!20, thin] (-0.7,-1.22) -- (0.7,1.22);
\draw[gray!20, thin] (-0.7,1.22) -- (0.7,-1.22);
\draw[gray!20, thin] (0,1.732) -- (1.7,-1.22);
\draw[gray!20, thin] (0,-1.732) -- (1.7,1.22);
\draw[gray!20, thin] (-1.25,1.732) -- (1.25,1.732);
\draw[gray!20, thin] (-1.25,-1.732) -- (1.25,-1.732);
\filldraw[black] (1,0) circle (2pt) node[right = 1pt, font=\scriptsize] {$(i+1,j)$};
\filldraw[black] (0.5,0.866) circle (2pt) node[above = 1pt, font=\scriptsize] {\quad\quad\quad$(i, j+1)$};
\filldraw[black] (-0.5,0.866) circle (2pt) node[above = 1pt, font=\scriptsize] {$(i-1, j+1)$\quad\quad\quad};
\filldraw[black] (-1,0) circle (2pt)  node[left = 1pt, font=\scriptsize] {$(i-1,j)$};
\filldraw[black] (-0.5,-0.866) circle (2pt) node[below = 1pt, font=\scriptsize] {$(i, j-1)$\quad\quad\quad};
\filldraw[black] (0.5,-0.866) circle (2pt) node[below = 1pt, font=\scriptsize] {\quad\quad\quad$(i+1, j-1)$};
\filldraw[gray!55] (1.5,0.866) circle (1.8pt);
\filldraw[gray!55] (-1.5,-0.866) circle (1.8pt);
\filldraw[gray!55] (-1.5,0.866) circle (1.8pt);
\filldraw[gray!55] (+1.5,-0.866) circle (1.8pt);
\filldraw[gray!55] (0,1.732) circle (1.8pt);
\filldraw[gray!55] (0,-1.732) circle (1.8pt);
\draw[->, very thick, color=blue] (0,0) -- (1,0);
\draw[->, very thick, color=red] (0,0) -- (0.5,0.866);
\draw[->, very thick, color=green] (0,0) -- (-0.5,0.866);
\draw[->, very thick, color=gray] (0,0) -- (-1,0);
\draw[->, very thick, color=gray] (0,0) -- (-0.5,-0.866);
\draw[->, very thick, color=gray] (0,0) -- (0.5,-0.866);
\filldraw[black] (0,0) circle (2pt) node[below left= 3pt, font=\scriptsize] {$(i,j)\,\,$};
\node at (2.2, 0)       {${\color{blue}\bf{1}}$};
\node at (0.85, 1.472)  {${\color{red}\bf{2}}$};
\node at (-0.85, 1.472) {${\color{green}\bf{3}}$};
\node at (-2.2, 0)      {$\bf{4}$};
\node at (-0.85,-1.472) {$\bf{5}$};
\node at (0.85,-1.472)  {$\bf{6}$};
\end{tikzpicture}
\captionof{figure}{Bond-direction convention on the triangular host $\triangle$. Directions $\{1,2,3\}$ (blue, red, green) are the independent directions; directions $\{4,5,6\}$ (grey) are their opposites.}
\label{fig:triangular-directions}
\end{center}
The labels are chosen so that directions \{1, 2, 4, 5\} coincide with the four directions of the square host of Section~\ref{sec:square-emulator}; directions \{3, 6\} are the third bond family absent from the square.
A triangular-lattice bond program is specified by three binary matrices
\begin{equation}
w^{(1)},\; w^{(2)},\; w^{(3)} \;\in\; \{0,1\}^{L\times L}\,.
\end{equation}
%
The three opposite matrices are fixed by the requirement that the two orientations of the same undirected bond agree:
\begin{equation}
w^{(4)}_{ij} \;=\; w^{(1)}_{i-1, j}
\qquad\qquad
w^{(5)}_{ij} \;=\; w^{(2)}_{i, j-1}
\qquad\qquad
w^{(6)}_{ij} \;=\; w^{(3)}_{i+1, j-1}\,.
\label{eq:triangular-consistency}
\end{equation}
The triangular program therefore has $6L^2$ binary entries, of which half are independent. 
When site dilution suffices, a single binary matrix $w \in \{0,1\}^{L\times L}$ determines all six directed masks via $w^{(a)}_{ij} = w_{ij}$. Cut bonds incident on absent sites contribute nothing to the determinant, exactly as in the square case.

The weighted graph description of the transition matrix $\mathbb{W}_\triangle$ is the straightforward generalization of the square case detailed above.
With $\alpha = \rme^{\rmi \frac \pi  3}$ and $\beta = 1/\alpha$, the thirty-six pairs of directions split into two blocks of eighteen, one for directions $a \in \{1, 2, 3\}$ and one for $a \in \{4, 5, 6\}$.
Directions $a \in \{1, 2, 3\}$ are:
\begin{equation}
\begin{array}{ccccccccccccc}
\underset{(i,j)}{1} & \overset{1\,w_{ij}^{(1)}\rme^{\rmi k_1}}{\longleftrightarrow} & \underset{(i+1,j)}{1} &  &  &
\underset{(i,j)}{2} & \overset{\sqrt{\beta}\,w_{ij}^{(2)}\rme^{\rmi k_2}}{\longleftrightarrow} & \underset{(i,j+1)}{1} &  &  &
\underset{(i,j)}{3} & \overset{\beta\,w_{ij}^{(3)}\rme^{\rmi k_3}}{\longleftrightarrow} & \underset{(i-1,j+1)}{1} \\[6pt]
\underset{(i,j)}{1} & \overset{\sqrt{\alpha}\,w_{ij}^{(1)}\rme^{\rmi k_1}}{\longleftrightarrow} & \underset{(i+1,j)}{2} &  &  &
\underset{(i,j)}{2} & \overset{1\,w_{ij}^{(2)}\rme^{\rmi k_2}}{\longleftrightarrow} & \underset{(i,j+1)}{2} &  &  &
\underset{(i,j)}{3} & \overset{\sqrt{\beta}\,w_{ij}^{(3)}\rme^{\rmi k_3}}{\longleftrightarrow} & \underset{(i-1,j+1)}{2} \\[6pt]
\underset{(i,j)}{1} & \overset{\alpha\,w_{ij}^{(1)}\rme^{\rmi k_1}}{\longleftrightarrow} & \underset{(i+1,j)}{3} &  &  &
\underset{(i,j)}{2} & \overset{\sqrt{\alpha}\,w_{ij}^{(2)}\rme^{\rmi k_2}}{\longleftrightarrow} & \underset{(i,j+1)}{3} &  &  &
\underset{(i,j)}{3} & \overset{1\,w_{ij}^{(3)}\rme^{\rmi k_3}}{\longleftrightarrow} & \underset{(i-1,j+1)}{3} \\[6pt]
\underset{(i,j)}{1} & \overset{0\,w_{ij}^{(1)}\rme^{\rmi k_1}}{\longleftrightarrow} & \underset{(i+1,j)}{4} &  &  &
\underset{(i,j)}{2} & \overset{\alpha\,w_{ij}^{(2)}\rme^{\rmi k_2}}{\longleftrightarrow} & \underset{(i,j+1)}{4} &  &  &
\underset{(i,j)}{3} & \overset{\sqrt{\alpha}\,w_{ij}^{(3)}\rme^{\rmi k_3}}{\longleftrightarrow} & \underset{(i-1,j+1)}{4} \\[6pt]
\underset{(i,j)}{1} & \overset{\beta\,w_{ij}^{(1)}\rme^{\rmi k_1}}{\longleftrightarrow} & \underset{(i+1,j)}{5} &  &  &
\underset{(i,j)}{2} & \overset{0\,w_{ij}^{(2)}\rme^{\rmi k_2}}{\longleftrightarrow} & \underset{(i,j+1)}{5} &  &  &
\underset{(i,j)}{3} & \overset{\alpha\,w_{ij}^{(3)}\rme^{\rmi k_3}}{\longleftrightarrow} & \underset{(i-1,j+1)}{5} \\[6pt]
\underset{(i,j)}{1} & \overset{\sqrt{\beta}\,w_{ij}^{(1)}\rme^{\rmi k_1}}{\longleftrightarrow} & \underset{(i+1,j)}{6} &  &  &
\underset{(i,j)}{2} & \overset{\beta\,w_{ij}^{(2)}\rme^{\rmi k_2}}{\longleftrightarrow} & \underset{(i,j+1)}{6} &  &  &
\underset{(i,j)}{3} & \overset{0\,w_{ij}^{(3)}\rme^{\rmi k_3}}{\longleftrightarrow} & \underset{(i-1,j+1)}{6}
\end{array}
\label{eq:triangular-graph-a}
\end{equation}
while directions $a \in \{4, 5, 6\}$ are:
\begin{equation}
\begin{array}{ccccccccccccc}
\underset{(i,j)}{4} & \overset{0\,w_{ij}^{(4)}\rme^{-\rmi k_1}}{\longleftrightarrow} & \underset{(i-1,j)}{1} &  &  &
\underset{(i,j)}{5} & \overset{\alpha\,w_{ij}^{(5)}\rme^{-\rmi k_2}}{\longleftrightarrow} & \underset{(i,j-1)}{1} &  &  &
\underset{(i,j)}{6} & \overset{\sqrt{\alpha}\,w_{ij}^{(6)}\rme^{-\rmi k_3}}{\longleftrightarrow} & \underset{(i+1,j-1)}{1} \\[6pt]
\underset{(i,j)}{4} & \overset{\beta\,w_{ij}^{(4)}\rme^{-\rmi k_1}}{\longleftrightarrow} & \underset{(i-1,j)}{2} &  &  &
\underset{(i,j)}{5} & \overset{0\,w_{ij}^{(5)}\rme^{-\rmi k_2}}{\longleftrightarrow} & \underset{(i,j-1)}{2} &  &  &
\underset{(i,j)}{6} & \overset{\alpha\,w_{ij}^{(6)}\rme^{-\rmi k_3}}{\longleftrightarrow} & \underset{(i+1,j-1)}{2} \\[6pt]
\underset{(i,j)}{4} & \overset{\sqrt{\beta}\,w_{ij}^{(4)}\rme^{-\rmi k_1}}{\longleftrightarrow} & \underset{(i-1,j)}{3} &  &  &
\underset{(i,j)}{5} & \overset{\beta\,w_{ij}^{(5)}\rme^{-\rmi k_2}}{\longleftrightarrow} & \underset{(i,j-1)}{3} &  &  &
\underset{(i,j)}{6} & \overset{0\,w_{ij}^{(6)}\rme^{-\rmi k_3}}{\longleftrightarrow} & \underset{(i+1,j-1)}{3} \\[6pt]
\underset{(i,j)}{4} & \overset{1\,w_{ij}^{(4)}\rme^{-\rmi k_1}}{\longleftrightarrow} & \underset{(i-1,j)}{4} &  &  &
\underset{(i,j)}{5} & \overset{\sqrt{\beta}\,w_{ij}^{(5)}\rme^{-\rmi k_2}}{\longleftrightarrow} & \underset{(i,j-1)}{4} &  &  &
\underset{(i,j)}{6} & \overset{\beta\,w_{ij}^{(6)}\rme^{-\rmi k_3}}{\longleftrightarrow} & \underset{(i+1,j-1)}{4} \\[6pt]
\underset{(i,j)}{4} & \overset{\sqrt{\alpha}\,w_{ij}^{(4)}\rme^{-\rmi k_1}}{\longleftrightarrow} & \underset{(i-1,j)}{5} &  &  &
\underset{(i,j)}{5} & \overset{1\,w_{ij}^{(5)}\rme^{-\rmi k_2}}{\longleftrightarrow} & \underset{(i,j-1)}{5} &  &  &
\underset{(i,j)}{6} & \overset{\sqrt{\beta}\,w_{ij}^{(6)}\rme^{-\rmi k_3}}{\longleftrightarrow} & \underset{(i+1,j-1)}{5} \\[6pt]
\underset{(i,j)}{4} & \overset{\alpha\,w_{ij}^{(4)}\rme^{-\rmi k_1}}{\longleftrightarrow} & \underset{(i-1,j)}{6} &  &  &
\underset{(i,j)}{5} & \overset{\sqrt{\alpha}\,w_{ij}^{(5)}\rme^{-\rmi k_2}}{\longleftrightarrow} & \underset{(i,j-1)}{6} &  &  &
\underset{(i,j)}{6} & \overset{1\,w_{ij}^{(6)}\rme^{-\rmi k_3}}{\longleftrightarrow} & \underset{(i+1,j-1)}{6}
\end{array}
\label{eq:triangular-graph-b}
\end{equation}
Here $k_3 =k_2-k_1$ and now the matrix has dimension $6 L^2$.
Finally, note that every square emulator program lifts to the corresponding triangular one just by setting $w^{(3)} \equiv 0$ (implying $w^{(6)}=0$).

\paragraph{Worked examples: kagome and snub-square.}
The kagome lattice $(3, 6, 3, 6)$ is the simplest target that requires the triangular host since it contains triangles, which the square host cannot accommodate.
Kagome admits both bond- and site-dilution realizations on the triangular host at supercell size $L = 2$, illustrating the redundancy mentioned in Section~\ref{sec:emulator-principle}.

\textit{Bond program.} Walking over the supercell in Fig.~\ref{fig:kagome-embedding} below, colour-coded by independent direction, we can read off the binary masks:
\begin{equation}
w^{(1)} = \begin{pmatrix} 0 & 1 \\ 0 & 1 \end{pmatrix}\,,
\qquad\qquad
w^{(2)} = \begin{pmatrix} 1 & 1 \\ 0 & 0 \end{pmatrix}\,,
\qquad\qquad
w^{(3)} = \begin{pmatrix} 1 & 0 \\ 0 & 1 \end{pmatrix}\,,
\label{eq:kagome-bond-program}
\end{equation}
while the other masks $w^{(4)}, w^{(5)}, w^{(6)}$ are determined by~\eqref{eq:triangular-consistency}.
\begin{center}
\begin{tikzpicture}[
    paneltitle/.style={
        font=\normalsize,
        anchor=south
    },
    panellabel/.style={
        font=\bfseries,
        anchor=north west
    },
    arrow/.style={
       ->,
        black!60,
        line width=0.8pt
    },
    matrixnode/.style={
    anchor=center,
    inner sep=0pt,
    font=\normalsize
    },
    vertexnumber/.style={
        font=\tiny,
        inner sep=1pt,
        fill=white,
        fill opacity=0.85,
        text opacity=1
    },
    coordarrow/.style={
    ->,
    >=stealth,
    thin,
    black!70
}
]

\def\PanelW{4.2cm}
\def\PanelGap{0.75cm}
\def\ArrowGap{0.18cm}
\def\PanelTop{4.25cm}

\coordinate (A0) at (0,0);
\coordinate (B0) at ($(A0)+(\PanelW+\PanelGap,0)$);
\coordinate (C0) at ($(B0)+(\PanelW+\PanelGap,0)$);

\node[
    anchor=south west,
    inner sep=0pt
] (panelAimg) at (A0) {
    \includegraphics[width=\PanelW]{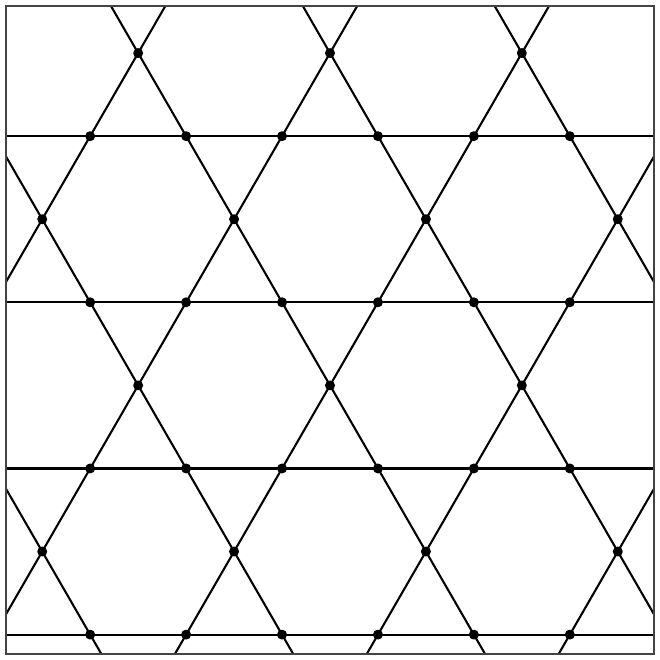}
};

\node[paneltitle] at ($(panelAimg.north)+(0,0.1cm)$)
    {$\Lambda = (3,6,3,6)$};

\node[
    anchor=south west,
    inner sep=0pt
] (panelBimg) at (B0) {
    \includegraphics[width=\PanelW]{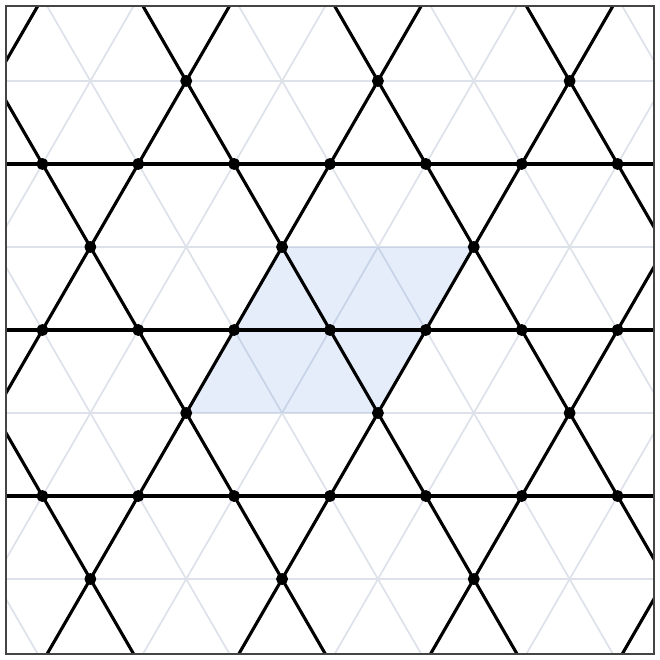}
};

\node[paneltitle] at ($(panelBimg.north)+(0,0.1cm)$)
    {$\Lambda \subset \triangle$};

\begin{scope}[
    shift={($(panelBimg.south west)$)},
    x={($(panelBimg.south east)-(panelBimg.south west)$)},
    y={($(panelBimg.north west)-(panelBimg.south west)$)}
]

\node[vertexnumber] at (0.43,0.50) {${\color{blue}1}$};
\node[vertexnumber] at (0.57,0.50) {${\color{blue}1}$};
\node[vertexnumber] at (0.355,0.375) {${\color{blue}0}$};
\node[vertexnumber] at (0.495,0.375) {${\color{blue}0}$};
\node[vertexnumber] at (0.4,0.57) {${\color{red}1}$};
\node[vertexnumber] at (0.54,0.57) {${\color{red}0}$};
\node[vertexnumber] at (0.325,0.445) {${\color{red}1}$};
\node[vertexnumber] at (0.465,0.445) {${\color{red}0}$};
\node[vertexnumber] at (0.32,0.57) {${\color{green}0}$};
\node[vertexnumber] at (0.46,0.57) {${\color{green}1}$};
\node[vertexnumber] at (0.245,0.445) {${\color{green}1}$};
\node[vertexnumber] at (0.385,0.445) {${\color{green}0}$};

\draw[coordarrow] (0.15,0.28) -- (0.15,0.42) node[pos=1.65, anchor=north,  font=\tiny] {$j$};
\draw[coordarrow] (0.15,0.28) -- (0.29,0.28) node[pos=1.45, anchor=east,  font=\tiny] {$i$};

\coordinate (triO) at (0.86,0.86);

\draw[arrow, color=blue]
    (0.75,0.78) -- (0.87,0.78)
    node[pos=1.85, anchor=east, font=\tiny] {${\color{blue}(1)}$};

\draw[arrow, color=red]
    (0.75,0.78) -- (0.81,0.885)
    node[pos=1.85, anchor=north east, font=\tiny] {${\color{red}(2)}$};

\draw[arrow, color=green]
    (0.75,0.78) -- (0.69,0.885)
    node[pos=0.5, anchor=south east, font=\tiny] {${\color{green}(3)}$};

\foreach \x/\y in {0.285/0.375,0.36/0.50,0.50/0.50}
  \fill[black] (\x,\y) circle (2.3pt);
\draw[black, line width=0.7pt, fill=white] (0.425,0.375) circle (2.3pt);
\end{scope}

\coordinate (Csw) at (C0);
\coordinate (Cse) at ($(Csw)+(\PanelW,0)$);
\coordinate (Cnw) at ($(Csw)+(0,\PanelW)$);
\coordinate (Cne) at ($(Csw)+(\PanelW,\PanelW)$);
\coordinate (Ccenter) at ($(Csw)!0.5!(Cne)$);

\node[
    anchor=center,
    inner sep=0pt,
    text width=\PanelW,
    align=center,
    font=\normalsize
] at (Ccenter) {%
{\color{blue}
\[
w^{(1)} \;=\; \begin{pmatrix} 0 & 1 \\ 0 & 1 \end{pmatrix}
\]
}
\vspace{0.35em}
{\color{red}
\[
w^{(2)} \;=\; \begin{pmatrix} 1 & 1 \\ 0 & 0 \end{pmatrix}
\]
}
\vspace{0.35em}
{\color{green}
\[
w^{(3)} \;=\; \begin{pmatrix} 1 & 0 \\ 0 & 1 \end{pmatrix}
\]
}
};

\draw[arrow]
    ($(panelAimg.east)+(\ArrowGap,0)$)
    --
    ($(panelBimg.west)-(\ArrowGap,0)$);

\draw[arrow]
    ($(panelBimg.east)+(\ArrowGap,0)$)
    --
    ($(Csw)+(0,\PanelW/2)-(\ArrowGap,0)$);

\end{tikzpicture}
\captionof{figure}{Kagome lattice $\Lambda=(3,6,3,6)$ bond program emulated on the triangular host. Left: the target graph. Middle: its embedding $\Lambda \subset \triangle$ at supercell size $L=2$, with the independent bond directions $\{1,2,3\}$ marked. The four host sites of the supercell are marked by enlarged dots: the three filled dots are the kagome sites, while the open dot is the host site $(i,j)=(2,1)$, which sits at the centre of a hexagon and is not a site of $\Lambda$; all bonds leaving and entering it carry weight $0$. Right: the resulting binary bond masks $w^{(1)}$, $w^{(2)}$, $w^{(3)}$ (transposes of the patterns drawn in the middle panel); the bottom-left entry $(2,1)$ vanishes in all three masks.}
\label{fig:kagome-embedding}
\end{center}
\vspace{5pt}

Note that the $2\times2$ supercell contains four host sites but only three kagome sites: the host site $(2,1)$ (open dot in Fig.~\ref{fig:kagome-embedding}) lies at the centre of a hexagonal face of $\Lambda$, so that all bonds leaving it vanish, $w^{(1)}_{21}=w^{(2)}_{21}=w^{(3)}_{21}=0$, and so do, by~\eqref{eq:triangular-consistency}, all bonds entering it. This empty site contributes nothing to the determinant and is precisely the site removed by the equivalent site program discussed next.

\textit{Site program.} Alternatively, a $2 \times 2$ triangular supercell contains four host sites; removing exactly one of them, the corner $(2,1)$, leaves three sites per supercell whose induced graph is precisely kagome on the surviving sublattice. The site mask is therefore
\begin{equation}
w \;=\; \begin{pmatrix} 1 & 1 \\ 0 & 1 \end{pmatrix}\,,
\qquad\qquad
|\mathbb{W}_\triangle^{(3,6,3,6)}| = 24\,,
\label{eq:kagome-site-program}
\end{equation}
with all six directed masks given by $w^{(a)}_{ij} = w_{ij}$.
Both programs reproduce the Kano--Naya critical temperature \cite{Kano_Naya_1953}
\begin{equation}
T_c \;=\; \frac{4}{\log(3 + 2\sqrt{3})} \;=\; 2.143319...\,,
\end{equation}
at the same parent-matrix dimension $|\mathbb{W}_\triangle^{(3,6,3,6)}| = 24$.

A less trivial example is the snub-square, or Shastry--Sutherland, $(3,3,4,3,4)$ lattice. A {\it bond} program of size $L=4$ is needed as shown in Fig.~\ref{fig:snub-square-embedding}:
\begin{center}
\begin{tikzpicture}[
    paneltitle/.style={
        font=\normalsize,
        anchor=south
    },
    panellabel/.style={
        font=\bfseries,
        anchor=north west
    },
    arrow/.style={
       ->,
        black!60,
        line width=0.8pt
    },
    matrixnode/.style={
    anchor=center,
    inner sep=0pt,
    font=\normalsize
    },
    vertexnumber/.style={
        font=\tiny,
        inner sep=1pt,
        fill=white,
        fill opacity=0.85,
        text opacity=1
    },
    coordarrow/.style={
    ->,
    >=stealth,
    thin,
    black!70
}
]

\def\PanelW{4.2cm}
\def\PanelGap{0.75cm}
\def\ArrowGap{0.18cm}
\def\PanelTop{4.25cm}

\coordinate (A0) at (0,0);
\coordinate (B0) at ($(A0)+(\PanelW+\PanelGap,0)$);
\coordinate (C0) at ($(B0)+(\PanelW+\PanelGap,0)$);

\node[
    anchor=south west,
    inner sep=0pt
] (panelAimg) at (A0) {
    \includegraphics[width=\PanelW]{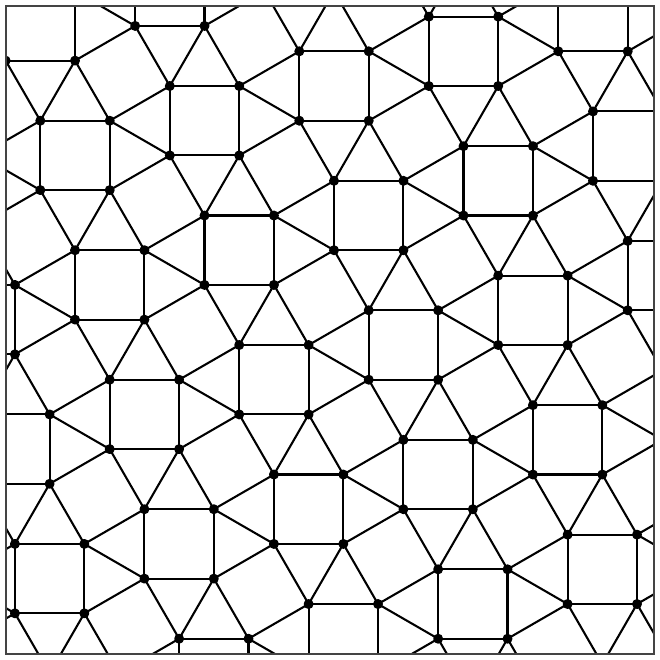}
};

\node[paneltitle] at ($(panelAimg.north)+(0,0.3cm)$)
    {$\Lambda = (3,3,4,3,4)$};

\node[
    anchor=south west,
    inner sep=0pt
] (panelBimg) at (B0) {
    \includegraphics[width=\PanelW]{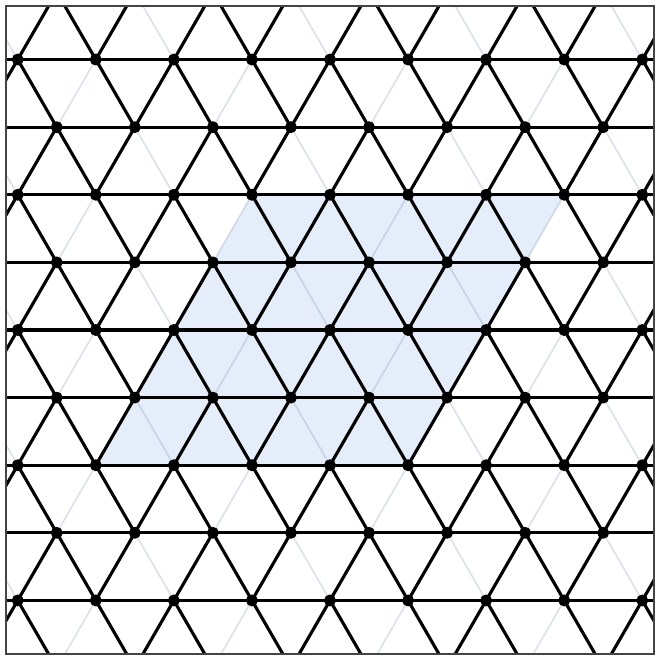}
};

\node[paneltitle] at ($(panelBimg.north)+(0,0.3cm)$)
    {$\Lambda \subset \triangle$};

\begin{scope}[
    shift={($(panelBimg.south west)$)},
    x={($(panelBimg.south east)-(panelBimg.south west)$)},
    y={($(panelBimg.north west)-(panelBimg.south west)$)}
]

\node[vertexnumber] at (0.115,0.35) {${\color{green}1}$};
\node[vertexnumber] at (0.175,0.45) {${\color{green}1}$};
\node[vertexnumber] at (0.235,0.55) {${\color{green}0}$};
\node[vertexnumber] at (0.295,0.65) {${\color{green}1}$};
\node[vertexnumber] at (0.235,0.35) {${\color{green}0}$};
\node[vertexnumber] at (0.295,0.45) {${\color{green}1}$};
\node[vertexnumber] at (0.355,0.55) {${\color{green}1}$};
\node[vertexnumber] at (0.415,0.65) {${\color{green}1}$};
\node[vertexnumber] at (0.355,0.35) {${\color{green}1}$};
\node[vertexnumber] at (0.415,0.45) {${\color{green}1}$};
\node[vertexnumber] at (0.475,0.55) {${\color{green}0}$};
\node[vertexnumber] at (0.535,0.65) {${\color{green}1}$};
\node[vertexnumber] at (0.475,0.35) {${\color{green}0}$};
\node[vertexnumber] at (0.535,0.45) {${\color{green}1}$};
\node[vertexnumber] at (0.595,0.55) {${\color{green}1}$};
\node[vertexnumber] at (0.655,0.65) {${\color{green}1}$};

\node[vertexnumber] at (0.175,0.35) {${\color{red}1}$};
\node[vertexnumber] at (0.235,0.45) {${\color{red}1}$};
\node[vertexnumber] at (0.295,0.55) {${\color{red}1}$};
\node[vertexnumber] at (0.345,0.65) {${\color{red}0}$};
\node[vertexnumber] at (0.295,0.35) {${\color{red}1}$};
\node[vertexnumber] at (0.355,0.45) {${\color{red}0}$};
\node[vertexnumber] at (0.405,0.55) {${\color{red}1}$};
\node[vertexnumber] at (0.465,0.65) {${\color{red}1}$};
\node[vertexnumber] at (0.415,0.35) {${\color{red}1}$};
\node[vertexnumber] at (0.475,0.45) {${\color{red}1}$};
\node[vertexnumber] at (0.525,0.55) {${\color{red}1}$};
\node[vertexnumber] at (0.585,0.65) {${\color{red}0}$};
\node[vertexnumber] at (0.535,0.35) {${\color{red}1}$};
\node[vertexnumber] at (0.595,0.45) {${\color{red}0}$};
\node[vertexnumber] at (0.645,0.55) {${\color{red}1}$};
\node[vertexnumber] at (0.705,0.65) {${\color{red}1}$};

\node[vertexnumber] at (0.205,0.3) {${\color{blue}1}$};
\node[vertexnumber] at (0.265,0.4) {${\color{blue}1}$};
\node[vertexnumber] at (0.325,0.5) {${\color{blue}1}$};
\node[vertexnumber] at (0.375,0.6) {${\color{blue}1}$};
\node[vertexnumber] at (0.315,0.3) {${\color{blue}1}$};
\node[vertexnumber] at (0.385,0.4) {${\color{blue}1}$};
\node[vertexnumber] at (0.435,0.5) {${\color{blue}1}$};
\node[vertexnumber] at (0.495,0.6) {${\color{blue}1}$};
\node[vertexnumber] at (0.435,0.3) {${\color{blue}1}$};
\node[vertexnumber] at (0.505,0.4) {${\color{blue}1}$};
\node[vertexnumber] at (0.555,0.5) {${\color{blue}1}$};
\node[vertexnumber] at (0.615,0.6) {${\color{blue}1}$};
\node[vertexnumber] at (0.565,0.3) {${\color{blue}1}$};
\node[vertexnumber] at (0.615,0.4) {${\color{blue}1}$};
\node[vertexnumber] at (0.675,0.5) {${\color{blue}1}$};
\node[vertexnumber] at (0.735,0.6) {${\color{blue}1}$};

\draw[coordarrow] (0.08,0.15) -- (0.08,0.29) node[pos=1.65, anchor=north,  font=\tiny] {$j$};
\draw[coordarrow] (0.08,0.15) -- (0.22,0.15) node[pos=1.45, anchor=east,  font=\tiny] {$i$};

\coordinate (triO) at (0.86,0.86);

\end{scope}

\coordinate (Csw) at (C0);
\coordinate (Cse) at ($(Csw)+(\PanelW,0)$);
\coordinate (Cnw) at ($(Csw)+(0,\PanelW)$);
\coordinate (Cne) at ($(Csw)+(\PanelW,\PanelW)$);
\coordinate (Ccenter) at ($(Csw)!0.5!(Cne)$);

\node[
    anchor=center,
    inner sep=0pt,
    text width=\PanelW,
    align=center,
    font=\normalsize
] at (Ccenter) {%
\resizebox{!}{2.8 cm}{%
\begin{minipage}{\PanelW}
\centering
{\color{blue}
\[
w^{(1)} \;=\;
\begin{pmatrix}
1&1&1&1\\
1&1&1&1\\
1&1&1&1\\
1&1&1&1
\end{pmatrix}
\]
}
\vspace{-0.4em}
{\color{red}
\[
w^{(2)} \;=\;
\begin{pmatrix}
1&1&1&0\\
1&0&1&1\\
1&1&1&0\\
1&0&1&1
\end{pmatrix}
\]
}
\vspace{-0.4em}
{\color{green}
\[
w^{(3)} \;=\;
\begin{pmatrix}
1&1&0&1\\
0&1&1&1\\
1&1&0&1\\
0&1&1&1
\end{pmatrix}
\]
}
\end{minipage}
}%
};

\draw[arrow]
    ($(panelAimg.east)+(\ArrowGap,0)$)
    --
    ($(panelBimg.west)-(\ArrowGap,0)$);

\draw[arrow]
    ($(panelBimg.east)+(\ArrowGap,0)$)
    --
    ($(Csw)+(0,\PanelW/2)-(\ArrowGap,0)$);

\end{tikzpicture}
\captionof{figure}{Snub-square (Shastry--Sutherland) lattice $\Lambda=(3,3,4,3,4)$ bond program emulated on the triangular host. Left: the target graph. Middle: its embedding $\Lambda \subset \triangle$ at supercell size $L=4$. Right: the resulting binary bond masks.}
\label{fig:snub-square-embedding}
\end{center}
\vspace{5pt}
The program correctly reproduces the Thompson--Wardrop critical temperature \cite{Thompson_Wardrop_1974}
\begin{equation}
T_c \;=\; 2.926261...\,.
\end{equation}
The parent matrix dimension is now $|\mathbb{W}_\triangle^{(3,3,4,3,4)}| = 96$ and one cannot easily obtain an analytical expression for the critical temperature (see \cite{Kassan-Ogly_2024} for an indirect closed formula).

\section{Thermodynamic observables}\label{sec:thermodynamics}

The square and triangular emulators share a common thermodynamic pipeline: once the parent transition matrix $\mathbb{W}^\Lambda_{\mathcal H}$ is fixed by a supercell program encoding the embedding $\Lambda \subset \mathcal{H}$, the per-spin free energy and all derived observables are obtained from a two-dimensional momentum integral of $\log P^\Lambda_{\mathcal H}(k_1,k_2)$, where $P^\Lambda_{\mathcal H}$ is the Kac--Ward determinant \eqref{eq:KW-determinant}. We collect the relations here in a form independent of which emulator is being used; the only emulator-specific input is the matrix $\mathbb{W}^\Lambda_{\mathcal H}$ itself.
It is clear that all thermodynamic observables only depend on the target lattice equivalence class $[\Lambda]$, which for simplicity we denote by $\Lambda$. 

\paragraph{Free energy.}
In the emulator framework, the per-spin free energy of the target lattice $\Lambda$ is
\begin{equation}
f_\Lambda(T) \;=\; -T \log 2 \;+\; T\,\frac{n_l}{2 n_v} \log(1 - v^2) \;-\; \frac{T}{2 n_v}\!\int\!\frac{{\rm d} k_1}{2\pi}\frac{{\rm d} k_2}{2\pi}\, \log P_\Lambda(v, k_1, k_2)\,,
\label{eq:free-energy}
\end{equation}
where $n_v$ and $n_l$ are respectively the number of sites and bonds in the supercell while the integration is over the range $0\leq k_1, k_2 \leq 2\pi$.
The FV/KW determinant $P_\Lambda(v,k_1,k_2)$ is defined in \eqref{eq:KW-determinant} and from now on we drop the host index ${\mathcal H}$ for clarity, so no confusion should arise.
Note that \eqref{eq:free-energy} is valid for both square and triangular hosts.
The temperature dependence enters through $v(T) = \tanh(\xi / T)$, with $\xi = +1$ for ferromagnetic couplings and $\xi = -1$ for antiferromagnetic ones.
\begin{table*}[t!]
\centering
\small
\caption{Thermodynamic quantities for the Archimedean lattices ($k=1$): ferromagnetic critical temperature $T_c$, energy density at the transition $\varepsilon_c$, and specific-heat amplitudes $A$ and $B$ of Eq.~\eqref{eq:specific-heat-asymptotic}; antiferromagnetic critical temperature $T_c^{\rm AF}$ (``none'' when there is no finite-temperature transition), ground-state energy $\varepsilon_0^{\rm AF}$, and residual entropy $s_0^{\rm AF}$. For the four bipartite lattices $T_c^{\rm AF}=T_c$ and $\varepsilon_0^{\rm AF}=\varepsilon_0^{\rm F}=-n_l/n_v$. Temperatures and energies are in units of $J$. The last column gives the atlas figure of Appendix~\ref{app:atlas} containing the program and the thermodynamic curves of each lattice.}
\label{tab:archimedean}
\renewcommand{\arraystretch}{1.25}
\setlength{\tabcolsep}{6pt}
\resizebox{\textwidth}{!}{%
\begin{tabular}{>{\centering\arraybackslash}m{2.3cm}
                >{\centering\arraybackslash}m{1.9cm}
                >{\centering\arraybackslash}m{1.55cm}
                >{\centering\arraybackslash}m{1.45cm}
                >{\centering\arraybackslash}m{1.2cm}
                >{\centering\arraybackslash}m{1.2cm}
                >{\centering\arraybackslash}m{1.45cm}
                >{\centering\arraybackslash}m{1.45cm}
                >{\centering\arraybackslash}m{1.2cm}
                >{\centering\arraybackslash}m{0.9cm}}
\toprule
\makecell{\textbf{Name}} &
\makecell{\textbf{Lattice} $\Lambda$} &
\multicolumn{4}{c}{\textbf{Ferromagnetic}} &
\multicolumn{3}{c}{\textbf{Antiferromagnetic}} &
\makecell{\textbf{Fig.}} \\
\cmidrule(lr){3-6} \cmidrule(lr){7-9}
&
&
$T_c$ &
$\varepsilon_c$ &
$A$ &
$B$ &
$T_c^{{\rm{AF}}}$ &
$\varepsilon_0^{{\rm{AF}}}$ &
$s_0^{{\rm{AF}}}$ &
\\
\midrule
\hyperref[atlas:t1011]{$(3,3,3,3,3,3)$} & \latimg{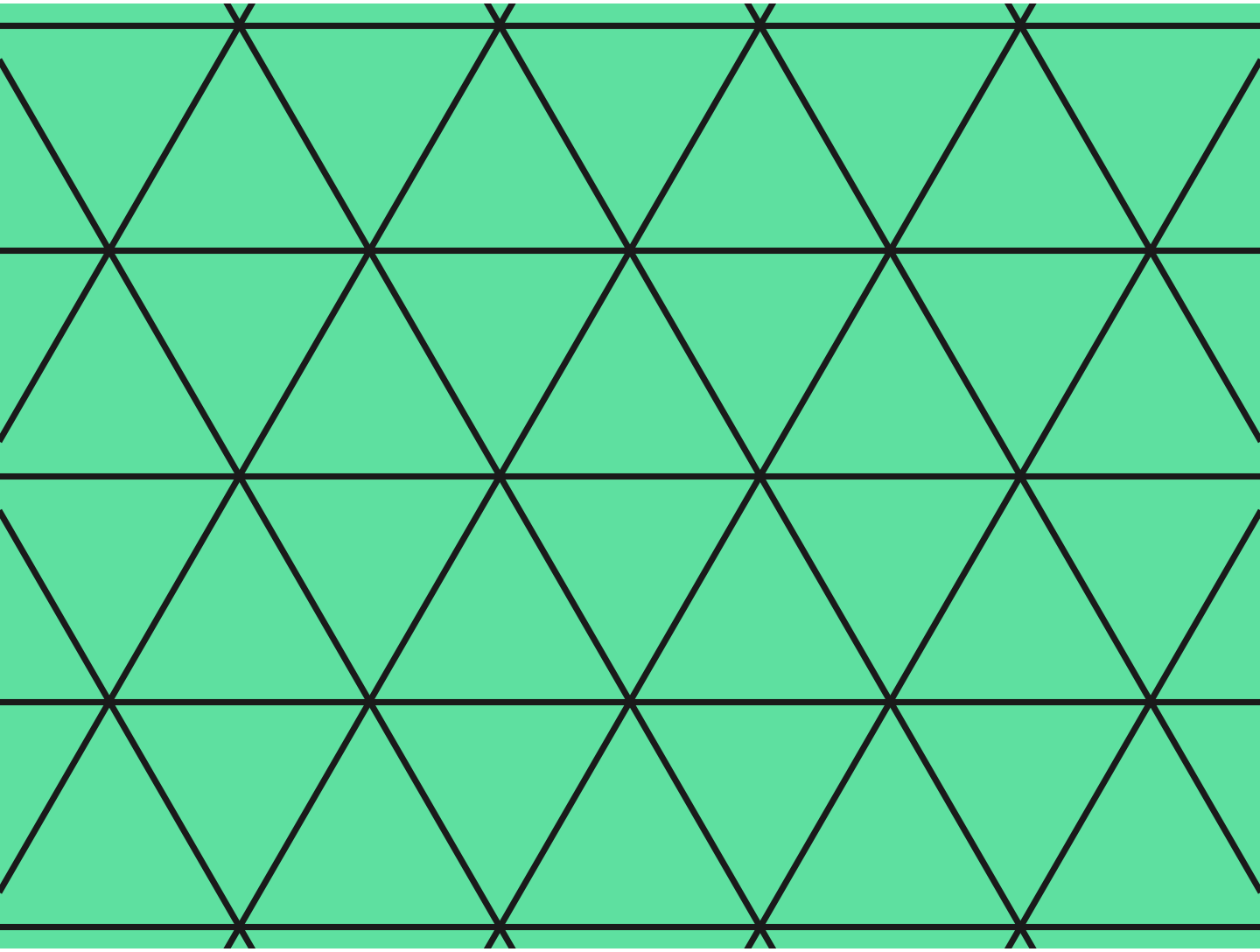} & $3.64096$ & $-2$ & $0.49907$ & $-0.307$ & none & $-1$ & $0.323$ & \ref{fig:atlas-t1011} \\
\hyperref[atlas:t1005]{$(4,4,4,4)$}     & \latimg{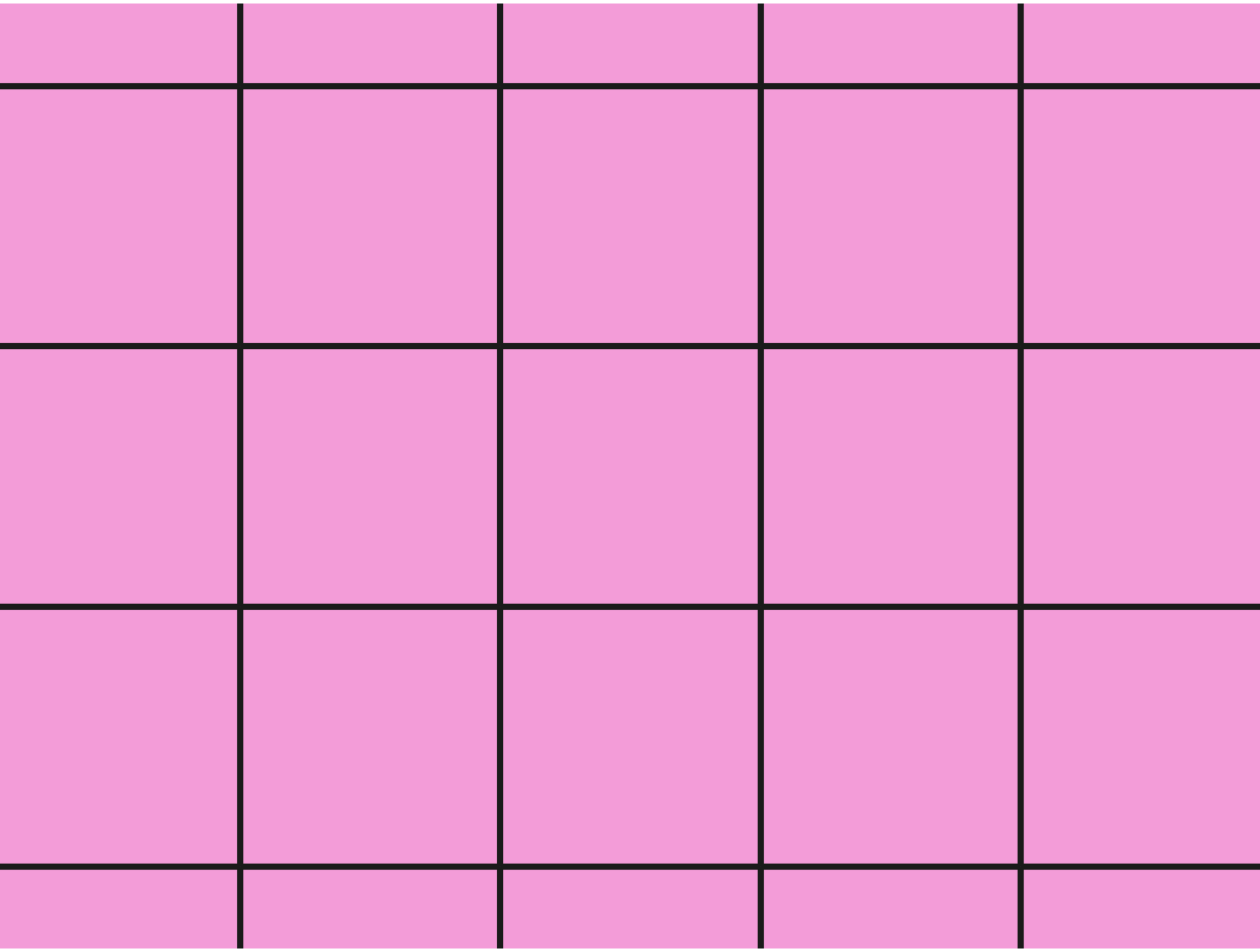} & $2.26919$ & $-1.41421$ & $0.49454$ & $-0.306$ & $2.26919$ & $-2$ & $0$ & \ref{fig:atlas-t1005} \\
\hyperref[atlas:t1001]{$(6,6,6)$}       & \latimg{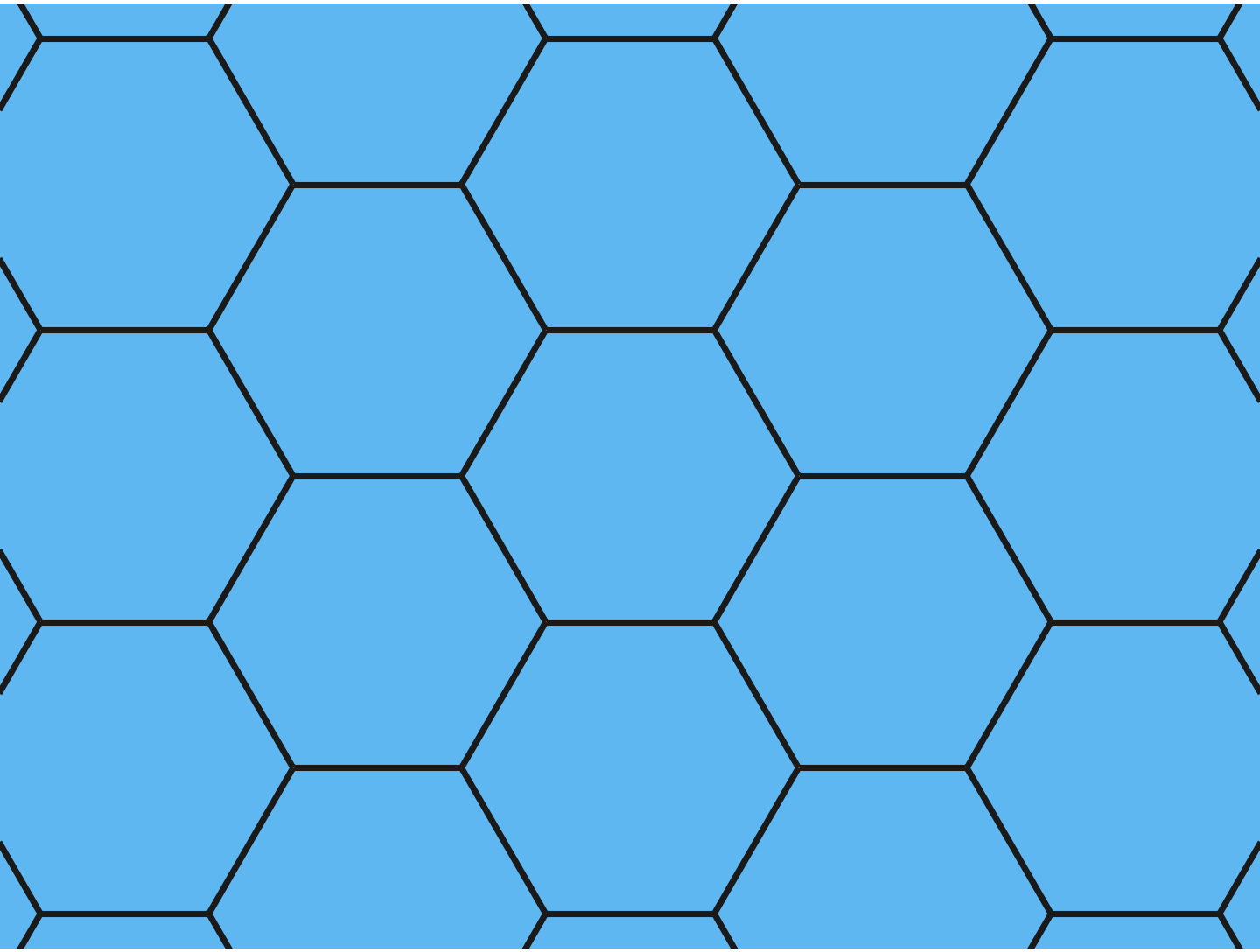} & $1.51865$ & $-1.15470$ & $0.47811$ & $-0.305$ & $1.51865$ & $-\tfrac{3}{2}$ & $0$ & \ref{fig:atlas-t1001} \\
\hyperref[atlas:t1007]{$(3,6,3,6)$}     & \latimg{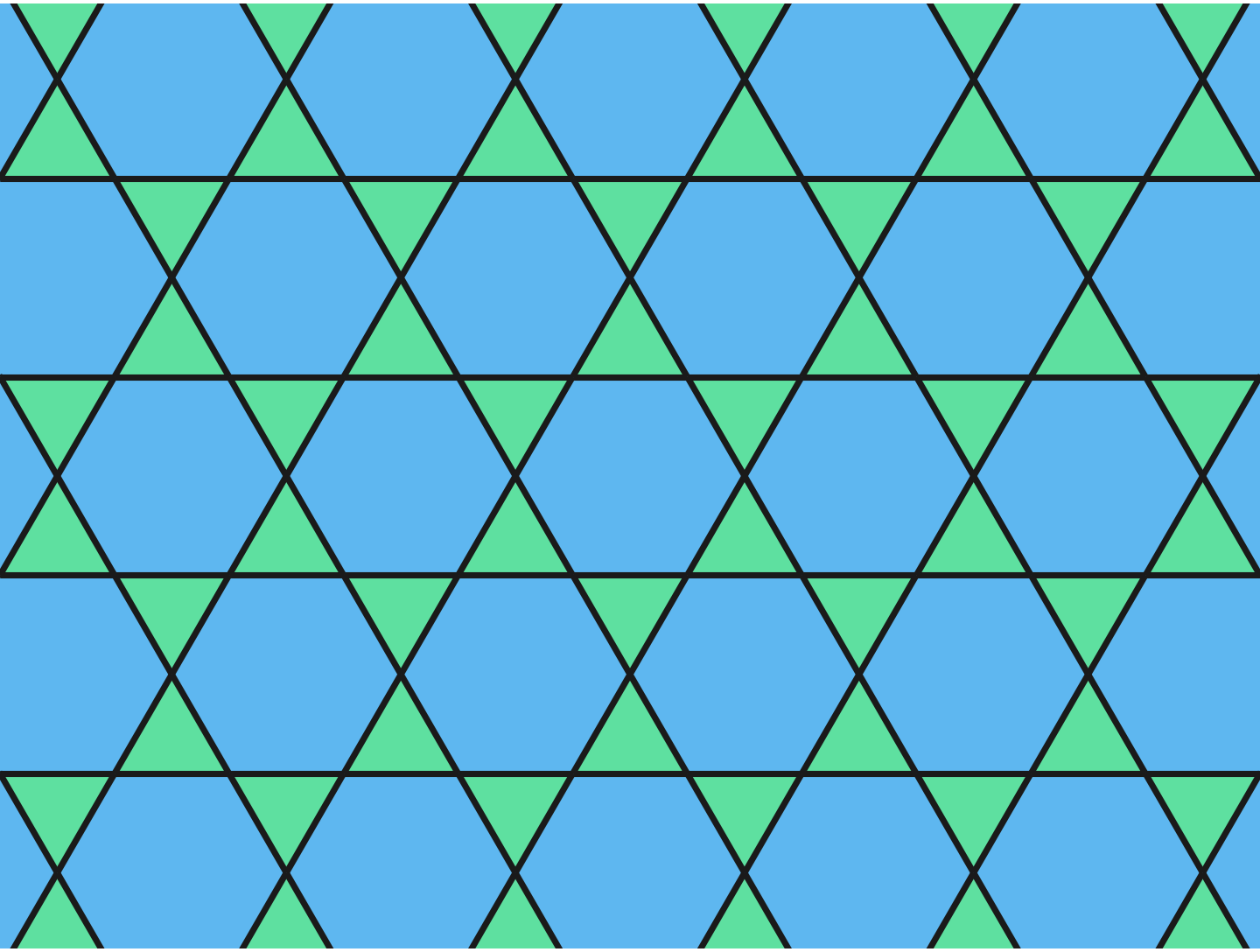} & $2.14332$ & $-1.48803$ & $0.48006$ & $-0.298$ & none & $-\tfrac{2}{3}$ & $0.502$ & \ref{fig:atlas-t1007} \\
\hyperref[atlas:t1010]{$(3,3,3,3,6)$}   & \latimg{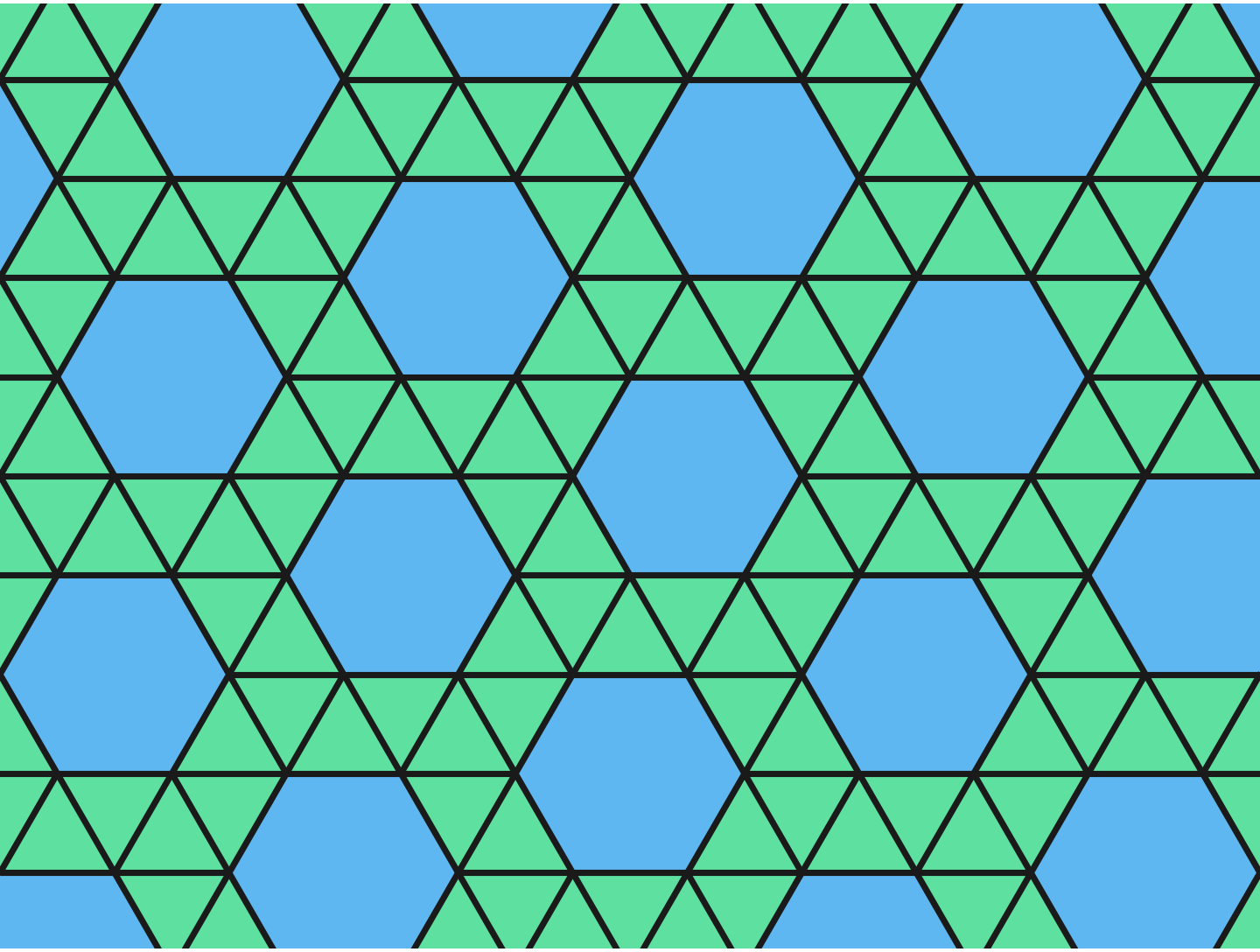} & $2.78584$ & $-1.82807$ & $0.46346$ & $-0.248$ & none & $-\tfrac{7}{6}$ & $0.054$ & \ref{fig:atlas-t1010} \\
\hyperref[atlas:t44_488]{$(4,8,8)$}     & \latimg{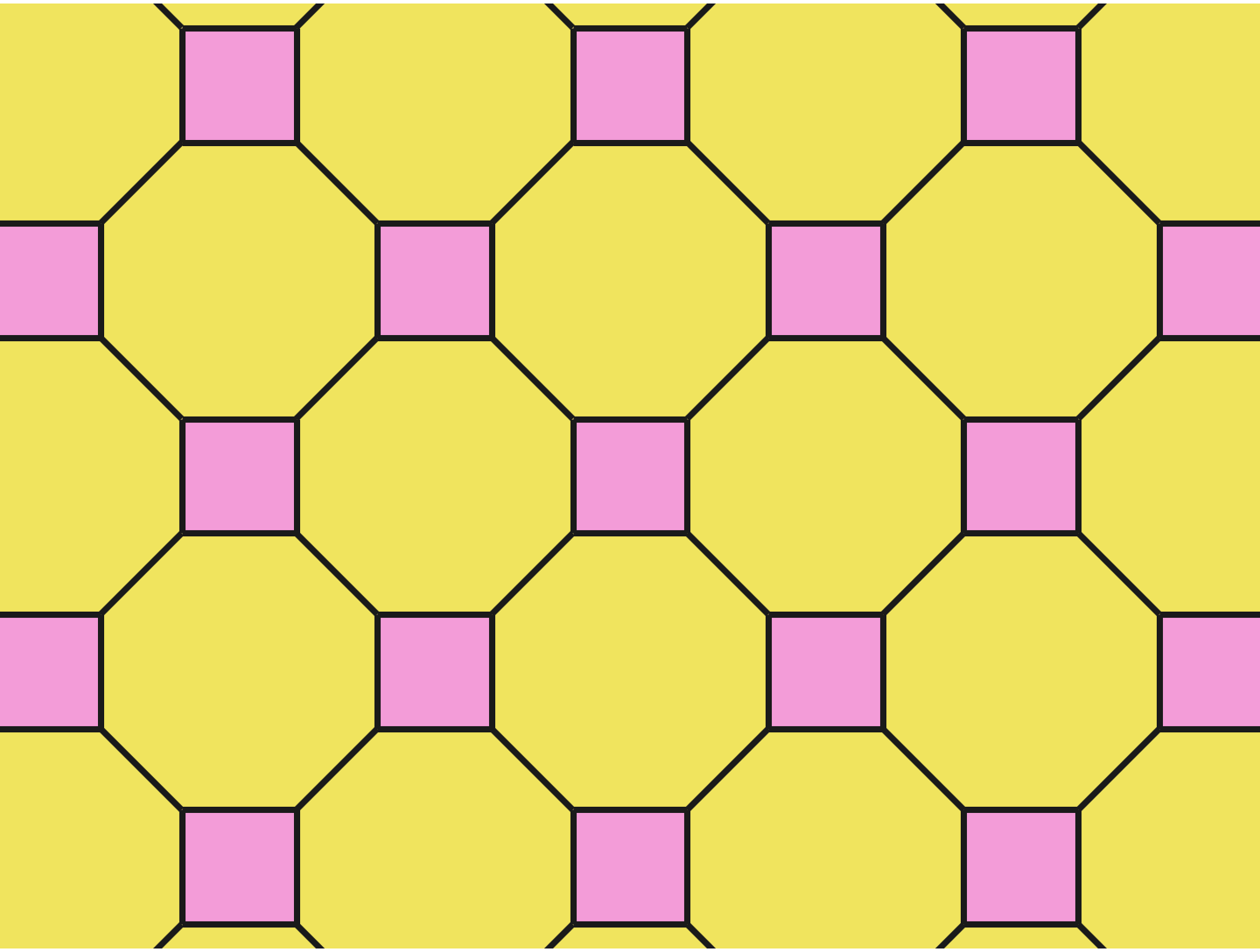} & $1.43870$ & $-1.20731$ & $0.43867$ & $-0.250$ & $1.43870$ & $-\tfrac{3}{2}$ & $0$ & \ref{fig:atlas-t44_488} \\
\hyperref[atlas:t1003]{$(4,6,12)$}      & \latimg{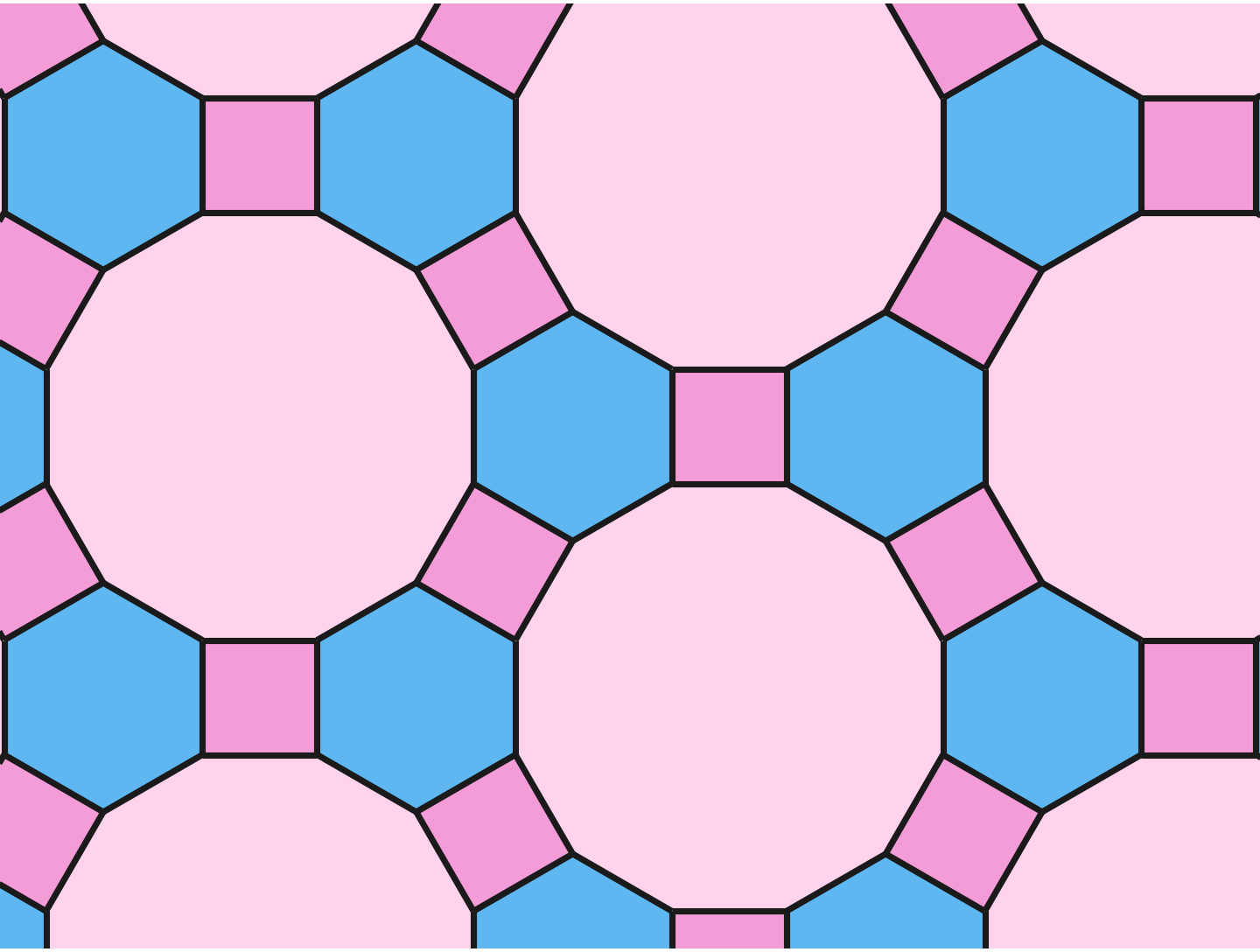} & $1.38983$ & $-1.24563$ & $0.40405$ & $-0.202$ & $1.38983$ & $-\tfrac{3}{2}$ & $0$ & \ref{fig:atlas-t1003} \\
\hyperref[atlas:t1009]{$(3,3,4,3,4)$}   & \latimg{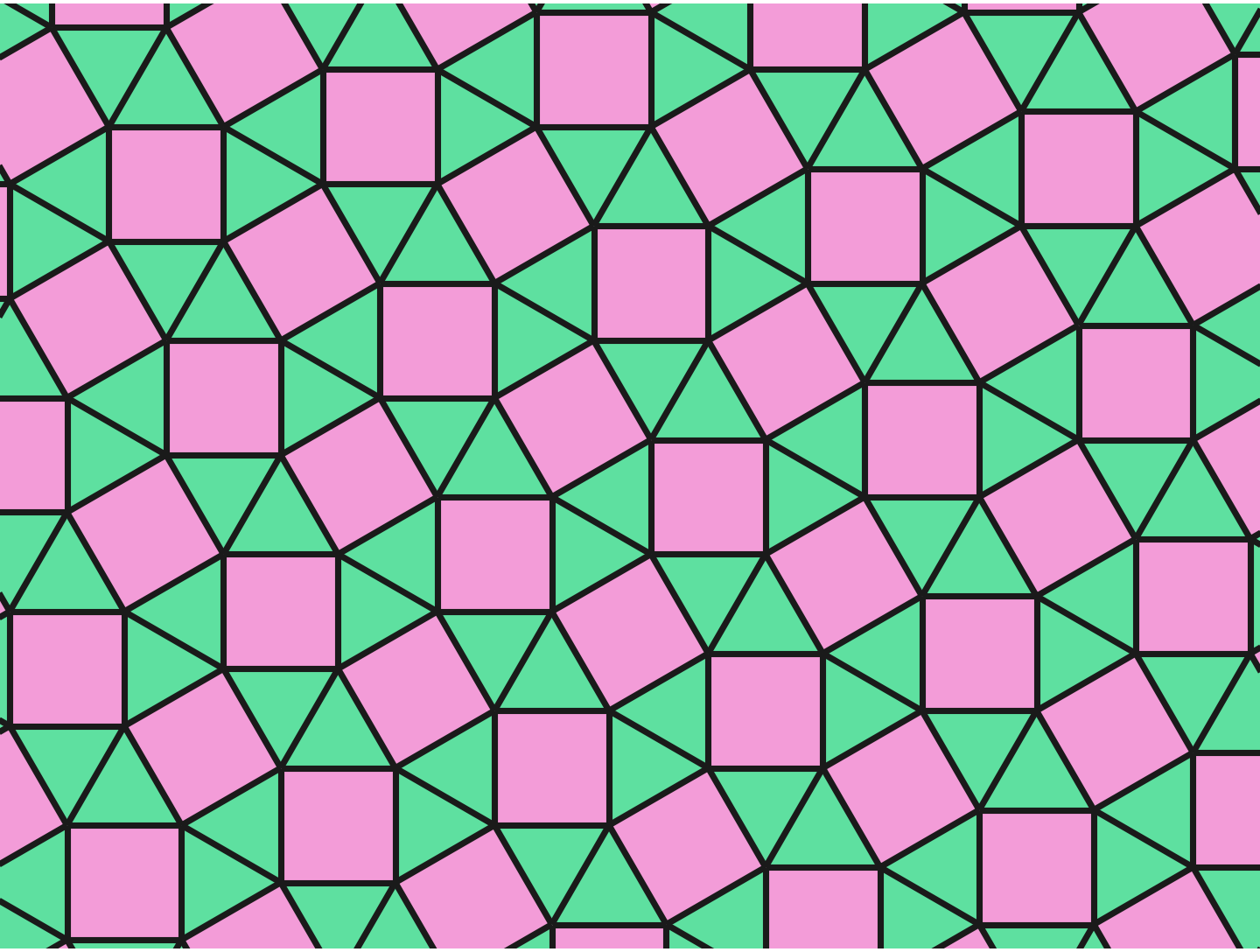} & $2.92626$ & $-1.72493$ & $0.49484$ & $-0.303$ & $1.26194$ & $-\tfrac{3}{2}$ & $0$ & \ref{fig:atlas-t1009} \\
\hyperref[atlas:t1006]{$(3,4,6,4)$}     & \latimg{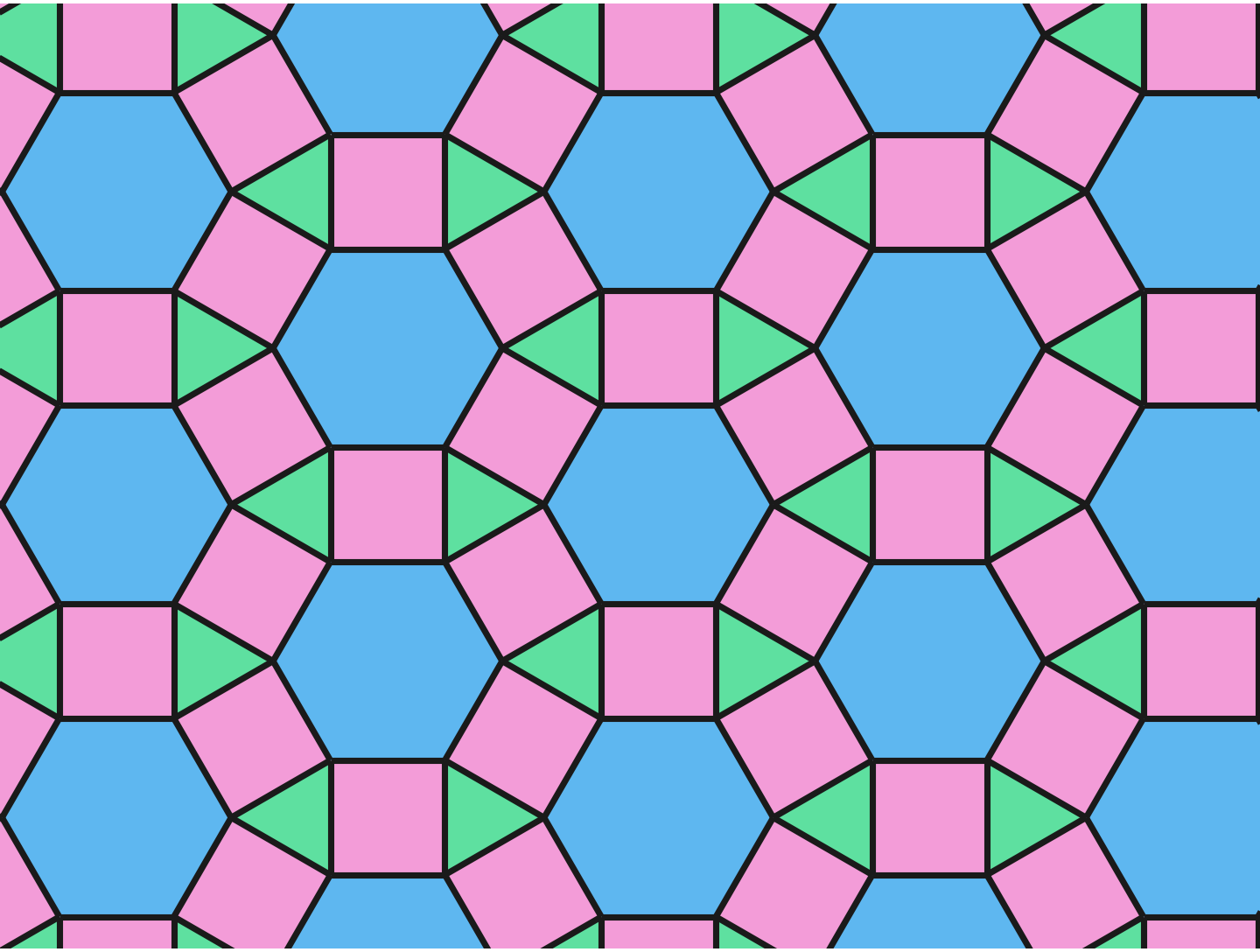} & $2.14332$ & $-1.50482$ & $0.44790$ & $-0.217$ & none & $-\tfrac{4}{3}$ & $0.054$ & \ref{fig:atlas-t1006} \\
\hyperref[atlas:t1008]{$(3,3,3,4,4)$}   & \latimg{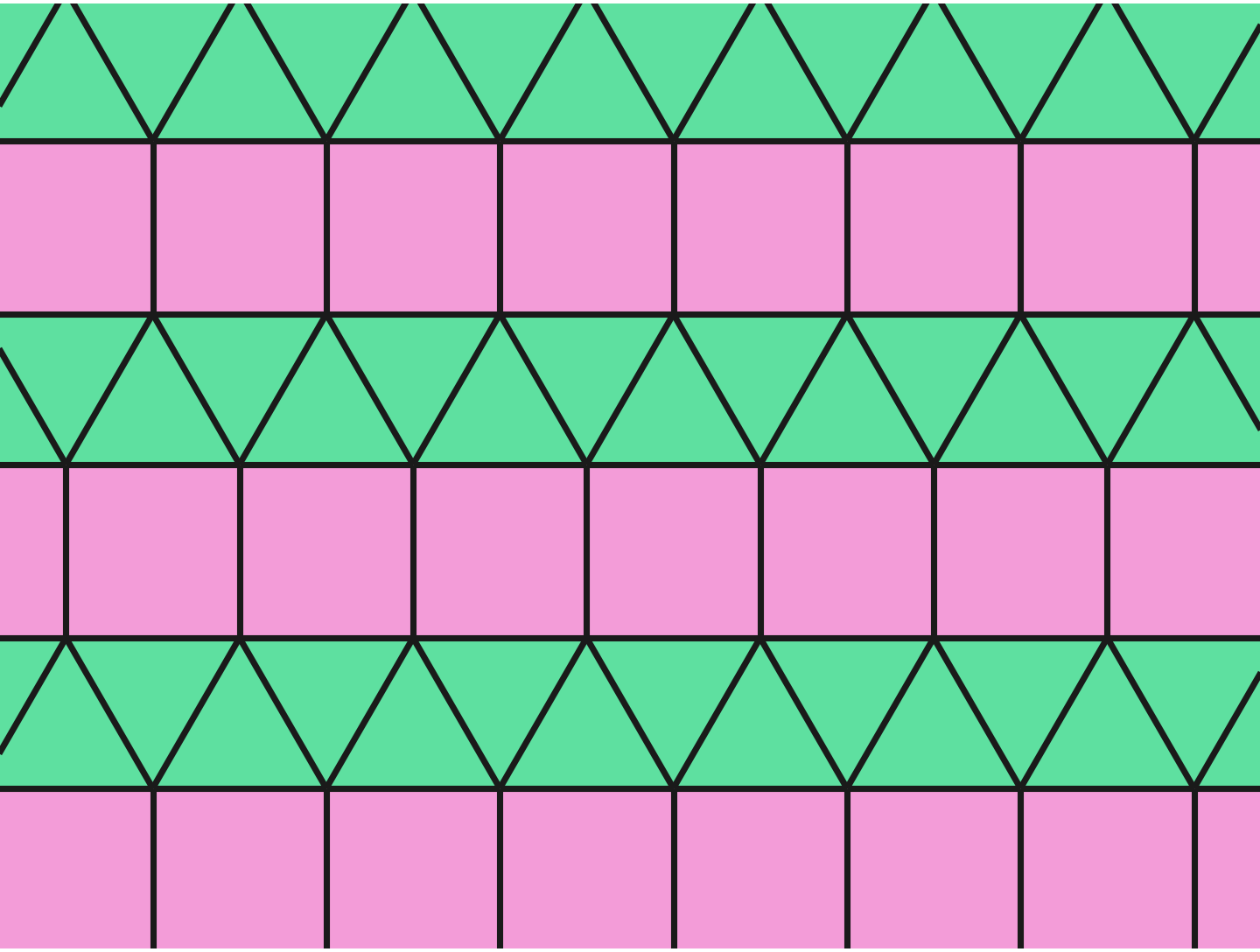} & $2.88539$ & $-1.75821$ & $0.47792$ & $-0.265$ & none & $-\tfrac{3}{2}$ & $0$ & \ref{fig:atlas-t1008} \\
\hyperref[atlas:t1004]{$(3,12,12)$}     & \latimg{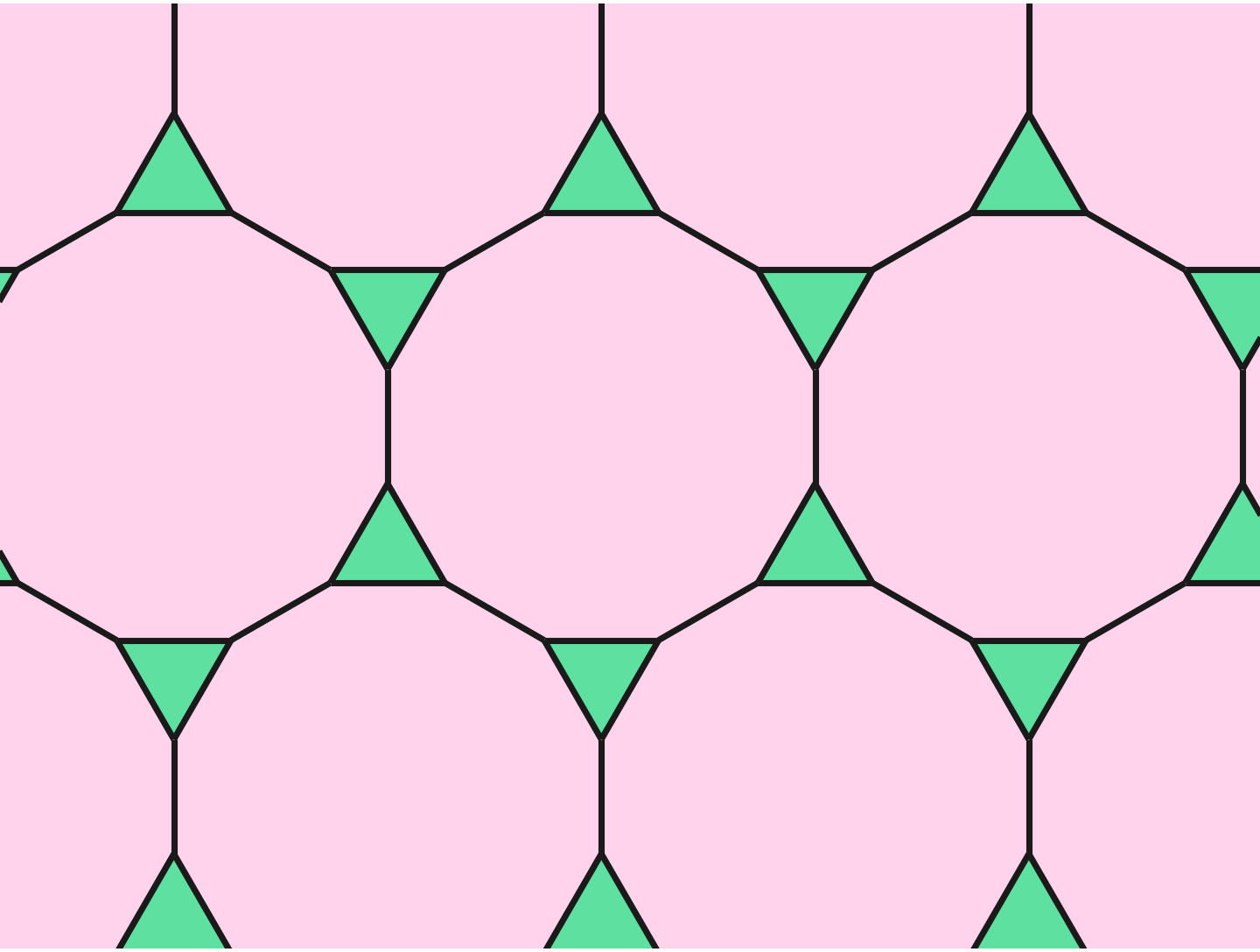} & $1.23151$ & $-1.31280$ & $0.35600$ & $-0.184$ & none & $-\tfrac{5}{6}$ & $0.251$ & \ref{fig:atlas-t1004} \\
\bottomrule
\end{tabular}
}
\end{table*}

\paragraph{Critical temperature.}
Phase transitions correspond to the singularities of the momentum integral in \eqref{eq:free-energy}, i.e.\ to the values of $v$ at which the determinant $P_\Lambda(v,k_1,k_2)$ vanishes for some $(k_1,k_2)$ in the integration domain $0\leq k_1, k_2 \leq 2\pi$. For ferromagnetic couplings, $0<v<1$, it is a rigorous result that on any planar doubly periodic graph the Kac--Ward determinant can vanish only at zero momentum, and only at the critical point \cite{Li_2012,Cimasoni_DuminilCopin_2013}. The ferromagnetic critical temperature is therefore determined by the singular point of the logarithm in \eqref{eq:free-energy} at $k_1 = k_2 = 0$. Writing
\begin{equation}
P_\Lambda(v) \;\equiv\; P_\Lambda(v, 0, 0)\,,
\label{eq:P-zero-momentum}
\end{equation}
for the zero-momentum reduction of the determinant, the ferromagnetic critical inverse temperature parameter $v_c$ is the unique root of
\begin{equation}
P_\Lambda(v_c) \;=\; 0\,,
\label{eq:Tc-condition}
\end{equation}
in the range $0 < v_c < 1$. The critical temperature is given by the standard relation
\begin{equation}
T_c = \frac{1}{\mathrm{arctanh}\, v_c}\,.\label{Tc}
\end{equation}
In the antiferromagnetic case we look for the solution in the negative range $- 1< v_c^{\rm AF} < 0$ and the critical temperature is given by $T_c^{\rm AF}= -1/ \mathrm{arctanh}\,v_c^{\rm AF}$.

The restriction to zero momentum requires some care in the antiferromagnetic case.
For $v<0$ the weights of the loop expansion are no longer positive, the results of \cite{Li_2012,Cimasoni_DuminilCopin_2013} do not apply, and frustrated lattices may in principle develop instabilities at non-zero momentum.
One thus has to check for zeros of $P_\Lambda(v,k_1,k_2)$ over the whole momentum range $0\leq k_1, k_2 \le 2\pi$ when $-1< v < 0$.
Among the lattices studied in this work, the $2$-uniform lattice t2011 is the only one displaying this behaviour: its antiferromagnetic transition $T_c^{\rm AF}=1.18195$ appears when $k_1=k_2=\pi/L=\pi/4$ of its $L=4$ program, i.e.\ at zero momentum on the doubled $L=8$ supercell: its ordered ground state does not fit in the $L=4$ supercell. For all the other lattices, a search for zeros of $P_\Lambda$ at non-zero momenta revealed no further transition.

For lattices small enough that the determinant~\eqref{eq:KW-determinant} can be evaluated symbolically, $T_c$ is obtained in closed form if the resulting polynomial is solvable by radicals. For larger supercells, $P_\Lambda(v)$ is a polynomial in $v$ of moderate degree whose root in $(0, 1)$ is found numerically to arbitrary precision. The critical temperatures collected in this paper are obtained by this second route.

It is clear that $T_c$ is the simplest invariant of the equivalence class of target lattices $[\Lambda]$. One might wonder if this invariant is able to classify equivalence classes; the answer is easily seen to be negative as counter examples show up already in the Archimedean family: lattices $(3,6,3,6)$ and $(3,4,6,4)$ are not topologically equivalent, yet they have the same $T_c = 2.14332 ...$.
The correspondence between $T_c$ and $[\Lambda]$ deserves a dedicated study, analogous to those in \cite{Joseph_2026, Joseph_2026b}.

\paragraph{Energy and specific heat.}
The per-spin internal energy and specific heat follow from standard thermodynamic relations
\begin{equation}
\varepsilon_\Lambda(T) = -T^2 \frac{\partial}{\partial T}\frac{ f_\Lambda(T)}{T}\,,
\qquad\qquad\qquad
c_\Lambda(T) = \frac{\partial }{\partial T}\varepsilon_\Lambda(T)\,.
\label{eq:thermo-relations}
\end{equation}
Carrying out the differentiations under the integral sign in~\eqref{eq:free-energy}, one finds
\begin{equation}
\varepsilon_\Lambda(T) \;=\; -\xi\, \frac{n_l}{n_v}\, v \;-\; \xi\frac{1-v^2}{2 n_v}\!\int\!\frac{{\rm d} k_1}{2\pi}\frac{{\rm d} k_2}{2\pi}\, \frac{P'_\Lambda}{P_\Lambda}\,,
\label{eq:energy}
\end{equation}
where 
the prime on $P_\Lambda$ denotes differentiation with respect to $v$, and
\begin{equation}
T^2 c_\Lambda(T) \;=\; 2 v \,\xi\, \varepsilon_\Lambda(T) \;+\; \frac{n_l}{n_v}(1 + v^2) \;+\; \frac{(1 - v^2)^2}{2 n_v}\!\int\!\frac{{\rm d} k_1}{2\pi}\frac{{\rm d} k_2}{2\pi}\!\left\{ \frac{P''_\Lambda}{P_\Lambda} - \left(\frac{P'_\Lambda}{P_\Lambda}\right)^{\!2}\, \right\}.
\label{eq:specific-heat}
\end{equation}
The per-spin entropy is then
\begin{equation}
s_\Lambda(T) = \frac{\varepsilon_\Lambda(T) - f_\Lambda(T)}{T}\,.
\label{eq:entropy}
\end{equation}
For supercells beyond a handful of sites the determinant $P_\Lambda$ is too large to evaluate symbolically, and direct symbolic differentiation in~\eqref{eq:energy} and~\eqref{eq:specific-heat} becomes impractical. The derivatives of $P_\Lambda$ are nevertheless directly accessible from $\mathbb{W}^\Lambda_{\mathcal H}$ through the trace identities
\begin{equation}
\frac{P'_\Lambda}{P_\Lambda} \;=\; -\,\mathrm{tr}\!\left[ \frac{\mathbb{W}^\Lambda_{\mathcal H}}{\mathbb{I} - v \mathbb{W}^\Lambda_{\mathcal H}} \right],
\label{eq:trace-Pprime}
\end{equation}
\begin{equation}
\frac{P''_\Lambda}{P_\Lambda} \;=\; \left\{\mathrm{tr}\!\left[ \frac{\mathbb{W}^\Lambda_{\mathcal H}}{\mathbb{I} - v \mathbb{W}^\Lambda_{\mathcal H}} \right]\right\}^{\!2} \;-\; \mathrm{tr}\!\left[ \left(\frac{\mathbb{W}^\Lambda_{\mathcal H}}{\mathbb{I} - v \mathbb{W}^\Lambda_{\mathcal H}}\right)^{\!2} \right].
\label{eq:trace-Pdoubleprime}
\end{equation}
These reduce~\eqref{eq:energy} and~\eqref{eq:specific-heat} to traces of resolvents of $\mathbb{W}^\Lambda_{\mathcal H}$, which are linear-algebraic operations on a matrix of fixed size $\nu L^2$ ($\nu = 4$ or $6$). The cost of evaluating energy and specific heat at a given temperature is therefore polynomial in $L$, independent of the structural complexity of the target lattice.

\paragraph{Critical amplitudes.}
Near the transition, the specific heat behaves as
\begin{equation}
c_\Lambda(T) \;\sim\; -A \log\!\left| 1 - \frac{T_c}{T} \right| \;+\; B\,,
\label{eq:specific-heat-asymptotic}
\end{equation}
with two lattice-dependent constants $A$ and $B$. The amplitude $A$ admits the closed-form expression
\begin{equation}
\rho\, T_c^2\, A \;=\; \frac{(1 - v_c^2)^2}{2\pi}\, \frac{P''_\Lambda(v_c)}{\sqrt{ P_{11}(v_c)\, P_{22}(v_c) - P_{12}^2(v_c) }},
\label{eq:A-amplitude}
\end{equation}
where $\rho = n_v / L^2$ is the site density of the supercell and the second-order momentum derivatives of $P_\Lambda$ are defined
\begin{equation}
P_{ij}(v) \;\equiv\; \left.\frac{\partial^2 P_\Lambda(v, k_1, k_2)}{\partial k_i\, \partial k_j}\right|_{k_1 = k_2 = 0}\,,
\label{eq:Pij-definition}
\end{equation}
%
which are given in terms of the FV/KW transition matrix by
%
%
%
\begin{equation}\label{Pij}
\frac{P_{ij}}{P}=\left({\rm tr}\frac{v\,\mathbb{W}_{i}}{\mathbb{I}-v\,\mathbb{W}}\right)\left({\rm tr}\frac{v\,\mathbb{W}_{j}}{\mathbb{I}-v\,\mathbb{W}}\right)-{\rm tr}\left(\frac{v\,\mathbb{W}_{ij}}{\mathbb{I}-v\,\mathbb{W}}+\frac{v^{2}\,\mathbb{W}_{i}\mathbb{W}_{j}}{(\mathbb{I}-v\,\mathbb{W})^{2}}\right)\,.
\end{equation}
A derivation of~\eqref{eq:A-amplitude} is given in Appendix~\ref{app:A-derivation}; the formula applies to both the square and triangular emulators, with the appropriate Jacobian of the momentum variables absorbed into the definition of $P_{ij}$.

The amplitude $B$ has no equally compact closed form for a general lattice $\Lambda$. It is obtained by subtracting the logarithmic divergence~\eqref{eq:specific-heat-asymptotic} from $c_\Lambda(T)$ evaluated near $T_c$ via~\eqref{eq:specific-heat} and~\eqref{eq:trace-Pdoubleprime}. The values of $B$ reported in this paper are obtained by this numerical subtraction and are given to three decimal digits.

\paragraph{Correlation length.}
The correlation length $\xi_\Lambda$ of the Ising model is directly related to the fermion lattice mass $m_\Lambda \sim 1/\xi_\Lambda$ which is derived from the fermion lattice propagator $P_\Lambda$ that diagonalizes the lattice Ising problem \cite{Itzykson:1989sx}.
One just needs to factor out the lattice wave-function renormalization in the $k_1,k_2\to 0$ limit from the lattice propagator.
More specifically, one finds the following general relation
\begin{equation}
\xi_\Lambda(T) = \frac{1}{2}\sqrt{\, \frac{P_{11}(v) + P_{22}(v)}{P_\Lambda(v)} \,} \,.
\label{eq:correlation-length}
\end{equation}
For any given lattice $\Lambda$, this expression diverges at $T_c$ with the expected $\xi_\Lambda(T)  \sim |T - T_c|^{-1}$ behavior characteristic of the two-dimensional Ising universality class.

\paragraph{Dual lattices.}
As is well known, the Ising models on a lattice and on its dual are closely related \cite{Kramers_Wannier_1941}.
In particular, using the critical values for the high temperature parameter $v_{c}$
found for the lattice $\Lambda$, we can calculate the critical
temperature of its dual $\Lambda^D$ from the relation
\begin{equation}
T_{c}^{D}= -\frac{2}{\log \tanh\frac{1}{T_c}} \label{TD}\,.
\end{equation}
In the ferromagnetic case one can also determine all thermodynamic quantities (see \cite{Laurent_2025} and references therein).
Duality thus gives free access to a large class of non-unit-edge-length lattices, dual to those that can be emulated in our framework.  Sometimes the dual lattice is more interesting than the original one; examples are the dice lattice, dual to the kagome lattice, and the Cairo pentagonal lattice, dual to the snub square.

\paragraph{Ground state $T\!\to\!0$ limit.}
For many applications the properties of the ground state are very important, specifically the energy $\varepsilon_0 \equiv \varepsilon(0)$ and the entropy $s_0\equiv s(0)$.
In the ferromagnetic case the limit is straightforward: the energy is just minus the density of links, $\varepsilon_0^{\rm F} = -\tfrac{n_l}{n_v}$, while the entropy per spin vanishes, $s_0^{\rm F} = 0$, since there are only two ground states (all spins up or all spins down), whose contribution $\log 2/N$ to the entropy per spin vanishes in the thermodynamic limit $N\to\infty$. This last fact follows directly from $\varepsilon_0^{\rm F} = f_0^{\rm F}$, where $f_0\equiv f(0)$, and from the definition of entropy.
The antiferromagnetic case is much more interesting, as non-trivial ground states are possible.
On bipartite lattices the sublattice spin flip maps the antiferromagnetic ground state onto the ferromagnetic one, so that $\varepsilon_0^{\rm AF}=\varepsilon_0^{\rm F}=-\tfrac{n_l}{n_v}$ and $s_0^{\rm AF}=0$.
On frustrated lattices, both quantities can be obtained directly from the programmed matrix at $v=-1$, without any extrapolation in temperature. Writing $v=-1+\delta$, each factor of $P_\Lambda=\prod_i\bigl(1-v\lambda_i(k_1,k_2)\bigr)$ associated with the eigenvalue $\lambda=-1$ of $\mathbb W^\Lambda_{\mathcal H}(k_1,k_2)$ equals $\delta$, so that
\begin{equation}
P_\Lambda(v,k_1,k_2)=\delta^{\,m}\,Q_\Lambda(k_1,k_2)\,\bigl[1+O(\delta)\bigr]\,,\qquad
Q_\Lambda(k_1,k_2)=\prod_{\lambda_i\neq-1}\bigl(1+\lambda_i(k_1,k_2)\bigr)\,,
\label{eq:P-ground-state}
\end{equation}
where $m$ is the multiplicity of the eigenvalue $-1$, which is the same for almost all $(k_1,k_2)$. Inserting~\eqref{eq:P-ground-state} into~\eqref{eq:free-energy}, \eqref{eq:energy}, \eqref{eq:entropy} and sending $T\to 0$
\begin{equation}
\varepsilon_0^{\rm AF}=-\frac{n_l-m}{n_v}\,,
\label{eq:e0-AF}
\end{equation}
\begin{equation}
s_0^{\rm AF}=\Bigl(1-\frac{2n_l-m}{2n_v}\Bigr)\log 2+\frac{1}{2n_v}\int\frac{{\rm d}k_1}{2\pi}\frac{{\rm d}k_2}{2\pi}\,\log Q_\Lambda(k_1,k_2)\,.
\label{eq:s0-AF}
\end{equation}
Equation~\eqref{eq:e0-AF} identifies $m=2n_X$ with twice the number $n_X$ of unsatisfied bonds per supercell in the antiferromagnetic ground state, i.e.\ the minimal number of bonds to be removed per supercell to leave a lattice with only even faces \cite{Laurent_2025}; since $m$ is an integer, $\varepsilon_0^{\rm AF}$ is always a rational number. On unfrustrated lattices $m=0$, $Q_\Lambda=P_\Lambda(-1,k_1,k_2)$, and~\eqref{eq:s0-AF} gives $s_0^{\rm AF}=0$. All the ground-state energies and residual entropies reported in this work are computed from~\eqref{eq:e0-AF} and~\eqref{eq:s0-AF}, which reproduce the known values for the triangular and kagome lattices \cite{Wannier_1950,Kano_Naya_1953}.\footnote{For some lattices $Q_\Lambda$ vanishes along lines of the integration domain. The integral in~\eqref{eq:s0-AF} remains finite, since the singularity of $\log Q_\Lambda$ is integrable, but its numerical evaluation on an $N\times N$ momentum grid converges as $1/N$, and grids containing points on these lines must be avoided.}
We stress that the zero-temperature energy $\varepsilon_0^{\rm AF}\equiv\varepsilon^{\rm AF}(0)$ should not be confused with the energy density at the antiferromagnetic transition, $\varepsilon_c^{\rm AF}\equiv\varepsilon^{\rm AF}(T_c^{\rm AF})$, the antiferromagnetic counterpart of $\varepsilon_c$, which is defined only for lattices with a finite-temperature antiferromagnetic transition and is reported in the atlas of Appendix~\ref{app:atlas}.

\paragraph{Summary.}
Equations~\eqref{eq:free-energy}--\eqref{eq:s0-AF} define the thermodynamic pipeline of the emulator framework.
Once an emulator and a supercell program are specified, the same set of determinants, traces, and momentum integrals delivers $f$, $T_c$, $\varepsilon$, $c$, $s$, $A$, $B$, $\xi$, $T^D_c$, $\varepsilon_0^{\rm AF}$ and $s_0^{\rm AF}$ for any target lattice $\Lambda$.
This is the operational content advocated in Section~\ref{sec:emulator-principle}: the analytic machinery is fixed once and for all, and changing target amounts to changing the program.
\begin{table*}[t!]
\centering
\small
\caption{Thermodynamic quantities for the first ten $2$-uniform lattices: ferromagnetic critical temperature $T_c$, energy density at the transition $\varepsilon_c$, and specific-heat amplitude $A$; antiferromagnetic critical temperature $T_c^{\rm AF}$ (``none'' when there is no finite-temperature transition), ground-state energy $\varepsilon_0^{\rm AF}$, and residual entropy $s_0^{\rm AF}$, computed from Eqs.~\eqref{eq:e0-AF} and~\eqref{eq:s0-AF}. The last column gives the corresponding atlas figure of Appendix~\ref{app:atlas}, which also reports the energy density $\varepsilon_c^{\rm AF}$ at the antiferromagnetic transition and the dual critical temperature $T_c^{\rm D}$.}
\label{galebach1}
\renewcommand{\arraystretch}{1.12}
\setlength{\tabcolsep}{5pt}
\begin{tabular}{>{\centering\arraybackslash}m{1.8cm}
                >{\centering\arraybackslash}m{1.65cm}
                >{\centering\arraybackslash}m{1.45cm}
                >{\centering\arraybackslash}m{1.35cm}
                >{\centering\arraybackslash}m{1.15cm}
                >{\centering\arraybackslash}m{1.3cm}
                >{\centering\arraybackslash}m{1.0cm}
                >{\centering\arraybackslash}m{1.0cm}
                >{\centering\arraybackslash}m{0.7cm}}
\toprule
\makecell{\textbf{Galebach}\\\textbf{name}} &
\makecell{\textbf{Lattice} $\Lambda$} &
\multicolumn{3}{c}{\textbf{Ferromagnetic}} &
\multicolumn{3}{c}{\textbf{Antiferromagnetic}} &
\makecell{\textbf{Fig.}} \\
\cmidrule(lr){3-5} \cmidrule(lr){6-8}
&
&
$T_c$ &
$\varepsilon_c$ &
$A$ &
$T_c^{{\rm{AF}}}$ &
$\varepsilon_0^{{\rm{AF}}}$ &
$s_0^{{\rm{AF}}}$ &
\\
\midrule
\hyperref[atlas:t2001]{t$2001$} & \latimgsmall{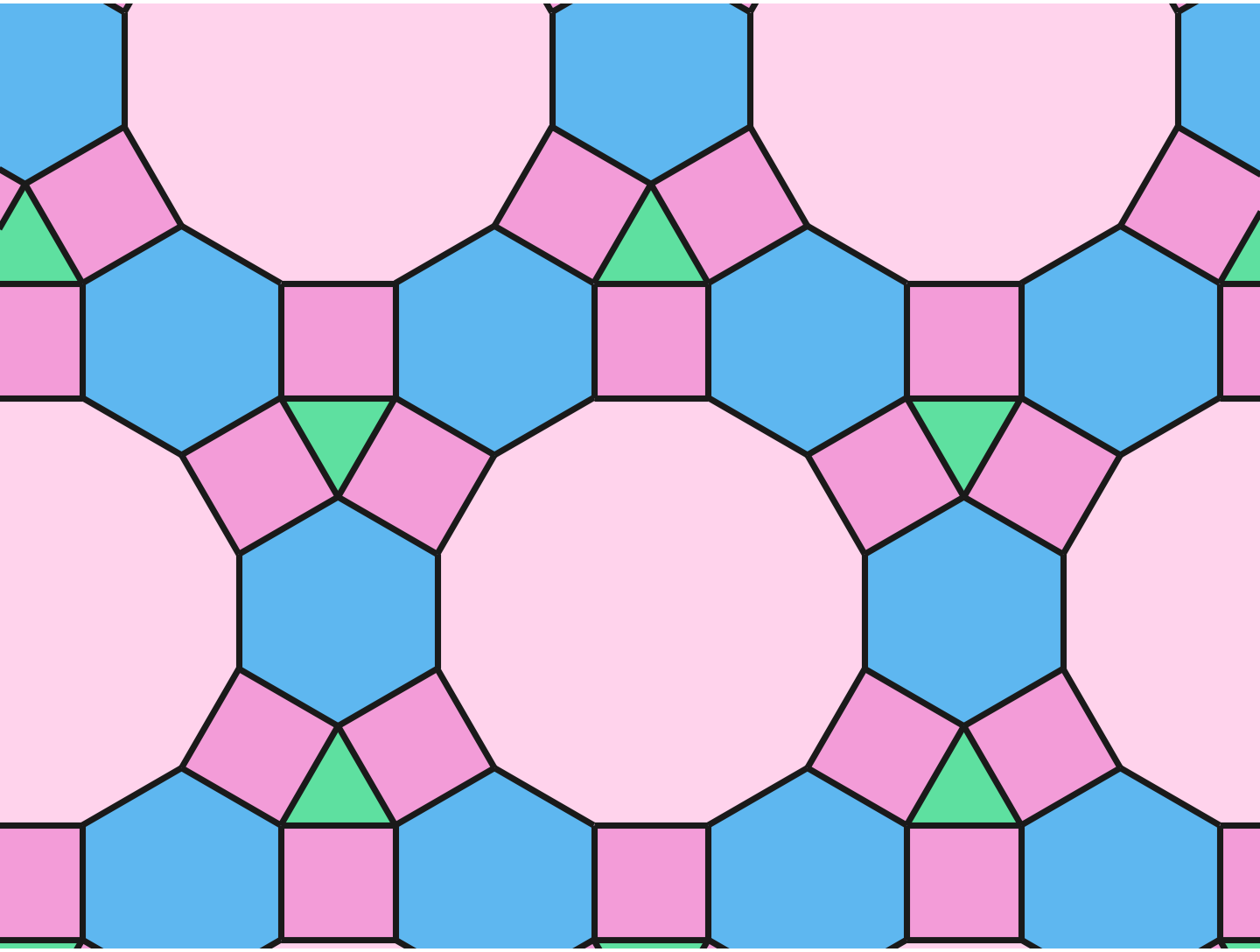} & $1.58017$ & $-1.38041$ & $0.36172$ & none & $-\tfrac{11}{9}$ & $0.129$ & \ref{fig:atlas-t2001} \\
\hyperref[atlas:t2002]{t$2002$} & \latimgsmall{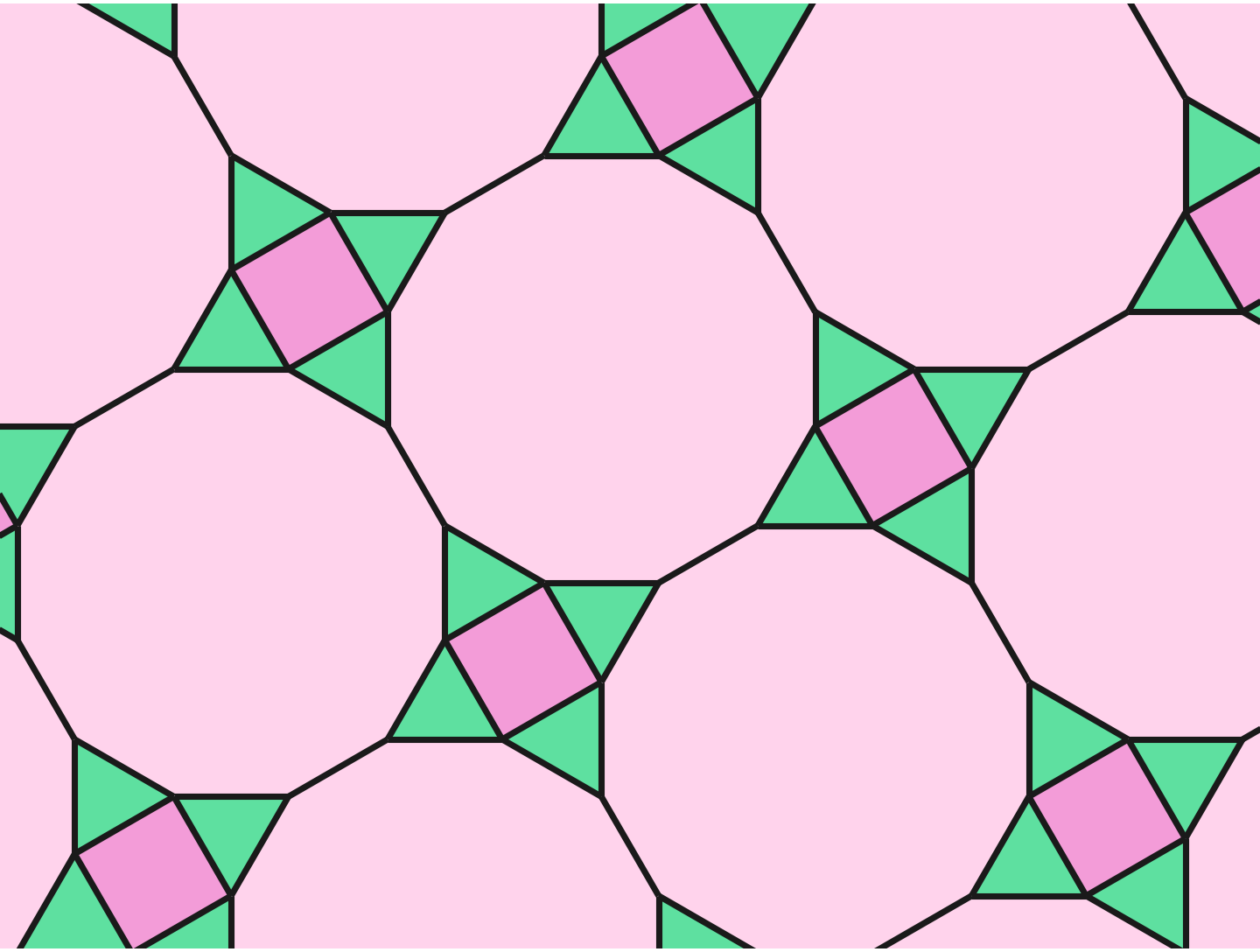} & $1.47614$ & $-1.54523$ & $0.28617$ & none & $-\tfrac{3}{4}$ & $0.378$ & \ref{fig:atlas-t2002} \\
\hyperref[atlas:t2003]{t$2003$} & \latimgsmall{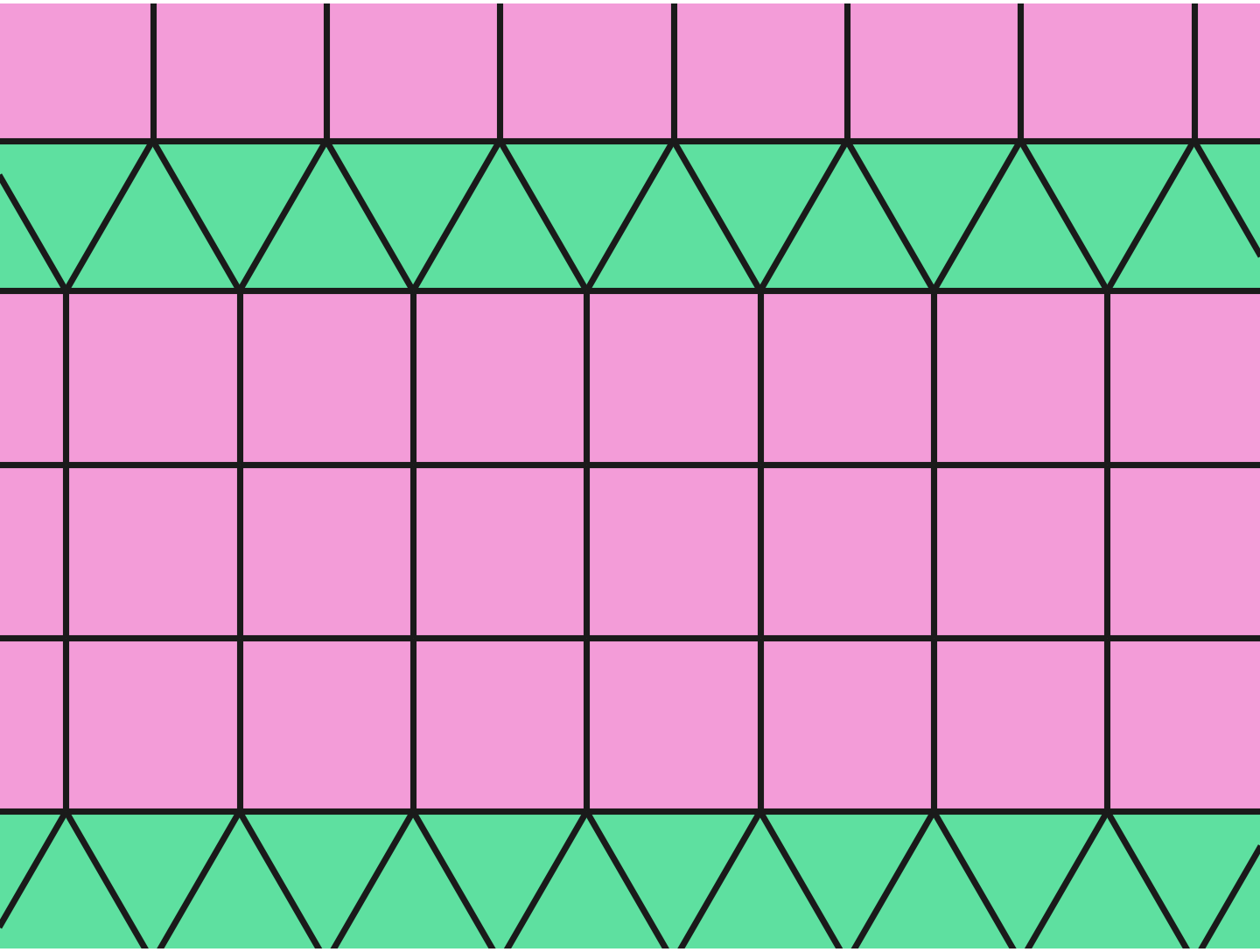} & $2.56381$ & $-1.59804$ & $0.46065$ & none & $-\tfrac{7}{4}$ & $0$ & \ref{fig:atlas-t2003} \\
\hyperref[atlas:t2004]{t$2004$} & \latimgsmall{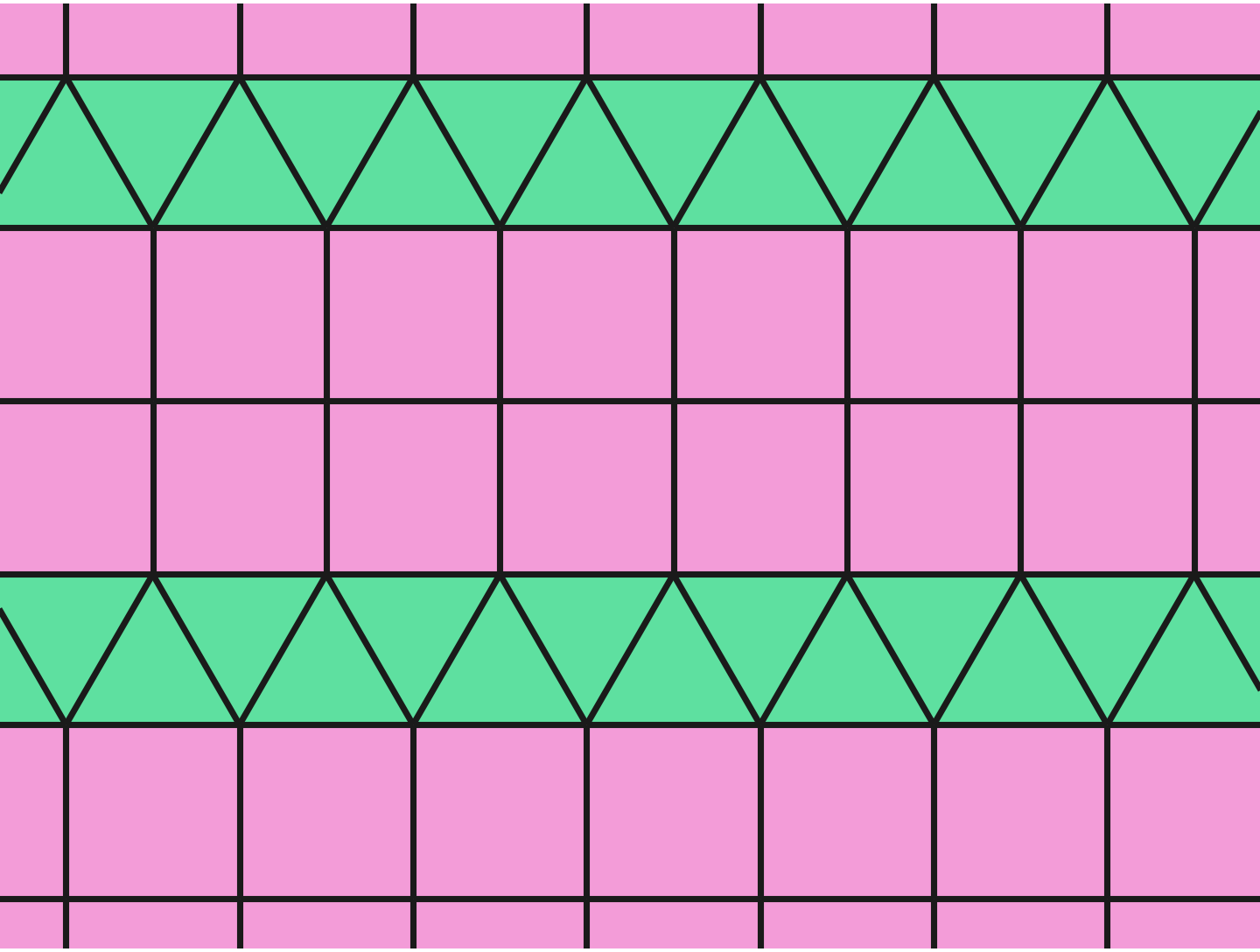} & $2.66764$ & $-1.65506$ & $0.46710$ & none & $-\tfrac{5}{3}$ & $0$ & \ref{fig:atlas-t2004} \\
\hyperref[atlas:t2005]{t$2005$} & \latimgsmall{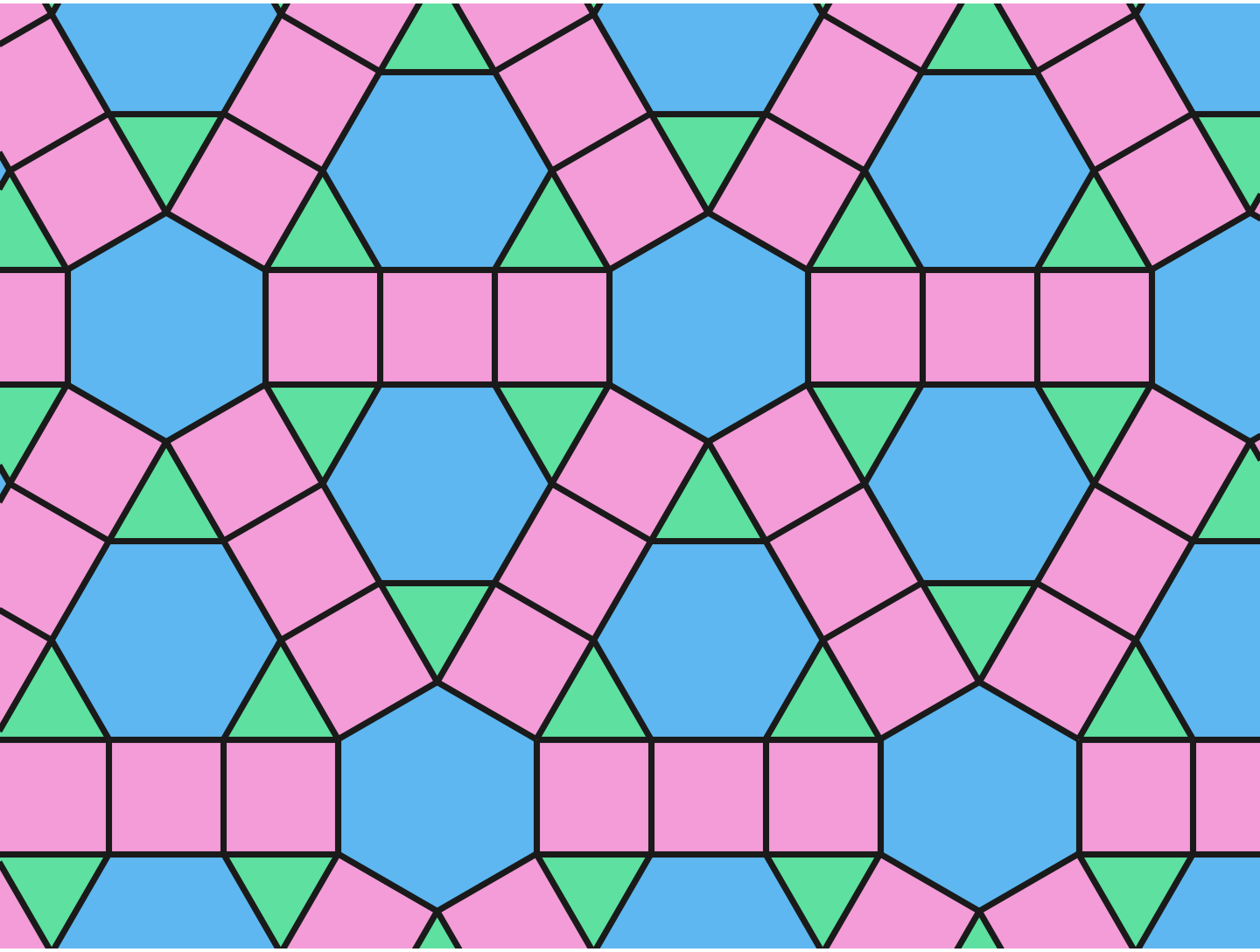} & $2.16795$ & $-1.48370$ & $0.46465$ & none & $-\tfrac{4}{3}$ & $0.084$ & \ref{fig:atlas-t2005} \\
\hyperref[atlas:t2006]{t$2006$} & \latimgsmall{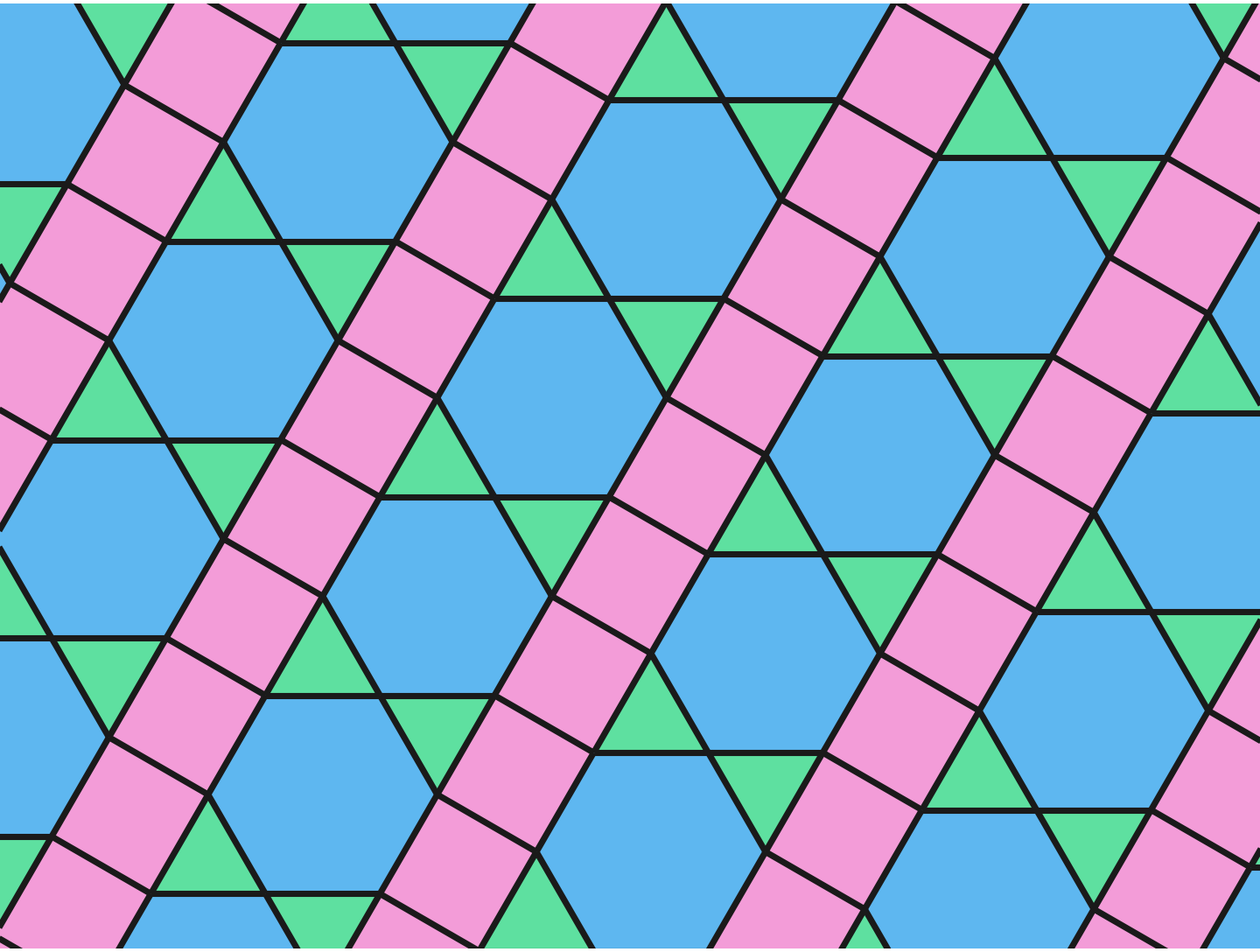} & $2.17788$ & $-1.47159$ & $0.47852$ & none & $-\tfrac{6}{5}$ & $0.139$ & \ref{fig:atlas-t2006} \\
\hyperref[atlas:t2007]{t$2007$} & \latimgsmall{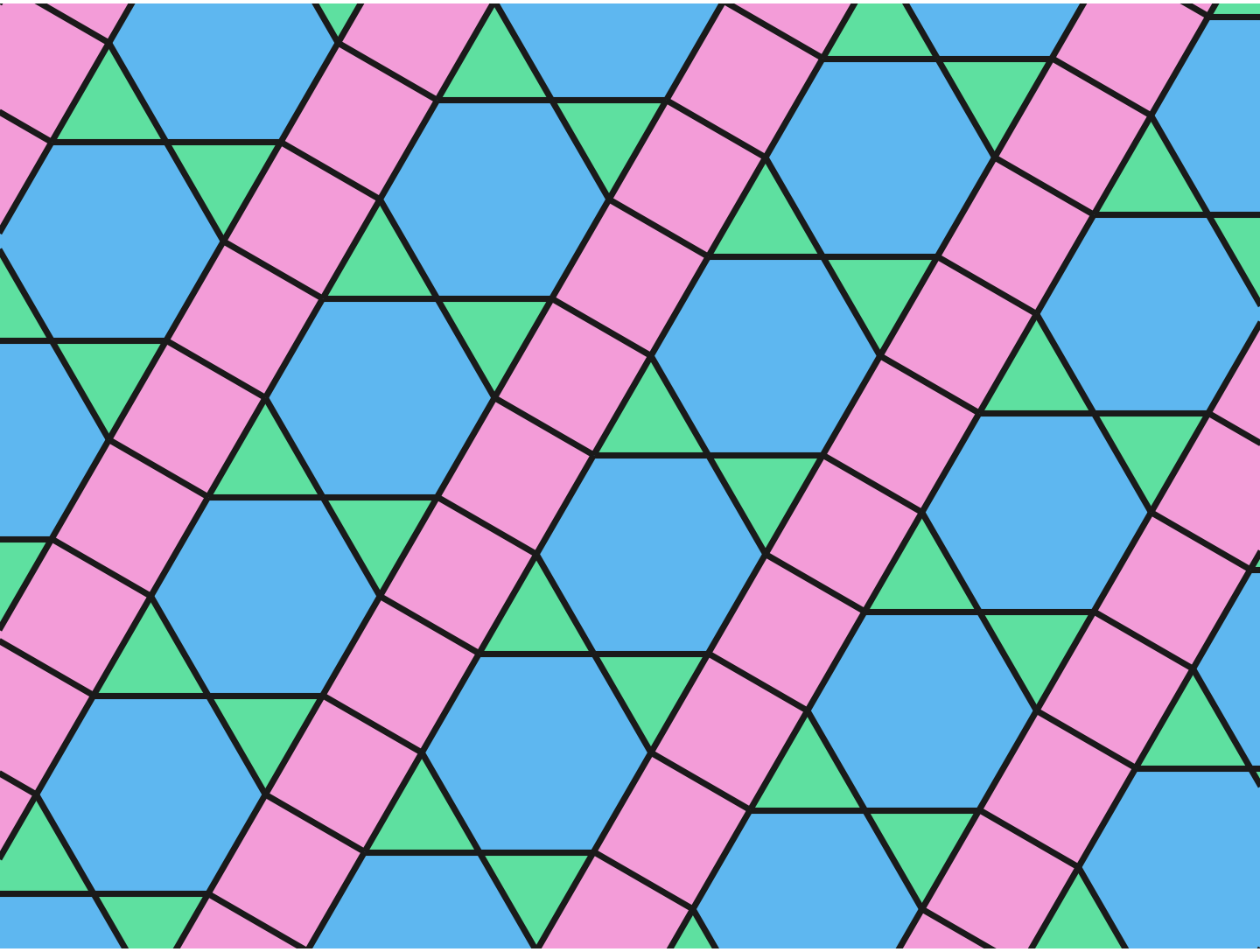} & $2.17171$ & $-1.47730$ & $0.47286$ & none & $-\tfrac{6}{5}$ & $0.184$ & \ref{fig:atlas-t2007} \\
\hyperref[atlas:t2008]{t$2008$} & \latimgsmall{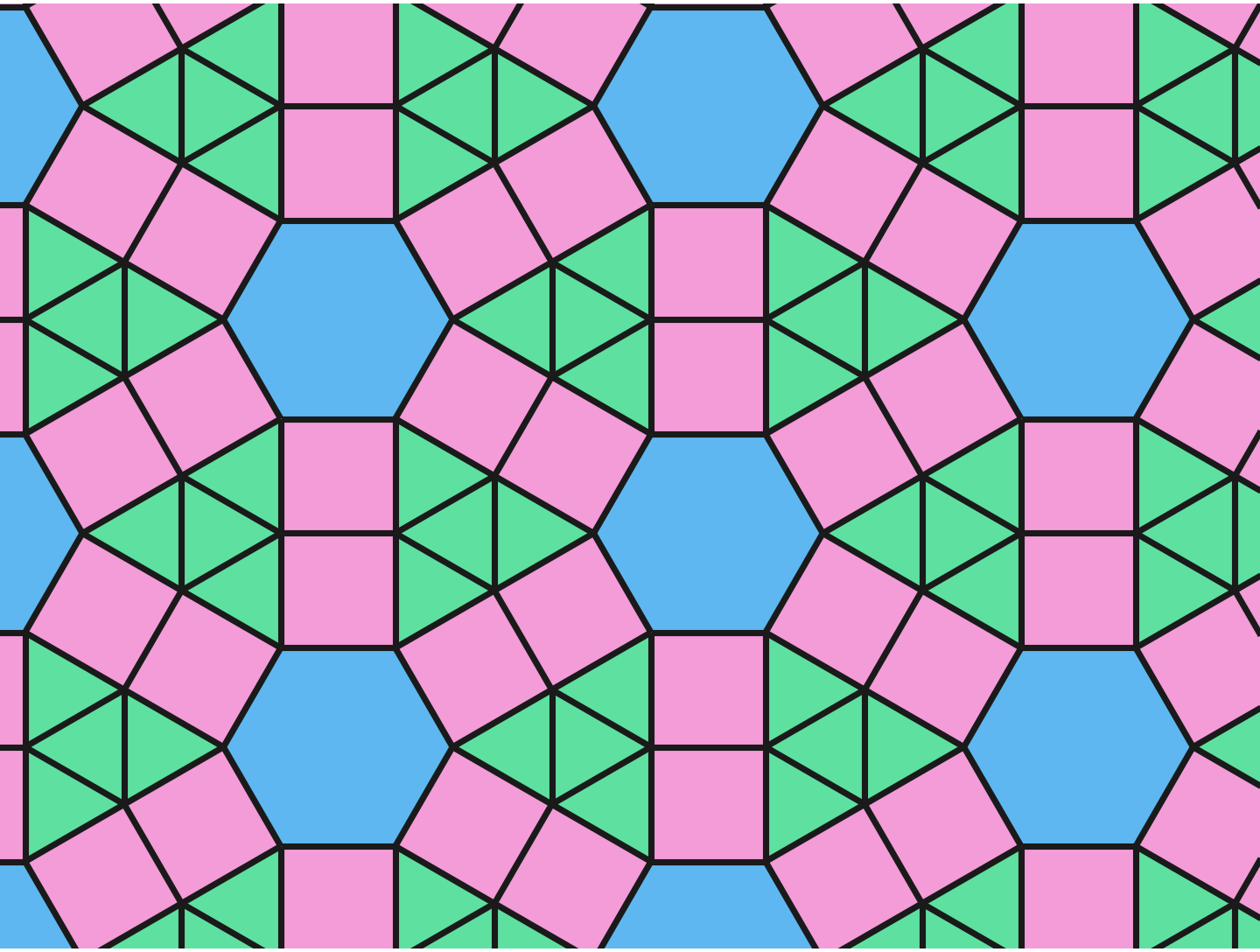} & $2.48264$ & $-1.66253$ & $0.43280$ & none & $-\tfrac{5}{4}$ & $0.126$ & \ref{fig:atlas-t2008} \\
\hyperref[atlas:t2009]{t$2009$} & \latimgsmall{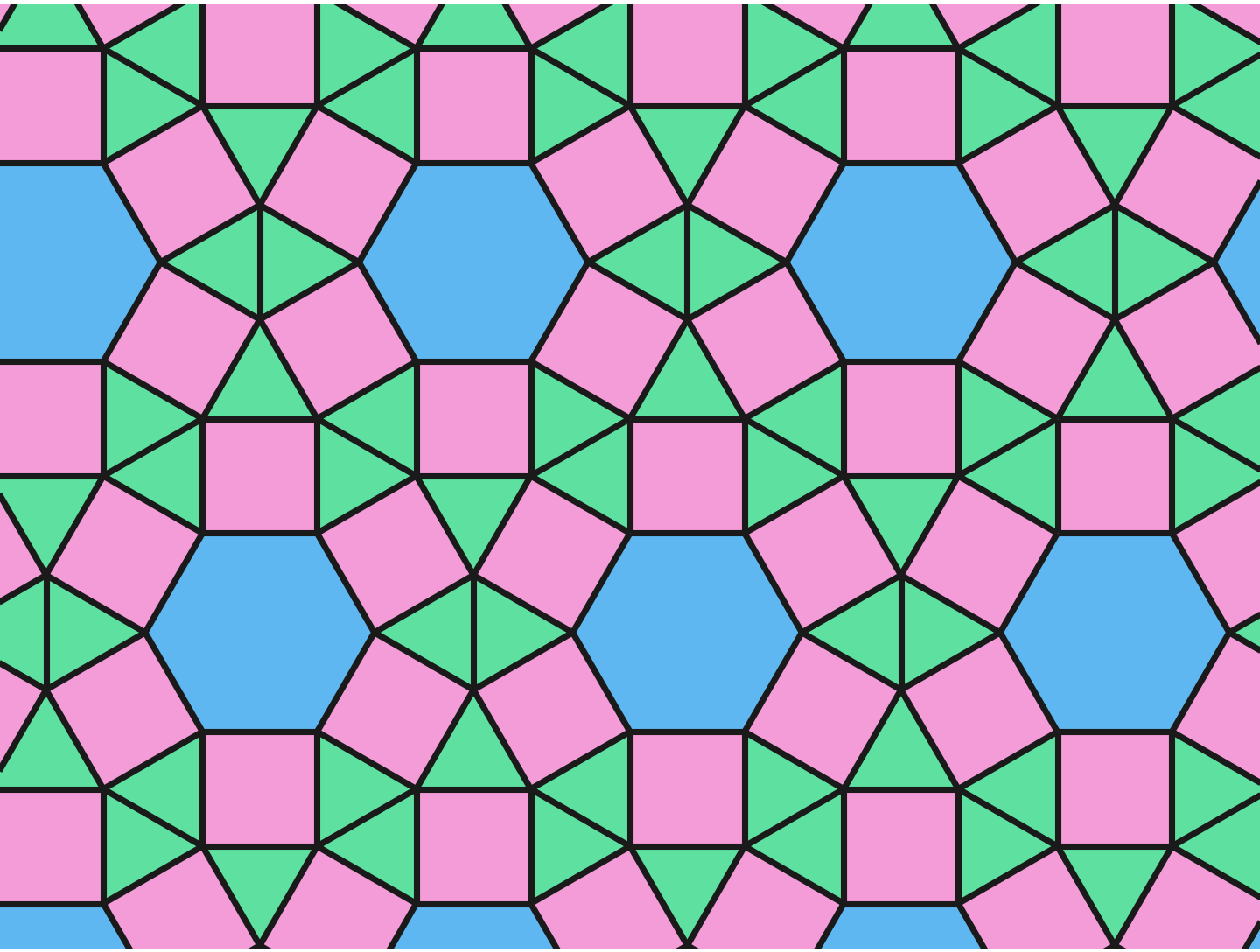} & $2.52372$ & $-1.62812$ & $0.45512$ & none & $-\tfrac{5}{4}$ & $0.085$ & \ref{fig:atlas-t2009} \\
\hyperref[atlas:t2010]{t$2010$} & \latimgsmall{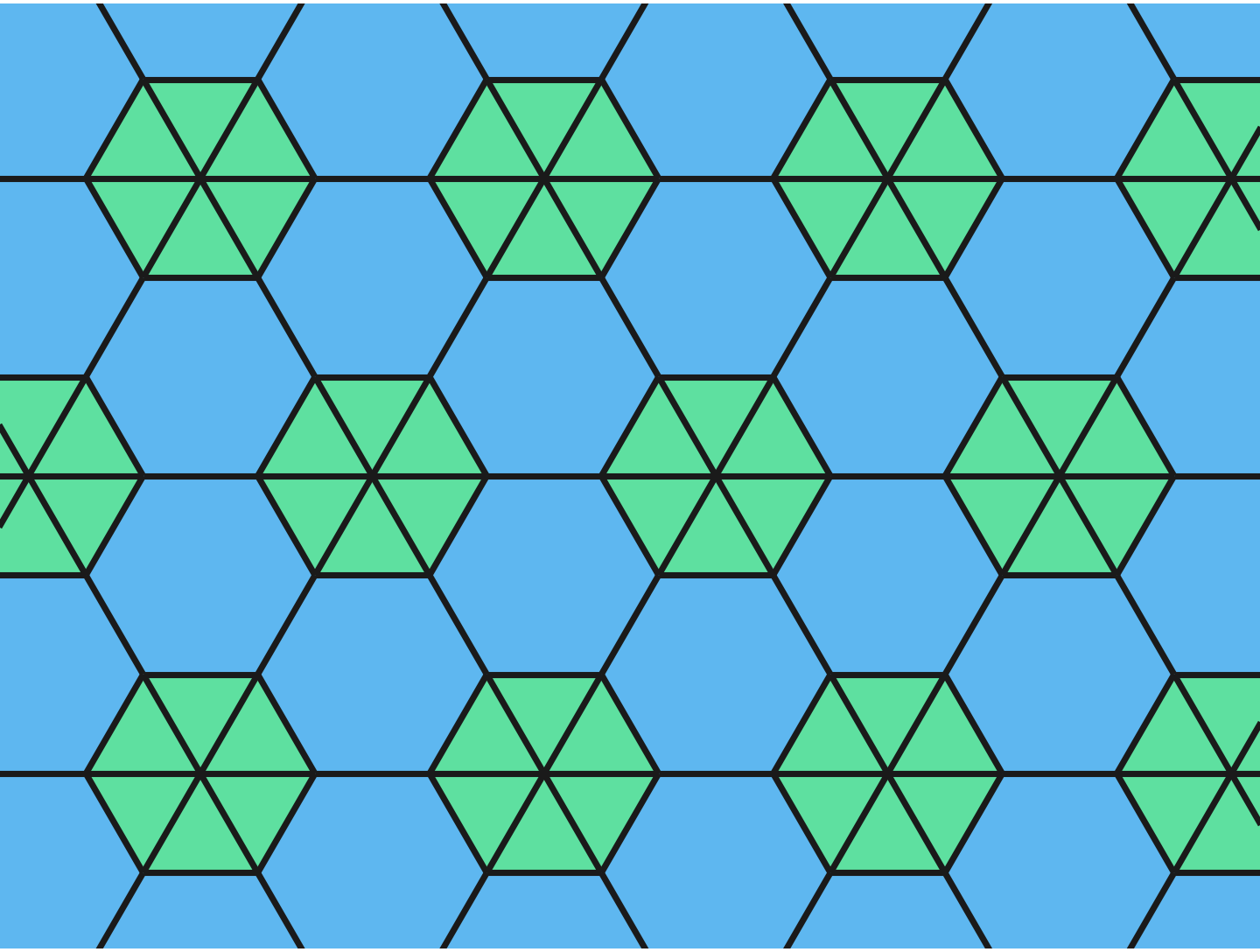} & $2.20203$ & $-1.68015$ & $0.38230$ & $1.32438$ & $-\tfrac{9}{7}$ & $0.099$ & \ref{fig:atlas-t2010} \\
\bottomrule
\end{tabular}
\end{table*}
\begin{table*}[t!]
\centering
\small
\caption{Thermodynamic quantities for the remaining ten $2$-uniform lattices; columns as in Table~\ref{galebach1}.}
\label{galebach2}
\renewcommand{\arraystretch}{1.12}
\setlength{\tabcolsep}{5pt}
\begin{tabular}{>{\centering\arraybackslash}m{1.8cm}
                >{\centering\arraybackslash}m{1.65cm}
                >{\centering\arraybackslash}m{1.45cm}
                >{\centering\arraybackslash}m{1.35cm}
                >{\centering\arraybackslash}m{1.15cm}
                >{\centering\arraybackslash}m{1.3cm}
                >{\centering\arraybackslash}m{1.0cm}
                >{\centering\arraybackslash}m{1.0cm}
                >{\centering\arraybackslash}m{0.7cm}}
\toprule
\makecell{\textbf{Galebach}\\\textbf{name}} &
\makecell{\textbf{Lattice} $\Lambda$} &
\multicolumn{3}{c}{\textbf{Ferromagnetic}} &
\multicolumn{3}{c}{\textbf{Antiferromagnetic}} &
\makecell{\textbf{Fig.}} \\
\cmidrule(lr){3-5} \cmidrule(lr){6-8}
&
&
$T_c$ &
$\varepsilon_c$ &
$A$ &
$T_c^{{\rm{AF}}}$ &
$\varepsilon_0^{{\rm{AF}}}$ &
$s_0^{{\rm{AF}}}$ &
\\
\midrule
\hyperref[atlas:t2011]{t$2011$} & \latimgsmall{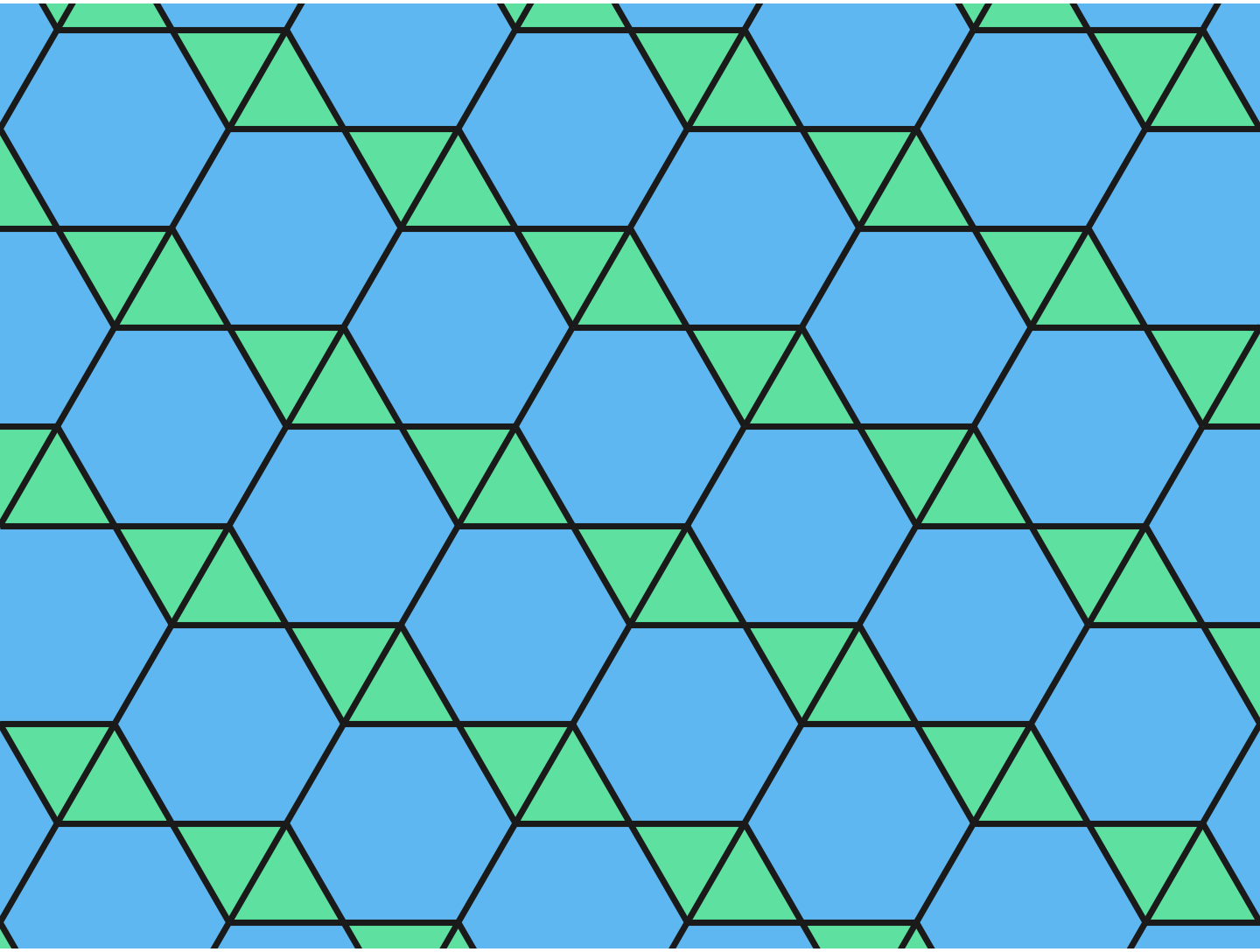} & $2.08372$ & $-1.53222$ & $0.44829$ & $1.18195$ & $-\tfrac{4}{3}$ & $0$ & \ref{fig:atlas-t2011} \\
\hyperref[atlas:t2012]{t$2012$} & \latimgsmall{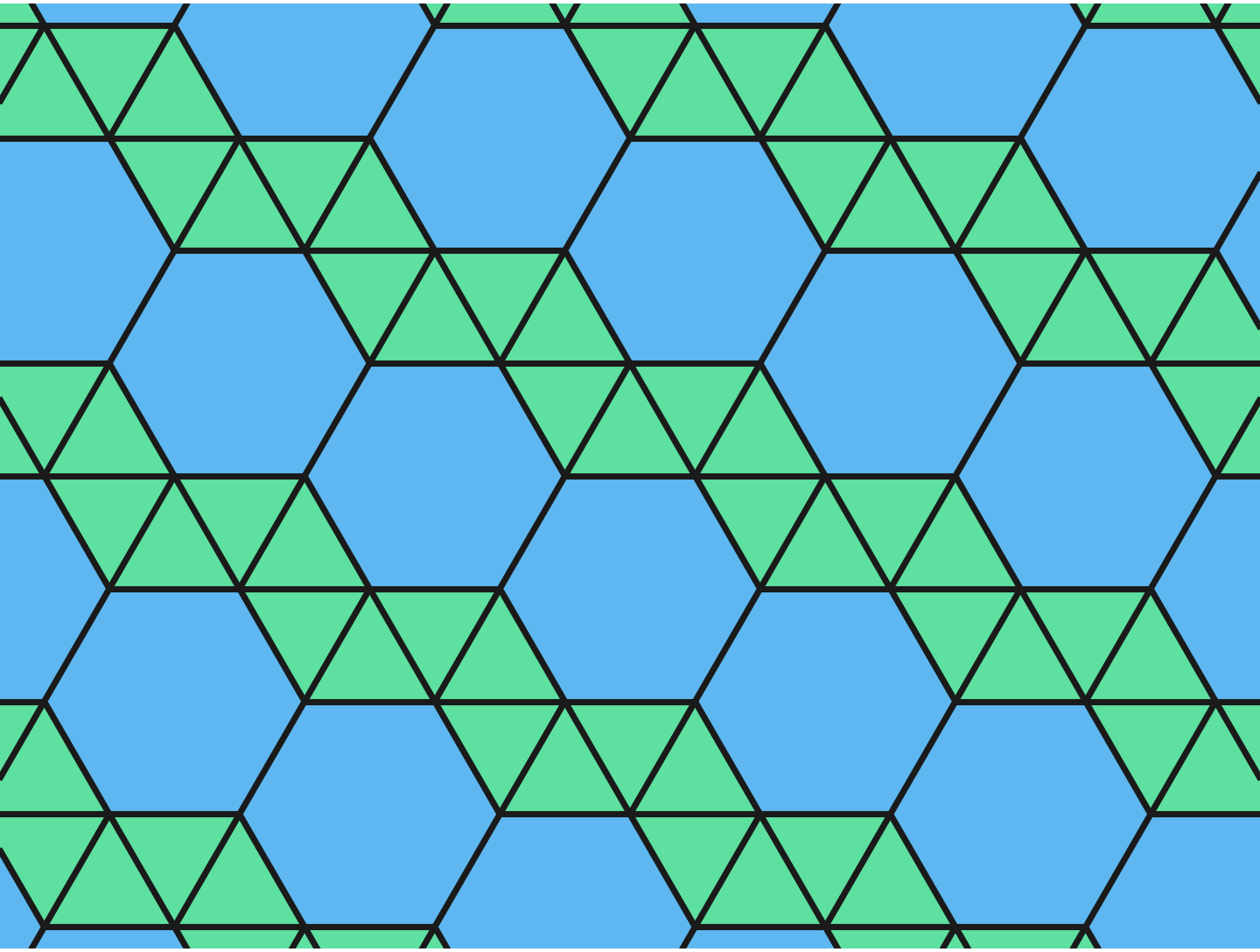} & $2.38312$ & $-1.72217$ & $0.42501$ & none & $-\tfrac{5}{4}$ & $0$ & \ref{fig:atlas-t2012} \\ 
\hyperref[atlas:t2013]{t$2013$} & \latimgsmall{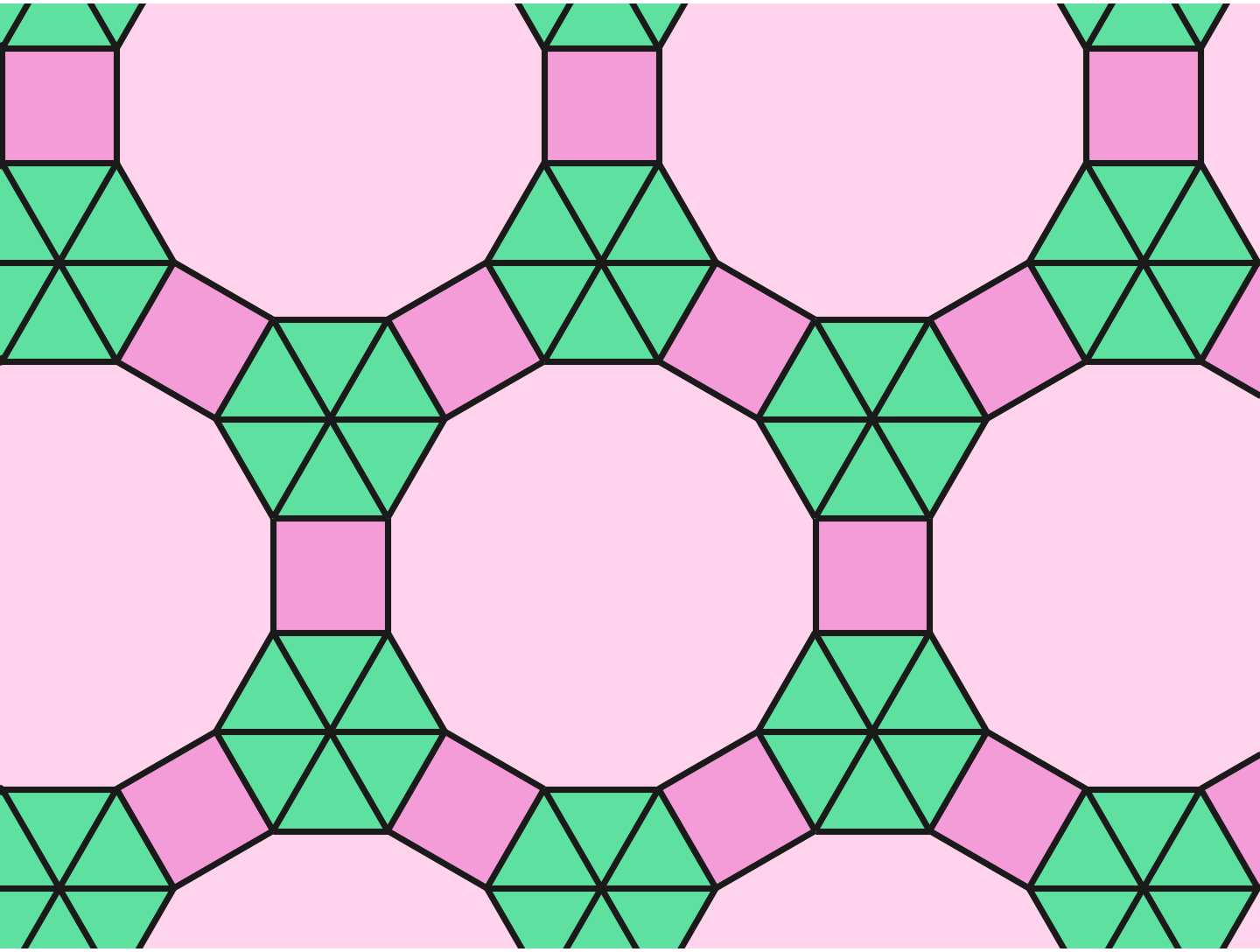} & $2.02213$ & $-1.81291$ & $0.31177$ & $1.21195$ & $-\tfrac{9}{7}$ & $0.099$ & \ref{fig:atlas-t2013} \\
\hyperref[atlas:t2014]{t$2014$} & \latimgsmall{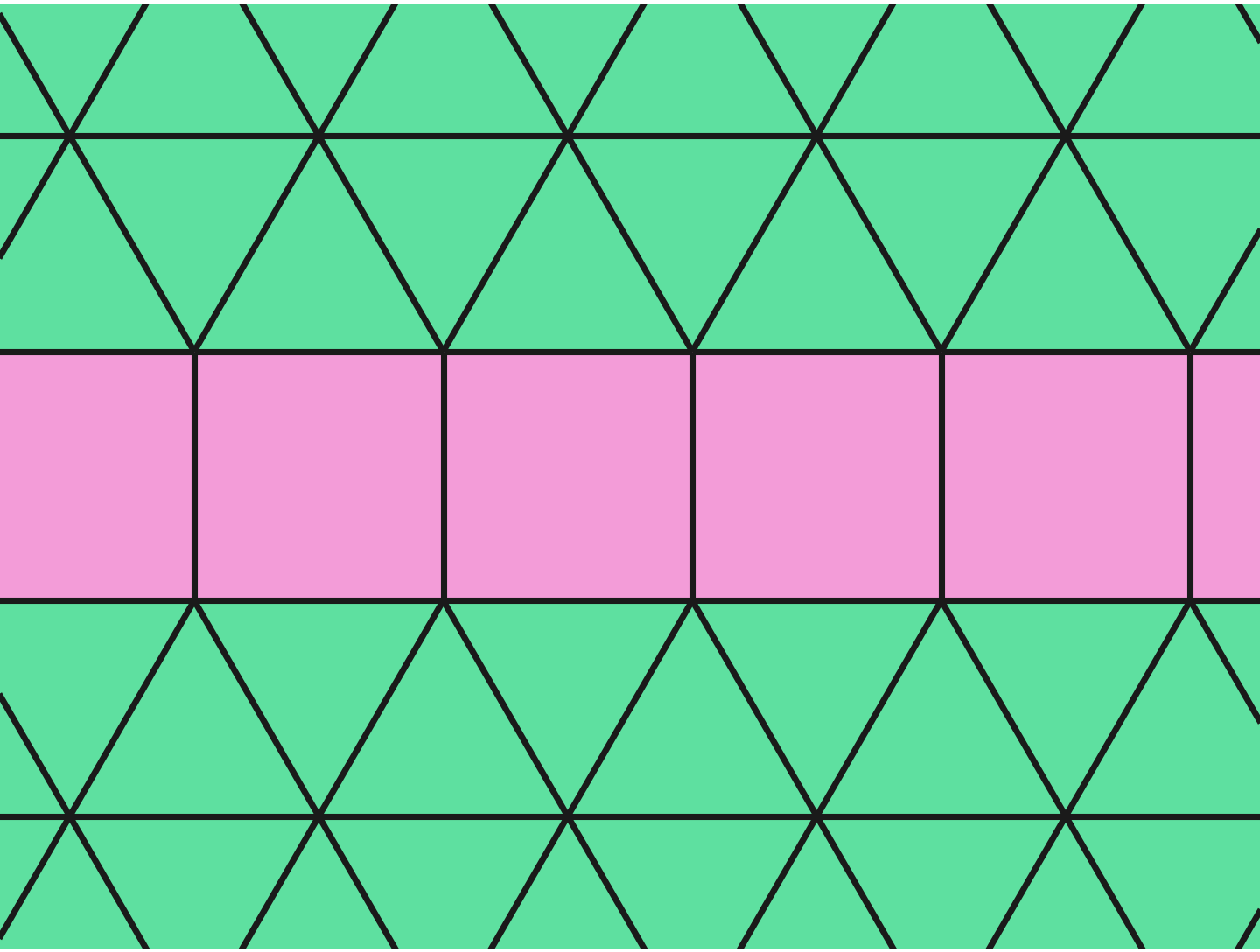} & $3.11852$ & $-1.85849$ & $0.46991$ & none & $-\tfrac{4}{3}$ & $0$ & \ref{fig:atlas-t2014} \\ 
\hyperref[atlas:t2015]{t$2015$} & \latimgsmall{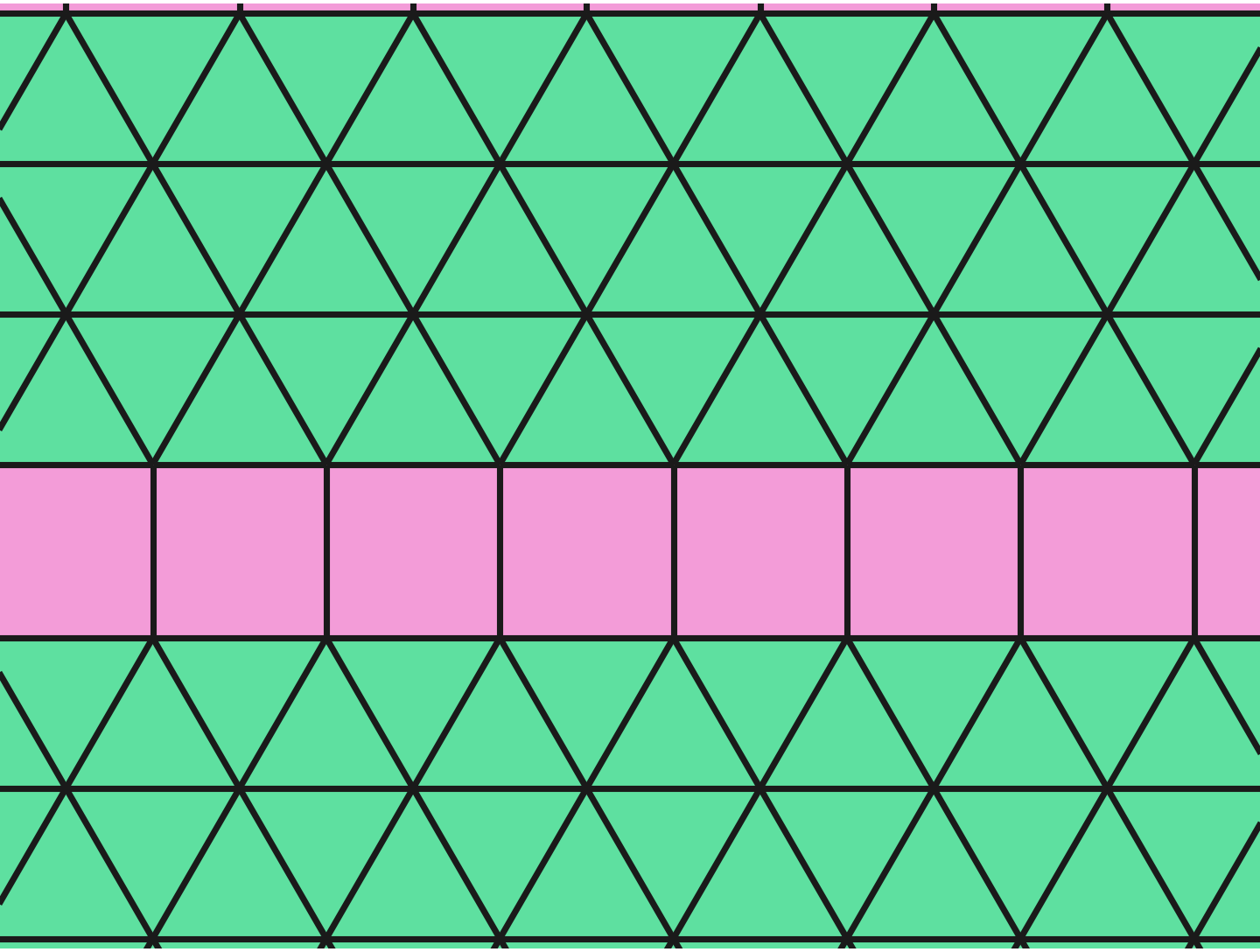} & $3.24161$ & $-1.90432$ & $0.46576$ & none & $-\tfrac{5}{4}$ & $0$ & \ref{fig:atlas-t2015} \\ 
\hyperref[atlas:t2016]{t$2016$} & \latimgsmall{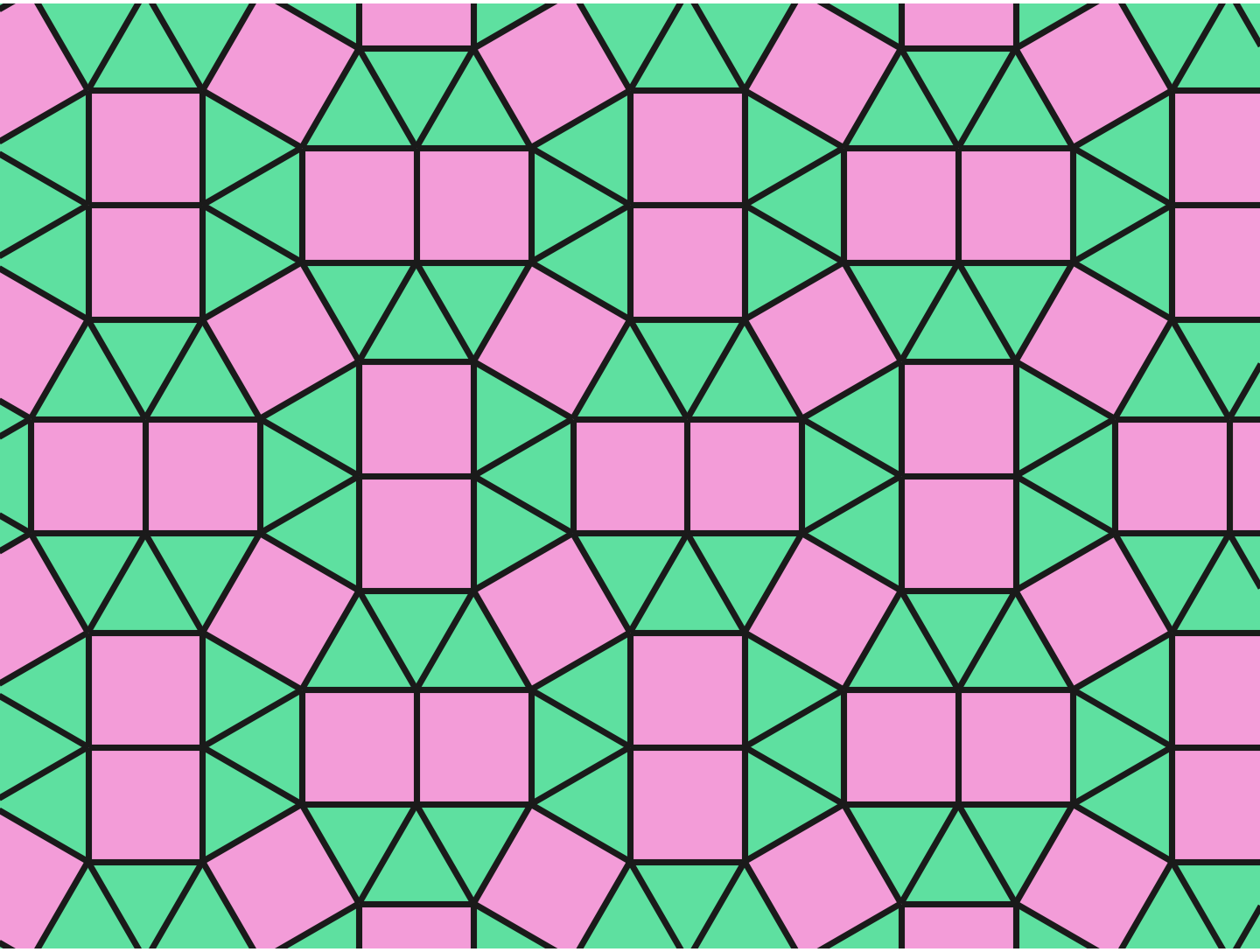} & $2.91611$ & $-1.73276$ & $0.49146$ & none & $-\tfrac{7}{6}$ & $0.185$ & \ref{fig:atlas-t2016} \\
\hyperref[atlas:t2017]{t$2017$} & \latimgsmall{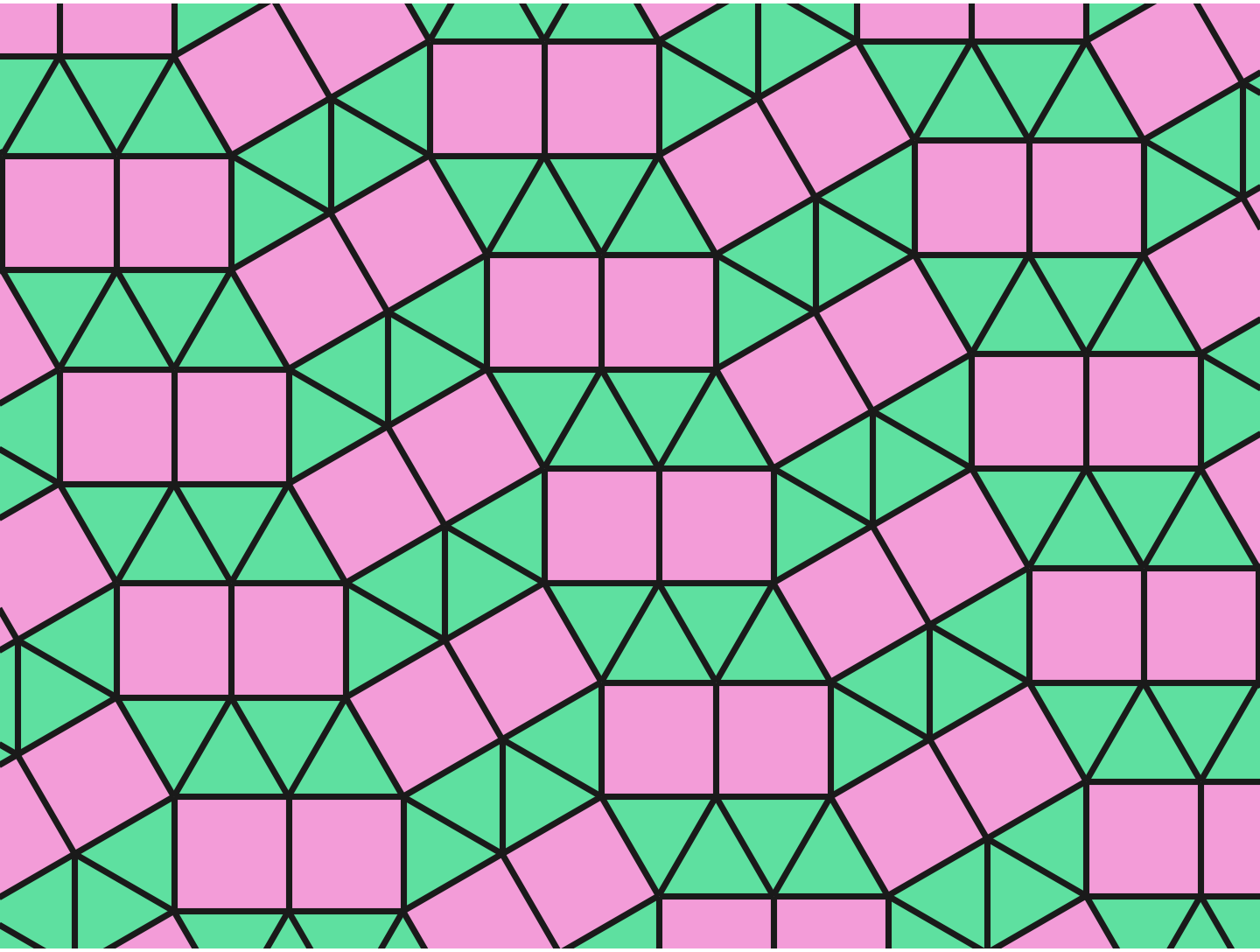} & $2.91019$ & $-1.73742$ & $0.48933$ & $1.19424$ & $-\tfrac{3}{2}$ & $0$ & \ref{fig:atlas-t2017} \\
\hyperref[atlas:t2018]{t$2018$} & \latimgsmall{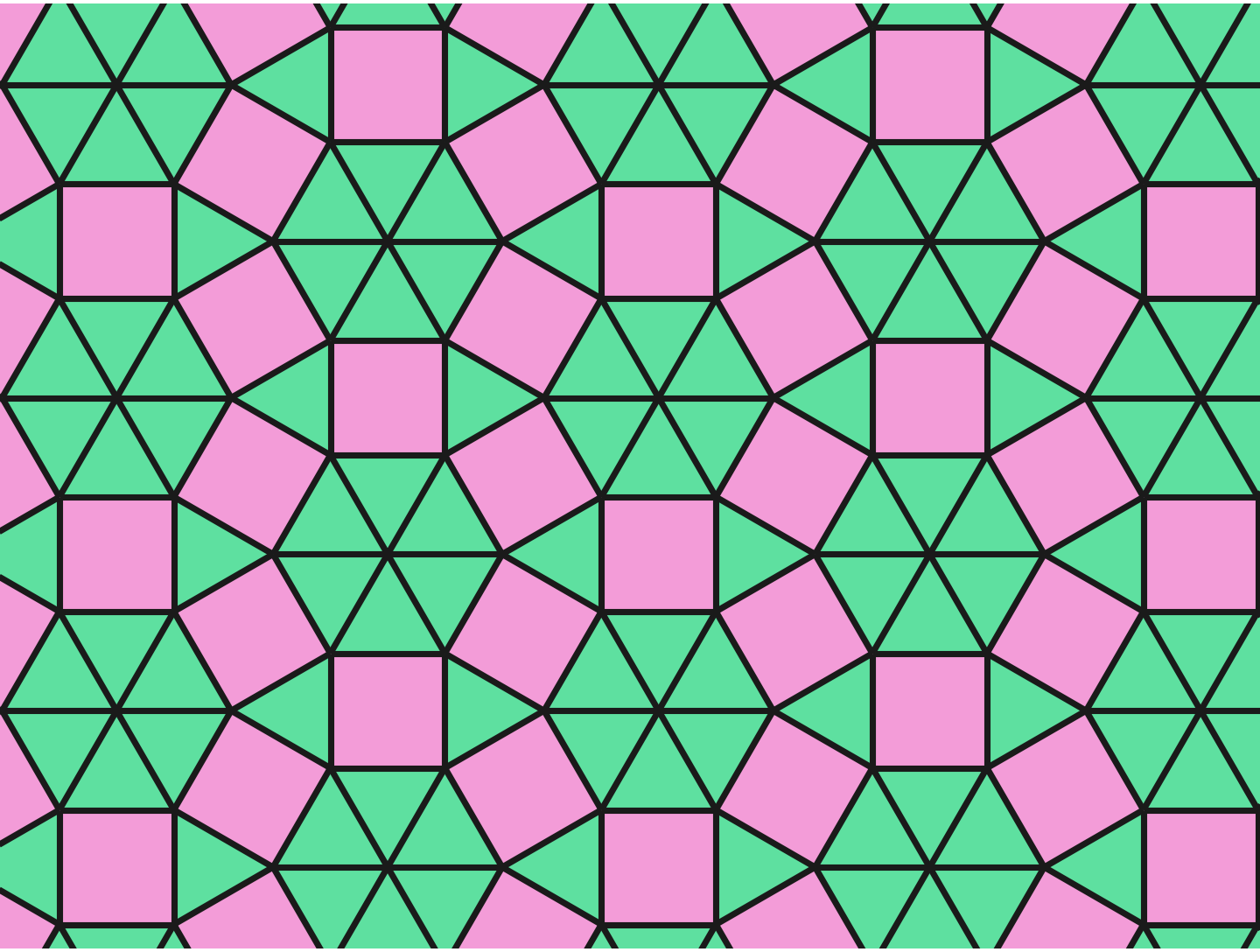} & $3.02267$ & $-1.76955$ & $0.48897$ & none & $-\tfrac{8}{7}$ & $0.145$ & \ref{fig:atlas-t2018} \\
\hyperref[atlas:t2019]{t$2019$} & \latimgsmall{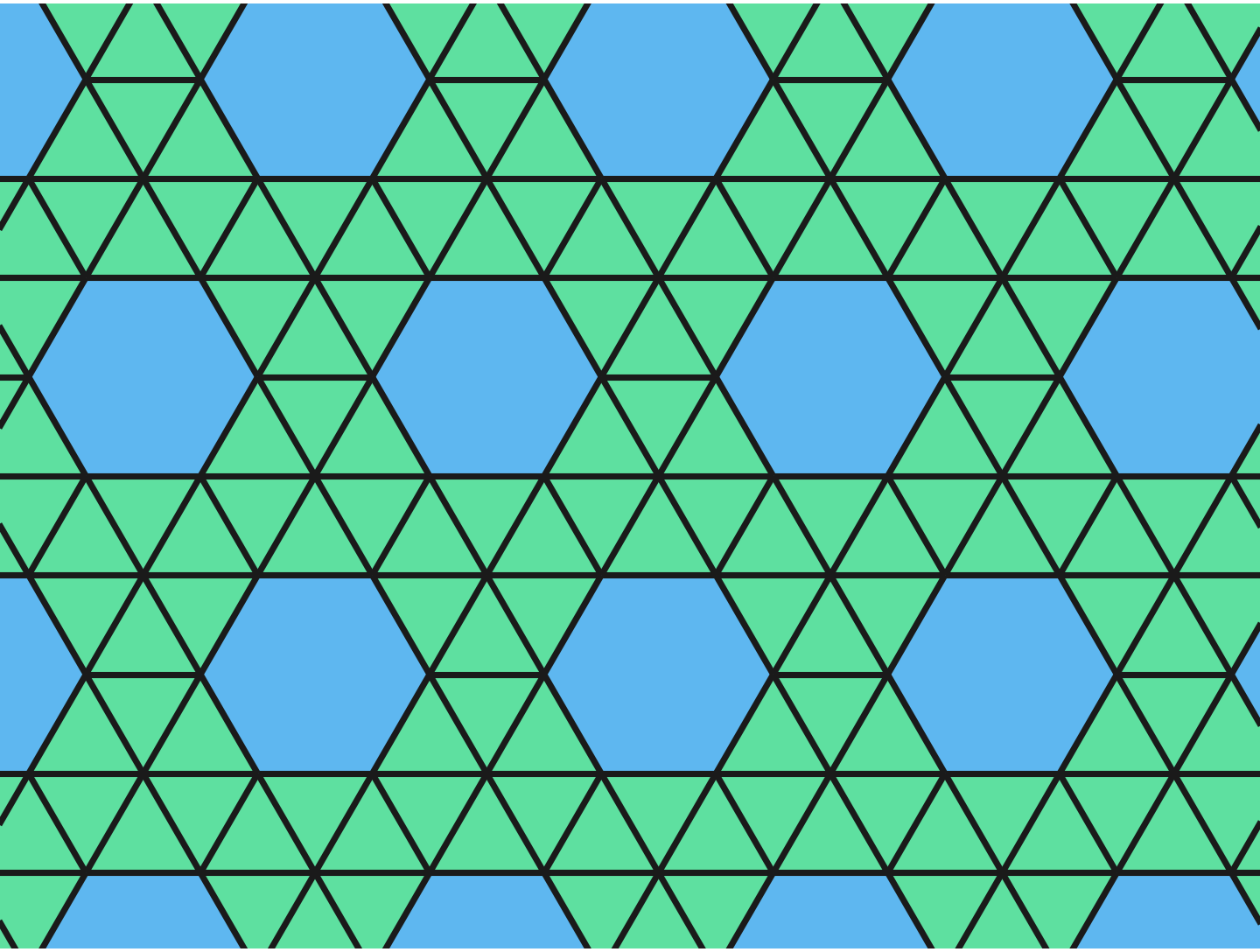} & $2.98574$ & $-1.88660$ & $0.46208$ & $1.07860$ & $-\tfrac{9}{8}$ & $0.173$ & \ref{fig:atlas-t2019} \\
\hyperref[atlas:t2020]{t$2020$} & \latimgsmall{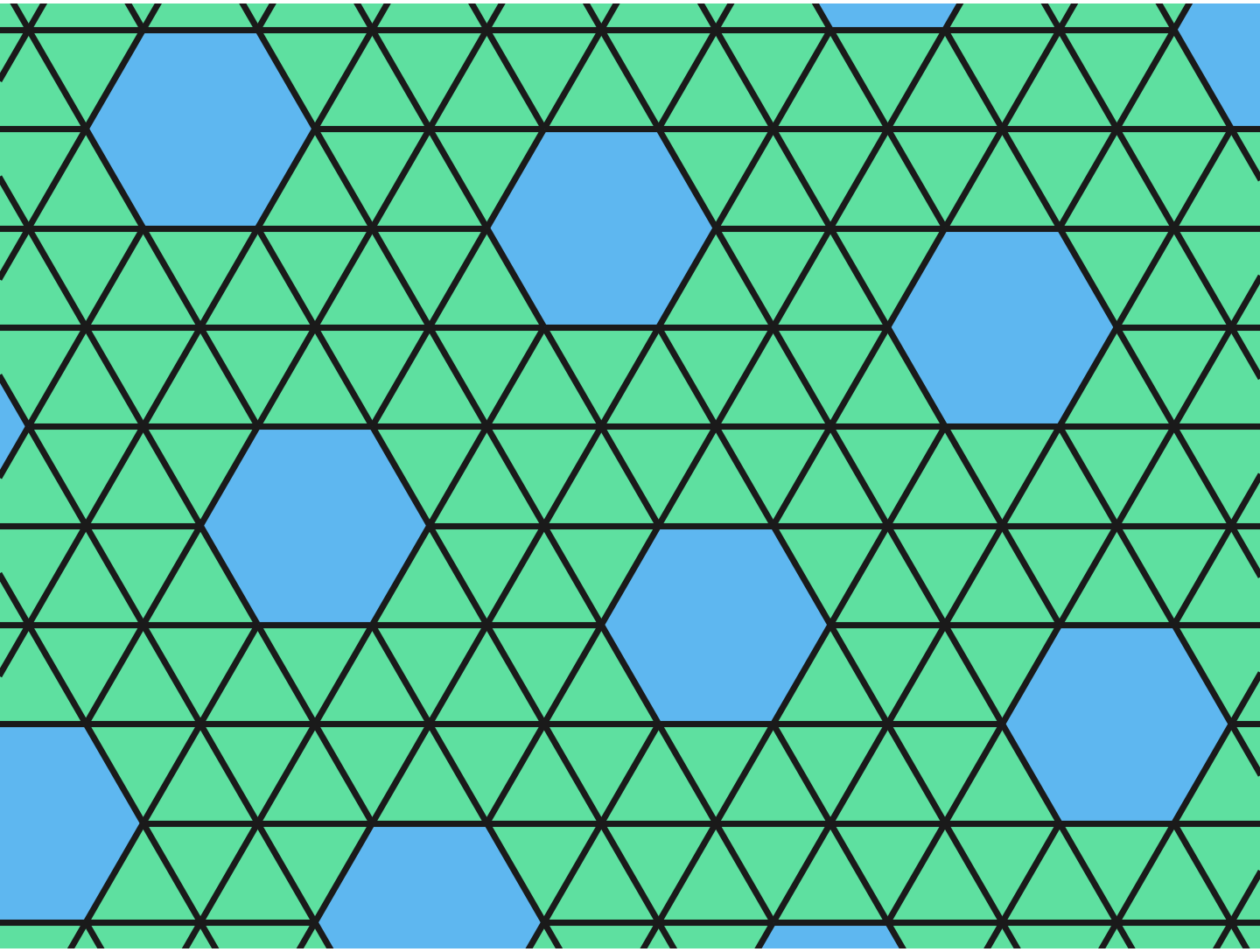} & $3.19352$ & $-1.93872$ & $0.46492$ & none & $-\tfrac{13}{12}$ & $0.173$ & \ref{fig:atlas-t2020} \\
\bottomrule
\end{tabular}
\end{table*}
\begin{table*}[t]
\centering
\small
\caption{Thermodynamic quantities for the five crystallographic pentagonal layer nets treated in this work; columns as in Table~\ref{galebach1}. None of these lattices has a finite-temperature antiferromagnetic transition.}
\label{tab:pentagonal-Tc}
\renewcommand{\arraystretch}{1.12}
\setlength{\tabcolsep}{5pt}
\resizebox{0.85\textwidth}{!}{%
\begin{tabular}{>{\centering\arraybackslash}m{1.8cm}
                >{\centering\arraybackslash}m{1.65cm}
                >{\centering\arraybackslash}m{1.45cm}
                >{\centering\arraybackslash}m{1.35cm}
                >{\centering\arraybackslash}m{1.15cm}
                >{\centering\arraybackslash}m{1.15cm}
                >{\centering\arraybackslash}m{1.0cm}
                >{\centering\arraybackslash}m{1.0cm}
                >{\centering\arraybackslash}m{0.7cm}}
\toprule
\makecell{\textbf{RCSR}\\\textbf{name}} &
\makecell{\textbf{Lattice} $\Lambda$} &
\multicolumn{3}{c}{\textbf{Ferromagnetic}} &
\multicolumn{3}{c}{\textbf{Antiferromagnetic}} &
\makecell{\textbf{Fig.}} \\
\cmidrule(lr){3-5} \cmidrule(lr){6-8}
&
&
$T_c$ &
$\varepsilon_c$ &
$A$ &
$T_c^{{\rm{AF}}}$ &
$\varepsilon_0^{{\rm{AF}}}$ &
$s_0^{{\rm{AF}}}$ &
\\
\midrule
\hyperref[atlas:mcm]{\texttt{mcm} (Cairo)} & \latimgsmall{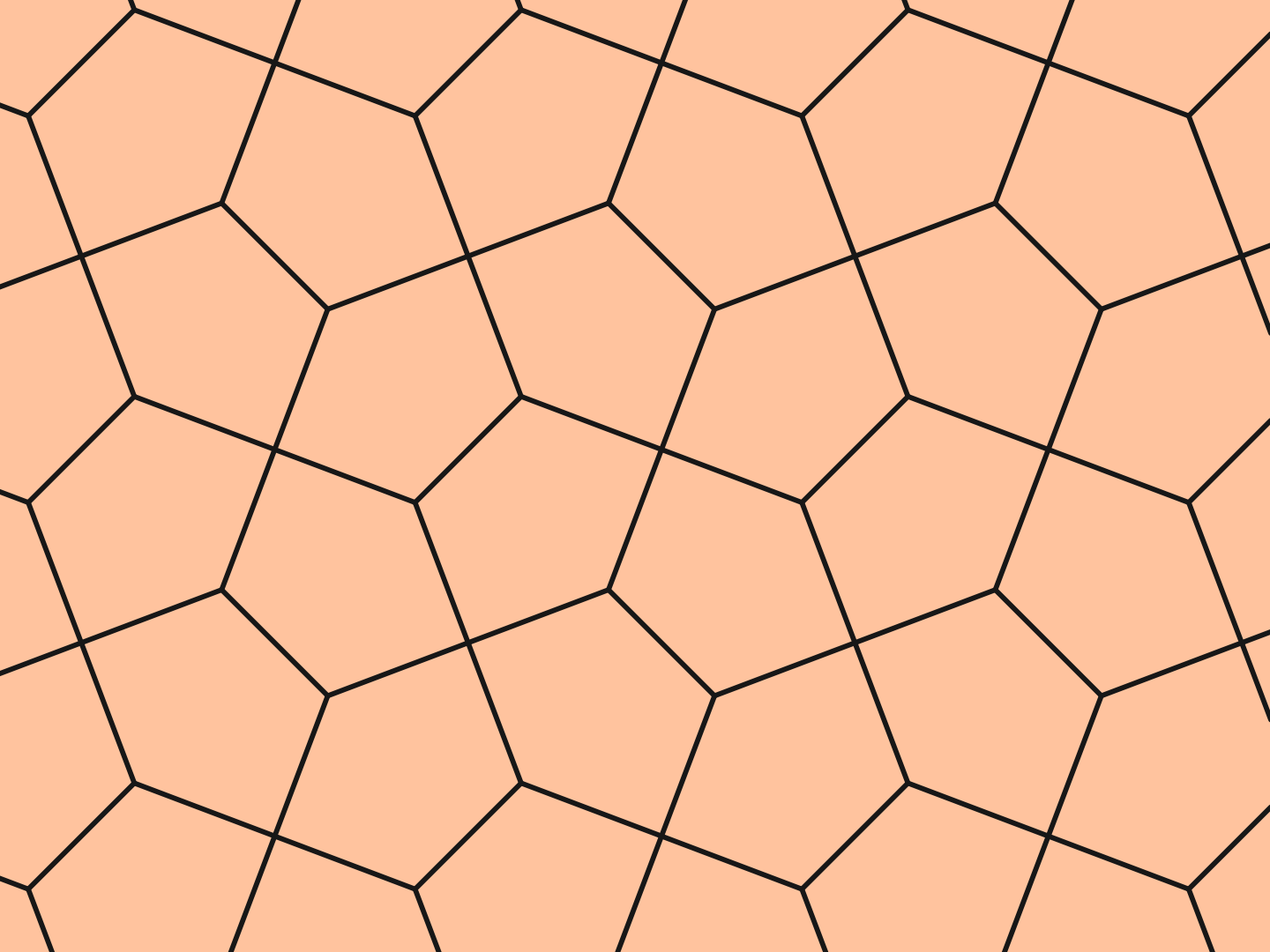} & $1.79917$ & $-1.22274$ & $0.47522$ & none & $-1$ & $0.232$ & \ref{fig:atlas-mcm} \\
\hyperref[atlas:krj-d]{\texttt{krj-d}} & \latimgsmall{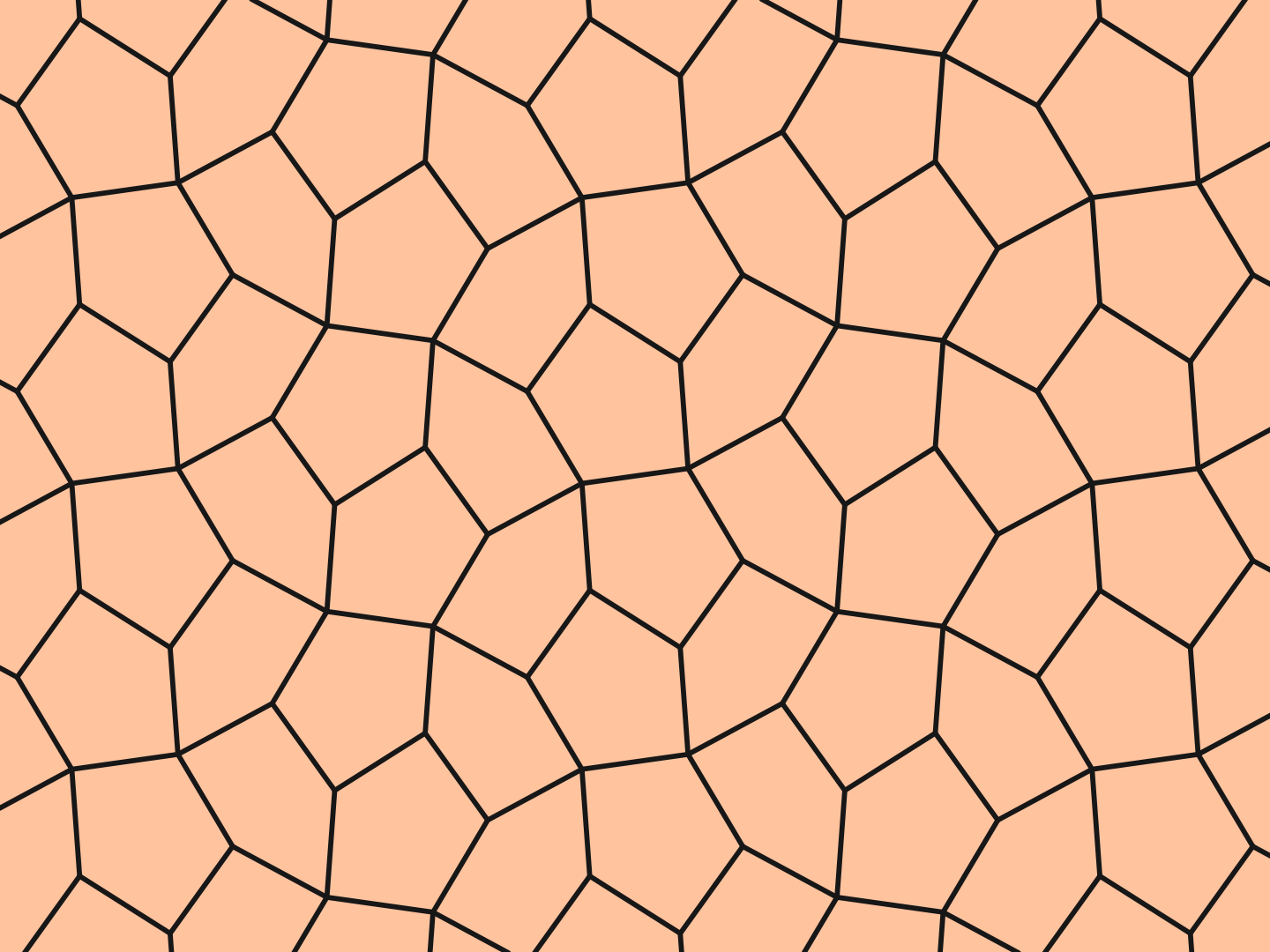} & $1.80746$ & $-1.21581$ & $0.46640$ & none & $-1$ & $0.232$ & \ref{fig:atlas-krj-d} \\
\hyperref[atlas:krv-d]{\texttt{krv-d}} & \latimgsmall{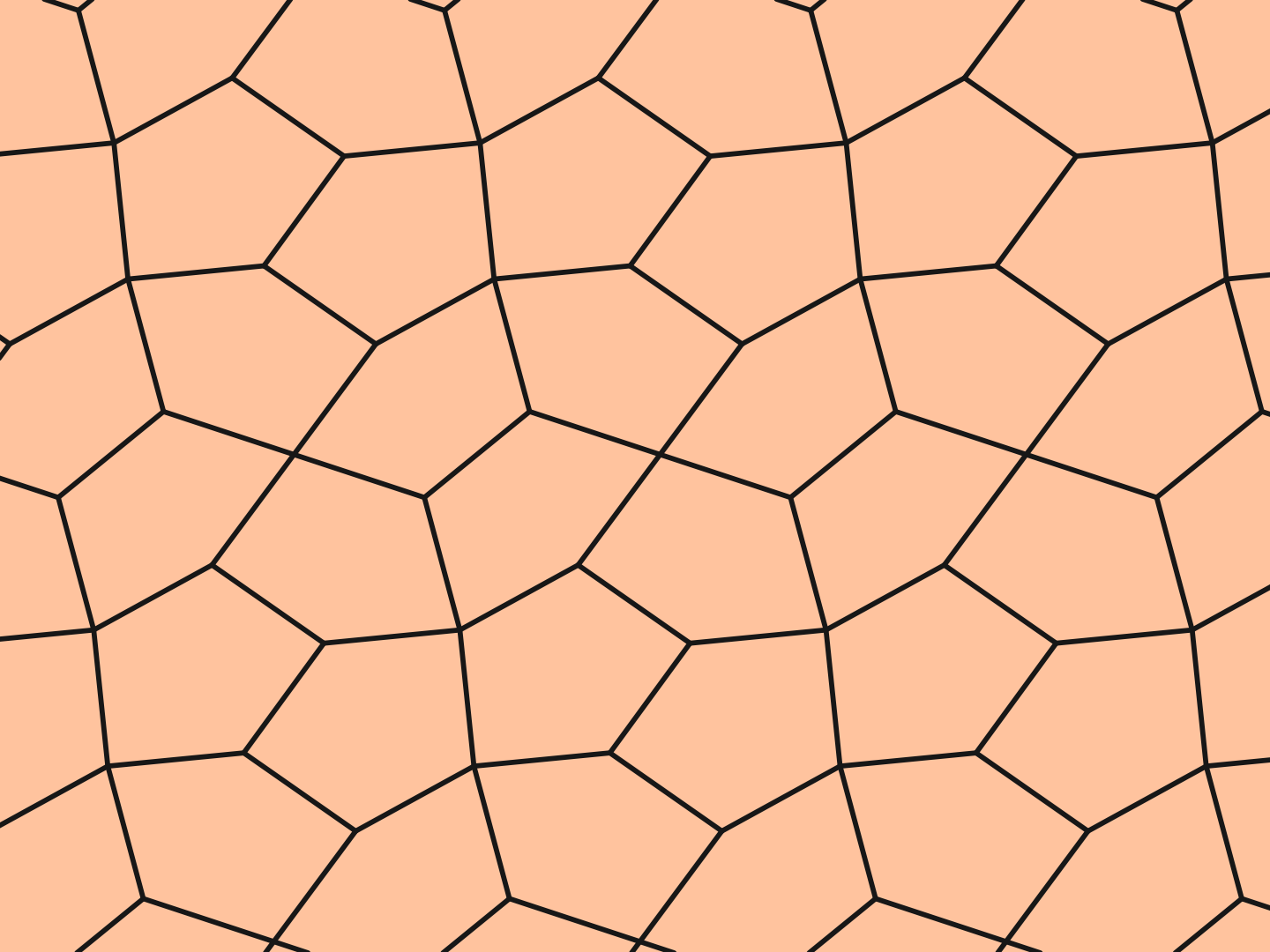} & $1.80477$ & $-1.21802$ & $0.46911$ & none & $-1$ & $0.232$ & \ref{fig:atlas-krv-d} \\
\hyperref[atlas:krw-d]{\texttt{krw-d}} & \latimgsmall{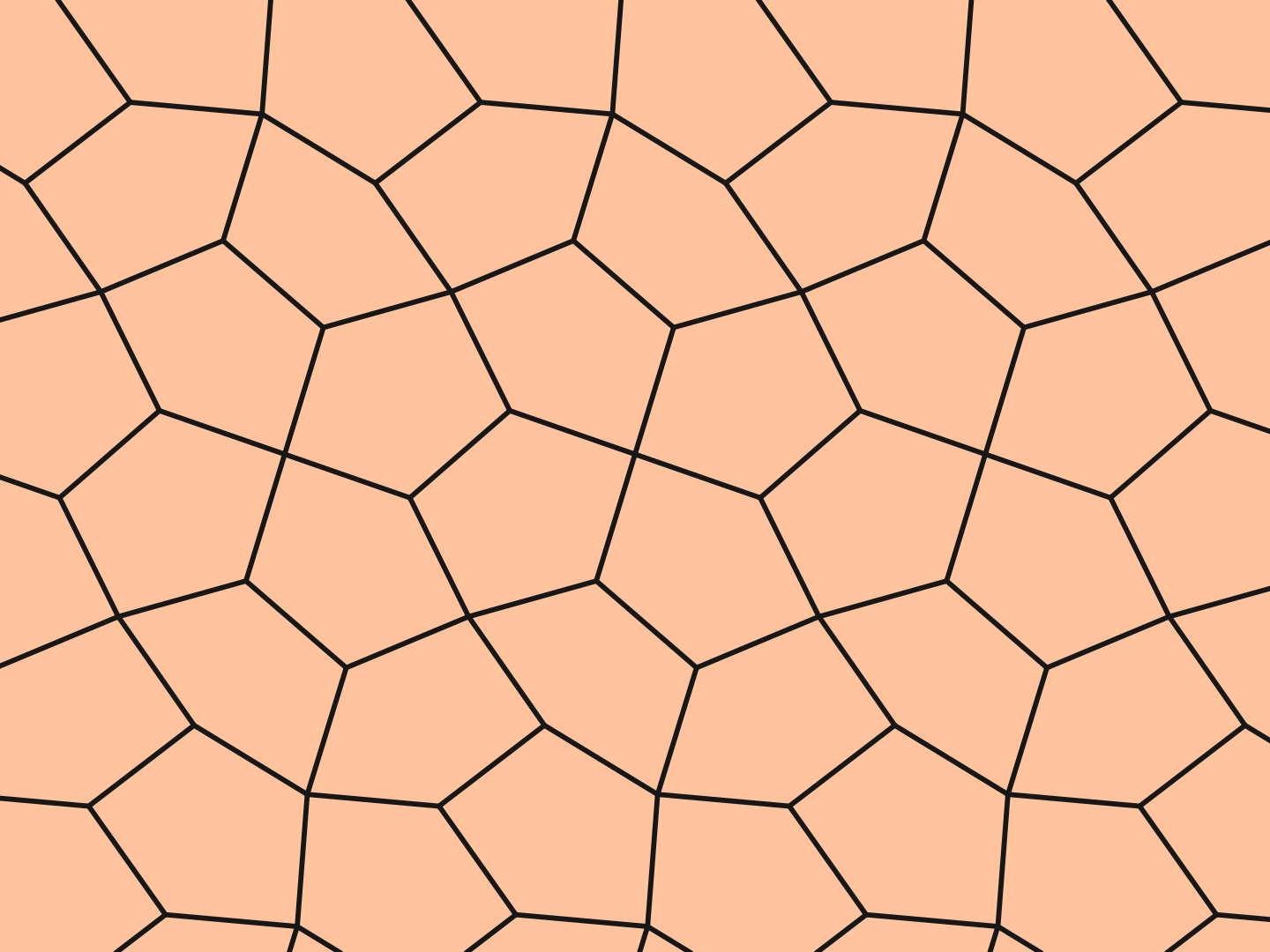} & $1.80250$ & $-1.21993$ & $0.47158$ & none & $-1$ & $0.232$ & \ref{fig:atlas-krw-d} \\
\hyperref[atlas:usm-d]{\texttt{usm-d}} & \latimgsmall{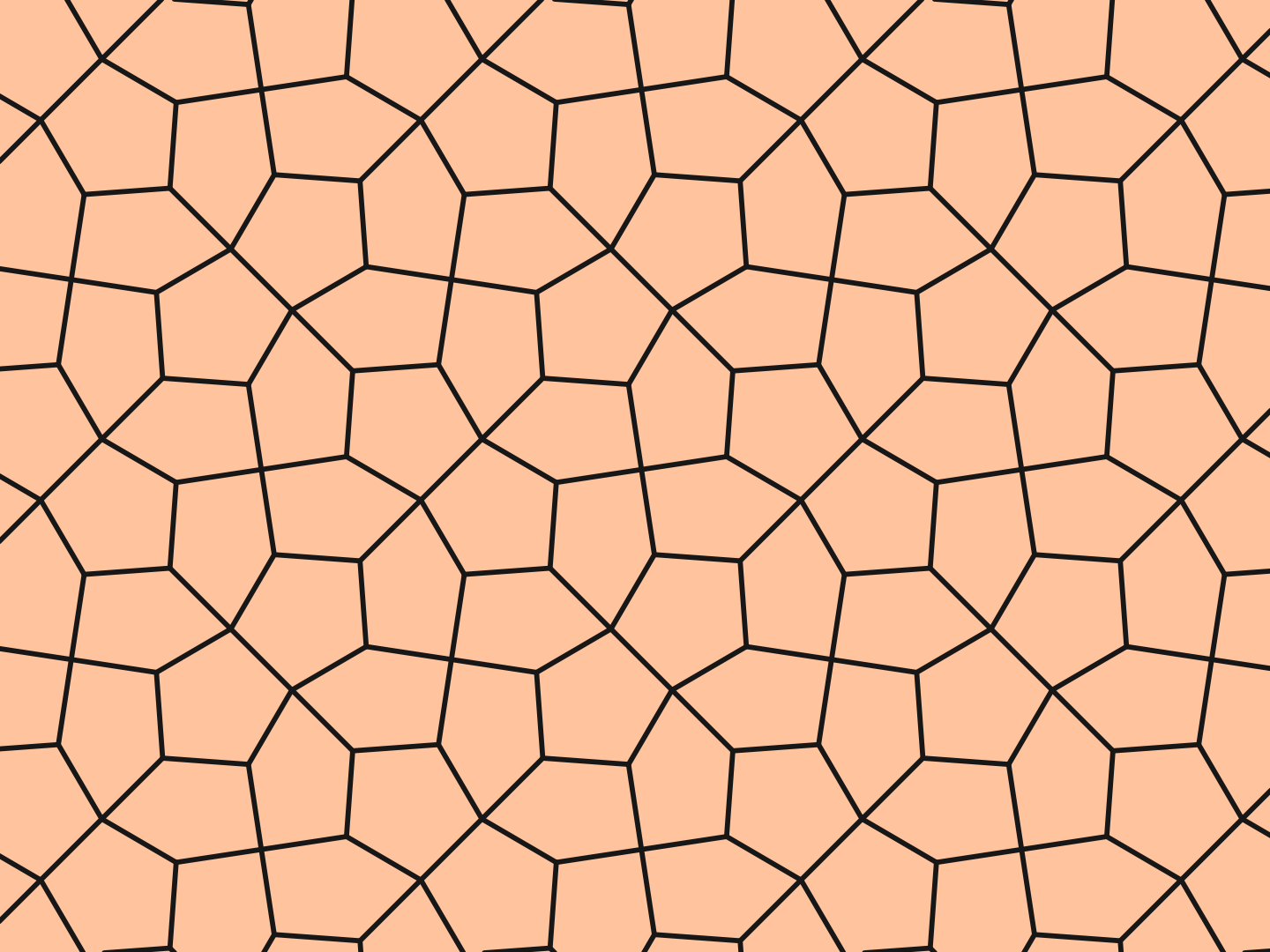} & $1.80439$ & $-1.21840$ & $0.46974$ & none & $-1$ & $0.231$ & \ref{fig:atlas-usm-d} \\
\bottomrule
\end{tabular}
}
\end{table*}

\section{Applications}\label{examples}
We now apply the square and triangular emulators to three finite families of planar lattices: the eleven Archimedean lattices \cite{Grunbaum_Shephard_1989}, the twenty two-uniform Galebach lattices \cite{galebach,chequesoto}, and five crystallographic pentagonal layer nets from the RCSR catalogue \cite{rcsr}.
For each lattice we construct a finite supercell program, insert it into the appropriate parent FV/KW matrix, and analyze both signs of the nearest-neighbour coupling.
%
While in the ferromagnetic cases all lattices behave qualitatively the same, in the antiferromagnetic cases, depending on the lattice topology, one finds different qualitative behaviour: a finite antiferromagnetic transition, no finite temperature transition, or the same critical temperature as the ferromagnetic model in the bipartite cases.

The purpose of this section is not to display the binary masks themselves.
Those are collected in Appendix~\ref{app:atlas}.
Instead, the tables below summarize the structural data of the emulator representation and the thermodynamic quantities obtained from it, for both ferro- and antiferromagnetic couplings and to provide an overall test of the emulator formalism.

\subsection{Archimedean lattices}
\label{sec:results-archimedean}
 
The eleven Archimedean lattices admit programs of supercell size $L \in \{1, 2, 4, 6, 7, 8, 12\}$, the smallest being the triangular lattice itself at $L=1$ and the largest the $(3,12,12)$ tiling at $L=12$. Table~\ref{tab:archimedean} collects the ferromagnetic critical temperature $T_c$, the energy density $\varepsilon_c$ at the transition, and the specific-heat amplitudes $A$ and $B$, together with the antiferromagnetic temperature $T_c^{\mathrm{AF}}$ and the zero-temperature antiferromagnetic energy $\varepsilon_0^{\mathrm{AF}}$ and entropy $s_0^{\mathrm{AF}}$.
All quantities agree with previous studies of this class of lattices \cite{Codello_2010,Yu_2015,Kassan-Ogly_2024,Laurent_2025},
providing a first non-trivial check of the emulator formalism.
 
Three regimes can be read off the antiferromagnetic column. The four bipartite lattices \! --- \!$(4,4,4,4)$, $(6,6,6)$, $(4,6,12)$, and $(4,8,8)$ --- have $T_c^{\mathrm{AF}} = T_c$, since on a bipartite graph the antiferromagnetic problem maps onto the ferromagnetic one by a sublattice spin flip. The six fully frustrated lattices --- $(3,3,3,3,3,3)$, $(3,6,3,6)$, $(3,3,3,3,6)$, $(3,4,6,4)$, $(3,3,3,4,4)$ and $(3,12,12)$ --- show no antiferromagnetic transition: the determinant $P_\Lambda(v, k_1, k_2)$ has no zero for $-1<v<0$ anywhere in the integration domain of \eqref{eq:free-energy} (see the discussion in Section~\ref{sec:thermodynamics}), and the specific heat exhibits only a broad maximum without a critical feature, indicative of a disordered ground state. The snub-square lattice $(3,3,4,3,4)$ is the intermediate case in which only some triangular plaquettes are frustrated, and a distinct antiferromagnetic transition appears at $T_c^{\mathrm{AF}} =1.2619...$, well below the ferromagnetic value.
Even though some lattices admit closed-form analytic determinants, all the results presented here are obtained numerically from the trace identities of Section~\ref{sec:thermodynamics}.
The supercell programs and the corresponding thermodynamic plots are collected in Appendix~\ref{app:atlas-archimedean}.

\subsection{2-uniform lattices}
\label{sec:results-galebach}
 
The twenty $2$-uniform tilings from the Galebach catalogue \cite{galebach,chequesoto, SOTOSANCHEZ202169} extend the Archimedean family by allowing two distinct vertex environments per unit cell. They admit triangular-emulator programs of supercell size $L \in \{3, 4, 5, 6, 7, 12, 13, 15, 20\}$, with the larger sizes reflecting the larger unit cells of these tilings. Tables~\ref{galebach1} and~\ref{galebach2} collect the ferromagnetic quantities $T_c$, $\varepsilon_c$ and $A$, together with the antiferromagnetic critical temperature $T_c^{\rm AF}$, ground-state energy $\varepsilon_0^{\rm AF}$ and residual entropy $s_0^{\rm AF}$; the remaining quantities (the energy density $\varepsilon_c^{\rm AF}$ at the antiferromagnetic transition and the dual critical temperature $T_c^{\rm D}$), together with the full thermodynamic curves, are reported in the corresponding atlas figures of Appendix~\ref{app:atlas-galebach}, referenced in the last column of the tables. The two-table split is for layout only and carries no taxonomic distinction between the first ten and the second ten lattices.
 
Of the three antiferromagnetic regimes identified for the Archimedean family --- bipartite ($T_c^{\mathrm{AF}} = T_c$), fully frustrated ($T_c^{\mathrm{AF}}$ absent), and partially frustrated ($T_c^{\mathrm{AF}} \neq T_c$ but finite) --- only the last two occur in the $2$-uniform family, since all twenty $2$-uniform tilings contain triangular faces and none of them is bipartite. Fifteen lattices show no antiferromagnetic transition, while five (t2010, t2011, t2013, t2017, and t2019) have a finite $0 < T_c^{\mathrm{AF}} < T_c$; for t2011 the transition is found at non-zero momentum, as discussed in Section~\ref{sec:thermodynamics}; to our best knowledge, this is the first example of such behaviour for a frustrated Ising antiferromagnet with nearest-neighbour couplings on a planar tiling. Another qualitatively new behaviour emerges for two-dimensional lattices: the lattices t2010, t2013 and t2019 have $0 < T_c^{\mathrm{AF}} < T_c$ together with a non-zero residual entropy $s_0^{\rm AF} > 0$, i.e.\ a finite-temperature antiferromagnetic transition coexisting with a macroscopically degenerate ground state.  To our knowledge, the present work provides the first systematic tabulation of these thermodynamic data for the $k=2$ uniform family, extending the tabulation of the critical temperatures presented in \cite{Portillo_2025, Joseph_2026} which are here fully confirmed.
The supercell programs and thermodynamic plots are collected in Appendix~\ref{app:atlas-galebach}.

\subsection{Pentagonal tilings}
Crystallographic pentagonal tilings provide a class of targets unreachable in practice by tailored FV/KW solutions: their unit cells contain several inequivalent vertices, their plane groups are nontrivial, and their topology is not built from a small number of regular polygons.
The RCSR layer-net catalog~\cite{rcsr} lists ten such tilings with all bonds of equal length; we have constructed triangular-emulator programs for five of them.
The pentagonal tilings have, to our knowledge, no prior systematic FV/KW treatment beyond the single isolated case of the Cairo lattice \cite{Urumov_2002}, dual to the snub square.
The emulator framework handles these lattices by exactly the same machinery used for the Archimedean and $2$-uniform families: one parent matrix, one program-substitution mechanism, one thermodynamic pipeline.
Only the fully frustrated antiferromagnetic regime ($T_c^{\mathrm{AF}}$ absent) is encountered. 
The pentagonal data in Table \ref{tab:pentagonal-Tc} constitute the most direct evidence in this paper that the triangular emulator reaches beyond the targets to which tailored solutions are conventionally applied.

\section{Range of applicability of the emulator framework}
\label{sec:targets}

The emulator framework introduced in this paper is useful well beyond the specific applications shown here for regular lattice classes.
Its power and strength lie in the fact that it can be applied to general classes of non-homogeneous lattices, such as fractal structures \cite{Codello:2015bia, Zinati:2025npw}, and -- more importantly -- to disordered systems as outlined in \cite{Zinati_Gori_Codello_2026}.
In the following, we review the basic ideas on how to use the Ising emulator to treat these problems (fractals and quenched disorder) and outline future directions of application.

\subsection{Deterministic fractals}
\label{sec:target-fractal}
\label{sec:fractal-targets}

The only periodicity the emulator framework requires is that of the supercell itself; the target lattice need not be periodic in the usual crystallographic sense.
This makes deterministic fractal constructions accessible through their finite-generation periodic approximants. 
For the generalized Sierpinski-carpet family $SC_k(a,b)$, one starts from an $a\times a$ block decomposition, removes a prescribed $b\times b$ subset of sub-blocks, and recursively applies the same rule on each surviving block.
At any fixed generation $k$, the resulting program is finite and can be used directly on the supercell of the square emulator, with supercell size $L = L_k = a^k$ in the notation of Section~\ref{sec:emulator-principle}. Increasing $k$ thus refines the program while preserving the finite-supercell structure the emulator requires.

As a concrete example, we consider the central-hole family $SC_k(3,1)$, the first few iterations are the following (Fig.~\ref{fig:beyond-crystalline}a):
The first nontrivial step, $SC_1(3,1)$, has supercell size $L=3$, with a single site removed from the centre; the corresponding site program is
\begin{equation}
w_{SC_1(3,1)} \;=\; \begin{pmatrix} 1 & 1 & 1 \\ 1 & 0 & 1 \\ 1 & 1 & 1 \end{pmatrix}\,.
\label{eq:program-SC1}
\end{equation}
The same finite approximant can equivalently be described by an explicit bond program, but in the present framework it is more natural to regard $SC_k(a,b)$ as a site-programmed supercell.
Both yield the same thermodynamics, as discussed in Section~\ref{sec:square-emulator}, but the site-programmed description is more compact.
At the next iterate $SC_2(3,1)$, the same rule is applied recursively to the eight surviving $3\times3$ blocks, producing a supercell of size $L=3^2=9$ with the site program~\eqref{eq:program-SC2}, which makes the recursive structure visible directly at the level of the emulator supercell.
\vspace{5pt}
\begin{equation}
w_{SC_2(3,1)} \;=\;
\begin{pmatrix}
1&1&1&1&1&1&1&1&1\\
1&0&1&1&0&1&1&0&1\\
1&1&1&1&1&1&1&1&1\\
1&1&1&0&0&0&1&1&1\\
1&0&1&0&0&0&1&0&1\\
1&1&1&0&0&0&1&1&1\\
1&1&1&1&1&1&1&1&1\\
1&0&1&1&0&1&1&0&1\\
1&1&1&1&1&1&1&1&1
\end{pmatrix},
\label{eq:program-SC2}
\end{equation}
\vspace{5pt}
The same logic extends directly to higher iterates: at fixed $k$, the finite-generation carpet defines a periodic supercell, the emulator treats that supercell exactly, and the infinite fractal is approached through the sequence $k=1,2,3,\dots$
This opens the framework to the broad class of finitely generated fractal lattices.
A dedicated study of these periodic fractal approximants for the Sierpinski $SC_k(a,b)$ family started in \cite{Codello:2015bia} and further developed in \cite{Zinati:2025npw}, enabling computations up to generation $k=10$ for $SC_k(3,1)$ and yielding the extrapolated estimate $T_c^{(3,1)} = 1.4782927(26)$; a 
similar analysis was performed on other members of the $SC_k(a,b)$ family.

\vspace{5pt}
\noindent\textbf{(a)} Finite-generation Sierpinski-carpet approximants $SC_k(3,1)$, $k=1,\dots,4$.\par
\vspace{3pt}
\fourpanelstrip
{\sierpinski[scale=1.5]{1}}
{\sierpinski[scale=1.5]{2}}
{\sierpinski[scale=1.5]{3}}
{\sierpinski[scale=1.5]{4}}
{$k=1$}
{$k=2$}
{$k=3$}
{$k=4$}
\vspace{10pt}

\noindent\textbf{(b)} Site-disorder realizations on a square supercell, $L=6$.\par
\vspace{3pt}
\fourpanelstrip
{\sitemask[scale=1.5]{6}{{1,0,0,1,0,0},{0,1,0,1,0,1},{0,0,1,0,0,0},{0,1,0,0,1,1},{0,1,0,1,1,1},{0,0,1,1,1,1}}}
{\sitemask[scale=1.5]{6}{{0,1,1,1,0,1},{1,1,0,1,1,1},{1,1,1,0,1,1},{1,1,0,1,1,0},{1,0,0,0,1,0},{1,0,0,1,0,0}}}
{\sitemask[scale=1.5]{6}{{1,1,1,0,1,1},{1,0,1,1,1,1},{1,1,1,1,1,1},{1,1,1,1,0,1},{1,1,1,0,1,1},{1,1,1,1,1,0}}}
{\sitemask[scale=1.5]{6}{{1,1,0,1,1,1},{1,1,1,1,1,1},{1,1,1,1,1,1},{1,1,1,0,1,1},{1,1,1,1,1,1},{1,1,1,1,1,1}}}
{$p=0.59$}
{$p=0.70$}
{$p=0.85$}
{$p=0.95$}
\vspace{10pt}

\noindent\textbf{(c)} Bond-disorder realizations on the snub-square $(3,3,4,3,4)$ supercell.\par
\vspace{3pt}
\begin{center}
\input{figures/random_t1009_supercell_strip.tex}
\end{center}
\vspace{5pt}

\captionof{figure}{Extensions beyond crystalline order. (a) Finite-generation approximants of the central-hole Sierpinski-carpet family $SC_k(3,1)$, used as periodic supercells of the square emulator with supercell size $L_k=3^k$. (b) Typical site-disorder realizations on a square supercell of size $L=6$, at occupation probabilities $p=0.59$, $0.70$, $0.85$, and $0.95$ (black tiles mark occupied sites). (c) Typical bond-disorder realizations on the snub-square $(3,3,4,3,4)$ supercell, at the same bond-occupation probabilities (left to right).}
\label{fig:beyond-crystalline}
\vspace{5pt}

\subsection{Quenched disorder}
\label{sec:target-random}

The targets considered so far are all deterministic. The emulator framework also accommodates quenched random site or bond dilution at no structural cost: each entry of the program (the site program $w$, or each bond program $w^{(a)}$) is drawn as an independent Bernoulli random variable with parameter $p$, the consistency conditions~\eqref{eq:triangular-consistency} are enforced on the resulting bond masks, and the thermodynamic pipeline of Section~\ref{sec:thermodynamics} is run on each realization.
Disorder-averaged observables are recovered by averaging over different realizations, with the supercell size $L$ playing the role of the disorder correlation scale.
As a minimal illustration, Fig.~\ref{fig:beyond-crystalline}b shows four typical site-disorder realizations on a square supercell of size $L = 6$ at occupation probabilities $p = 0.59$, $0.70$, $0.85$, and $0.95$.
Neither the supercell-program data structure nor the thermodynamic pipeline is modified; only the input data is.
This construction has been used to treat the site-diluted Ising model on the square lattice in~\cite{Zinati_Gori_Codello_2026}. There, the full phase boundary $T_c(p)$ is resolved from the pure-Ising point at $p = 1$ down to the percolation limit $T_c(p_c) = 0$ at $p_c = 0.592746...$. The technical content of that work is the extension of the combinatorial solution to randomized supercells; conceptually, it is one instance of the framework presented here.
 
The instance treated in~\cite{Zinati_Gori_Codello_2026} is one of many. The emulator framework is host- and disorder-agnostic: any of the lattices catalogued in Section~\ref{examples} can be randomly diluted by the same one-line modification of its program. Three directions stand out as immediately accessible and physically interesting:
(1) bond-diluted Ising replaces the per-site Bernoulli with a per-bond Bernoulli on each of the three directed mask families $w^{(1)}, w^{(2)}, w^{(3)}$; on the square host this maps to the bond-percolation problem with $p_c = \tfrac{1}{2}$, on the triangular host to $p_c = 2\sin\tfrac{\pi}{18}$, and on each of the Archimedean and $k$-uniform targets to a distinct non-universal critical-line problem. Four typical bond-disorder realizations on the snub-square  $(3,3,4,3,4)$ supercell are shown in Fig.~\ref{fig:beyond-crystalline}c at the same dilution values;
(2) site-diluted Ising models on intrinsically triangular targets --- kagome, $(3,4,6,4)$, snub-square --- are still an open problem: the percolation thresholds of these lattices are known numerically \cite{Suding_Ziff_1999, Ziff_Suding_1997}, but the diluted Ising phase boundary has not been resolved exactly, whereas the emulator framework applies to them with the same machinery used for the square;
(3) allowing signed Bernoulli entries $J_{ij} \in \{+J, -J\}$ at probability $p$ extends the framework to the random-bond Ising model and, along the Nishimori line $\tanh \beta J = 2p - 1$, to gauge-invariant disorder; this connects the framework to a long line of work on the Nishimori point \cite{Nishimori_1981}  and the random-bond Ising universality class, opening access to lattices beyond the square where the analytic tools have historically been sparse.
 
These extensions share the structural feature that no modification of the construction is required: the parent matrix is unchanged as is the thermodynamic pipeline of Section \ref{sec:thermodynamics}. What changes is only the binary input data, and the outer average over realizations. This is the operational content of the emulator picture in its broadest form: one parent matrix, many programs --- deterministic, fractal, and random alike --- across the full atlas of flat unit-edge-length planar targets.

\section{Conclusions and Outlook}\vspace{-0.2cm}
We have presented a unified emulator-based approach to Ising models on
two-dimensional flat lattices, based on the fact 
that a broad family of planar lattices can be embedded as site- or bond-depleted
supercells of a universal host lattice. In this sense, the square
and triangular constructions behave as an {\it Ising processing unit (IPU)}: once
the emulator is fixed, changing the lattice amounts just to changing a
small binary supercell program rather than rebuilding the analytic
machinery from scratch.
The triangular emulator is the main outcome of this perspective. It
subsumes the square construction, reproduces the Archimedean
benchmark cases, and naturally extends to $k$-uniform, fractal and more general non-homogeneous
examples. The same workflow also gives access to antiferromagnetic
sectors, residual entropies, and correlation-length data within a
single computational pipeline.
\enlargethispage{3\baselineskip}%
This opens several directions: the first is the systematic completion
of the atlas for higher $k$-uniform and pentagonal families; the
second is the study of quenched site and bond disorder, as well as
aperiodic or substitution-generated lattices. A third direction is to
generalize the emulator concept beyond equal-edge-length planar
families with binary weights, with the longer-term goal of treating more general coupling
patterns and random geometric realizations.
\subsection*{Acknowledgments}
The authors acknowledge support from the CSIS grant I+D-2022-22520220100174UD. \\
Special thanks to S. Sanchez Majos for precious help with manipulating lattice datasets.

\contacts{%
  \safeemail{riccardo.ben.ali.zinati}{usi.ch} \\
  \safeemail{alessandro.codello}{unive.it}%
}

\spcolophon


\newpage
\bibliography{ref}

\newpage
\appendix
\section{Computing $A$ from the Free Energy}
\label{app:A-derivation}

We want to determine the singular part of the free energy per-spin. This comes from the singularity of the logarithm in the double integral
\begin{equation}
I \equiv \int\frac{{\rm d}k_{1}}{2\pi}\frac{{\rm d}k_{2}}{2\pi}\log P_{\Lambda}(v,k_{1},k_{2})\,,
\end{equation}
in the limit $k_1\to 0$ and $k_2\to 0$.
First we expand the determinant function around zero momenta
\begin{equation}\label{I}
P(v,k_{1},k_{2})=P(v)+\tfrac{1}{2}k_{1}^{2}P_{11}(v)+\tfrac{1}{2}k_{2}^{2}P_{22}(v)+k_{1}k_{2}P_{12}(v)+O(k^{4})\,.
\end{equation}
Then we also expand $P(v)=P(v,0,0)$ around the critical point 
\begin{equation}
P(v)=\underbrace{P(v_{c})}_{=0}+(v-v_{c})\underbrace{P'(v_{c})}_{=0}+\tfrac{1}{2}(v-v_{c})^{2}P''(v_{c})+...
\end{equation}
since $P(v_c)=0$ defines $v_c$, while $P'(v_c)=0$ holds in general, $P(v)$ having a double zero at $v_c$.
The integral \eqref{I} then becomes 
\begin{eqnarray}
I &=&\int\frac{{\rm d}k_{1}}{2\pi}\frac{{\rm d}k_{2}}{2\pi}\log\left\{ \tfrac{1}{2}(v-v_{c})^{2}P''(v_{c})+\tfrac{1}{2}k_{1}^{2}P_{11}(v_{c})+\tfrac{1}{2}k_{2}^{2}P_{22}(v_{c})+k_{1}k_{2}P_{12}(v_{c})+...\right\} \nonumber \\
&=&\tfrac{1}{(2\pi)^{2}}\int_{0}^{2\pi}{\rm d}\theta\int_{0}^{\infty}{\rm d}k\,k\log\left\{ \tfrac{1}{2}(v-v_{c})^{2}P''(v_{c})+\tfrac{1}{2}F(\theta)k^{2}+...\right\}\,.
\end{eqnarray}
Following \cite{Laurent_2025} we define the function $F(\theta)$ that takes the following form (on the square lattice) 
\begin{equation}
F_{\square}(\theta)=\cos^{2}\theta\, P_{11}(v_c)+\sin^{2}\theta\, P_{22}(v_c)+\sin 2\theta\, P_{12}(v_c)
\end{equation}
where $k_{1}=k_x=k\cos\theta$ and $k_{2}=k_y=k\sin\theta$.
We need the finite part of $k$-integral which is
\begin{equation}
\int_{0}^{\infty}{\rm d}k\,k\log\left\{ a+b\, k^{2}+...\right\} =\infty+\tfrac{a}{2b}\log a+\epsilon\,.
\end{equation}
Using this relation and performing the angular integration gives
\begin{eqnarray}
I &\sim&-\frac{(v-v_{c})^{2}P''(v_{c})}{2(2\pi)^{2}}\log(v-v_{c})^{2}\int_{0}^{2\pi}\frac{{\rm d}\theta}{F_{\square}(\theta)}+...\\
&=&-\frac{(v-v_{c})^{2}P''(v_{c})}{(2\pi)^{2}}\frac{2\pi}{\sqrt{P_{11}(v_c)P_{22}(v_c)-P_{12}^{2}(v_c)}}\log(v-v_{c})\,,
\end{eqnarray}
where we used
\begin{equation}
\int_{0}^{2\pi}\frac{{\rm d}\theta}{F_{\square}(\theta)}
= \frac{2\pi}{\sqrt{P_{11}(v_c)P_{22}(v_c)-P_{12}^{2}(v_c)}}\,.
\end{equation}
Since
\begin{equation}
(v-v_{c})^{2}=\frac{(1-v_{c}^{2})^{2}}{T_{c}^{4}}(T-T_{c})^{2}+...
\end{equation}
we finally get the singular part of the free energy
\begin{equation}
f_{sing}(T)\sim-\frac{T_{c}}{2n_{v}}I\sim \frac{(1-v_{c}^{2})^{2}}{4\pi n_{v}T_{c}^{3}}\frac{P''(v_{c})}{\sqrt{P_{11}P_{22}-P_{12}^{2}}}(T-T_{c})^{2}\log|T-T_{c}|\,.
\end{equation}
This is the expected $\alpha = 0$ singularity of the free energy $f_{sing} \sim t^2 \log t$ with $t= 1-T_{c}/T$ the reduced temperature.
We compute now $c_{sing}(T)$ to find
\begin{equation}
c_{sing}(T)\sim -\frac{(1-v_{c}^{2})^{2}}{2\pi n_{v}T_{c}^{2}}\frac{P''(v_{c})}{\sqrt{P_{11}P_{22}-P_{12}^{2}}}\log|1-\tfrac{T_{c}}{T}|\,,
\end{equation}
from the definition $c_{sing} \sim -A\log|1-\tfrac{T_{c}}{T}|$
we extract our final relation for $A$
\begin{equation}\label{Aapp}
\rho \,T_{c}^{2}A=\frac{(1-v_{c}^{2})^{2}}{2\pi}\frac{P''(v_{c})}{\sqrt{P_{11}(v_{c})P_{22}(v_{c})-P_{12}^{2}(v_{c})}}\,.
\end{equation}
Note the density correction $\rho=\frac{n_{v}}{L^{2}}$, necessary for both emulators.
Finally if we are on the triangle emulator we would have $k_{1} =k_{x}-\tfrac{1}{\sqrt{3}}k_{y}=k\cos\theta-\tfrac{1}{\sqrt{3}}k\sin\theta$ and $k_{2}=\tfrac{2}{\sqrt{3}}k_{y}=\tfrac{2}{\sqrt{3}}k\sin\theta$ which would give
\begin{eqnarray}
F_{\triangle}(\theta)& =&\left(\frac{1}{3}\sin^{2}\theta+\cos^{2}\theta-\frac{2\sin\theta\cos\theta}{\sqrt{3}}\right)P_{11}^{\triangle}(v_c)
+\tfrac{4}{3}\sin^{2}\theta \,P_{22}^{\triangle}(v_c)\nonumber\\&&+2\left(\tfrac{2}{\sqrt{3}}\cos\theta\sin\theta-\tfrac{2}{3}\sin^{2}\theta\right)P_{12}^{\triangle}(v_c)\,.
\end{eqnarray}
The angular integral will now result in
\begin{equation}
\int_{0}^{2\pi}\frac{{\rm d}\theta}{F_{\triangle}(\theta)}
= \frac{2\pi}{\sqrt{\tfrac{4}{3}\big[P_{11}(v_c)P_{22}(v_c)-P_{12}^{2}(v_c)\big]}}\,.
\end{equation}
The extra factor $\tfrac{2}{\sqrt{3}}$ in the denominator exactly cancels the Jacobian
\begin{equation}
\tfrac{{\rm d}k_{1}}{2\pi}\tfrac{{\rm d}k_{2}}{2\pi}=\left|\begin{array}{cc}
1 & -\tfrac{1}{\sqrt{3}}\\
0 & \tfrac{2}{\sqrt{3}}
\end{array}\right|\tfrac{{\rm d}k_{x}}{2\pi}\tfrac{{\rm d}k_{y}}{2\pi}=\tfrac{2}{\sqrt{3}}\tfrac{{\rm d}k_{x}}{2\pi}\tfrac{{\rm d}k_{y}}{2\pi}\,,
\end{equation}
resulting in the same formula \eqref{Aapp}  for both square and triangle emulators.

\newpage
\input{atlas_thermo/atlas}

\end{document}

%% file: tikz_macros.tex
\usetikzlibrary{calc}
\usetikzlibrary{matrix,positioning,fit}
\usetikzlibrary{arrows.meta}

\definecolor{maskOne}{HTML}{0000FF}
\definecolor{maskTwo}{HTML}{FF0000}
\colorlet{maskThree}{green}
\definecolor{cellOn}{HTML}{2B2B2B}
\definecolor{cellMuted}{HTML}{B6B3A8}
\definecolor{cellOff}{HTML}{F3F1EA}
\definecolor{cellBorder}{HTML}{C9C6BC}
\definecolor{coralPink}{HTML}{FF858D}

\newlength{\masktilesize}
\newcommand{\masktileLref}{4}
\newcommand{\masktile}[3]{%
\begin{minipage}[t]{\masktilesize}\centering
{\footnotesize $#1$}\par\vspace{0.30em}
\pgfmathsetlengthmacro{\cellsize}{min(\masktilesize/#2, \masktilesize/\masktileLref)}%
\pgfmathsetmacro{\digitfontpt}{0.55*\cellsize/1pt}%
\begin{tikzpicture}[x=\cellsize, y=\cellsize, baseline=0]
\foreach \row [count=\j from 0] in {#3} {
\foreach \v [count=\i from 0] in \row {
\pgfmathsetmacro{\yy}{(#2-1)-\j}%
\draw[cellBorder, line width=0.18pt] (\i,\yy) rectangle ++(1,1);
\ifnum\v=1
\node[font=\fontsize{\digitfontpt pt}{\digitfontpt pt}\selectfont,
text=cellOn, inner sep=0] at (\i+0.5, \yy+0.5) {1};
\else
\node[font=\fontsize{\digitfontpt pt}{\digitfontpt pt}\selectfont,
text=cellMuted, inner sep=0] at (\i+0.5, \yy+0.5) {0};
\fi
}
}
\draw[cellBorder, line width=0.5pt] (0,0) rectangle (#2,#2);
\end{tikzpicture}
\end{minipage}%
}

\newsavebox{\masktilematbox}
\newtoks\masktiletoks
\newcommand{\masktilematrows}[1]{%
\global\masktiletoks={}%
\foreach \row [count=\j from 1] in {#1} {%
  \ifnum\j>1
    \global\masktiletoks=\expandafter{\the\masktiletoks\\}%
  \fi
  \foreach \v [count=\i from 1] in \row {%
    \ifnum\i>1
      \global\masktiletoks=\expandafter{\the\masktiletoks&}%
    \fi
    \ifnum\v=1
      \global\masktiletoks=\expandafter{\the\masktiletoks\textcolor{cellOn}{1}}%
    \else
      \global\masktiletoks=\expandafter{\the\masktiletoks\textcolor{cellMuted}{0}}%
    \fi
  }%
}%
\the\masktiletoks
}
\newcommand{\masktilemat}[3]{%
\begin{minipage}[t]{\masktilesize}\centering
{\footnotesize $#1$}\par\vspace{0.30em}
\sbox{\masktilematbox}{%
  {\scriptsize$\setlength{\arraycolsep}{1.6pt}%
  \renewcommand{\arraystretch}{0.95}%
  \begin{array}{*{#2}{c}}
  \masktilematrows{#3}%
  \end{array}$}%
}%
\ifdim\wd\masktilematbox>\masktilesize
  \resizebox{\masktilesize}{!}{\usebox{\masktilematbox}}%
\else
  \usebox{\masktilematbox}%
\fi
\end{minipage}%
}

\newcommand{\sierpinski}[2][]{\tikz[baseline=(current bounding box.center),transform shape,#1]{
\draw[fill=black] rectangle(1,1);
\foreach \n[
evaluate=\n as \m using \n-1,
evaluate=\n as \s using 1/3^\n,
evaluate=\m as \p using 3^\m] in {1,...,#2}{
\foreach \k[evaluate=\k as \x using (2*\k-1)/2/3^\m] in {1,...,\p}{
\foreach \j[evaluate=\j as \y using (2*\j-1)/2/3^\m] in {1,...,\p}{
\node[fill=white, minimum size=\s cm, inner sep=0] at (\x,\y){};
}}}}}

\newcommand{\sitemask}[3][]{\tikz[baseline=(current bounding box.center),transform shape,#1]{
\def\L{#2}
\draw[fill=black] rectangle(1,1);
\foreach \row [count=\j from 0] in {#3} {
\foreach \v [count=\i from 0] in \row {
\ifnum\v=0
\pgfmathsetmacro{\xx}{\i/\L}
\pgfmathsetmacro{\yy}{1-(\j+1)/\L}
\pgfmathsetmacro{\ss}{1/\L}
\fill[white] (\xx,\yy) rectangle ++(\ss,\ss);
\fi
}
}
\draw[line width=0.35pt] (0,0) rectangle(1,1);
}}

\newlength{\fourpanelcellwidth}
\newcommand{\fourpanelstrip}[8]{%
\begin{center}
\noindent\hbox to \linewidth{%
\hss\begin{minipage}[t]{\fourpanelcellwidth}\centering #1\par\vspace{0.55em}#5\end{minipage}%
\hss\begin{minipage}[t]{\fourpanelcellwidth}\centering #2\par\vspace{0.55em}#6\end{minipage}%
\hss\begin{minipage}[t]{\fourpanelcellwidth}\centering #3\par\vspace{0.55em}#7\end{minipage}%
\hss\begin{minipage}[t]{\fourpanelcellwidth}\centering #4\par\vspace{0.55em}#8\end{minipage}\hss}%
\end{center}}


%% file: figures/random_t1009_supercell_strip.tex
\begin{tikzpicture}[baseline=(current bounding box.center),x=0.18cm,y=0.18cm]
  \begin{scope}[xshift=0cm]
    \coordinate (su1) at (0.000000,0.000000);
    \coordinate (su2) at (0.500000,-0.866025);
    \coordinate (su3) at (1.366025,-1.366025);
    \coordinate (su4) at (1.366025,-0.366025);
    \coordinate (su5) at (1.866025,0.500000);
    \coordinate (su6) at (2.366025,-0.366025);
    \coordinate (su7) at (3.232051,-0.866025);
    \coordinate (su8) at (3.232051,0.133975);
    \coordinate (su9) at (3.732051,1.000000);
    \coordinate (su10) at (4.232051,0.133975);
    \coordinate (su11) at (5.098076,-0.366025);
    \coordinate (su12) at (5.098076,0.633975);
    \coordinate (su13) at (5.598076,1.500000);
    \coordinate (su14) at (6.098076,0.633975);
    \coordinate (su15) at (6.964102,0.133975);
    \coordinate (su16) at (6.964102,1.133975);
    \coordinate (su17) at (0.500000,-1.866025);
    \coordinate (su18) at (1.000000,-2.732051);
    \coordinate (su19) at (1.866025,-3.232051);
    \coordinate (su20) at (1.866025,-2.232051);
    \coordinate (su21) at (2.366025,-1.366025);
    \coordinate (su22) at (2.866025,-2.232051);
    \coordinate (su23) at (3.732051,-2.732051);
    \coordinate (su24) at (3.732051,-1.732051);
    \coordinate (su25) at (4.232051,-0.866025);
    \coordinate (su26) at (4.732051,-1.732051);
    \coordinate (su27) at (5.598076,-2.232051);
    \coordinate (su28) at (5.598076,-1.232051);
    \coordinate (su29) at (6.098076,-0.366025);
    \coordinate (su30) at (6.598076,-1.232051);
    \coordinate (su31) at (7.464102,-1.732051);
    \coordinate (su32) at (7.464102,-0.732051);
    \coordinate (su33) at (1.000000,-3.732051);
    \coordinate (su34) at (1.500000,-4.598076);
    \coordinate (su35) at (2.366025,-5.098076);
    \coordinate (su36) at (2.366025,-4.098076);
    \coordinate (su37) at (2.866025,-3.232051);
    \coordinate (su38) at (3.366025,-4.098076);
    \coordinate (su39) at (4.232051,-4.598076);
    \coordinate (su40) at (4.232051,-3.598076);
    \coordinate (su41) at (4.732051,-2.732051);
    \coordinate (su42) at (5.232051,-3.598076);
    \coordinate (su43) at (6.098076,-4.098076);
    \coordinate (su44) at (6.098076,-3.098076);
    \coordinate (su45) at (6.598076,-2.232051);
    \coordinate (su46) at (7.098076,-3.098076);
    \coordinate (su47) at (7.964102,-3.598076);
    \coordinate (su48) at (7.964102,-2.598076);
    \coordinate (su49) at (1.500000,-5.598076);
    \coordinate (su50) at (2.000000,-6.464102);
    \coordinate (su51) at (2.866025,-6.964102);
    \coordinate (su52) at (2.866025,-5.964102);
    \coordinate (su53) at (3.366025,-5.098076);
    \coordinate (su54) at (3.866025,-5.964102);
    \coordinate (su55) at (4.732051,-6.464102);
    \coordinate (su56) at (4.732051,-5.464102);
    \coordinate (su57) at (5.232051,-4.598076);
    \coordinate (su58) at (5.732051,-5.464102);
    \coordinate (su59) at (6.598076,-5.964102);
    \coordinate (su60) at (6.598076,-4.964102);
    \coordinate (su61) at (7.098076,-4.098076);
    \coordinate (su62) at (7.598076,-4.964102);
    \coordinate (su63) at (8.464102,-5.464102);
    \coordinate (su64) at (8.464102,-4.464102);
    \draw[line width=0.48pt] (su1) -- (su2);
    \draw[line width=0.48pt] (su2) -- (su3);
    \draw[line width=0.48pt] (su2) -- (su4);
    \draw[line width=0.48pt] (su3) -- (su17);
    \draw[line width=0.48pt] (su3) -- (su20);
    \draw[line width=0.48pt] (su4) -- (su5);
    \draw[line width=0.48pt] (su6) -- (su7);
    \draw[line width=0.48pt] (su6) -- (su21);
    \draw[line width=0.48pt] (su7) -- (su8);
    \draw[line width=0.48pt] (su7) -- (su25);
    \draw[line width=0.48pt] (su8) -- (su9);
    \draw[line width=0.48pt] (su8) -- (su10);
    \draw[line width=0.48pt] (su10) -- (su12);
    \draw[line width=0.48pt] (su10) -- (su25);
    \draw[line width=0.48pt] (su11) -- (su28);
    \draw[line width=0.48pt] (su12) -- (su13);
    \draw[line width=0.48pt] (su14) -- (su16);
    \draw[line width=0.48pt] (su14) -- (su29);
    \draw[line width=0.48pt] (su15) -- (su29);
    \draw[line width=0.48pt] (su15) -- (su32);
    \draw[line width=0.48pt] (su19) -- (su33);
    \draw[line width=0.48pt] (su19) -- (su36);
    \draw[line width=0.48pt] (su23) -- (su24);
    \draw[line width=0.48pt] (su23) -- (su37);
    \draw[line width=0.48pt] (su23) -- (su40);
    \draw[line width=0.48pt] (su23) -- (su41);
    \draw[line width=0.48pt] (su24) -- (su25);
    \draw[line width=0.48pt] (su24) -- (su26);
    \draw[line width=0.48pt] (su25) -- (su26);
    \draw[line width=0.48pt] (su26) -- (su27);
    \draw[line width=0.48pt] (su26) -- (su28);
    \draw[line width=0.48pt] (su26) -- (su41);
    \draw[line width=0.48pt] (su27) -- (su28);
    \draw[line width=0.48pt] (su28) -- (su29);
    \draw[line width=0.48pt] (su28) -- (su30);
    \draw[line width=0.48pt] (su30) -- (su32);
    \draw[line width=0.48pt] (su30) -- (su45);
    \draw[line width=0.48pt] (su31) -- (su45);
    \draw[line width=0.48pt] (su31) -- (su48);
    \draw[line width=0.48pt] (su34) -- (su49);
    \draw[line width=0.48pt] (su35) -- (su49);
    \draw[line width=0.48pt] (su35) -- (su52);
    \draw[line width=0.48pt] (su38) -- (su39);
    \draw[line width=0.48pt] (su38) -- (su40);
    \draw[line width=0.48pt] (su38) -- (su53);
    \draw[line width=0.48pt] (su39) -- (su40);
    \draw[line width=0.48pt] (su39) -- (su57);
    \draw[line width=0.48pt] (su41) -- (su42);
    \draw[line width=0.48pt] (su42) -- (su43);
    \draw[line width=0.48pt] (su43) -- (su44);
    \draw[line width=0.48pt] (su43) -- (su57);
    \draw[line width=0.48pt] (su43) -- (su61);
    \draw[line width=0.48pt] (su44) -- (su45);
    \draw[line width=0.48pt] (su44) -- (su46);
    \draw[line width=0.48pt] (su45) -- (su46);
    \draw[line width=0.48pt] (su46) -- (su48);
    \draw[line width=0.48pt] (su46) -- (su61);
    \draw[line width=0.48pt] (su47) -- (su48);
    \draw[line width=0.48pt] (su47) -- (su64);
    \draw[line width=0.48pt] (su50) -- (su51);
    \draw[line width=0.48pt] (su51) -- (su52);
    \draw[line width=0.48pt] (su52) -- (su53);
    \draw[line width=0.48pt] (su52) -- (su54);
    \draw[line width=0.48pt] (su54) -- (su56);
    \draw[line width=0.48pt] (su56) -- (su58);
    \draw[line width=0.48pt] (su57) -- (su58);
    \draw[line width=0.48pt] (su58) -- (su60);
    \draw[line width=0.48pt] (su59) -- (su60);
    \draw[line width=0.48pt] (su60) -- (su61);
    \draw[line width=0.48pt] (su60) -- (su62);
    \draw[line width=0.48pt] (su61) -- (su62);
    \draw[line width=0.48pt] (su62) -- (su63);
    \draw[line width=0.48pt] (su63) -- (su64);
    \fill (su1) circle (0.58pt);
    \fill (su2) circle (0.58pt);
    \fill (su3) circle (0.58pt);
    \fill (su4) circle (0.58pt);
    \fill (su5) circle (0.58pt);
    \fill (su6) circle (0.58pt);
    \fill (su7) circle (0.58pt);
    \fill (su8) circle (0.58pt);
    \fill (su9) circle (0.58pt);
    \fill (su10) circle (0.58pt);
    \fill (su11) circle (0.58pt);
    \fill (su12) circle (0.58pt);
    \fill (su13) circle (0.58pt);
    \fill (su14) circle (0.58pt);
    \fill (su15) circle (0.58pt);
    \fill (su16) circle (0.58pt);
    \fill (su17) circle (0.58pt);
    \fill (su18) circle (0.58pt);
    \fill (su19) circle (0.58pt);
    \fill (su20) circle (0.58pt);
    \fill (su21) circle (0.58pt);
    \fill (su22) circle (0.58pt);
    \fill (su23) circle (0.58pt);
    \fill (su24) circle (0.58pt);
    \fill (su25) circle (0.58pt);
    \fill (su26) circle (0.58pt);
    \fill (su27) circle (0.58pt);
    \fill (su28) circle (0.58pt);
    \fill (su29) circle (0.58pt);
    \fill (su30) circle (0.58pt);
    \fill (su31) circle (0.58pt);
    \fill (su32) circle (0.58pt);
    \fill (su33) circle (0.58pt);
    \fill (su34) circle (0.58pt);
    \fill (su35) circle (0.58pt);
    \fill (su36) circle (0.58pt);
    \fill (su37) circle (0.58pt);
    \fill (su38) circle (0.58pt);
    \fill (su39) circle (0.58pt);
    \fill (su40) circle (0.58pt);
    \fill (su41) circle (0.58pt);
    \fill (su42) circle (0.58pt);
    \fill (su43) circle (0.58pt);
    \fill (su44) circle (0.58pt);
    \fill (su45) circle (0.58pt);
    \fill (su46) circle (0.58pt);
    \fill (su47) circle (0.58pt);
    \fill (su48) circle (0.58pt);
    \fill (su49) circle (0.58pt);
    \fill (su50) circle (0.58pt);
    \fill (su51) circle (0.58pt);
    \fill (su52) circle (0.58pt);
    \fill (su53) circle (0.58pt);
    \fill (su54) circle (0.58pt);
    \fill (su55) circle (0.58pt);
    \fill (su56) circle (0.58pt);
    \fill (su57) circle (0.58pt);
    \fill (su58) circle (0.58pt);
    \fill (su59) circle (0.58pt);
    \fill (su60) circle (0.58pt);
    \fill (su61) circle (0.58pt);
    \fill (su62) circle (0.58pt);
    \fill (su63) circle (0.58pt);
    \fill (su64) circle (0.58pt);
    \node at (4.232051,-8.35) {$p=0.59$};
  \end{scope}
  \begin{scope}[xshift=3.8cm]
    \coordinate (su1) at (0.000000,0.000000);
    \coordinate (su2) at (0.500000,-0.866025);
    \coordinate (su3) at (1.366025,-1.366025);
    \coordinate (su4) at (1.366025,-0.366025);
    \coordinate (su5) at (1.866025,0.500000);
    \coordinate (su6) at (2.366025,-0.366025);
    \coordinate (su7) at (3.232051,-0.866025);
    \coordinate (su8) at (3.232051,0.133975);
    \coordinate (su9) at (3.732051,1.000000);
    \coordinate (su10) at (4.232051,0.133975);
    \coordinate (su11) at (5.098076,-0.366025);
    \coordinate (su12) at (5.098076,0.633975);
    \coordinate (su13) at (5.598076,1.500000);
    \coordinate (su14) at (6.098076,0.633975);
    \coordinate (su15) at (6.964102,0.133975);
    \coordinate (su16) at (6.964102,1.133975);
    \coordinate (su17) at (0.500000,-1.866025);
    \coordinate (su18) at (1.000000,-2.732051);
    \coordinate (su19) at (1.866025,-3.232051);
    \coordinate (su20) at (1.866025,-2.232051);
    \coordinate (su21) at (2.366025,-1.366025);
    \coordinate (su22) at (2.866025,-2.232051);
    \coordinate (su23) at (3.732051,-2.732051);
    \coordinate (su24) at (3.732051,-1.732051);
    \coordinate (su25) at (4.232051,-0.866025);
    \coordinate (su26) at (4.732051,-1.732051);
    \coordinate (su27) at (5.598076,-2.232051);
    \coordinate (su28) at (5.598076,-1.232051);
    \coordinate (su29) at (6.098076,-0.366025);
    \coordinate (su30) at (6.598076,-1.232051);
    \coordinate (su31) at (7.464102,-1.732051);
    \coordinate (su32) at (7.464102,-0.732051);
    \coordinate (su33) at (1.000000,-3.732051);
    \coordinate (su34) at (1.500000,-4.598076);
    \coordinate (su35) at (2.366025,-5.098076);
    \coordinate (su36) at (2.366025,-4.098076);
    \coordinate (su37) at (2.866025,-3.232051);
    \coordinate (su38) at (3.366025,-4.098076);
    \coordinate (su39) at (4.232051,-4.598076);
    \coordinate (su40) at (4.232051,-3.598076);
    \coordinate (su41) at (4.732051,-2.732051);
    \coordinate (su42) at (5.232051,-3.598076);
    \coordinate (su43) at (6.098076,-4.098076);
    \coordinate (su44) at (6.098076,-3.098076);
    \coordinate (su45) at (6.598076,-2.232051);
    \coordinate (su46) at (7.098076,-3.098076);
    \coordinate (su47) at (7.964102,-3.598076);
    \coordinate (su48) at (7.964102,-2.598076);
    \coordinate (su49) at (1.500000,-5.598076);
    \coordinate (su50) at (2.000000,-6.464102);
    \coordinate (su51) at (2.866025,-6.964102);
    \coordinate (su52) at (2.866025,-5.964102);
    \coordinate (su53) at (3.366025,-5.098076);
    \coordinate (su54) at (3.866025,-5.964102);
    \coordinate (su55) at (4.732051,-6.464102);
    \coordinate (su56) at (4.732051,-5.464102);
    \coordinate (su57) at (5.232051,-4.598076);
    \coordinate (su58) at (5.732051,-5.464102);
    \coordinate (su59) at (6.598076,-5.964102);
    \coordinate (su60) at (6.598076,-4.964102);
    \coordinate (su61) at (7.098076,-4.098076);
    \coordinate (su62) at (7.598076,-4.964102);
    \coordinate (su63) at (8.464102,-5.464102);
    \coordinate (su64) at (8.464102,-4.464102);
    \draw[line width=0.48pt] (su2) -- (su3);
    \draw[line width=0.48pt] (su2) -- (su4);
    \draw[line width=0.48pt] (su2) -- (su17);
    \draw[line width=0.48pt] (su3) -- (su4);
    \draw[line width=0.48pt] (su3) -- (su17);
    \draw[line width=0.48pt] (su3) -- (su20);
    \draw[line width=0.48pt] (su3) -- (su21);
    \draw[line width=0.48pt] (su4) -- (su6);
    \draw[line width=0.48pt] (su5) -- (su6);
    \draw[line width=0.48pt] (su6) -- (su7);
    \draw[line width=0.48pt] (su6) -- (su21);
    \draw[line width=0.48pt] (su7) -- (su8);
    \draw[line width=0.48pt] (su7) -- (su21);
    \draw[line width=0.48pt] (su7) -- (su24);
    \draw[line width=0.48pt] (su8) -- (su9);
    \draw[line width=0.48pt] (su8) -- (su10);
    \draw[line width=0.48pt] (su10) -- (su11);
    \draw[line width=0.48pt] (su10) -- (su25);
    \draw[line width=0.48pt] (su11) -- (su12);
    \draw[line width=0.48pt] (su11) -- (su28);
    \draw[line width=0.48pt] (su11) -- (su29);
    \draw[line width=0.48pt] (su12) -- (su13);
    \draw[line width=0.48pt] (su14) -- (su15);
    \draw[line width=0.48pt] (su14) -- (su16);
    \draw[line width=0.48pt] (su14) -- (su29);
    \draw[line width=0.48pt] (su15) -- (su16);
    \draw[line width=0.48pt] (su15) -- (su29);
    \draw[line width=0.48pt] (su17) -- (su18);
    \draw[line width=0.48pt] (su18) -- (su19);
    \draw[line width=0.48pt] (su18) -- (su20);
    \draw[line width=0.48pt] (su19) -- (su20);
    \draw[line width=0.48pt] (su19) -- (su33);
    \draw[line width=0.48pt] (su20) -- (su21);
    \draw[line width=0.48pt] (su22) -- (su24);
    \draw[line width=0.48pt] (su22) -- (su37);
    \draw[line width=0.48pt] (su23) -- (su24);
    \draw[line width=0.48pt] (su23) -- (su37);
    \draw[line width=0.48pt] (su23) -- (su40);
    \draw[line width=0.48pt] (su23) -- (su41);
    \draw[line width=0.48pt] (su24) -- (su26);
    \draw[line width=0.48pt] (su25) -- (su26);
    \draw[line width=0.48pt] (su26) -- (su28);
    \draw[line width=0.48pt] (su26) -- (su41);
    \draw[line width=0.48pt] (su27) -- (su28);
    \draw[line width=0.48pt] (su27) -- (su44);
    \draw[line width=0.48pt] (su27) -- (su45);
    \draw[line width=0.48pt] (su28) -- (su29);
    \draw[line width=0.48pt] (su28) -- (su30);
    \draw[line width=0.48pt] (su29) -- (su30);
    \draw[line width=0.48pt] (su30) -- (su31);
    \draw[line width=0.48pt] (su30) -- (su32);
    \draw[line width=0.48pt] (su30) -- (su45);
    \draw[line width=0.48pt] (su31) -- (su32);
    \draw[line width=0.48pt] (su31) -- (su45);
    \draw[line width=0.48pt] (su31) -- (su48);
    \draw[line width=0.48pt] (su33) -- (su34);
    \draw[line width=0.48pt] (su34) -- (su35);
    \draw[line width=0.48pt] (su34) -- (su36);
    \draw[line width=0.48pt] (su35) -- (su36);
    \draw[line width=0.48pt] (su35) -- (su52);
    \draw[line width=0.48pt] (su35) -- (su53);
    \draw[line width=0.48pt] (su36) -- (su37);
    \draw[line width=0.48pt] (su36) -- (su38);
    \draw[line width=0.48pt] (su37) -- (su38);
    \draw[line width=0.48pt] (su38) -- (su39);
    \draw[line width=0.48pt] (su38) -- (su40);
    \draw[line width=0.48pt] (su38) -- (su53);
    \draw[line width=0.48pt] (su39) -- (su56);
    \draw[line width=0.48pt] (su39) -- (su57);
    \draw[line width=0.48pt] (su40) -- (su41);
    \draw[line width=0.48pt] (su40) -- (su42);
    \draw[line width=0.48pt] (su41) -- (su42);
    \draw[line width=0.48pt] (su42) -- (su43);
    \draw[line width=0.48pt] (su42) -- (su44);
    \draw[line width=0.48pt] (su42) -- (su57);
    \draw[line width=0.48pt] (su43) -- (su44);
    \draw[line width=0.48pt] (su43) -- (su57);
    \draw[line width=0.48pt] (su45) -- (su46);
    \draw[line width=0.48pt] (su46) -- (su47);
    \draw[line width=0.48pt] (su46) -- (su48);
    \draw[line width=0.48pt] (su47) -- (su48);
    \draw[line width=0.48pt] (su47) -- (su61);
    \draw[line width=0.48pt] (su49) -- (su50);
    \draw[line width=0.48pt] (su50) -- (su51);
    \draw[line width=0.48pt] (su51) -- (su52);
    \draw[line width=0.48pt] (su52) -- (su54);
    \draw[line width=0.48pt] (su53) -- (su54);
    \draw[line width=0.48pt] (su54) -- (su56);
    \draw[line width=0.48pt] (su56) -- (su57);
    \draw[line width=0.48pt] (su56) -- (su58);
    \draw[line width=0.48pt] (su57) -- (su58);
    \draw[line width=0.48pt] (su58) -- (su59);
    \draw[line width=0.48pt] (su58) -- (su60);
    \draw[line width=0.48pt] (su59) -- (su60);
    \draw[line width=0.48pt] (su60) -- (su61);
    \draw[line width=0.48pt] (su60) -- (su62);
    \draw[line width=0.48pt] (su61) -- (su62);
    \draw[line width=0.48pt] (su62) -- (su64);
    \draw[line width=0.48pt] (su63) -- (su64);
    \fill (su1) circle (0.58pt);
    \fill (su2) circle (0.58pt);
    \fill (su3) circle (0.58pt);
    \fill (su4) circle (0.58pt);
    \fill (su5) circle (0.58pt);
    \fill (su6) circle (0.58pt);
    \fill (su7) circle (0.58pt);
    \fill (su8) circle (0.58pt);
    \fill (su9) circle (0.58pt);
    \fill (su10) circle (0.58pt);
    \fill (su11) circle (0.58pt);
    \fill (su12) circle (0.58pt);
    \fill (su13) circle (0.58pt);
    \fill (su14) circle (0.58pt);
    \fill (su15) circle (0.58pt);
    \fill (su16) circle (0.58pt);
    \fill (su17) circle (0.58pt);
    \fill (su18) circle (0.58pt);
    \fill (su19) circle (0.58pt);
    \fill (su20) circle (0.58pt);
    \fill (su21) circle (0.58pt);
    \fill (su22) circle (0.58pt);
    \fill (su23) circle (0.58pt);
    \fill (su24) circle (0.58pt);
    \fill (su25) circle (0.58pt);
    \fill (su26) circle (0.58pt);
    \fill (su27) circle (0.58pt);
    \fill (su28) circle (0.58pt);
    \fill (su29) circle (0.58pt);
    \fill (su30) circle (0.58pt);
    \fill (su31) circle (0.58pt);
    \fill (su32) circle (0.58pt);
    \fill (su33) circle (0.58pt);
    \fill (su34) circle (0.58pt);
    \fill (su35) circle (0.58pt);
    \fill (su36) circle (0.58pt);
    \fill (su37) circle (0.58pt);
    \fill (su38) circle (0.58pt);
    \fill (su39) circle (0.58pt);
    \fill (su40) circle (0.58pt);
    \fill (su41) circle (0.58pt);
    \fill (su42) circle (0.58pt);
    \fill (su43) circle (0.58pt);
    \fill (su44) circle (0.58pt);
    \fill (su45) circle (0.58pt);
    \fill (su46) circle (0.58pt);
    \fill (su47) circle (0.58pt);
    \fill (su48) circle (0.58pt);
    \fill (su49) circle (0.58pt);
    \fill (su50) circle (0.58pt);
    \fill (su51) circle (0.58pt);
    \fill (su52) circle (0.58pt);
    \fill (su53) circle (0.58pt);
    \fill (su54) circle (0.58pt);
    \fill (su55) circle (0.58pt);
    \fill (su56) circle (0.58pt);
    \fill (su57) circle (0.58pt);
    \fill (su58) circle (0.58pt);
    \fill (su59) circle (0.58pt);
    \fill (su60) circle (0.58pt);
    \fill (su61) circle (0.58pt);
    \fill (su62) circle (0.58pt);
    \fill (su63) circle (0.58pt);
    \fill (su64) circle (0.58pt);
    \node at (4.232051,-8.35) {$p=0.70$};
  \end{scope}
  \begin{scope}[xshift=7.6cm]
    \coordinate (su1) at (0.000000,0.000000);
    \coordinate (su2) at (0.500000,-0.866025);
    \coordinate (su3) at (1.366025,-1.366025);
    \coordinate (su4) at (1.366025,-0.366025);
    \coordinate (su5) at (1.866025,0.500000);
    \coordinate (su6) at (2.366025,-0.366025);
    \coordinate (su7) at (3.232051,-0.866025);
    \coordinate (su8) at (3.232051,0.133975);
    \coordinate (su9) at (3.732051,1.000000);
    \coordinate (su10) at (4.232051,0.133975);
    \coordinate (su11) at (5.098076,-0.366025);
    \coordinate (su12) at (5.098076,0.633975);
    \coordinate (su13) at (5.598076,1.500000);
    \coordinate (su14) at (6.098076,0.633975);
    \coordinate (su15) at (6.964102,0.133975);
    \coordinate (su16) at (6.964102,1.133975);
    \coordinate (su17) at (0.500000,-1.866025);
    \coordinate (su18) at (1.000000,-2.732051);
    \coordinate (su19) at (1.866025,-3.232051);
    \coordinate (su20) at (1.866025,-2.232051);
    \coordinate (su21) at (2.366025,-1.366025);
    \coordinate (su22) at (2.866025,-2.232051);
    \coordinate (su23) at (3.732051,-2.732051);
    \coordinate (su24) at (3.732051,-1.732051);
    \coordinate (su25) at (4.232051,-0.866025);
    \coordinate (su26) at (4.732051,-1.732051);
    \coordinate (su27) at (5.598076,-2.232051);
    \coordinate (su28) at (5.598076,-1.232051);
    \coordinate (su29) at (6.098076,-0.366025);
    \coordinate (su30) at (6.598076,-1.232051);
    \coordinate (su31) at (7.464102,-1.732051);
    \coordinate (su32) at (7.464102,-0.732051);
    \coordinate (su33) at (1.000000,-3.732051);
    \coordinate (su34) at (1.500000,-4.598076);
    \coordinate (su35) at (2.366025,-5.098076);
    \coordinate (su36) at (2.366025,-4.098076);
    \coordinate (su37) at (2.866025,-3.232051);
    \coordinate (su38) at (3.366025,-4.098076);
    \coordinate (su39) at (4.232051,-4.598076);
    \coordinate (su40) at (4.232051,-3.598076);
    \coordinate (su41) at (4.732051,-2.732051);
    \coordinate (su42) at (5.232051,-3.598076);
    \coordinate (su43) at (6.098076,-4.098076);
    \coordinate (su44) at (6.098076,-3.098076);
    \coordinate (su45) at (6.598076,-2.232051);
    \coordinate (su46) at (7.098076,-3.098076);
    \coordinate (su47) at (7.964102,-3.598076);
    \coordinate (su48) at (7.964102,-2.598076);
    \coordinate (su49) at (1.500000,-5.598076);
    \coordinate (su50) at (2.000000,-6.464102);
    \coordinate (su51) at (2.866025,-6.964102);
    \coordinate (su52) at (2.866025,-5.964102);
    \coordinate (su53) at (3.366025,-5.098076);
    \coordinate (su54) at (3.866025,-5.964102);
    \coordinate (su55) at (4.732051,-6.464102);
    \coordinate (su56) at (4.732051,-5.464102);
    \coordinate (su57) at (5.232051,-4.598076);
    \coordinate (su58) at (5.732051,-5.464102);
    \coordinate (su59) at (6.598076,-5.964102);
    \coordinate (su60) at (6.598076,-4.964102);
    \coordinate (su61) at (7.098076,-4.098076);
    \coordinate (su62) at (7.598076,-4.964102);
    \coordinate (su63) at (8.464102,-5.464102);
    \coordinate (su64) at (8.464102,-4.464102);
    \draw[line width=0.48pt] (su1) -- (su2);
    \draw[line width=0.48pt] (su2) -- (su3);
    \draw[line width=0.48pt] (su2) -- (su4);
    \draw[line width=0.48pt] (su2) -- (su17);
    \draw[line width=0.48pt] (su3) -- (su4);
    \draw[line width=0.48pt] (su3) -- (su20);
    \draw[line width=0.48pt] (su3) -- (su21);
    \draw[line width=0.48pt] (su4) -- (su5);
    \draw[line width=0.48pt] (su4) -- (su6);
    \draw[line width=0.48pt] (su5) -- (su6);
    \draw[line width=0.48pt] (su6) -- (su7);
    \draw[line width=0.48pt] (su6) -- (su21);
    \draw[line width=0.48pt] (su7) -- (su8);
    \draw[line width=0.48pt] (su7) -- (su21);
    \draw[line width=0.48pt] (su7) -- (su24);
    \draw[line width=0.48pt] (su8) -- (su9);
    \draw[line width=0.48pt] (su8) -- (su10);
    \draw[line width=0.48pt] (su9) -- (su10);
    \draw[line width=0.48pt] (su10) -- (su12);
    \draw[line width=0.48pt] (su10) -- (su25);
    \draw[line width=0.48pt] (su11) -- (su25);
    \draw[line width=0.48pt] (su11) -- (su28);
    \draw[line width=0.48pt] (su11) -- (su29);
    \draw[line width=0.48pt] (su12) -- (su13);
    \draw[line width=0.48pt] (su12) -- (su14);
    \draw[line width=0.48pt] (su13) -- (su14);
    \draw[line width=0.48pt] (su14) -- (su15);
    \draw[line width=0.48pt] (su14) -- (su16);
    \draw[line width=0.48pt] (su14) -- (su29);
    \draw[line width=0.48pt] (su15) -- (su16);
    \draw[line width=0.48pt] (su15) -- (su29);
    \draw[line width=0.48pt] (su15) -- (su32);
    \draw[line width=0.48pt] (su17) -- (su18);
    \draw[line width=0.48pt] (su18) -- (su19);
    \draw[line width=0.48pt] (su18) -- (su20);
    \draw[line width=0.48pt] (su18) -- (su33);
    \draw[line width=0.48pt] (su19) -- (su20);
    \draw[line width=0.48pt] (su19) -- (su33);
    \draw[line width=0.48pt] (su19) -- (su36);
    \draw[line width=0.48pt] (su19) -- (su37);
    \draw[line width=0.48pt] (su20) -- (su21);
    \draw[line width=0.48pt] (su20) -- (su22);
    \draw[line width=0.48pt] (su21) -- (su22);
    \draw[line width=0.48pt] (su22) -- (su23);
    \draw[line width=0.48pt] (su22) -- (su24);
    \draw[line width=0.48pt] (su22) -- (su37);
    \draw[line width=0.48pt] (su23) -- (su24);
    \draw[line width=0.48pt] (su23) -- (su37);
    \draw[line width=0.48pt] (su23) -- (su40);
    \draw[line width=0.48pt] (su23) -- (su41);
    \draw[line width=0.48pt] (su24) -- (su25);
    \draw[line width=0.48pt] (su24) -- (su26);
    \draw[line width=0.48pt] (su25) -- (su26);
    \draw[line width=0.48pt] (su26) -- (su41);
    \draw[line width=0.48pt] (su27) -- (su28);
    \draw[line width=0.48pt] (su27) -- (su41);
    \draw[line width=0.48pt] (su27) -- (su44);
    \draw[line width=0.48pt] (su27) -- (su45);
    \draw[line width=0.48pt] (su28) -- (su29);
    \draw[line width=0.48pt] (su28) -- (su30);
    \draw[line width=0.48pt] (su29) -- (su30);
    \draw[line width=0.48pt] (su30) -- (su32);
    \draw[line width=0.48pt] (su30) -- (su45);
    \draw[line width=0.48pt] (su31) -- (su32);
    \draw[line width=0.48pt] (su31) -- (su45);
    \draw[line width=0.48pt] (su31) -- (su48);
    \draw[line width=0.48pt] (su33) -- (su34);
    \draw[line width=0.48pt] (su34) -- (su35);
    \draw[line width=0.48pt] (su34) -- (su36);
    \draw[line width=0.48pt] (su34) -- (su49);
    \draw[line width=0.48pt] (su35) -- (su36);
    \draw[line width=0.48pt] (su35) -- (su49);
    \draw[line width=0.48pt] (su35) -- (su52);
    \draw[line width=0.48pt] (su35) -- (su53);
    \draw[line width=0.48pt] (su36) -- (su37);
    \draw[line width=0.48pt] (su36) -- (su38);
    \draw[line width=0.48pt] (su37) -- (su38);
    \draw[line width=0.48pt] (su38) -- (su39);
    \draw[line width=0.48pt] (su38) -- (su40);
    \draw[line width=0.48pt] (su38) -- (su53);
    \draw[line width=0.48pt] (su39) -- (su40);
    \draw[line width=0.48pt] (su39) -- (su53);
    \draw[line width=0.48pt] (su39) -- (su56);
    \draw[line width=0.48pt] (su39) -- (su57);
    \draw[line width=0.48pt] (su40) -- (su42);
    \draw[line width=0.48pt] (su41) -- (su42);
    \draw[line width=0.48pt] (su42) -- (su43);
    \draw[line width=0.48pt] (su42) -- (su44);
    \draw[line width=0.48pt] (su42) -- (su57);
    \draw[line width=0.48pt] (su43) -- (su44);
    \draw[line width=0.48pt] (su43) -- (su57);
    \draw[line width=0.48pt] (su43) -- (su60);
    \draw[line width=0.48pt] (su43) -- (su61);
    \draw[line width=0.48pt] (su44) -- (su45);
    \draw[line width=0.48pt] (su44) -- (su46);
    \draw[line width=0.48pt] (su45) -- (su46);
    \draw[line width=0.48pt] (su46) -- (su48);
    \draw[line width=0.48pt] (su47) -- (su48);
    \draw[line width=0.48pt] (su47) -- (su64);
    \draw[line width=0.48pt] (su49) -- (su50);
    \draw[line width=0.48pt] (su50) -- (su51);
    \draw[line width=0.48pt] (su50) -- (su52);
    \draw[line width=0.48pt] (su51) -- (su52);
    \draw[line width=0.48pt] (su52) -- (su53);
    \draw[line width=0.48pt] (su52) -- (su54);
    \draw[line width=0.48pt] (su53) -- (su54);
    \draw[line width=0.48pt] (su54) -- (su55);
    \draw[line width=0.48pt] (su54) -- (su56);
    \draw[line width=0.48pt] (su56) -- (su57);
    \draw[line width=0.48pt] (su57) -- (su58);
    \draw[line width=0.48pt] (su58) -- (su59);
    \draw[line width=0.48pt] (su59) -- (su60);
    \draw[line width=0.48pt] (su60) -- (su61);
    \draw[line width=0.48pt] (su60) -- (su62);
    \draw[line width=0.48pt] (su61) -- (su62);
    \draw[line width=0.48pt] (su62) -- (su63);
    \draw[line width=0.48pt] (su63) -- (su64);
    \fill (su1) circle (0.58pt);
    \fill (su2) circle (0.58pt);
    \fill (su3) circle (0.58pt);
    \fill (su4) circle (0.58pt);
    \fill (su5) circle (0.58pt);
    \fill (su6) circle (0.58pt);
    \fill (su7) circle (0.58pt);
    \fill (su8) circle (0.58pt);
    \fill (su9) circle (0.58pt);
    \fill (su10) circle (0.58pt);
    \fill (su11) circle (0.58pt);
    \fill (su12) circle (0.58pt);
    \fill (su13) circle (0.58pt);
    \fill (su14) circle (0.58pt);
    \fill (su15) circle (0.58pt);
    \fill (su16) circle (0.58pt);
    \fill (su17) circle (0.58pt);
    \fill (su18) circle (0.58pt);
    \fill (su19) circle (0.58pt);
    \fill (su20) circle (0.58pt);
    \fill (su21) circle (0.58pt);
    \fill (su22) circle (0.58pt);
    \fill (su23) circle (0.58pt);
    \fill (su24) circle (0.58pt);
    \fill (su25) circle (0.58pt);
    \fill (su26) circle (0.58pt);
    \fill (su27) circle (0.58pt);
    \fill (su28) circle (0.58pt);
    \fill (su29) circle (0.58pt);
    \fill (su30) circle (0.58pt);
    \fill (su31) circle (0.58pt);
    \fill (su32) circle (0.58pt);
    \fill (su33) circle (0.58pt);
    \fill (su34) circle (0.58pt);
    \fill (su35) circle (0.58pt);
    \fill (su36) circle (0.58pt);
    \fill (su37) circle (0.58pt);
    \fill (su38) circle (0.58pt);
    \fill (su39) circle (0.58pt);
    \fill (su40) circle (0.58pt);
    \fill (su41) circle (0.58pt);
    \fill (su42) circle (0.58pt);
    \fill (su43) circle (0.58pt);
    \fill (su44) circle (0.58pt);
    \fill (su45) circle (0.58pt);
    \fill (su46) circle (0.58pt);
    \fill (su47) circle (0.58pt);
    \fill (su48) circle (0.58pt);
    \fill (su49) circle (0.58pt);
    \fill (su50) circle (0.58pt);
    \fill (su51) circle (0.58pt);
    \fill (su52) circle (0.58pt);
    \fill (su53) circle (0.58pt);
    \fill (su54) circle (0.58pt);
    \fill (su55) circle (0.58pt);
    \fill (su56) circle (0.58pt);
    \fill (su57) circle (0.58pt);
    \fill (su58) circle (0.58pt);
    \fill (su59) circle (0.58pt);
    \fill (su60) circle (0.58pt);
    \fill (su61) circle (0.58pt);
    \fill (su62) circle (0.58pt);
    \fill (su63) circle (0.58pt);
    \fill (su64) circle (0.58pt);
    \node at (4.232051,-8.35) {$p=0.85$};
  \end{scope}
  \begin{scope}[xshift=11.4cm]
    \coordinate (su1) at (0.000000,0.000000);
    \coordinate (su2) at (0.500000,-0.866025);
    \coordinate (su3) at (1.366025,-1.366025);
    \coordinate (su4) at (1.366025,-0.366025);
    \coordinate (su5) at (1.866025,0.500000);
    \coordinate (su6) at (2.366025,-0.366025);
    \coordinate (su7) at (3.232051,-0.866025);
    \coordinate (su8) at (3.232051,0.133975);
    \coordinate (su9) at (3.732051,1.000000);
    \coordinate (su10) at (4.232051,0.133975);
    \coordinate (su11) at (5.098076,-0.366025);
    \coordinate (su12) at (5.098076,0.633975);
    \coordinate (su13) at (5.598076,1.500000);
    \coordinate (su14) at (6.098076,0.633975);
    \coordinate (su15) at (6.964102,0.133975);
    \coordinate (su16) at (6.964102,1.133975);
    \coordinate (su17) at (0.500000,-1.866025);
    \coordinate (su18) at (1.000000,-2.732051);
    \coordinate (su19) at (1.866025,-3.232051);
    \coordinate (su20) at (1.866025,-2.232051);
    \coordinate (su21) at (2.366025,-1.366025);
    \coordinate (su22) at (2.866025,-2.232051);
    \coordinate (su23) at (3.732051,-2.732051);
    \coordinate (su24) at (3.732051,-1.732051);
    \coordinate (su25) at (4.232051,-0.866025);
    \coordinate (su26) at (4.732051,-1.732051);
    \coordinate (su27) at (5.598076,-2.232051);
    \coordinate (su28) at (5.598076,-1.232051);
    \coordinate (su29) at (6.098076,-0.366025);
    \coordinate (su30) at (6.598076,-1.232051);
    \coordinate (su31) at (7.464102,-1.732051);
    \coordinate (su32) at (7.464102,-0.732051);
    \coordinate (su33) at (1.000000,-3.732051);
    \coordinate (su34) at (1.500000,-4.598076);
    \coordinate (su35) at (2.366025,-5.098076);
    \coordinate (su36) at (2.366025,-4.098076);
    \coordinate (su37) at (2.866025,-3.232051);
    \coordinate (su38) at (3.366025,-4.098076);
    \coordinate (su39) at (4.232051,-4.598076);
    \coordinate (su40) at (4.232051,-3.598076);
    \coordinate (su41) at (4.732051,-2.732051);
    \coordinate (su42) at (5.232051,-3.598076);
    \coordinate (su43) at (6.098076,-4.098076);
    \coordinate (su44) at (6.098076,-3.098076);
    \coordinate (su45) at (6.598076,-2.232051);
    \coordinate (su46) at (7.098076,-3.098076);
    \coordinate (su47) at (7.964102,-3.598076);
    \coordinate (su48) at (7.964102,-2.598076);
    \coordinate (su49) at (1.500000,-5.598076);
    \coordinate (su50) at (2.000000,-6.464102);
    \coordinate (su51) at (2.866025,-6.964102);
    \coordinate (su52) at (2.866025,-5.964102);
    \coordinate (su53) at (3.366025,-5.098076);
    \coordinate (su54) at (3.866025,-5.964102);
    \coordinate (su55) at (4.732051,-6.464102);
    \coordinate (su56) at (4.732051,-5.464102);
    \coordinate (su57) at (5.232051,-4.598076);
    \coordinate (su58) at (5.732051,-5.464102);
    \coordinate (su59) at (6.598076,-5.964102);
    \coordinate (su60) at (6.598076,-4.964102);
    \coordinate (su61) at (7.098076,-4.098076);
    \coordinate (su62) at (7.598076,-4.964102);
    \coordinate (su63) at (8.464102,-5.464102);
    \coordinate (su64) at (8.464102,-4.464102);
    \draw[line width=0.48pt] (su1) -- (su2);
    \draw[line width=0.48pt] (su2) -- (su3);
    \draw[line width=0.48pt] (su2) -- (su4);
    \draw[line width=0.48pt] (su2) -- (su17);
    \draw[line width=0.48pt] (su3) -- (su4);
    \draw[line width=0.48pt] (su3) -- (su17);
    \draw[line width=0.48pt] (su3) -- (su20);
    \draw[line width=0.48pt] (su4) -- (su5);
    \draw[line width=0.48pt] (su4) -- (su6);
    \draw[line width=0.48pt] (su5) -- (su6);
    \draw[line width=0.48pt] (su6) -- (su7);
    \draw[line width=0.48pt] (su6) -- (su8);
    \draw[line width=0.48pt] (su6) -- (su21);
    \draw[line width=0.48pt] (su7) -- (su8);
    \draw[line width=0.48pt] (su7) -- (su21);
    \draw[line width=0.48pt] (su7) -- (su24);
    \draw[line width=0.48pt] (su7) -- (su25);
    \draw[line width=0.48pt] (su8) -- (su9);
    \draw[line width=0.48pt] (su8) -- (su10);
    \draw[line width=0.48pt] (su9) -- (su10);
    \draw[line width=0.48pt] (su10) -- (su11);
    \draw[line width=0.48pt] (su10) -- (su12);
    \draw[line width=0.48pt] (su10) -- (su25);
    \draw[line width=0.48pt] (su11) -- (su12);
    \draw[line width=0.48pt] (su11) -- (su25);
    \draw[line width=0.48pt] (su11) -- (su28);
    \draw[line width=0.48pt] (su11) -- (su29);
    \draw[line width=0.48pt] (su12) -- (su13);
    \draw[line width=0.48pt] (su12) -- (su14);
    \draw[line width=0.48pt] (su13) -- (su14);
    \draw[line width=0.48pt] (su14) -- (su15);
    \draw[line width=0.48pt] (su14) -- (su16);
    \draw[line width=0.48pt] (su14) -- (su29);
    \draw[line width=0.48pt] (su15) -- (su16);
    \draw[line width=0.48pt] (su15) -- (su29);
    \draw[line width=0.48pt] (su15) -- (su32);
    \draw[line width=0.48pt] (su17) -- (su18);
    \draw[line width=0.48pt] (su18) -- (su19);
    \draw[line width=0.48pt] (su18) -- (su20);
    \draw[line width=0.48pt] (su18) -- (su33);
    \draw[line width=0.48pt] (su19) -- (su20);
    \draw[line width=0.48pt] (su19) -- (su33);
    \draw[line width=0.48pt] (su19) -- (su36);
    \draw[line width=0.48pt] (su19) -- (su37);
    \draw[line width=0.48pt] (su20) -- (su21);
    \draw[line width=0.48pt] (su20) -- (su22);
    \draw[line width=0.48pt] (su21) -- (su22);
    \draw[line width=0.48pt] (su22) -- (su23);
    \draw[line width=0.48pt] (su22) -- (su24);
    \draw[line width=0.48pt] (su22) -- (su37);
    \draw[line width=0.48pt] (su23) -- (su24);
    \draw[line width=0.48pt] (su23) -- (su37);
    \draw[line width=0.48pt] (su23) -- (su40);
    \draw[line width=0.48pt] (su23) -- (su41);
    \draw[line width=0.48pt] (su24) -- (su25);
    \draw[line width=0.48pt] (su24) -- (su26);
    \draw[line width=0.48pt] (su25) -- (su26);
    \draw[line width=0.48pt] (su26) -- (su27);
    \draw[line width=0.48pt] (su26) -- (su28);
    \draw[line width=0.48pt] (su26) -- (su41);
    \draw[line width=0.48pt] (su27) -- (su28);
    \draw[line width=0.48pt] (su27) -- (su41);
    \draw[line width=0.48pt] (su27) -- (su44);
    \draw[line width=0.48pt] (su27) -- (su45);
    \draw[line width=0.48pt] (su28) -- (su29);
    \draw[line width=0.48pt] (su28) -- (su30);
    \draw[line width=0.48pt] (su30) -- (su31);
    \draw[line width=0.48pt] (su30) -- (su32);
    \draw[line width=0.48pt] (su30) -- (su45);
    \draw[line width=0.48pt] (su31) -- (su45);
    \draw[line width=0.48pt] (su31) -- (su48);
    \draw[line width=0.48pt] (su34) -- (su35);
    \draw[line width=0.48pt] (su34) -- (su36);
    \draw[line width=0.48pt] (su34) -- (su49);
    \draw[line width=0.48pt] (su35) -- (su36);
    \draw[line width=0.48pt] (su35) -- (su49);
    \draw[line width=0.48pt] (su35) -- (su52);
    \draw[line width=0.48pt] (su35) -- (su53);
    \draw[line width=0.48pt] (su36) -- (su37);
    \draw[line width=0.48pt] (su36) -- (su38);
    \draw[line width=0.48pt] (su37) -- (su38);
    \draw[line width=0.48pt] (su38) -- (su39);
    \draw[line width=0.48pt] (su38) -- (su40);
    \draw[line width=0.48pt] (su38) -- (su53);
    \draw[line width=0.48pt] (su39) -- (su40);
    \draw[line width=0.48pt] (su39) -- (su53);
    \draw[line width=0.48pt] (su39) -- (su56);
    \draw[line width=0.48pt] (su39) -- (su57);
    \draw[line width=0.48pt] (su40) -- (su41);
    \draw[line width=0.48pt] (su40) -- (su42);
    \draw[line width=0.48pt] (su41) -- (su42);
    \draw[line width=0.48pt] (su42) -- (su43);
    \draw[line width=0.48pt] (su42) -- (su44);
    \draw[line width=0.48pt] (su42) -- (su57);
    \draw[line width=0.48pt] (su43) -- (su44);
    \draw[line width=0.48pt] (su43) -- (su57);
    \draw[line width=0.48pt] (su43) -- (su60);
    \draw[line width=0.48pt] (su43) -- (su61);
    \draw[line width=0.48pt] (su44) -- (su46);
    \draw[line width=0.48pt] (su45) -- (su46);
    \draw[line width=0.48pt] (su46) -- (su47);
    \draw[line width=0.48pt] (su46) -- (su48);
    \draw[line width=0.48pt] (su46) -- (su61);
    \draw[line width=0.48pt] (su47) -- (su48);
    \draw[line width=0.48pt] (su47) -- (su61);
    \draw[line width=0.48pt] (su47) -- (su64);
    \draw[line width=0.48pt] (su49) -- (su50);
    \draw[line width=0.48pt] (su50) -- (su51);
    \draw[line width=0.48pt] (su50) -- (su52);
    \draw[line width=0.48pt] (su52) -- (su53);
    \draw[line width=0.48pt] (su52) -- (su54);
    \draw[line width=0.48pt] (su53) -- (su54);
    \draw[line width=0.48pt] (su54) -- (su55);
    \draw[line width=0.48pt] (su54) -- (su56);
    \draw[line width=0.48pt] (su55) -- (su56);
    \draw[line width=0.48pt] (su56) -- (su57);
    \draw[line width=0.48pt] (su56) -- (su58);
    \draw[line width=0.48pt] (su57) -- (su58);
    \draw[line width=0.48pt] (su58) -- (su59);
    \draw[line width=0.48pt] (su58) -- (su60);
    \draw[line width=0.48pt] (su59) -- (su60);
    \draw[line width=0.48pt] (su60) -- (su61);
    \draw[line width=0.48pt] (su60) -- (su62);
    \draw[line width=0.48pt] (su61) -- (su62);
    \draw[line width=0.48pt] (su62) -- (su63);
    \draw[line width=0.48pt] (su62) -- (su64);
    \draw[line width=0.48pt] (su63) -- (su64);
    \fill (su1) circle (0.58pt);
    \fill (su2) circle (0.58pt);
    \fill (su3) circle (0.58pt);
    \fill (su4) circle (0.58pt);
    \fill (su5) circle (0.58pt);
    \fill (su6) circle (0.58pt);
    \fill (su7) circle (0.58pt);
    \fill (su8) circle (0.58pt);
    \fill (su9) circle (0.58pt);
    \fill (su10) circle (0.58pt);
    \fill (su11) circle (0.58pt);
    \fill (su12) circle (0.58pt);
    \fill (su13) circle (0.58pt);
    \fill (su14) circle (0.58pt);
    \fill (su15) circle (0.58pt);
    \fill (su16) circle (0.58pt);
    \fill (su17) circle (0.58pt);
    \fill (su18) circle (0.58pt);
    \fill (su19) circle (0.58pt);
    \fill (su20) circle (0.58pt);
    \fill (su21) circle (0.58pt);
    \fill (su22) circle (0.58pt);
    \fill (su23) circle (0.58pt);
    \fill (su24) circle (0.58pt);
    \fill (su25) circle (0.58pt);
    \fill (su26) circle (0.58pt);
    \fill (su27) circle (0.58pt);
    \fill (su28) circle (0.58pt);
    \fill (su29) circle (0.58pt);
    \fill (su30) circle (0.58pt);
    \fill (su31) circle (0.58pt);
    \fill (su32) circle (0.58pt);
    \fill (su33) circle (0.58pt);
    \fill (su34) circle (0.58pt);
    \fill (su35) circle (0.58pt);
    \fill (su36) circle (0.58pt);
    \fill (su37) circle (0.58pt);
    \fill (su38) circle (0.58pt);
    \fill (su39) circle (0.58pt);
    \fill (su40) circle (0.58pt);
    \fill (su41) circle (0.58pt);
    \fill (su42) circle (0.58pt);
    \fill (su43) circle (0.58pt);
    \fill (su44) circle (0.58pt);
    \fill (su45) circle (0.58pt);
    \fill (su46) circle (0.58pt);
    \fill (su47) circle (0.58pt);
    \fill (su48) circle (0.58pt);
    \fill (su49) circle (0.58pt);
    \fill (su50) circle (0.58pt);
    \fill (su51) circle (0.58pt);
    \fill (su52) circle (0.58pt);
    \fill (su53) circle (0.58pt);
    \fill (su54) circle (0.58pt);
    \fill (su55) circle (0.58pt);
    \fill (su56) circle (0.58pt);
    \fill (su57) circle (0.58pt);
    \fill (su58) circle (0.58pt);
    \fill (su59) circle (0.58pt);
    \fill (su60) circle (0.58pt);
    \fill (su61) circle (0.58pt);
    \fill (su62) circle (0.58pt);
    \fill (su63) circle (0.58pt);
    \fill (su64) circle (0.58pt);
    \node at (4.232051,-8.35) {$p=0.95$};
  \end{scope}
\end{tikzpicture}

%% file: atlas_thermo/atlas.tex
\section{Atlas of programs and thermodynamic data}
\label{app:atlas}

This appendix collects, for each of the 36 lattices treated in Section~\ref{examples}, the triangular-emulator program and the corresponding structural and thermodynamic data, organised into three families --- Archimedean (Appendix~\ref{app:atlas-archimedean}), $2$-uniform Galebach (Appendix~\ref{app:atlas-galebach}), and pentagonal RCSR layer nets (Appendix~\ref{app:atlas-pentagonal}) --- and within each family by increasing supercell size $L$. Each lattice is presented as an individually captioned and labeled figure with a uniform layout: the lattice identifier and a thumbnail, the structural parameters (host, program type, supercell size $L$, parent-matrix dimension $|\mathbb{W}|=6L^2$, number of supercell sites $n_v$ and bonds $n_l$), the thermodynamic parameters (critical temperature $T_c$, critical energy density $\varepsilon_c$, and amplitudes $A$, $B$ for the ferromagnetic branch; $T_c^{\rm AF}$, $\varepsilon_c^{\rm AF}$, $s_0^{\rm AF}$ for the antiferromagnetic branch; and the dual critical temperature $T_c^{\rm D}$), the binary program mask(s), and the free energy, entropy, energy, and specific heat versus temperature for both couplings. Here $\varepsilon_c^{\rm AF}=\varepsilon^{\rm AF}(T_c^{\rm AF})$ denotes the energy density at the antiferromagnetic transition, the counterpart of the ferromagnetic $\varepsilon_c$; it is defined only when a finite-temperature antiferromagnetic transition exists (``---'' otherwise) and should not be confused with the ground-state energy $\varepsilon_0^{\rm AF}$ reported in Tables~\ref{tab:archimedean}--\ref{tab:pentagonal-Tc}. As in the main text, the masks are displayed with the index $i$ labelling rows and $j$ labelling columns. Numerical thermodynamic data for the same lattices are also collected in the summary Tables~\ref{tab:archimedean}--\ref{tab:pentagonal-Tc} of Section~\ref{examples}, whose last column refers to the atlas figure of each lattice: Figs.~\ref{fig:atlas-t1011}--\ref{fig:atlas-t1004} for the Archimedean lattices, Figs.~\ref{fig:atlas-t2004}--\ref{fig:atlas-t2013} for the $2$-uniform lattices, and Figs.~\ref{fig:atlas-mcm}--\ref{fig:atlas-usm-d} for the pentagonal lattices.

\subsection{Archimedean lattices}
\label{app:atlas-archimedean}

\begin{atlascard}{$(3,3,3,3,3,3)$}{table_icons_static/t1011.png}{\carddataarch{1}{6}{3.640956907}{-2.000000244}{0.49906938}{-0.30673}{none}{\NA}{0.323}\cardtype{site}\cardcounts{1}{3}}
\atlasplotobject{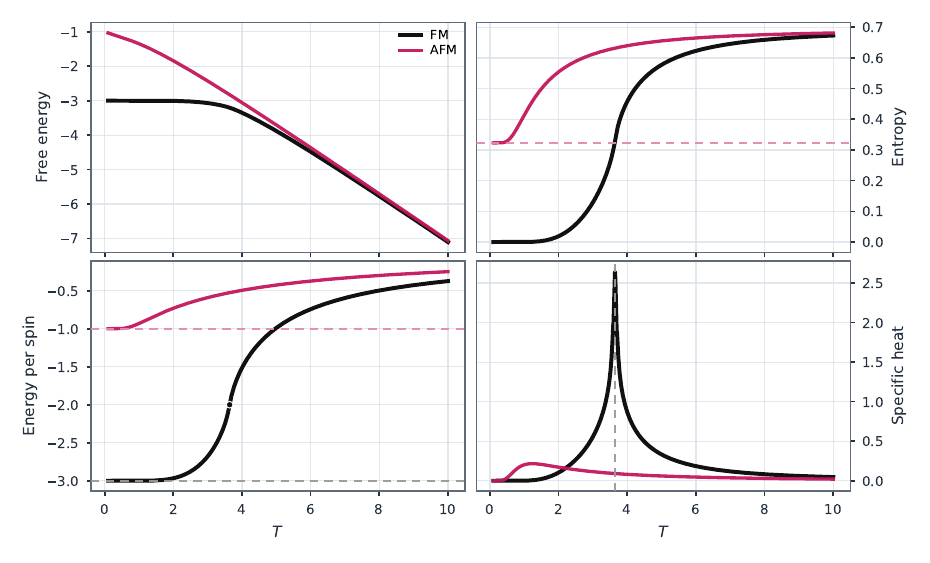}

\vspace{0.6em}
\hbox to \linewidth{\hss
\masktilemat{{\color{maskOne}w}}{1}{{1}}%
\hss}
\end{atlascard}

\begin{atlascard}{$(4,4,4,4)$}{table_icons_static/t1005.png}{\carddataarch{1}{6}{2.269185314}{-1.414212787}{0.49453859}{-0.30632}{\sameasfm}{-1.414212787}{-0}\cardtype{bond}\cardcounts{1}{2}}
\atlasplotobject{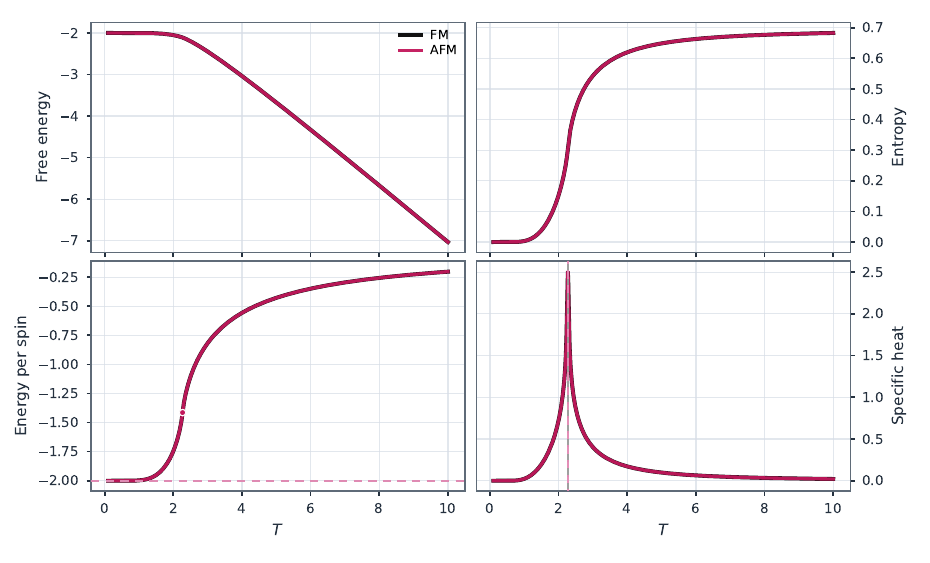}

\vspace{0.6em}
\hbox to \linewidth{%
  \hss
  \masktilemat{{\color{maskOne}w^{(1)}}}{1}{{1}}
  \hss
  \masktilemat{{\color{maskTwo}w^{(2)}}}{1}{{1}}
  \hss
  \masktilemat{{\color{maskThree}w^{(3)}}}{1}{{0}}%
  \hss
}
\end{atlascard}

\begin{atlascard}{$(6,6,6)$}{table_icons_static/t1001.png}{\carddataarch{2}{24}{1.518651435}{-1.154699664}{0.47810638}{-0.30457}{\sameasfm}{-1.154699664}{0}\cardtype{bond}\cardcounts{4}{6}}
\atlasplotobject{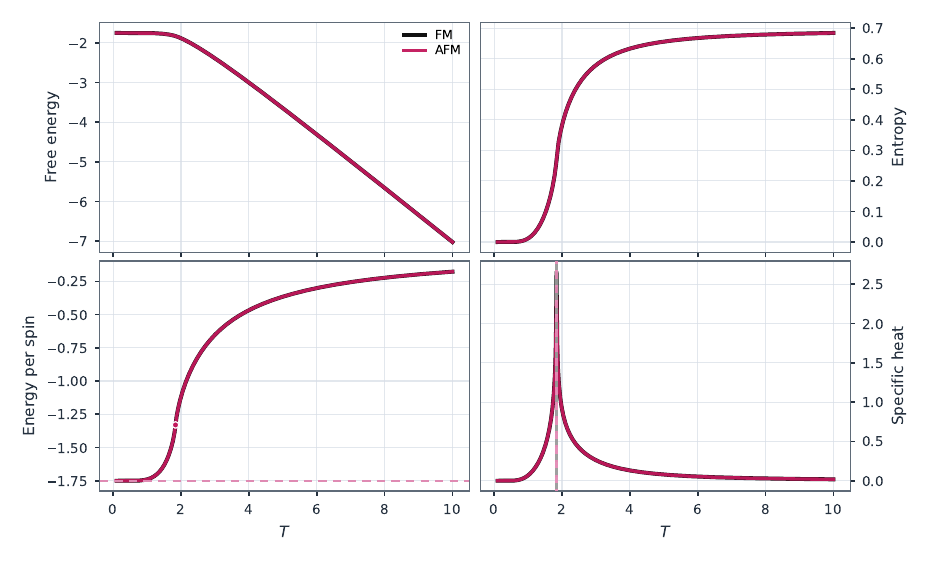}

\vspace{0.6em}
\hbox to \linewidth{%
  \hss
  \masktilemat{{\color{maskOne}w^{(1)}}}{2}{{1,1},{1,1}}
  \hss
  \masktilemat{{\color{maskTwo}w^{(2)}}}{2}{{0,1},{1,0}}
  \hss
  \masktilemat{{\color{maskThree}w^{(3)}}}{2}{{0,0},{0,0}}%
  \hss
}
\end{atlascard}

\begin{atlascard}{$(3,6,3,6)$}{table_icons_static/t1007.png}{\carddataarch{2}{24}{2.143319440}{-1.488032937}{0.48006157}{-0.29799}{none}{\NA}{0.502}\cardtype{bond}\cardcounts{3}{6}}
\atlasplotobject{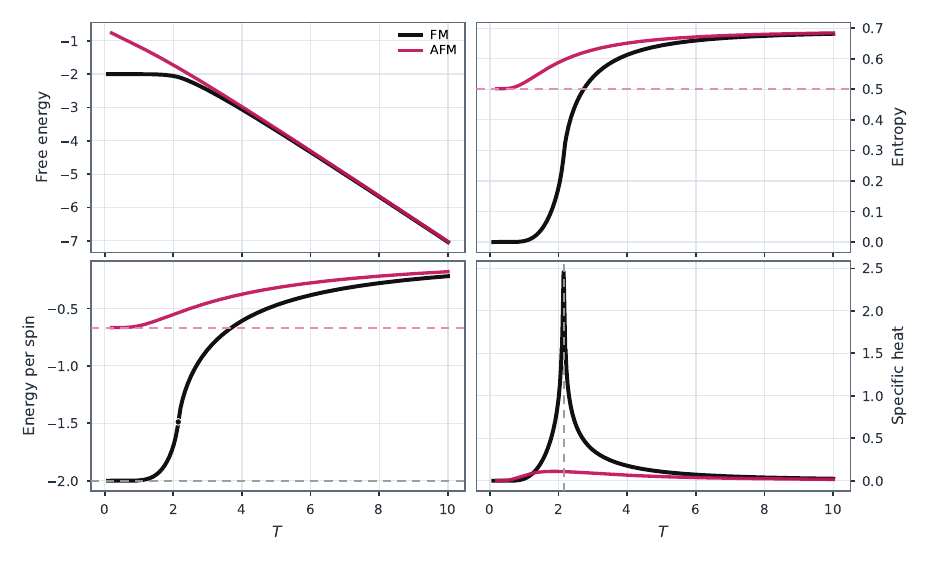}

\vspace{0.6em}
\hbox to \linewidth{%
  \hss
  \masktilemat{{\color{maskOne}w^{(1)}}}{2}{{1,0},{1,0}}
  \hss
  \masktilemat{{\color{maskTwo}w^{(2)}}}{2}{{1,1},{0,0}}
  \hss
  \masktilemat{{\color{maskThree}w^{(3)}}}{2}{{0,1},{1,0}}%
  \hss
}
\end{atlascard}

\begin{atlascard}{$(3,3,3,4,4)$}{table_icons_static/t1008.png}{\carddataarch{2}{24}{2.885390082}{-1.758213445}{0.47791545}{-0.26540}{none}{\NA}{0}\cardtype{bond}\cardcounts{4}{10}}
\atlasplotobject{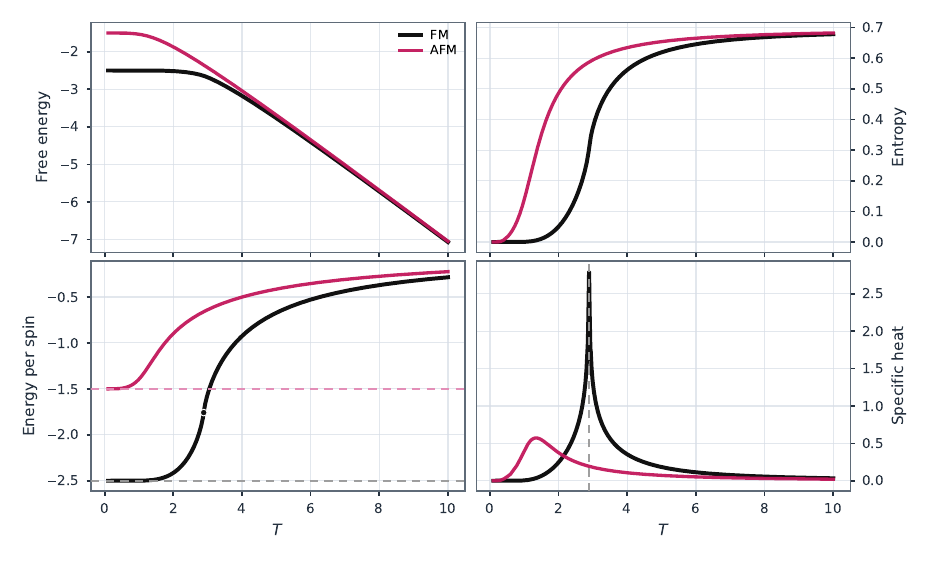}

\vspace{0.6em}
\hbox to \linewidth{%
  \hss
  \masktilemat{{\color{maskOne}w^{(1)}}}{2}{{1,1},{1,1}}
  \hss
  \masktilemat{{\color{maskTwo}w^{(2)}}}{2}{{1,1},{1,1}}
  \hss
  \masktilemat{{\color{maskThree}w^{(3)}}}{2}{{1,1},{0,0}}%
  \hss
}
\end{atlascard}

\begin{atlascard}{$(4,8,8)$}{table_icons_static/t44_488.png}{\carddataarch{4}{96}{1.438695454}{-1.20731}{0.43867371}{-0.25000}{\sameasfm}{-1.20731}{0}\cardtype{bond}\cardcounts{16}{24}}
\atlasplotobject{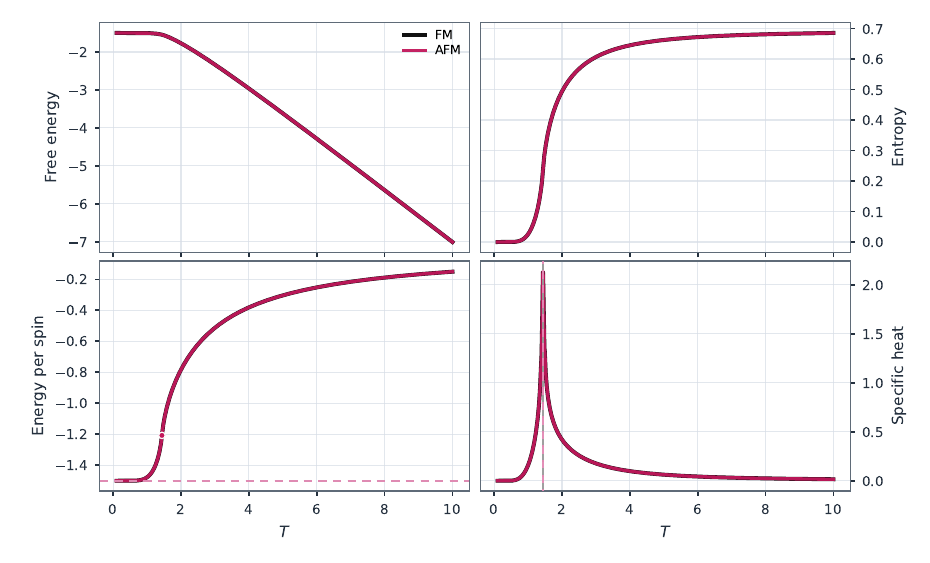}

\vspace{0.6em}
\hbox to \linewidth{%
  \hss
  \masktilemat{{\color{maskOne}w^{(1)}}}{4}{%
    {1,1,1,1},{1,1,1,1},{1,1,1,1},{1,1,1,1}}
  \hss
  \masktilemat{{\color{maskTwo}w^{(2)}}}{4}{%
    {0,1,0,1},{0,1,0,1},{1,0,1,0},{1,0,1,0}}
  \hss
  \masktilemat{{\color{maskThree}w^{(3)}}}{4}{%
    {0,0,0,0},{0,0,0,0},{0,0,0,0},{0,0,0,0}}%
  \hss
}
\end{atlascard}

\begin{atlascard}{$(3,3,4,3,4)$}{table_icons_static/t1009.png}{\carddataarch{4}{96}{2.926261573}{-1.724927185}{0.49484450}{-0.30308}{1.261938877}{-1.264705315}{-0.00000006}\cardtype{bond}\cardcounts{16}{40}}
\atlasplotobject{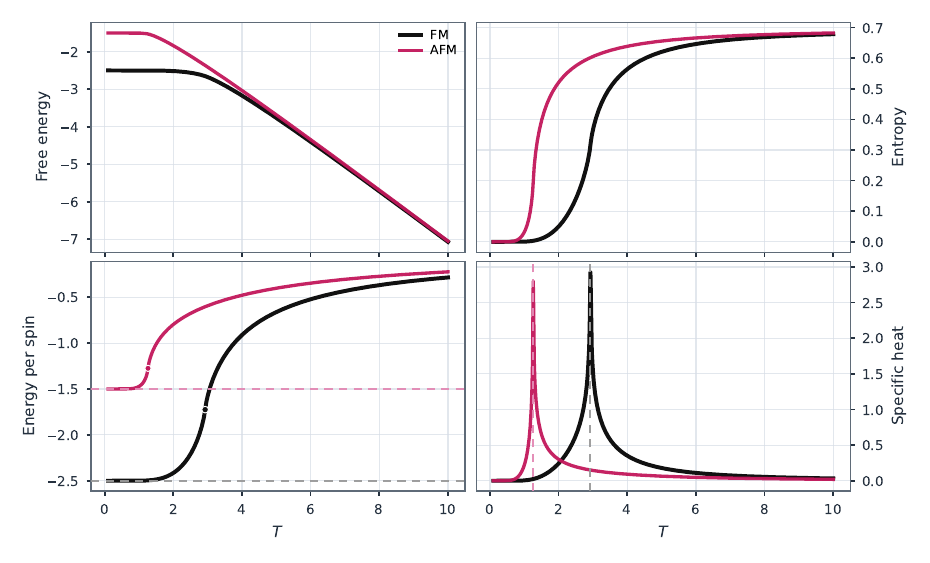}

\vspace{0.6em}
\hbox to \linewidth{%
  \hss
  \masktilemat{{\color{maskOne}w^{(1)}}}{4}{%
    {1,1,1,1},{1,1,1,1},{1,1,1,1},{1,1,1,1}}
  \hss
  \masktilemat{{\color{maskTwo}w^{(2)}}}{4}{%
    {1,1,1,0},{1,0,1,1},{1,1,1,0},{1,0,1,1}}
  \hss
  \masktilemat{{\color{maskThree}w^{(3)}}}{4}{%
    {1,1,0,1},{0,1,1,1},{1,1,0,1},{0,1,1,1}}%
  \hss
}
\end{atlascard}

\begin{atlascard}{$(3,4,6,4)$}{table_icons_static/t1006.png}{\carddataarch{6}{216}{2.143319440}{-1.504824610}{0.44790355}{-0.21705}{none}{\NA}{0.05378546}\cardtype{bond}\cardcounts{36}{72}}
\atlasplotobject{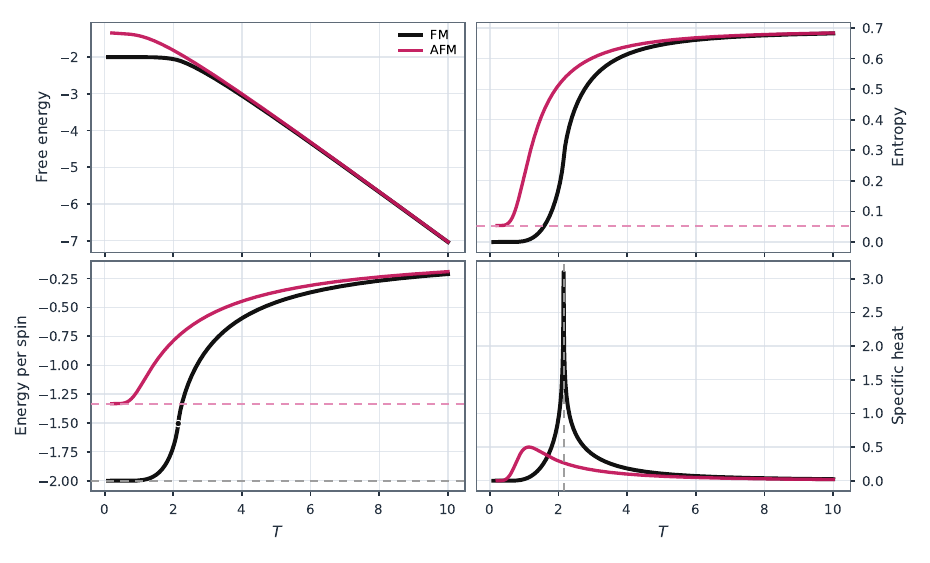}

\vspace{0.6em}
\hbox to \linewidth{%
  \hss
  \masktilemat{{\color{maskOne}w^{(1)}}}{6}{%
    {0,1,1,0,1,1},%
    {1,1,0,1,1,0},%
    {1,0,1,1,0,1},%
    {0,1,1,0,1,1},%
    {1,1,0,1,1,0},%
    {1,0,1,1,0,1}}
  \hss
  \masktilemat{{\color{maskTwo}w^{(2)}}}{6}{%
    {1,1,1,1,1,1},%
    {1,1,1,1,1,1},%
    {1,1,1,1,1,1},%
    {1,1,1,1,1,1},%
    {1,1,1,1,1,1},%
    {1,1,1,1,1,1}}
  \hss
  \masktilemat{{\color{maskThree}w^{(3)}}}{6}{%
    {1,1,0,1,1,0},%
    {0,0,0,0,0,0},%
    {0,1,1,0,1,1},%
    {0,0,0,0,0,0},%
    {1,0,1,1,0,1},%
    {0,0,0,0,0,0}}%
  \hss
}
\end{atlascard}

\begin{atlascard}{$(3,3,3,3,6)$}{table_icons_static/t1010.png}{\carddataarch{7}{294}{2.785843005}{-1.828066616}{0.46346373}{-0.24838}{none}{\NA}{0.054}\cardtype{site}\cardcounts{42}{105}}
\atlasplotobject{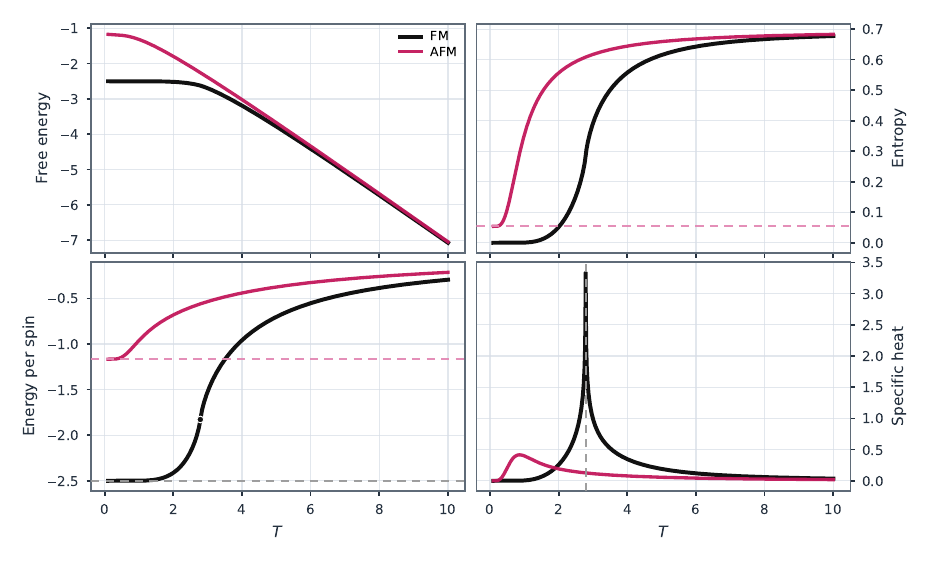}

\vspace{0.6em}
\hbox to \linewidth{\hss
\masktilemat{{\color{maskOne}w}}{7}{%
{1,0,1,1,1,1,1},%
{1,1,1,1,1,0,1},%
{1,1,0,1,1,1,1},%
{1,1,1,1,1,1,0},%
{1,1,1,0,1,1,1},%
{0,1,1,1,1,1,1},%
{1,1,1,1,0,1,1}}%
\hss}
\end{atlascard}

\begin{atlascard}{$(4,6,12)$}{table_icons_static/t1003.png}{\carddataarch{8}{384}{1.389834391}{-1.245633247}{0.40404504}{-0.20200}{\sameasfm}{-1.245633247}{0}\cardtype{bond}\cardcounts{48}{72}}
\atlasplotobject{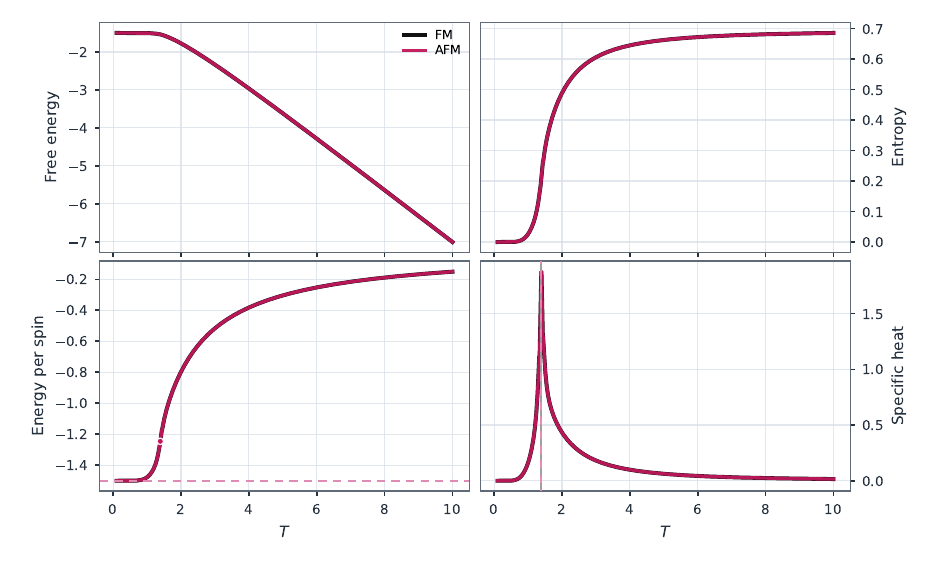}

\vspace{0.6em}
\hbox to \linewidth{%
  \hss
  \masktilemat{{\color{maskOne}w^{(1)}}}{8}{%
    {1,1,1,1,1,1,1,1},%
    {1,1,0,0,1,1,0,0},%
    {1,1,0,0,1,1,0,0},%
    {1,1,0,0,1,1,0,0},%
    {1,1,1,1,1,1,1,1},%
    {0,0,1,1,0,0,1,1},%
    {0,0,1,1,0,0,1,1},%
    {0,0,1,1,0,0,1,1}}
  \hss
  \masktilemat{{\color{maskTwo}w^{(2)}}}{8}{%
    {1,1,0,1,1,1,0,1},%
    {0,1,1,1,0,1,1,1},%
    {1,0,0,0,1,0,0,0},%
    {1,0,0,0,1,0,0,0},%
    {0,1,1,1,0,1,1,1},%
    {1,1,0,1,1,1,0,1},%
    {0,0,1,0,0,0,1,0},%
    {0,0,1,0,0,0,1,0}}
  \hss
  \masktilemat{{\color{maskThree}w^{(3)}}}{8}{%
    {0,0,0,0,0,0,0,0},%
    {0,0,0,0,0,0,0,0},%
    {0,0,0,0,0,0,0,0},%
    {0,0,0,0,0,0,0,0},%
    {0,0,0,0,0,0,0,0},%
    {0,0,0,0,0,0,0,0},%
    {0,0,0,0,0,0,0,0},%
    {0,0,0,0,0,0,0,0}}%
  \hss
}
\end{atlascard}

\begin{atlascard}{$(3,12,12)$}{table_icons_static/t1004.png}{\carddataarch{12}{864}{1.231511701}{-1.312795364}{0.35600372}{-0.184119}{none}{\NA}{0.25116515}\cardtype{site}\cardcounts{72}{108}}
\atlasplotobject{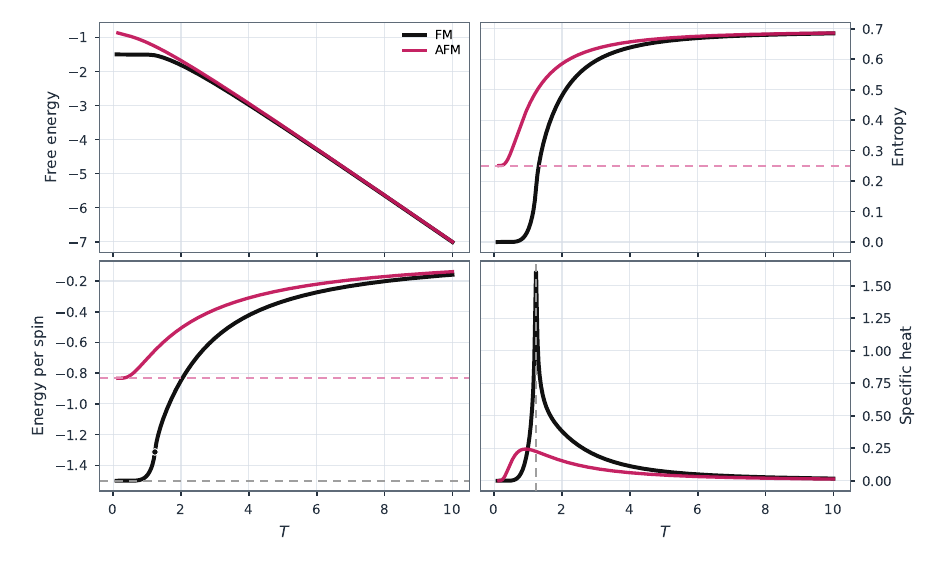}

\vspace{0.6em}
\hbox to \linewidth{\hss
\masktilemat{{\color{maskOne}w}}{12}{%
{1,1,0,1,1,0,1,1,0,1,1,0},%
{1,0,0,1,0,0,1,0,0,1,0,0},%
{1,0,0,1,0,0,1,0,0,1,0,0},%
{1,0,1,1,0,1,1,0,1,1,0,1},%
{1,1,0,1,1,0,1,1,0,1,1,0},%
{1,0,0,1,0,0,1,0,0,1,0,0},%
{1,0,0,1,0,0,1,0,0,1,0,0},%
{1,0,1,1,0,1,1,0,1,1,0,1},%
{1,1,0,1,1,0,1,1,0,1,1,0},%
{1,0,0,1,0,0,1,0,0,1,0,0},%
{1,0,0,1,0,0,1,0,0,1,0,0},%
{1,0,1,1,0,1,1,0,1,1,0,1}}%
\hss}
\end{atlascard}

\clearpage
\subsection{$2$-uniform Galebach lattices}
\label{app:atlas-galebach}

\begin{atlascard}{$t2004$}{table_icons_static/t2004.png}{\carddatastruct{3}{54}{9}{21}{bond}\cardphysicsgeneric{2.667641963}{-1.655059514}{0.467103485}{none}{\NA}{0}}
\atlasplotobject{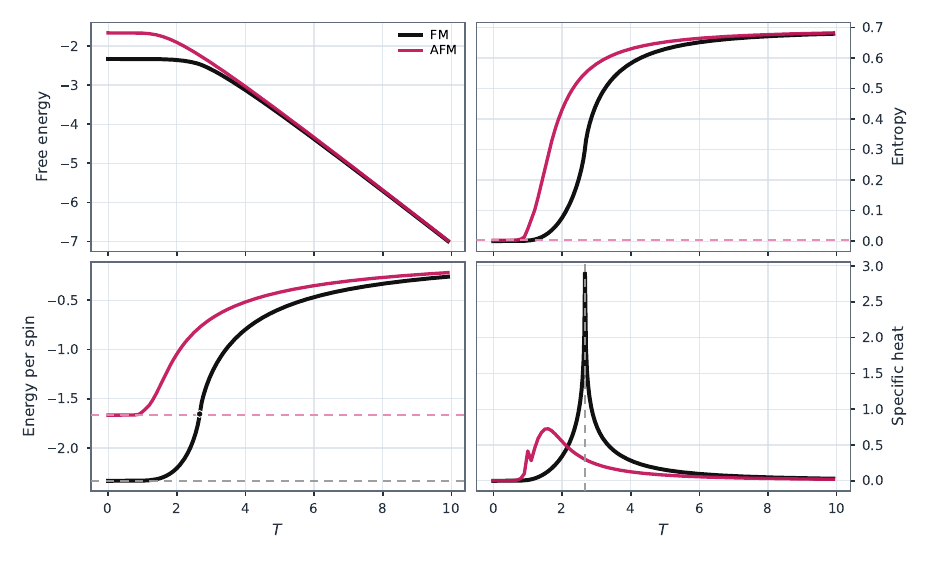}

\vspace{0.6em}
\hbox to \linewidth{%
  \hss
  \masktilemat{{\color{maskOne}w^{(1)}}}{3}{{1,1,1},{1,1,1},{1,1,1}}
  \hss
  \masktilemat{{\color{maskTwo}w^{(2)}}}{3}{{1,1,1},{1,1,1},{1,1,1}}
  \hss
  \masktilemat{{\color{maskThree}w^{(3)}}}{3}{{1,1,1},{0,0,0},{0,0,0}}%
  \hss
}
\end{atlascard}

\begin{atlascard}{$t2010$}{table_icons_static/t2010.png}{\carddatastruct{3}{54}{7}{15}{site}\cardphysicsgeneric{2.202025850}{-1.680149827}{0.38230442}{1.324375296}{-1.047040907}{0.09902105}}
\atlasplotobject{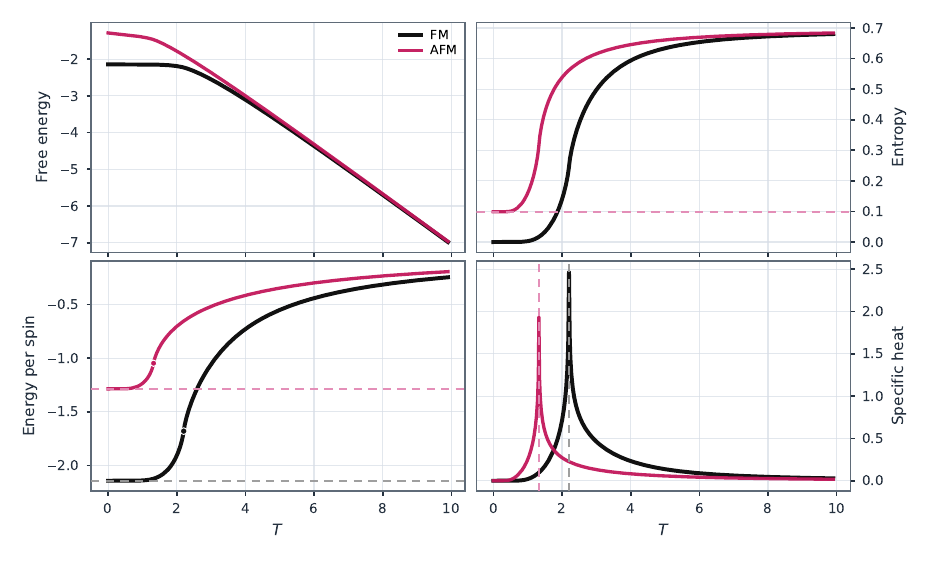}

\vspace{0.6em}
\hbox to \linewidth{\hss
\masktilemat{w}{3}{{1,1,1},{1,0,1},{1,1,0}}%
\hss}
\end{atlascard}

\begin{atlascard}{$t2014$}{table_icons_static/t2014.png}{\carddatastruct{3}{54}{9}{24}{bond}\cardphysicsgeneric{3.118520977}{-1.858494172}{0.46991309}{none}{\NA}{0}}
\atlasplotobject{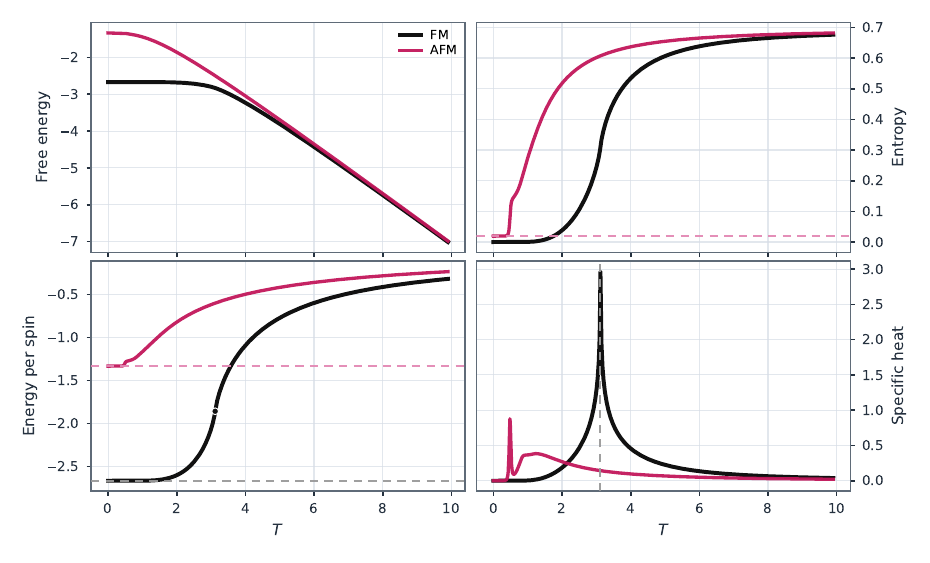}

\vspace{0.6em}
\hbox to \linewidth{%
  \hss
  \masktilemat{{\color{maskOne}w^{(1)}}}{3}{{1,1,1},{1,1,1},{1,1,1}}
  \hss
  \masktilemat{{\color{maskTwo}w^{(2)}}}{3}{{1,1,0},{1,1,0},{1,1,0}}
  \hss
  \masktilemat{{\color{maskThree}w^{(3)}}}{3}{{1,1,1},{1,1,1},{1,1,1}}%
  \hss
}
\end{atlascard}

\begin{atlascard}{$t2019$}{table_icons_static/t2019.png}{\carddatastruct{3}{54}{8}{21}{site}\cardphysicsgeneric{2.985743598}{-1.886598362}{0.46208195}{1.078600300}{-0.979233164}{0.17338044}}
\atlasplotobject{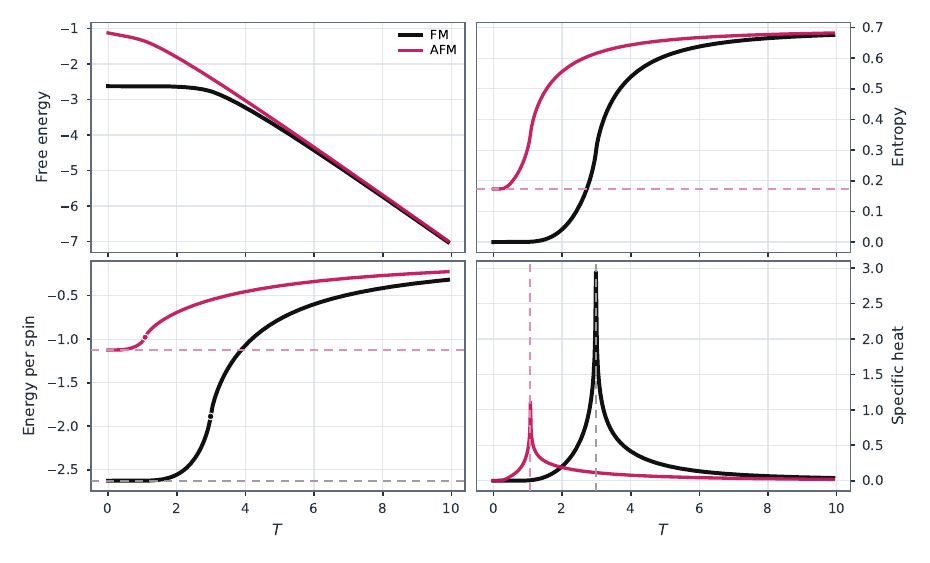}

\vspace{0.6em}
\hbox to \linewidth{\hss
\masktilemat{w}{3}{{1,1,1},{1,0,1},{1,1,1}}%
\hss}
\end{atlascard}

\begin{atlascard}{$t2003$}{table_icons_static/t2003.png}{\carddatastruct{4}{96}{16}{36}{bond}\cardphysicsgeneric{2.563814512}{-1.598037866}{0.460647212}{none}{\NA}{0}}
\atlasplotobject{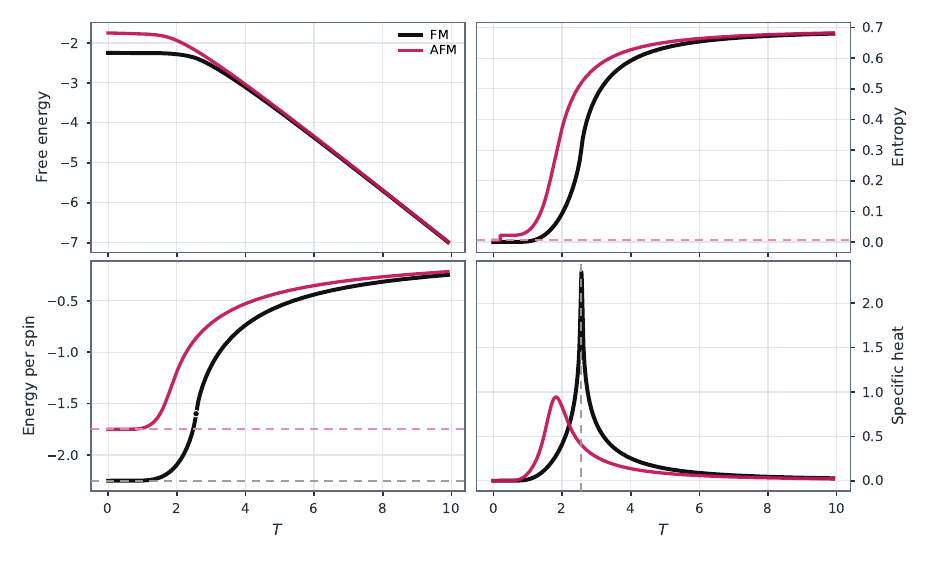}

\vspace{0.6em}
\hbox to \linewidth{%
  \hss
  \masktilemat{{\color{maskOne}w^{(1)}}}{4}{{1,1,1,1},{1,1,1,1},{1,1,1,1},{1,1,1,1}}
  \hss
  \masktilemat{{\color{maskTwo}w^{(2)}}}{4}{{1,1,1,1},{1,1,1,1},{1,1,1,1},{1,1,1,1}}
  \hss
  \masktilemat{{\color{maskThree}w^{(3)}}}{4}{{1,1,1,1},{0,0,0,0},{0,0,0,0},{0,0,0,0}}%
  \hss
}
\end{atlascard}

\begin{atlascard}{$t2011$}{table_icons_static/t2011.png}{\carddatastruct{4}{96}{12}{24}{site}\cardphysicsgeneric{2.083724154}{-1.532224196}{0.44828951}{1.181950572}{-1.079180}{0}}
\atlasplotobject{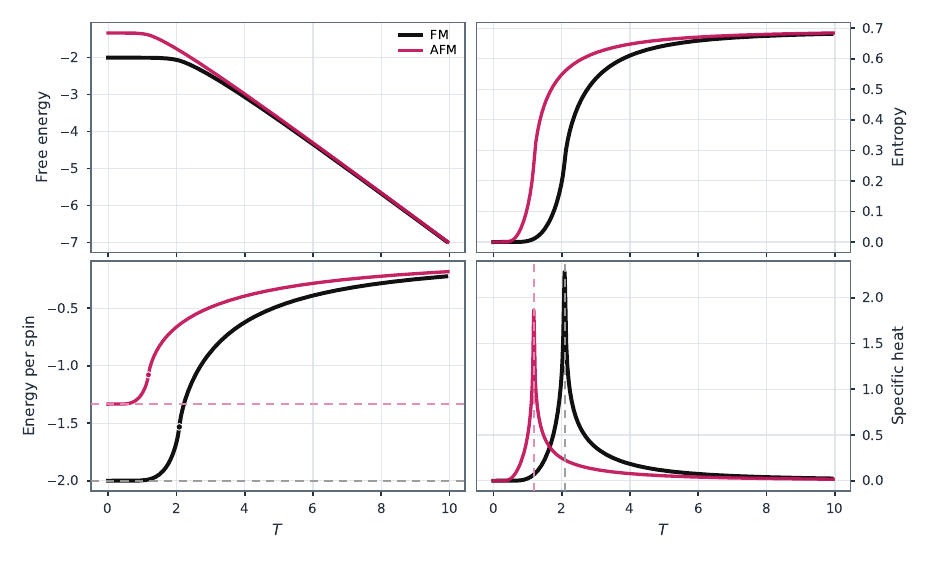}

\vspace{0.6em}
\hbox to \linewidth{\hss
\masktilemat{w}{4}{{1,0,1,1},{1,1,0,1},{1,1,1,0},{0,1,1,1}}%
\hss}
\end{atlascard}

\begin{atlascard}{$t2015$}{table_icons_static/t2015.png}{\carddatastruct{4}{96}{16}{44}{bond}\cardphysicsgeneric{3.241613389}{-1.904323730}{0.465757612}{none}{\NA}{0}}
\atlasplotobject{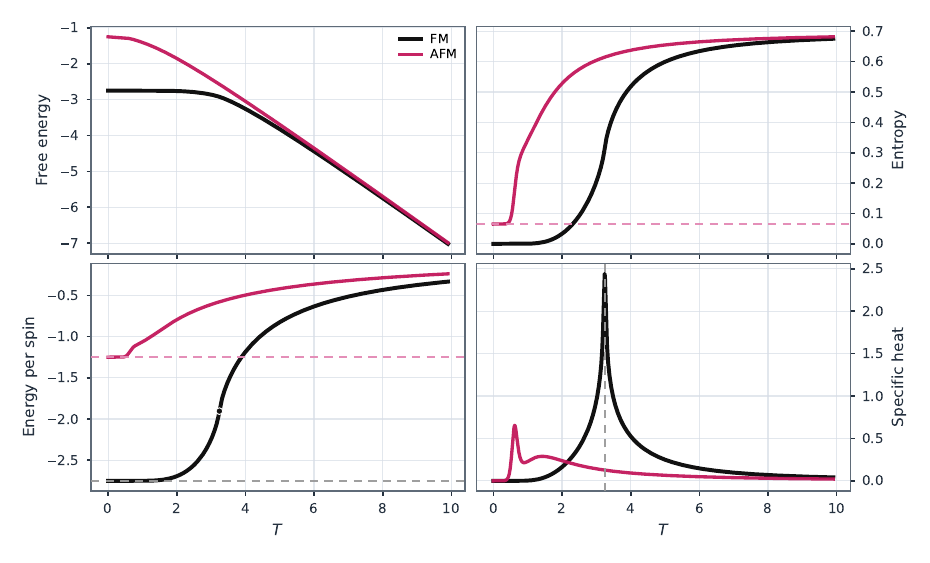}

\vspace{0.6em}
\hbox to \linewidth{%
  \hss
  \masktilemat{{\color{maskOne}w^{(1)}}}{4}{{1,1,1,1},{1,1,1,1},{1,1,1,1},{1,1,1,1}}
  \hss
  \masktilemat{{\color{maskTwo}w^{(2)}}}{4}{{0,1,1,1},{0,1,1,1},{0,1,1,1},{0,1,1,1}}
  \hss
  \masktilemat{{\color{maskThree}w^{(3)}}}{4}{{1,1,1,1},{1,1,1,1},{1,1,1,1},{1,1,1,1}}%
  \hss
}
\end{atlascard}

\begin{atlascard}{$t2017$}{table_icons_static/t2017.png}{\carddatastruct{4}{96}{16}{40}{bond}\cardphysicsgeneric{2.910190121}{-1.737420824}{0.48932689}{1.194238932}{-1.307265056}{0.00000000}}
\atlasplotobject{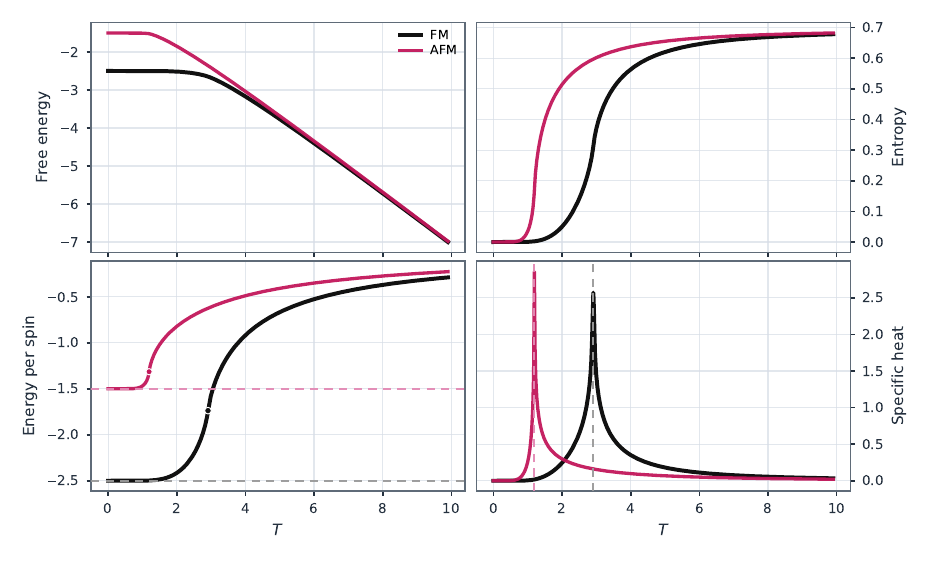}

\vspace{0.6em}
\hbox to \linewidth{%
  \hss
  \masktilemat{{\color{maskOne}w^{(1)}}}{4}{{1,1,1,1},{1,1,1,1},{1,1,1,1},{1,1,1,1}}
  \hss
  \masktilemat{{\color{maskTwo}w^{(2)}}}{4}{{1,0,1,1},{1,1,0,1},{1,0,1,1},{1,1,0,1}}
  \hss
  \masktilemat{{\color{maskThree}w^{(3)}}}{4}{{0,1,1,0},{1,1,1,1},{0,1,1,0},{1,1,1,1}}%
  \hss
}
\end{atlascard}

\begin{atlascard}{$t2012$}{table_icons_static/t2012.png}{\carddatastruct{5}{150}{20}{45}{site}\cardphysicsgeneric{2.383122015}{-1.722165274}{0.42500728}{none}{\NA}{0}}
\atlasplotobject{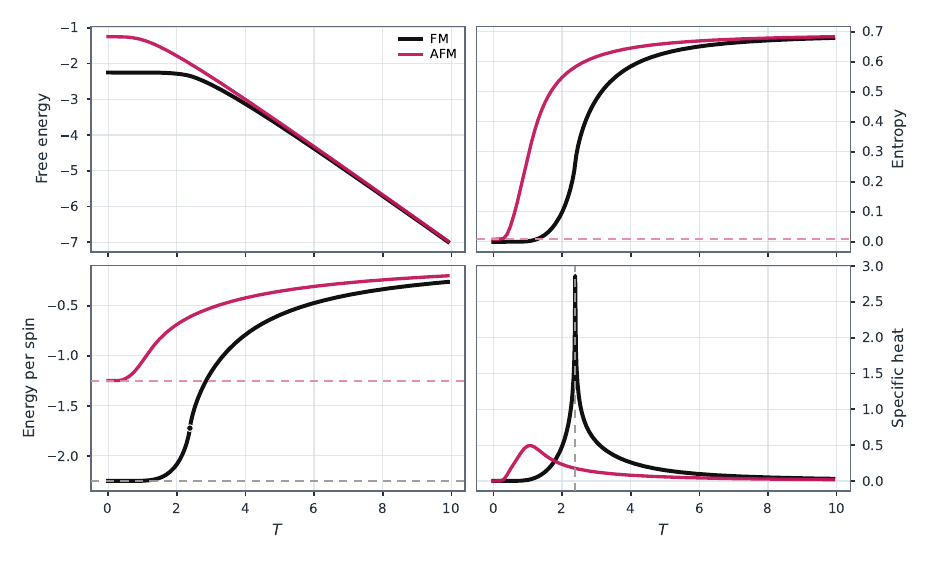}

\vspace{0.6em}
\hbox to \linewidth{\hss
\masktilemat{w}{5}{{1,1,1,1,0},{0,1,1,1,1},{1,0,1,1,1},{1,1,0,1,1},{1,1,1,0,1}}%
\hss}
\end{atlascard}

\begin{atlascard}{$t2006$}{table_icons_static/t2006.png}{\carddatastruct{6}{216}{30}{60}{bond}\cardphysicsgeneric{2.177876317}{-1.471592}{0.478520}{none}{\NA}{0.1385}}
\atlasplotobject{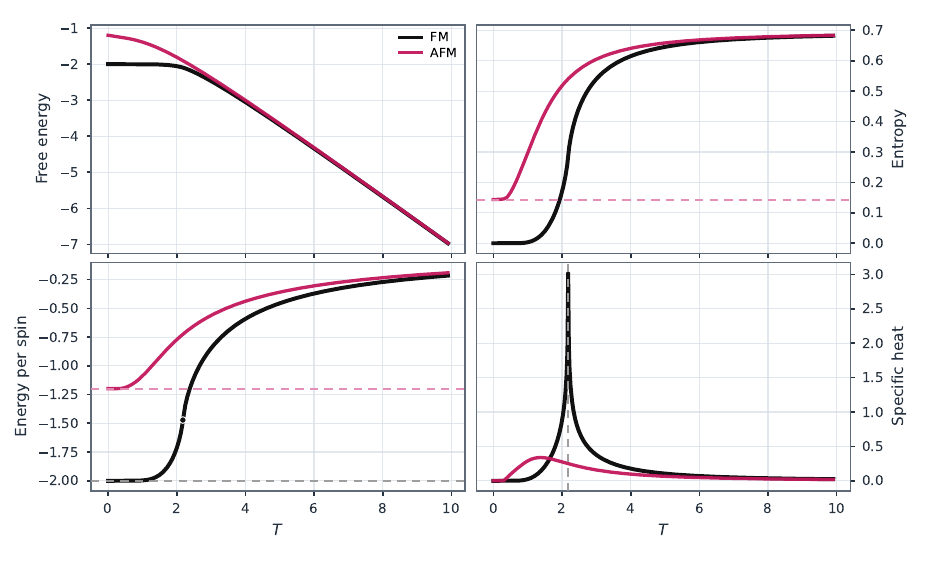}

\vspace{0.6em}
\hbox to \linewidth{%
  \hss
  \masktilemat{{\color{maskOne}w^{(1)}}}{6}{{1, 0, 0, 1, 0, 0}, {1, 0, 0, 1, 0, 0}, {0, 1, 0, 0, 1, 0}, {0, 1, 0,
   0, 1, 0}, {0, 0, 1, 0, 0, 1}, {0, 0, 1, 0, 0, 1}}
  \hss
  \masktilemat{{\color{maskTwo}w^{(2)}}}{6}{{0, 0, 1, 0, 0, 1}, {1, 1, 1, 1, 1, 1}, {1, 0, 0, 1, 0, 0}, {1, 1, 1,
   1, 1, 1}, {0, 1, 0, 0, 1, 0}, {1, 1, 1, 1, 1, 1}}
  \hss
  \masktilemat{{\color{maskThree}w^{(3)}}}{6}{{1,0,1,1,0,1},{0,1,1,0,1,1},{1,1,0,1,1,0},{1,0,1,1,0,1},{0,1,1,0,1,1},{1,1,0,1,1,0}}%
  \hss
}
\end{atlascard}

\begin{atlascard}{$t2007$}{table_icons_static/t2007.png}{\carddatastruct{6}{216}{30}{60}{bond}\cardphysicsgeneric{2.171709941}{-1.477304}{0.47286218}{none}{\NA}{0.1836}}
\atlasplotobject{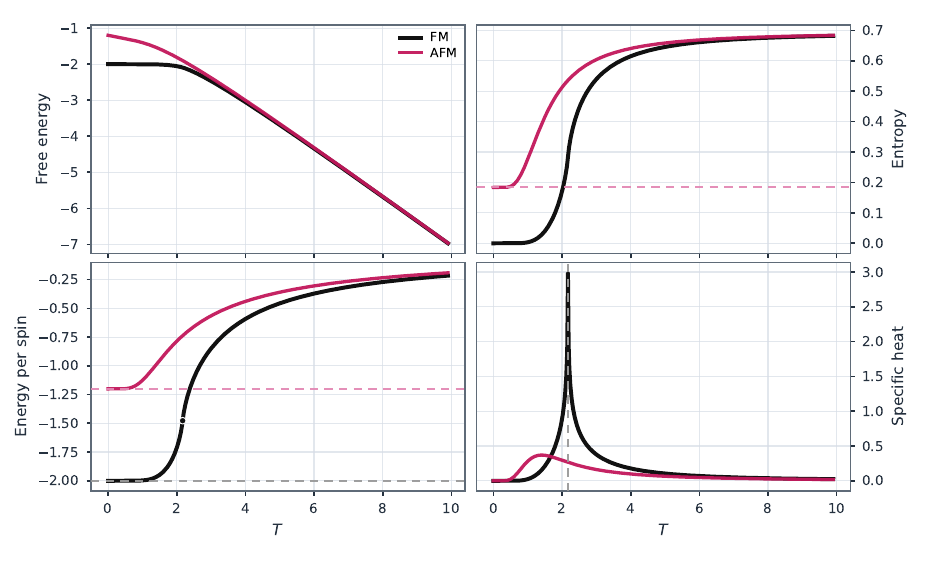}

\vspace{0.6em}
\hbox to \linewidth{%
  \hss
  \masktilemat{{\color{maskOne}w^{(1)}}}{6}{{0, 1, 0, 1, 0, 0}, {0, 0, 0, 1, 0, 1}, {0, 1, 0, 0, 0, 1}, {0, 1, 0,
   1, 0, 0}, {0, 0, 0, 1, 0, 1}, {0, 1, 0, 0, 0, 1}}
  \hss
  \masktilemat{{\color{maskTwo}w^{(2)}}}{6}{{1, 1, 1, 0, 0, 1}, {0, 1, 1, 1, 1, 0}, {1, 0, 0, 1, 1, 1}, {1, 1, 1,
   0, 0, 1}, {0, 1, 1, 1, 1, 0}, {1, 0, 0, 1, 1, 1}}
  \hss
  \masktilemat{{\color{maskThree}w^{(3)}}}{6}{{1,0,1,1,0,1},{0,1,1,0,1,1},{1,1,0,1,1,0},{1,0,1,1,0,1},{0,1,1,0,1,1},{1,1,0,1,1,0}}%
  \hss
}
\end{atlascard}

\begin{atlascard}{$t2018$}{table_icons_static/t2018.png}{\carddatastruct{7}{294}{49}{126}{bond}\cardphysicsgeneric{3.022669833}{-1.769554274}{0.48897101}{none}{\NA}{0.14537675}}
\atlasplotobject{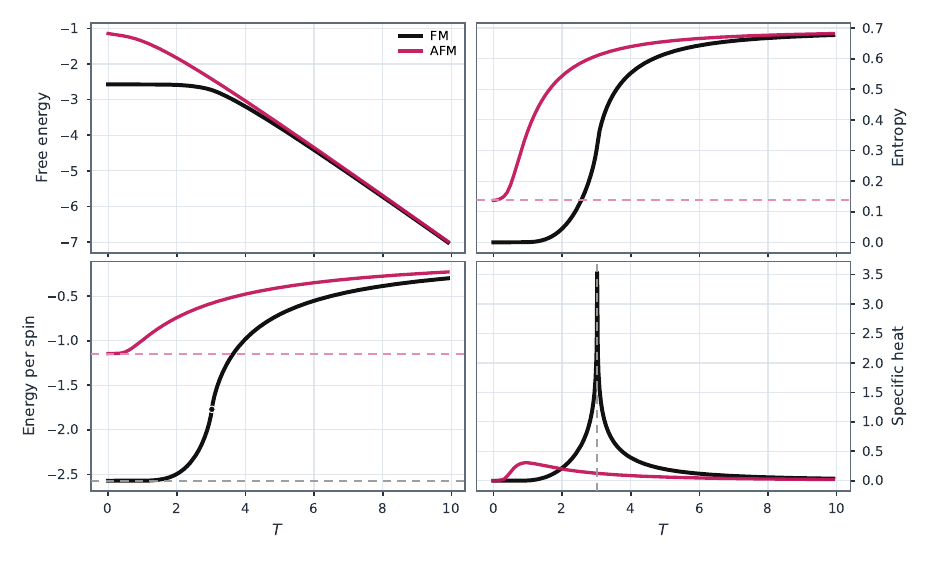}

\vspace{0.6em}
\hbox to \linewidth{%
  \hss
  \masktilemat{{\color{maskOne}w^{(1)}}}{7}{{1, 1, 0, 1, 1, 1, 1}, {1, 1, 1, 1, 1, 1, 0}, {1, 1, 1, 0, 1, 1, 
  1}, {0, 1, 1, 1, 1, 1, 1}, {1, 1, 1, 1, 0, 1, 1}, {1, 0, 1, 1, 1, 
  1, 1}, {1, 1, 1, 1, 1, 0, 1}}
  \hss
  \masktilemat{{\color{maskTwo}w^{(2)}}}{7}{{1, 1, 1, 0, 1, 1, 1}, {0, 1, 1, 1, 1, 1, 1}, {1, 1, 1, 1, 0, 1, 
  1}, {1, 0, 1, 1, 1, 1, 1}, {1, 1, 1, 1, 1, 0, 1}, {1, 1, 0, 1, 1, 
  1, 1}, {1, 1, 1, 1, 1, 1, 0}}
  \hss
  \masktilemat{{\color{maskThree}w^{(3)}}}{7}{{1, 0, 1, 1, 1, 1, 1}, {1, 1, 1, 1, 1, 0, 1}, {1, 1, 0, 1, 1, 1, 
  1}, {1, 1, 1, 1, 1, 1, 0}, {1, 1, 1, 0, 1, 1, 1}, {0, 1, 1, 1, 1, 
  1, 1}, {1, 1, 1, 1, 0, 1, 1}}%
  \hss
}
\end{atlascard}

\begin{atlascard}{$t2001$}{table_icons_static/t2001.png}{\carddatastruct{12}{864}{108}{180}{bond}\cardphysicsgeneric{1.580173228}{-1.380405611}{0.36171627}{none}{\NA}{0.1288}}
\atlasplotobject{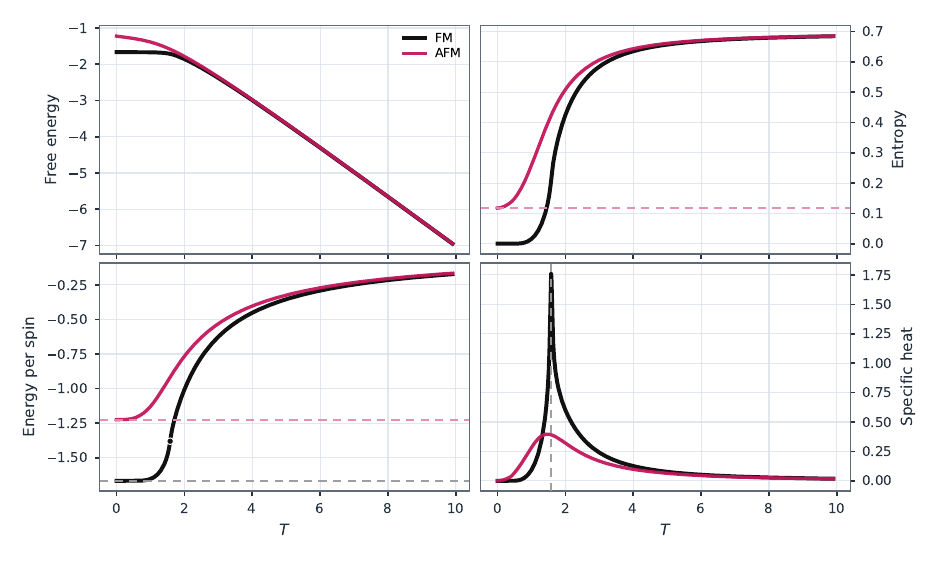}

\vspace{0.6em}
\hbox to \linewidth{%
  \hss
  \masktilemat{{\color{maskOne}w^{(1)}}}{12}{{0, 0, 0, 0, 1, 1, 0, 0, 0, 0, 1, 1}, {0, 0, 0, 0, 0, 0, 0, 0, 0, 0, 
  0, 0}, {0, 0, 0, 0, 1, 1, 0, 0, 0, 0, 1, 1}, {0, 0, 0, 0, 0, 0, 0, 
  0, 0, 0, 0, 0}, {0, 0, 0, 0, 1, 1, 0, 0, 0, 0, 1, 1}, {0, 0, 0, 0, 
  0, 0, 0, 0, 0, 0, 0, 0}, {0, 0, 0, 0, 1, 1, 0, 0, 0, 0, 1, 1}, {0, 
  0, 0, 0, 0, 0, 0, 0, 0, 0, 0, 0}, {0, 0, 0, 0, 1, 1, 0, 0, 0, 0, 1, 
  1}, {0, 0, 0, 0, 0, 0, 0, 0, 0, 0, 0, 0}, {0, 0, 0, 0, 1, 1, 0, 0, 
  0, 0, 1, 1}, {0, 0, 0, 0, 0, 0, 0, 0, 0, 0, 0, 0}}
  \hss
  \masktilemat{{\color{maskTwo}w^{(2)}}}{12}{{1, 1, 1, 1, 1, 1, 1, 1, 1, 1, 1, 1}, {0, 0, 0, 0, 1, 1, 0, 0, 0, 0, 
  1, 1}, {0, 0, 0, 1, 1, 0, 0, 0, 0, 1, 1, 0}, {1, 1, 1, 1, 1, 1, 1, 
  1, 1, 1, 1, 1}, {1, 1, 1, 1, 1, 1, 1, 1, 1, 1, 1, 1}, {0, 0, 0, 0, 
  1, 1, 0, 0, 0, 0, 1, 1}, {0, 0, 0, 1, 1, 0, 0, 0, 0, 1, 1, 0}, {1, 
  1, 1, 1, 1, 1, 1, 1, 1, 1, 1, 1}, {1, 1, 1, 1, 1, 1, 1, 1, 1, 1, 1, 
  1}, {0, 0, 0, 0, 1, 1, 0, 0, 0, 0, 1, 1}, {0, 0, 0, 1, 1, 0, 0, 0, 
  0, 1, 1, 0}, {1, 1, 1, 1, 1, 1, 1, 1, 1, 1, 1, 1}}
  \hss
  \masktilemat{{\color{maskThree}w^{(3)}}}{12}{{1, 0, 1, 1, 0, 1, 1, 0, 1, 1, 0, 1}, {1, 0, 0, 0, 0, 1, 1, 0, 0, 0, 
  0, 1}, {0, 0, 0, 1, 0, 1, 0, 0, 0, 1, 0, 1}, {0, 0, 1, 1, 0, 0, 0, 
  0, 1, 1, 0, 0}, {1, 0, 1, 1, 0, 1, 1, 0, 1, 1, 0, 1}, {1, 0, 0, 0, 
  0, 1, 1, 0, 0, 0, 0, 1}, {0, 0, 0, 1, 0, 1, 0, 0, 0, 1, 0, 1}, {0, 
  0, 1, 1, 0, 0, 0, 0, 1, 1, 0, 0}, {1, 0, 1, 1, 0, 1, 1, 0, 1, 1, 0, 
  1}, {1, 0, 0, 0, 0, 1, 1, 0, 0, 0, 0, 1}, {0, 0, 0, 1, 0, 1, 0, 0, 
  0, 1, 0, 1}, {0, 0, 1, 1, 0, 0, 0, 0, 1, 1, 0, 0}}%
  \hss
}
\end{atlascard}

\begin{atlascard}{$t2008$}{table_icons_static/t2008.png}{\carddatastruct{12}{864}{144}{324}{bond}\cardphysicsgeneric{2.482643364}{-1.662532959}{0.43280277}{none}{\NA}{0.12584308}}
\atlasplotobject{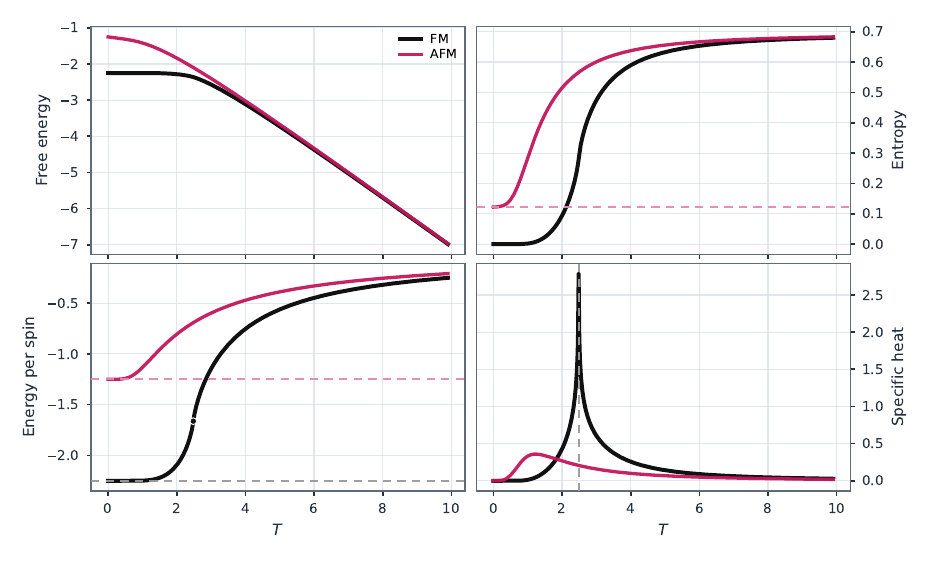}

\vspace{0.6em}
\hbox to \linewidth{%
  \hss
  \masktilemat{{\color{maskOne}w^{(1)}}}{12}{{0, 0, 0, 0, 0, 0, 0, 0, 0, 0, 0, 0}, {1, 1, 1, 0, 1, 1, 1, 0, 1, 1, 
  1, 0}, {1, 1, 0, 1, 1, 1, 0, 1, 1, 1, 0, 1}, {0, 0, 0, 0, 0, 0, 0, 
  0, 0, 0, 0, 0}, {1, 1, 1, 0, 1, 1, 1, 0, 1, 1, 1, 0}, {1, 1, 0, 1, 
  1, 1, 0, 1, 1, 1, 0, 1}, {0, 0, 0, 0, 0, 0, 0, 0, 0, 0, 0, 0}, {1, 
  1, 1, 0, 1, 1, 1, 0, 1, 1, 1, 0}, {1, 1, 0, 1, 1, 1, 0, 1, 1, 1, 0, 
  1}, {0, 0, 0, 0, 0, 0, 0, 0, 0, 0, 0, 0}, {1, 1, 1, 0, 1, 1, 1, 0, 
  1, 1, 1, 0}, {1, 1, 0, 1, 1, 1, 0, 1, 1, 1, 0, 1}}
  \hss
  \masktilemat{{\color{maskTwo}w^{(2)}}}{12}{{1, 1, 1, 1, 1, 1, 1, 1, 1, 1, 1, 1}, {1, 1, 1, 1, 1, 1, 1, 1, 1, 1, 
  1, 1}, {1, 1, 1, 1, 1, 1, 1, 1, 1, 1, 1, 1}, {1, 1, 1, 1, 1, 1, 1, 
  1, 1, 1, 1, 1}, {1, 1, 1, 1, 1, 1, 1, 1, 1, 1, 1, 1}, {1, 1, 1, 1, 
  1, 1, 1, 1, 1, 1, 1, 1}, {1, 1, 1, 1, 1, 1, 1, 1, 1, 1, 1, 1}, {1, 
  1, 1, 1, 1, 1, 1, 1, 1, 1, 1, 1}, {1, 1, 1, 1, 1, 1, 1, 1, 1, 1, 1, 
  1}, {1, 1, 1, 1, 1, 1, 1, 1, 1, 1, 1, 1}, {1, 1, 1, 1, 1, 1, 1, 1, 
  1, 1, 1, 1}, {1, 1, 1, 1, 1, 1, 1, 1, 1, 1, 1, 1}}
  \hss
  \masktilemat{{\color{maskThree}w^{(3)}}}{12}{{0, 1, 1, 1, 0, 1, 1, 1, 0, 1, 1, 1}, {0, 1, 1, 1, 0, 1, 1, 1, 0, 1, 
  1, 1}, {0, 1, 1, 1, 0, 1, 1, 1, 0, 1, 1, 1}, {0, 1, 1, 1, 0, 1, 1, 
  1, 0, 1, 1, 1}, {0, 1, 1, 1, 0, 1, 1, 1, 0, 1, 1, 1}, {0, 1, 1, 1, 
  0, 1, 1, 1, 0, 1, 1, 1}, {0, 1, 1, 1, 0, 1, 1, 1, 0, 1, 1, 1}, {0, 
  1, 1, 1, 0, 1, 1, 1, 0, 1, 1, 1}, {0, 1, 1, 1, 0, 1, 1, 1, 0, 1, 1, 
  1}, {0, 1, 1, 1, 0, 1, 1, 1, 0, 1, 1, 1}, {0, 1, 1, 1, 0, 1, 1, 1, 
  0, 1, 1, 1}, {0, 1, 1, 1, 0, 1, 1, 1, 0, 1, 1, 1}}%
  \hss
}
\end{atlascard}

\begin{atlascard}{$t2016$}{table_icons_static/t2016.png}{\carddatastruct{12}{864}{144}{360}{bond}\cardphysicsgeneric{2.916114715}{-1.732764232}{0.49145900}{none}{\NA}{0.18486575}}
\atlasplotobject{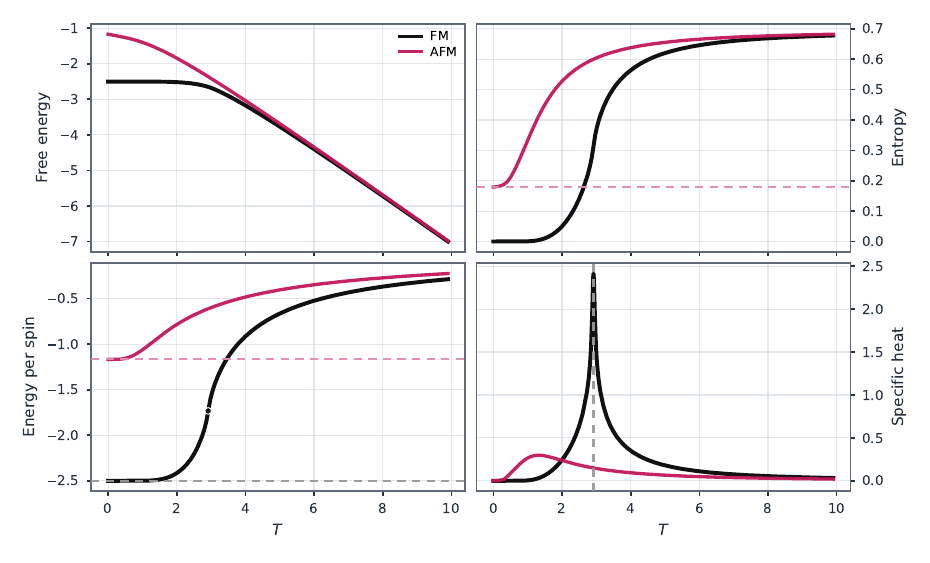}

\vspace{0.6em}
\hbox to \linewidth{%
  \hss
  \masktilemat{{\color{maskOne}w^{(1)}}}{12}{{1, 1, 1, 1, 1, 1, 1, 1, 1, 1, 0, 1}, {1, 1, 1, 1, 1, 0, 1, 1, 1, 1, 
  1, 1}, {0, 1, 1, 1, 1, 1, 1, 1, 1, 1, 1, 1}, {1, 1, 1, 1, 1, 1, 1, 
  0, 1, 1, 1, 1}, {1, 1, 0, 1, 1, 1, 1, 1, 1, 1, 1, 1}, {1, 1, 1, 1, 
  1, 1, 1, 1, 1, 0, 1, 1}, {1, 1, 1, 1, 0, 1, 1, 1, 1, 1, 1, 1}, {1, 
  1, 1, 1, 1, 1, 1, 1, 1, 1, 1, 0}, {1, 1, 1, 1, 1, 1, 0, 1, 1, 1, 1, 
  1}, {1, 0, 1, 1, 1, 1, 1, 1, 1, 1, 1, 1}, {1, 1, 1, 1, 1, 1, 1, 1, 
  0, 1, 1, 1}, {1, 1, 1, 0, 1, 1, 1, 1, 1, 1, 1, 1}}
  \hss
  \masktilemat{{\color{maskTwo}w^{(2)}}}{12}{{0, 1, 1, 1, 1, 1, 1, 1, 1, 1, 1, 1}, {1, 1, 1, 1, 1, 1, 1, 0, 1, 1, 
  1, 1}, {1, 1, 0, 1, 1, 1, 1, 1, 1, 1, 1, 1}, {1, 1, 1, 1, 1, 1, 1, 
  1, 1, 0, 1, 1}, {1, 1, 1, 1, 0, 1, 1, 1, 1, 1, 1, 1}, {1, 1, 1, 1, 
  1, 1, 1, 1, 1, 1, 1, 0}, {1, 1, 1, 1, 1, 1, 0, 1, 1, 1, 1, 1}, {1, 
  0, 1, 1, 1, 1, 1, 1, 1, 1, 1, 1}, {1, 1, 1, 1, 1, 1, 1, 1, 0, 1, 1, 
  1}, {1, 1, 1, 0, 1, 1, 1, 1, 1, 1, 1, 1}, {1, 1, 1, 1, 1, 1, 1, 1, 
  1, 1, 0, 1}, {1, 1, 1, 1, 1, 0, 1, 1, 1, 1, 1, 1}}
  \hss
  \masktilemat{{\color{maskThree}w^{(3)}}}{12}{{1, 1, 1, 1, 1, 1, 0, 1, 0, 0, 1, 0}, {1, 0, 1, 0, 0, 1, 0, 1, 1, 1, 
  1, 1}, {1, 0, 1, 1, 1, 1, 1, 1, 0, 1, 0, 0}, {1, 1, 1, 0, 1, 0, 0, 
  1, 0, 1, 1, 1}, {0, 0, 1, 0, 1, 1, 1, 1, 1, 1, 0, 1}, {1, 1, 1, 1, 
  1, 0, 1, 0, 0, 1, 0, 1}, {0, 1, 0, 0, 1, 0, 1, 1, 1, 1, 1, 1}, {0, 
  1, 1, 1, 1, 1, 1, 0, 1, 0, 0, 1}, {1, 1, 0, 1, 0, 0, 1, 0, 1, 1, 1, 
  1}, {0, 1, 0, 1, 1, 1, 1, 1, 1, 0, 1, 0}, {1, 1, 1, 1, 0, 1, 0, 0, 
  1, 0, 1, 1}, {1, 0, 0, 1, 0, 1, 1, 1, 1, 1, 1, 0}}%
  \hss
}
\end{atlascard}

\begin{atlascard}{$t2009$}{table_icons_static/t2009.png}{\carddatastruct{13}{1014}{156}{351}{bond}\cardphysicsgeneric{2.523724013}{-1.62812}{0.455121}{none}{\NA}{0.0847}}
\atlasplotobject{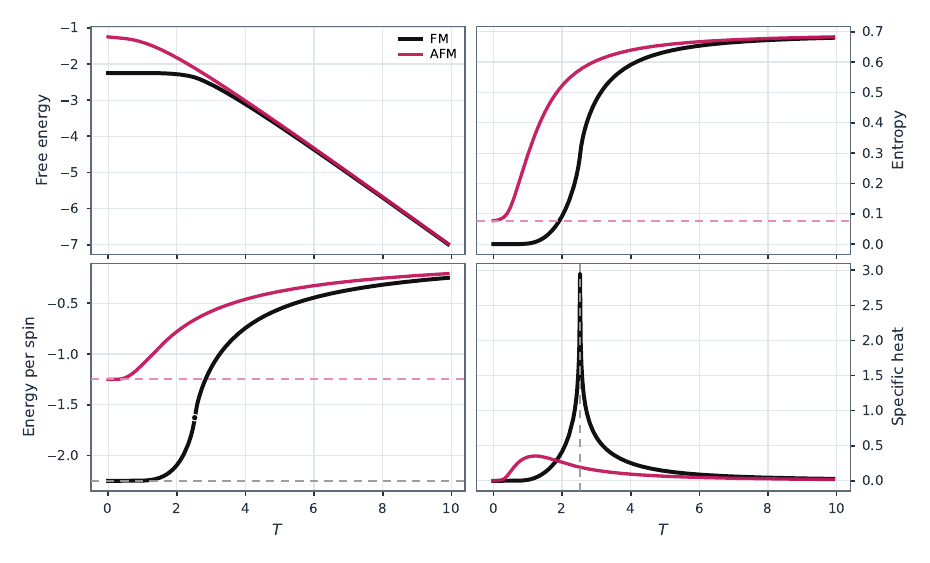}

\vspace{0.6em}
\hbox to \linewidth{%
  \hss
  \masktilemat{{\color{maskOne}w^{(1)}}}{13}{{0,0,1,1,1,1,1,1,0,0,1,1,1},{1,1,1,1,0,0,1,1,1,0,0,1,1},{0,0,1,1,1,0,0,1,1,1,1,1,1},{1,0,0,1,1,1,1,1,1,0,0,1,1},{1,1,1,1,1,0,0,1,1,1,0,0,1},{1,0,0,1,1,1,0,0,1,1,1,1,1},{1,1,0,0,1,1,1,1,1,1,0,0,1},{1,1,1,1,1,1,0,0,1,1,1,0,0},{1,1,0,0,1,1,1,0,0,1,1,1,1},{1,1,1,0,0,1,1,1,1,1,1,0,0},{0,1,1,1,1,1,1,0,0,1,1,1,0},{1,1,1,0,0,1,1,1,0,0,1,1,1},{0,1,1,1,0,0,1,1,1,1,1,1,0}}
  \hss
  \masktilemat{{\color{maskTwo}w^{(2)}}}{13}{{0, 1, 0, 1, 1, 1, 1, 1, 1, 1, 0, 1, 0}, {1, 1, 1, 1, 1, 1, 0, 1, 0, 
  0, 1, 0, 1}, {1, 1, 0, 1, 0, 0, 1, 0, 1, 1, 1, 1, 1}, {0, 0, 1, 0, 
  1, 1, 1, 1, 1, 1, 1, 0, 1}, {1, 1, 1, 1, 1, 1, 1, 0, 1, 0, 0, 1, 
  0}, {1, 1, 1, 0, 1, 0, 0, 1, 0, 1, 1, 1, 1}, {1, 0, 0, 1, 0, 1, 1, 
  1, 1, 1, 1, 1, 0}, {0, 1, 1, 1, 1, 1, 1, 1, 0, 1, 0, 0, 1}, {1, 1, 
  1, 1, 0, 1, 0, 0, 1, 0, 1, 1, 1}, {0, 1, 0, 0, 1, 0, 1, 1, 1, 1, 1, 
  1, 1}, {1, 0, 1, 1, 1, 1, 1, 1, 1, 0, 1, 0, 0}, {1, 1, 1, 1, 1, 0, 
  1, 0, 0, 1, 0, 1, 1}, {1, 0, 1, 0, 0, 1, 0, 1, 1, 1, 1, 1, 1}}
  \hss
  \masktilemat{{\color{maskThree}w^{(3)}}}{13}{{0, 1, 1, 0, 1, 1, 1, 0, 1, 0, 1, 1, 1}, {1, 1, 1, 0, 1, 0, 1, 1, 1, 
  0, 1, 1, 0}, {1, 0, 1, 1, 1, 0, 1, 1, 0, 1, 1, 1, 0}, {1, 0, 1, 1, 
  0, 1, 1, 1, 0, 1, 0, 1, 1}, {0, 1, 1, 1, 0, 1, 0, 1, 1, 1, 0, 1, 
  1}, {0, 1, 0, 1, 1, 1, 0, 1, 1, 0, 1, 1, 1}, {1, 1, 0, 1, 1, 0, 1, 
  1, 1, 0, 1, 0, 1}, {1, 0, 1, 1, 1, 0, 1, 0, 1, 1, 1, 0, 1}, {1, 0, 
  1, 0, 1, 1, 1, 0, 1, 1, 0, 1, 1}, {1, 1, 1, 0, 1, 1, 0, 1, 1, 1, 0, 
  1, 0}, {1, 1, 0, 1, 1, 1, 0, 1, 0, 1, 1, 1, 0}, {1, 1, 0, 1, 0, 1, 
  1, 1, 0, 1, 1, 0, 1}, {0, 1, 1, 1, 0, 1, 1, 0, 1, 1, 1, 0, 1}}%
  \hss
}
\end{atlascard}

\begin{atlascard}{$t2020$}{table_icons_static/t2020.png}{\carddatastruct{13}{1014}{156}{429}{site}\cardphysicsgeneric{3.193522437}{-1.938717187}{0.46491640}{none}{\NA}{0.17280643}}
\atlasplotobject{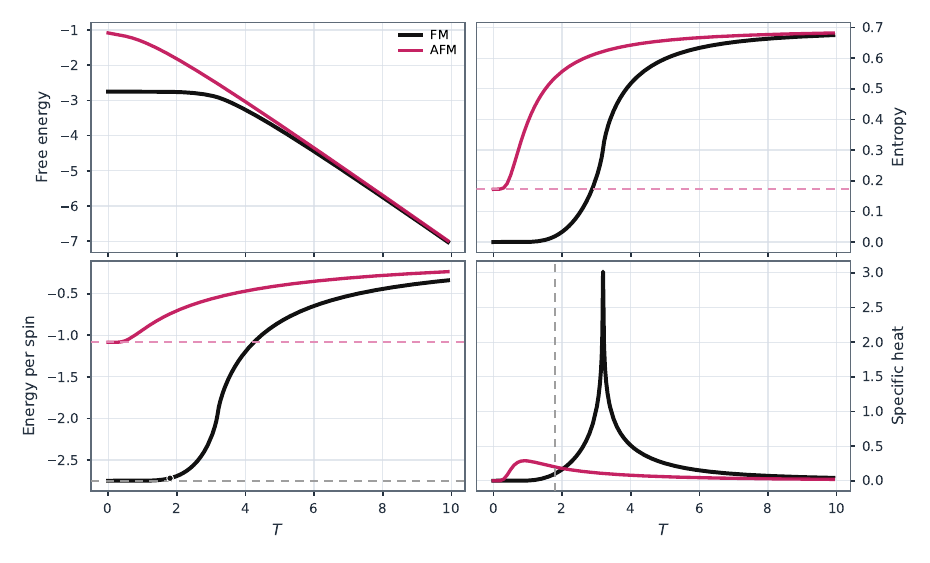}

\vspace{0.6em}
\hbox to \linewidth{\hss
\masktilemat{w}{13}{{1, 0, 1, 1, 1, 1, 1, 1, 1, 1, 1, 1, 1},{1, 1, 1, 1, 0, 1, 1, 1, 1, 1, 1, 1, 1},{1, 1, 1, 1, 1, 1, 1, 0, 1, 1, 1, 1, 1},{1, 1, 1, 1, 1, 1, 1, 1, 1, 1, 0, 1, 1},{0, 1, 1, 1, 1, 1, 1, 1, 1, 1, 1, 1, 1},{1, 1, 1, 0, 1, 1, 1, 1, 1, 1, 1, 1, 1},{1, 1, 1, 1, 1, 1, 0, 1, 1, 1, 1, 1, 1},{1, 1, 1, 1, 1, 1, 1, 1, 1, 0, 1, 1, 1},{1, 1, 1, 1, 1, 1, 1, 1, 1, 1, 1, 1, 0},{1, 1, 0, 1, 1, 1, 1, 1, 1, 1, 1, 1, 1},{1, 1, 1, 1, 1, 0, 1, 1, 1, 1, 1, 1, 1},{1, 1, 1, 1, 1, 1, 1, 1, 0, 1, 1, 1, 1},{1, 1, 1, 1, 1, 1, 1, 1, 1, 1, 1, 0, 1}}%
\hss}
\end{atlascard}

\begin{atlascard}{$t2002$}{table_icons_static/t2002.png}{\carddatastruct{15}{1350}{120}{210}{bond}\cardphysicsgeneric{1.476141516}{-1.545228576}{0.28617311}{none}{\NA}{0.37770580}}
\atlasplotobject{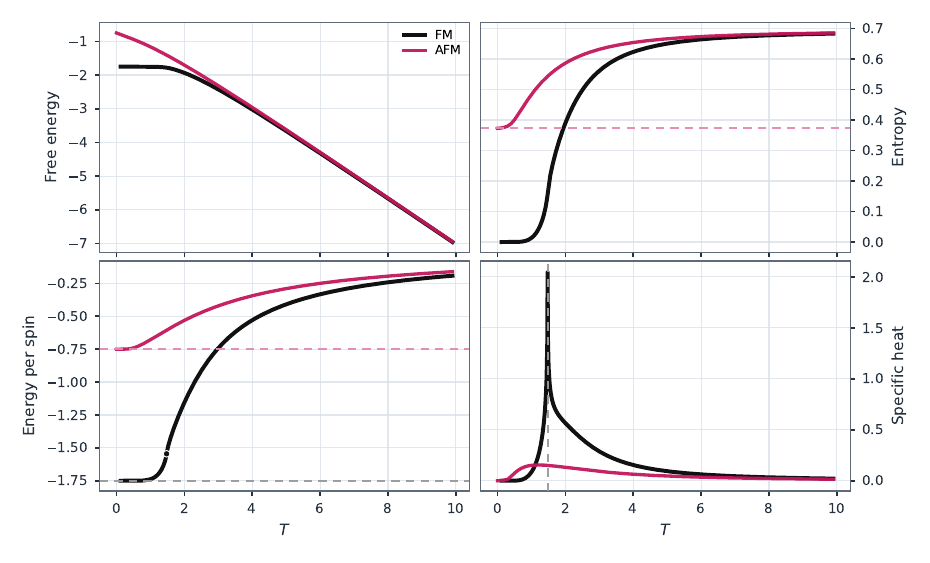}

\vspace{0.6em}
\hbox to \linewidth{%
  \hss
  \masktilemat{{\color{maskOne}w^{(1)}}}{15}{{0, 0, 0, 0, 1, 0, 0, 1, 0, 1, 0, 0, 1, 0, 0}, {1, 0, 0, 0, 0, 0, 0, 
  1, 0, 0, 1, 0, 1, 0, 0}, {1, 0, 0, 1, 0, 0, 0, 0, 0, 0, 1, 0, 0, 1, 
  0}, {0, 1, 0, 1, 0, 0, 1, 0, 0, 0, 0, 0, 0, 1, 0}, {0, 1, 0, 0, 1, 
  0, 1, 0, 0, 1, 0, 0, 0, 0, 0}, {0, 0, 0, 0, 1, 0, 0, 1, 0, 1, 0, 0, 
  1, 0, 0}, {1, 0, 0, 0, 0, 0, 0, 1, 0, 0, 1, 0, 1, 0, 0}, {1, 0, 0, 
  1, 0, 0, 0, 0, 0, 0, 1, 0, 0, 1, 0}, {0, 1, 0, 1, 0, 0, 1, 0, 0, 0, 
  0, 0, 0, 1, 0}, {0, 1, 0, 0, 1, 0, 1, 0, 0, 1, 0, 0, 0, 0, 0}, {0, 
  0, 0, 0, 1, 0, 0, 1, 0, 1, 0, 0, 1, 0, 0}, {1, 0, 0, 0, 0, 0, 0, 1, 
  0, 0, 1, 0, 1, 0, 0}, {1, 0, 0, 1, 0, 0, 0, 0, 0, 0, 1, 0, 0, 1, 
  0}, {0, 1, 0, 1, 0, 0, 1, 0, 0, 0, 0, 0, 0, 1, 0}, {0, 1, 0, 0, 1, 
  0, 1, 0, 0, 1, 0, 0, 0, 0, 0}}
  \hss
  \masktilemat{{\color{maskTwo}w^{(2)}}}{15}{{0, 0, 0, 0, 1, 1, 1, 1, 1, 0, 0, 0, 0, 0, 0}, {0, 0, 0, 0, 0, 0, 0, 
  1, 1, 1, 1, 1, 0, 0, 0}, {0, 0, 0, 0, 0, 0, 0, 0, 0, 0, 1, 1, 1, 1, 
  1}, {1, 1, 1, 0, 0, 0, 0, 0, 0, 0, 0, 0, 0, 1, 1}, {0, 1, 1, 1, 1, 
  1, 0, 0, 0, 0, 0, 0, 0, 0, 0}, {0, 0, 0, 0, 1, 1, 1, 1, 1, 0, 0, 0, 
  0, 0, 0}, {0, 0, 0, 0, 0, 0, 0, 1, 1, 1, 1, 1, 0, 0, 0}, {0, 0, 0, 
  0, 0, 0, 0, 0, 0, 0, 1, 1, 1, 1, 1}, {1, 1, 1, 0, 0, 0, 0, 0, 0, 0, 
  0, 0, 0, 1, 1}, {0, 1, 1, 1, 1, 1, 0, 0, 0, 0, 0, 0, 0, 0, 0}, {0, 
  0, 0, 0, 1, 1, 1, 1, 1, 0, 0, 0, 0, 0, 0}, {0, 0, 0, 0, 0, 0, 0, 1, 
  1, 1, 1, 1, 0, 0, 0}, {0, 0, 0, 0, 0, 0, 0, 0, 0, 0, 1, 1, 1, 1, 
  1}, {1, 1, 1, 0, 0, 0, 0, 0, 0, 0, 0, 0, 0, 1, 1}, {0, 1, 1, 1, 1, 
  1, 0, 0, 0, 0, 0, 0, 0, 0, 0}}
  \hss
  \masktilemat{{\color{maskThree}w^{(3)}}}{15}{{0, 1, 0, 0, 1, 1, 0, 0, 1, 0, 0, 0, 1, 0, 0}, {1, 0, 0, 0, 1, 0, 0, 
  1, 1, 0, 0, 1, 0, 0, 0}, {0, 0, 0, 1, 0, 0, 0, 1, 0, 0, 1, 1, 0, 0, 
  1}, {0, 0, 1, 0, 0, 0, 1, 0, 0, 0, 1, 0, 0, 1, 1}, {0, 1, 1, 0, 0, 
  1, 0, 0, 0, 1, 0, 0, 0, 1, 0}, {0, 1, 0, 0, 1, 1, 0, 0, 1, 0, 0, 0, 
  1, 0, 0}, {1, 0, 0, 0, 1, 0, 0, 1, 1, 0, 0, 1, 0, 0, 0}, {0, 0, 0, 
  1, 0, 0, 0, 1, 0, 0, 1, 1, 0, 0, 1}, {0, 0, 1, 0, 0, 0, 1, 0, 0, 0, 
  1, 0, 0, 1, 1}, {0, 1, 1, 0, 0, 1, 0, 0, 0, 1, 0, 0, 0, 1, 0}, {0, 
  1, 0, 0, 1, 1, 0, 0, 1, 0, 0, 0, 1, 0, 0}, {1, 0, 0, 0, 1, 0, 0, 1, 
  1, 0, 0, 1, 0, 0, 0}, {0, 0, 0, 1, 0, 0, 0, 1, 0, 0, 1, 1, 0, 0, 
  1}, {0, 0, 1, 0, 0, 0, 1, 0, 0, 0, 1, 0, 0, 1, 1}, {0, 1, 1, 0, 0, 
  1, 0, 0, 0, 1, 0, 0, 0, 1, 0}}%
  \hss
}
\end{atlascard}

\begin{atlascard}{$t2005$}{table_icons_static/t2005.png}{\carddatastruct{20}{2400}{360}{720}{bond}\cardphysicsgeneric{2.167953816}{-1.483695261}{0.46465291}{none}{\NA}{0.08389828}}
\atlasplotobject{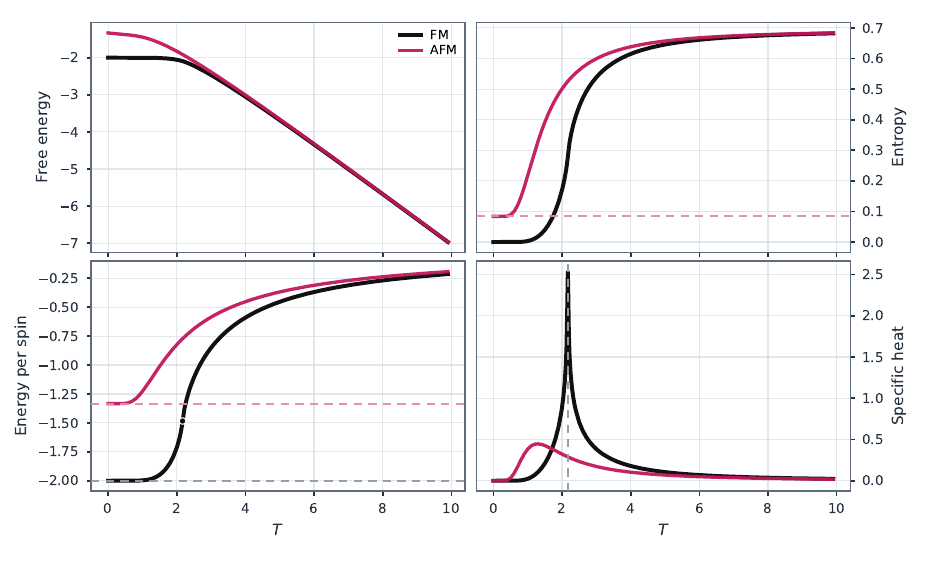}

\vspace{0.6em}
\hbox to \linewidth{%
  \hss
  \masktilemat{{\color{maskOne}w^{(1)}}}{20}{{1, 0, 1, 1, 0, 1, 0, 1, 1, 0, 1, 0, 1, 1, 0, 1, 0, 1, 1, 0}, {0, 0, 
  0, 0, 0, 0, 0, 0, 0, 0, 0, 0, 0, 0, 0, 0, 0, 0, 0, 0}, {1, 0, 1, 1, 
  0, 1, 0, 1, 1, 0, 1, 0, 1, 1, 0, 1, 0, 1, 1, 0}, {0, 0, 1, 1, 0, 0, 
  0, 1, 1, 0, 0, 0, 1, 1, 0, 0, 0, 1, 1, 0}, {1, 0, 1, 1, 0, 1, 0, 1, 
  1, 0, 1, 0, 1, 1, 0, 1, 0, 1, 1, 0}, {0, 0, 0, 0, 0, 0, 0, 0, 0, 0, 
  0, 0, 0, 0, 0, 0, 0, 0, 0, 0}, {1, 0, 1, 1, 0, 1, 0, 1, 1, 0, 1, 0, 
  1, 1, 0, 1, 0, 1, 1, 0}, {0, 0, 1, 1, 0, 0, 0, 1, 1, 0, 0, 0, 1, 1, 
  0, 0, 0, 1, 1, 0}, {1, 0, 1, 1, 0, 1, 0, 1, 1, 0, 1, 0, 1, 1, 0, 1, 
  0, 1, 1, 0}, {0, 0, 0, 0, 0, 0, 0, 0, 0, 0, 0, 0, 0, 0, 0, 0, 0, 0, 
  0, 0}, {1, 0, 1, 1, 0, 1, 0, 1, 1, 0, 1, 0, 1, 1, 0, 1, 0, 1, 1, 
  0}, {0, 0, 1, 1, 0, 0, 0, 1, 1, 0, 0, 0, 1, 1, 0, 0, 0, 1, 1, 
  0}, {1, 0, 1, 1, 0, 1, 0, 1, 1, 0, 1, 0, 1, 1, 0, 1, 0, 1, 1, 
  0}, {0, 0, 0, 0, 0, 0, 0, 0, 0, 0, 0, 0, 0, 0, 0, 0, 0, 0, 0, 
  0}, {1, 0, 1, 1, 0, 1, 0, 1, 1, 0, 1, 0, 1, 1, 0, 1, 0, 1, 1, 
  0}, {0, 0, 1, 1, 0, 0, 0, 1, 1, 0, 0, 0, 1, 1, 0, 0, 0, 1, 1, 
  0}, {1, 0, 1, 1, 0, 1, 0, 1, 1, 0, 1, 0, 1, 1, 0, 1, 0, 1, 1, 
  0}, {0, 0, 0, 0, 0, 0, 0, 0, 0, 0, 0, 0, 0, 0, 0, 0, 0, 0, 0, 
  0}, {1, 0, 1, 1, 0, 1, 0, 1, 1, 0, 1, 0, 1, 1, 0, 1, 0, 1, 1, 
  0}, {0, 0, 1, 1, 0, 0, 0, 1, 1, 0, 0, 0, 1, 1, 0, 0, 0, 1, 1, 0}}
  \hss
  \masktilemat{{\color{maskTwo}w^{(2)}}}{20}{{0, 0, 1, 1, 1, 0, 0, 1, 1, 1, 0, 0, 1, 1, 1, 0, 0, 1, 1, 1}, {1, 1, 
  1, 1, 1, 1, 1, 1, 1, 1, 1, 1, 1, 1, 1, 1, 1, 1, 1, 1}, {1, 1, 1, 1, 
  1, 1, 1, 1, 1, 1, 1, 1, 1, 1, 1, 1, 1, 1, 1, 1}, {1, 1, 1, 0, 0, 1, 
  1, 1, 0, 0, 1, 1, 1, 0, 0, 1, 1, 1, 0, 0}, {0, 0, 1, 1, 1, 0, 0, 1, 
  1, 1, 0, 0, 1, 1, 1, 0, 0, 1, 1, 1}, {1, 1, 1, 1, 1, 1, 1, 1, 1, 1, 
  1, 1, 1, 1, 1, 1, 1, 1, 1, 1}, {1, 1, 1, 1, 1, 1, 1, 1, 1, 1, 1, 1, 
  1, 1, 1, 1, 1, 1, 1, 1}, {1, 1, 1, 0, 0, 1, 1, 1, 0, 0, 1, 1, 1, 0, 
  0, 1, 1, 1, 0, 0}, {0, 0, 1, 1, 1, 0, 0, 1, 1, 1, 0, 0, 1, 1, 1, 0, 
  0, 1, 1, 1}, {1, 1, 1, 1, 1, 1, 1, 1, 1, 1, 1, 1, 1, 1, 1, 1, 1, 1, 
  1, 1}, {1, 1, 1, 1, 1, 1, 1, 1, 1, 1, 1, 1, 1, 1, 1, 1, 1, 1, 1, 
  1}, {1, 1, 1, 0, 0, 1, 1, 1, 0, 0, 1, 1, 1, 0, 0, 1, 1, 1, 0, 
  0}, {0, 0, 1, 1, 1, 0, 0, 1, 1, 1, 0, 0, 1, 1, 1, 0, 0, 1, 1, 
  1}, {1, 1, 1, 1, 1, 1, 1, 1, 1, 1, 1, 1, 1, 1, 1, 1, 1, 1, 1, 
  1}, {1, 1, 1, 1, 1, 1, 1, 1, 1, 1, 1, 1, 1, 1, 1, 1, 1, 1, 1, 
  1}, {1, 1, 1, 0, 0, 1, 1, 1, 0, 0, 1, 1, 1, 0, 0, 1, 1, 1, 0, 
  0}, {0, 0, 1, 1, 1, 0, 0, 1, 1, 1, 0, 0, 1, 1, 1, 0, 0, 1, 1, 
  1}, {1, 1, 1, 1, 1, 1, 1, 1, 1, 1, 1, 1, 1, 1, 1, 1, 1, 1, 1, 
  1}, {1, 1, 1, 1, 1, 1, 1, 1, 1, 1, 1, 1, 1, 1, 1, 1, 1, 1, 1, 
  1}, {1, 1, 1, 0, 0, 1, 1, 1, 0, 0, 1, 1, 1, 0, 0, 1, 1, 1, 0, 0}}
  \hss
  \masktilemat{{\color{maskThree}w^{(3)}}}{20}{{1, 0, 0, 0, 1, 1, 0, 0, 0, 1, 1, 0, 0, 0, 1, 1, 0, 0, 0, 1}, {0, 1, 
  0, 1, 1, 0, 1, 0, 1, 1, 0, 1, 0, 1, 1, 0, 1, 0, 1, 1}, {1, 1, 0, 1, 
  1, 1, 1, 0, 1, 1, 1, 1, 0, 1, 1, 1, 1, 0, 1, 1}, {1, 1, 0, 1, 0, 1, 
  1, 0, 1, 0, 1, 1, 0, 1, 0, 1, 1, 0, 1, 0}, {1, 0, 0, 0, 1, 1, 0, 0, 
  0, 1, 1, 0, 0, 0, 1, 1, 0, 0, 0, 1}, {0, 1, 0, 1, 1, 0, 1, 0, 1, 1, 
  0, 1, 0, 1, 1, 0, 1, 0, 1, 1}, {1, 1, 0, 1, 1, 1, 1, 0, 1, 1, 1, 1, 
  0, 1, 1, 1, 1, 0, 1, 1}, {1, 1, 0, 1, 0, 1, 1, 0, 1, 0, 1, 1, 0, 1, 
  0, 1, 1, 0, 1, 0}, {1, 0, 0, 0, 1, 1, 0, 0, 0, 1, 1, 0, 0, 0, 1, 1, 
  0, 0, 0, 1}, {0, 1, 0, 1, 1, 0, 1, 0, 1, 1, 0, 1, 0, 1, 1, 0, 1, 0, 
  1, 1}, {1, 1, 0, 1, 1, 1, 1, 0, 1, 1, 1, 1, 0, 1, 1, 1, 1, 0, 1, 
  1}, {1, 1, 0, 1, 0, 1, 1, 0, 1, 0, 1, 1, 0, 1, 0, 1, 1, 0, 1, 
  0}, {1, 0, 0, 0, 1, 1, 0, 0, 0, 1, 1, 0, 0, 0, 1, 1, 0, 0, 0, 
  1}, {0, 1, 0, 1, 1, 0, 1, 0, 1, 1, 0, 1, 0, 1, 1, 0, 1, 0, 1, 
  1}, {1, 1, 0, 1, 1, 1, 1, 0, 1, 1, 1, 1, 0, 1, 1, 1, 1, 0, 1, 
  1}, {1, 1, 0, 1, 0, 1, 1, 0, 1, 0, 1, 1, 0, 1, 0, 1, 1, 0, 1, 
  0}, {1, 0, 0, 0, 1, 1, 0, 0, 0, 1, 1, 0, 0, 0, 1, 1, 0, 0, 0, 
  1}, {0, 1, 0, 1, 1, 0, 1, 0, 1, 1, 0, 1, 0, 1, 1, 0, 1, 0, 1, 
  1}, {1, 1, 0, 1, 1, 1, 1, 0, 1, 1, 1, 1, 0, 1, 1, 1, 1, 0, 1, 
  1}, {1, 1, 0, 1, 0, 1, 1, 0, 1, 0, 1, 1, 0, 1, 0, 1, 1, 0, 1, 0}}%
  \hss
}
\end{atlascard}

\begin{atlascard}{$t2013$}{table_icons_static/t2013.png}{\carddatastruct{20}{2400}{280}{600}{bond}\cardphysicsgeneric{2.022131356}{-1.812909278}{0.31177070}{1.211948098}{-1.102027514}{0.09902105}}
\atlasplotobject{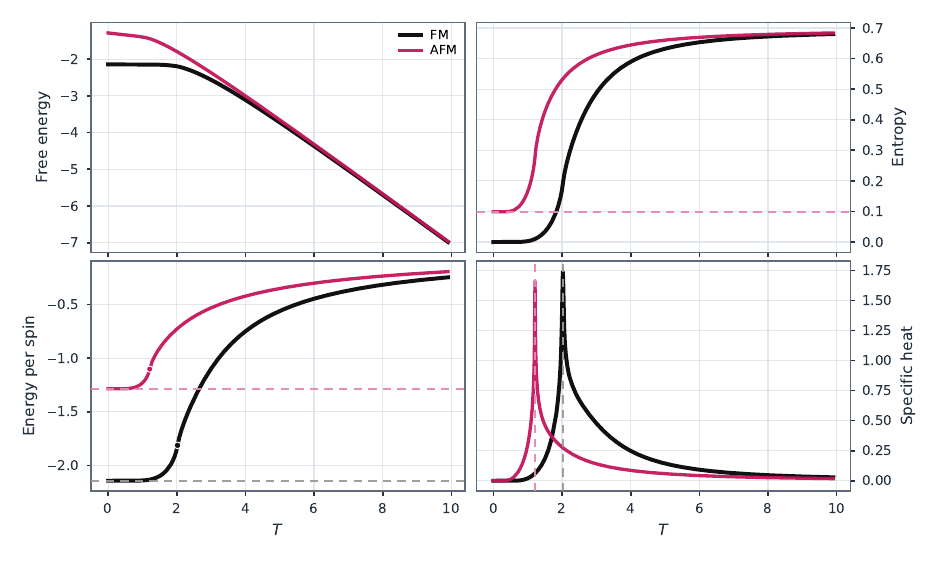}

\vspace{0.6em}
\hbox to \linewidth{%
  \hss
  \masktilemat{{\color{maskOne}w^{(1)}}}{20}{{1, 0, 0, 0, 1, 1, 0, 0, 0, 1, 1, 0, 0, 0, 1, 1, 0, 0, 0, 1}, {0, 0, 
  0, 0, 0, 0, 0, 0, 0, 0, 0, 0, 0, 0, 0, 0, 0, 0, 0, 0}, {0, 0, 0, 1, 
  1, 0, 0, 0, 1, 1, 0, 0, 0, 1, 1, 0, 0, 0, 1, 1}, {1, 1, 1, 1, 0, 1, 
  1, 1, 1, 0, 1, 1, 1, 1, 0, 1, 1, 1, 1, 0}, {1, 0, 0, 0, 1, 1, 0, 0, 
  0, 1, 1, 0, 0, 0, 1, 1, 0, 0, 0, 1}, {0, 0, 0, 0, 0, 0, 0, 0, 0, 0, 
  0, 0, 0, 0, 0, 0, 0, 0, 0, 0}, {0, 0, 0, 1, 1, 0, 0, 0, 1, 1, 0, 0, 
  0, 1, 1, 0, 0, 0, 1, 1}, {1, 1, 1, 1, 0, 1, 1, 1, 1, 0, 1, 1, 1, 1, 
  0, 1, 1, 1, 1, 0}, {1, 0, 0, 0, 1, 1, 0, 0, 0, 1, 1, 0, 0, 0, 1, 1, 
  0, 0, 0, 1}, {0, 0, 0, 0, 0, 0, 0, 0, 0, 0, 0, 0, 0, 0, 0, 0, 0, 0, 
  0, 0}, {0, 0, 0, 1, 1, 0, 0, 0, 1, 1, 0, 0, 0, 1, 1, 0, 0, 0, 1, 
  1}, {1, 1, 1, 1, 0, 1, 1, 1, 1, 0, 1, 1, 1, 1, 0, 1, 1, 1, 1, 
  0}, {1, 0, 0, 0, 1, 1, 0, 0, 0, 1, 1, 0, 0, 0, 1, 1, 0, 0, 0, 
  1}, {0, 0, 0, 0, 0, 0, 0, 0, 0, 0, 0, 0, 0, 0, 0, 0, 0, 0, 0, 
  0}, {0, 0, 0, 1, 1, 0, 0, 0, 1, 1, 0, 0, 0, 1, 1, 0, 0, 0, 1, 
  1}, {1, 1, 1, 1, 0, 1, 1, 1, 1, 0, 1, 1, 1, 1, 0, 1, 1, 1, 1, 
  0}, {1, 0, 0, 0, 1, 1, 0, 0, 0, 1, 1, 0, 0, 0, 1, 1, 0, 0, 0, 
  1}, {0, 0, 0, 0, 0, 0, 0, 0, 0, 0, 0, 0, 0, 0, 0, 0, 0, 0, 0, 
  0}, {0, 0, 0, 1, 1, 0, 0, 0, 1, 1, 0, 0, 0, 1, 1, 0, 0, 0, 1, 
  1}, {1, 1, 1, 1, 0, 1, 1, 1, 1, 0, 1, 1, 1, 1, 0, 1, 1, 1, 1, 0}}
  \hss
  \masktilemat{{\color{maskTwo}w^{(2)}}}{20}{{1, 1, 1, 1, 1, 1, 1, 1, 1, 1, 1, 1, 1, 1, 1, 1, 1, 1, 1, 1}, {0, 0, 
  0, 0, 1, 0, 0, 0, 0, 1, 0, 0, 0, 0, 1, 0, 0, 0, 0, 1}, {0, 0, 0, 1, 
  0, 0, 0, 0, 1, 0, 0, 0, 0, 1, 0, 0, 0, 0, 1, 0}, {1, 1, 1, 1, 1, 1, 
  1, 1, 1, 1, 1, 1, 1, 1, 1, 1, 1, 1, 1, 1}, {1, 1, 1, 1, 1, 1, 1, 1, 
  1, 1, 1, 1, 1, 1, 1, 1, 1, 1, 1, 1}, {0, 0, 0, 0, 1, 0, 0, 0, 0, 1, 
  0, 0, 0, 0, 1, 0, 0, 0, 0, 1}, {0, 0, 0, 1, 0, 0, 0, 0, 1, 0, 0, 0, 
  0, 1, 0, 0, 0, 0, 1, 0}, {1, 1, 1, 1, 1, 1, 1, 1, 1, 1, 1, 1, 1, 1, 
  1, 1, 1, 1, 1, 1}, {1, 1, 1, 1, 1, 1, 1, 1, 1, 1, 1, 1, 1, 1, 1, 1, 
  1, 1, 1, 1}, {0, 0, 0, 0, 1, 0, 0, 0, 0, 1, 0, 0, 0, 0, 1, 0, 0, 0, 
  0, 1}, {0, 0, 0, 1, 0, 0, 0, 0, 1, 0, 0, 0, 0, 1, 0, 0, 0, 0, 1, 
  0}, {1, 1, 1, 1, 1, 1, 1, 1, 1, 1, 1, 1, 1, 1, 1, 1, 1, 1, 1, 
  1}, {1, 1, 1, 1, 1, 1, 1, 1, 1, 1, 1, 1, 1, 1, 1, 1, 1, 1, 1, 
  1}, {0, 0, 0, 0, 1, 0, 0, 0, 0, 1, 0, 0, 0, 0, 1, 0, 0, 0, 0, 
  1}, {0, 0, 0, 1, 0, 0, 0, 0, 1, 0, 0, 0, 0, 1, 0, 0, 0, 0, 1, 
  0}, {1, 1, 1, 1, 1, 1, 1, 1, 1, 1, 1, 1, 1, 1, 1, 1, 1, 1, 1, 
  1}, {1, 1, 1, 1, 1, 1, 1, 1, 1, 1, 1, 1, 1, 1, 1, 1, 1, 1, 1, 
  1}, {0, 0, 0, 0, 1, 0, 0, 0, 0, 1, 0, 0, 0, 0, 1, 0, 0, 0, 0, 
  1}, {0, 0, 0, 1, 0, 0, 0, 0, 1, 0, 0, 0, 0, 1, 0, 0, 0, 0, 1, 
  0}, {1, 1, 1, 1, 1, 1, 1, 1, 1, 1, 1, 1, 1, 1, 1, 1, 1, 1, 1, 1}}
  \hss
  \masktilemat{{\color{maskThree}w^{(3)}}}{20}{{1, 0, 1, 1, 1, 1, 0, 1, 1, 1, 1, 0, 1, 1, 1, 1, 0, 1, 1, 1}, {1, 0, 
  0, 0, 1, 1, 0, 0, 0, 1, 1, 0, 0, 0, 1, 1, 0, 0, 0, 1}, {0, 0, 0, 1, 
  1, 0, 0, 0, 1, 1, 0, 0, 0, 1, 1, 0, 0, 0, 1, 1}, {0, 0, 1, 1, 0, 0, 
  0, 1, 1, 0, 0, 0, 1, 1, 0, 0, 0, 1, 1, 0}, {1, 0, 1, 1, 1, 1, 0, 1, 
  1, 1, 1, 0, 1, 1, 1, 1, 0, 1, 1, 1}, {1, 0, 0, 0, 1, 1, 0, 0, 0, 1, 
  1, 0, 0, 0, 1, 1, 0, 0, 0, 1}, {0, 0, 0, 1, 1, 0, 0, 0, 1, 1, 0, 0, 
  0, 1, 1, 0, 0, 0, 1, 1}, {0, 0, 1, 1, 0, 0, 0, 1, 1, 0, 0, 0, 1, 1, 
  0, 0, 0, 1, 1, 0}, {1, 0, 1, 1, 1, 1, 0, 1, 1, 1, 1, 0, 1, 1, 1, 1, 
  0, 1, 1, 1}, {1, 0, 0, 0, 1, 1, 0, 0, 0, 1, 1, 0, 0, 0, 1, 1, 0, 0, 
  0, 1}, {0, 0, 0, 1, 1, 0, 0, 0, 1, 1, 0, 0, 0, 1, 1, 0, 0, 0, 1, 
  1}, {0, 0, 1, 1, 0, 0, 0, 1, 1, 0, 0, 0, 1, 1, 0, 0, 0, 1, 1, 
  0}, {1, 0, 1, 1, 1, 1, 0, 1, 1, 1, 1, 0, 1, 1, 1, 1, 0, 1, 1, 
  1}, {1, 0, 0, 0, 1, 1, 0, 0, 0, 1, 1, 0, 0, 0, 1, 1, 0, 0, 0, 
  1}, {0, 0, 0, 1, 1, 0, 0, 0, 1, 1, 0, 0, 0, 1, 1, 0, 0, 0, 1, 
  1}, {0, 0, 1, 1, 0, 0, 0, 1, 1, 0, 0, 0, 1, 1, 0, 0, 0, 1, 1, 
  0}, {1, 0, 1, 1, 1, 1, 0, 1, 1, 1, 1, 0, 1, 1, 1, 1, 0, 1, 1, 
  1}, {1, 0, 0, 0, 1, 1, 0, 0, 0, 1, 1, 0, 0, 0, 1, 1, 0, 0, 0, 
  1}, {0, 0, 0, 1, 1, 0, 0, 0, 1, 1, 0, 0, 0, 1, 1, 0, 0, 0, 1, 
  1}, {0, 0, 1, 1, 0, 0, 0, 1, 1, 0, 0, 0, 1, 1, 0, 0, 0, 1, 1, 0}}%
  \hss
}
\end{atlascard}

\clearpage
\subsection{Pentagonal RCSR layer nets}
\label{app:atlas-pentagonal}

\begin{atlascard}{\texttt{mcm} (Cairo)}{table_icons_static/mcm.png}{\carddatastruct{6}{216}{36}{60}{bond}\cardphysicsgeneric{1.79917}{-1.22274}{0.47522}{none}{none}{0.23198}}
\atlasplotobject{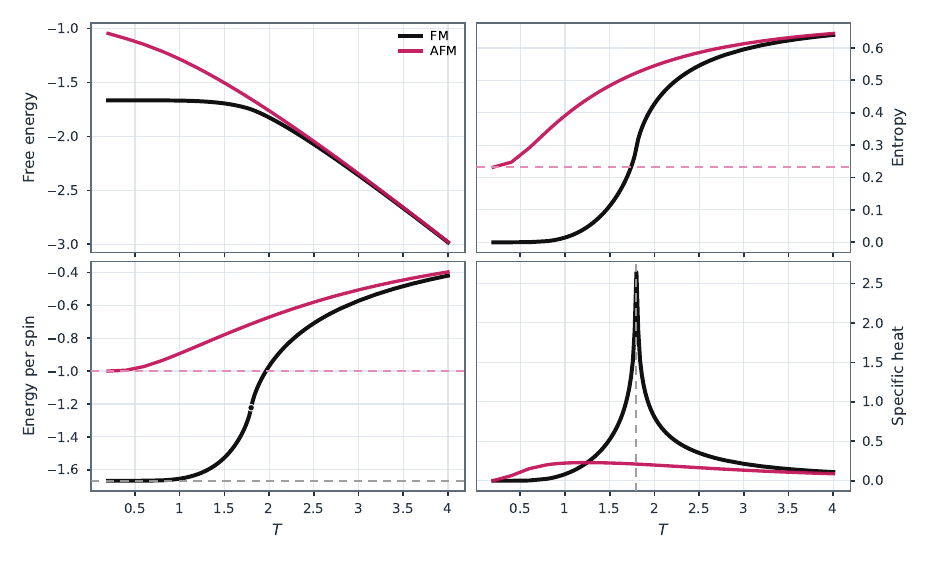}

\vspace{0.6em}
\hbox to \linewidth{%
  \hss
  \masktilemat{{\color{maskOne}w^{(1)}}}{6}{{0,1,0,1,0,1},{1,1,1,1,1,1},{1,1,1,1,1,1},{1,0,1,0,1,0},{1,1,1,1,1,1},{1,1,1,1,1,1}}
  \hss
  \masktilemat{{\color{maskTwo}w^{(2)}}}{6}{{0,0,0,0,0,0},{0,1,0,1,0,1},{1,0,1,0,1,0},{0,0,0,0,0,0},{1,0,1,0,1,0},{0,1,0,1,0,1}}
  \hss
  \masktilemat{{\color{maskThree}w^{(3)}}}{6}{{1,0,1,0,1,0},{1,1,1,1,1,1},{0,0,0,0,0,0},{0,1,0,1,0,1},{1,1,1,1,1,1},{0,0,0,0,0,0}}%
  \hss
}
\end{atlascard}

\begin{atlascard}{\texttt{krv-d}}{table_icons_static/krv-d.png}{\carddatastruct{9}{486}{81}{135}{bond}\cardphysicsgeneric{1.80477}{-1.21802}{0.46911}{none}{none}{0.23230}}
\atlasplotobject{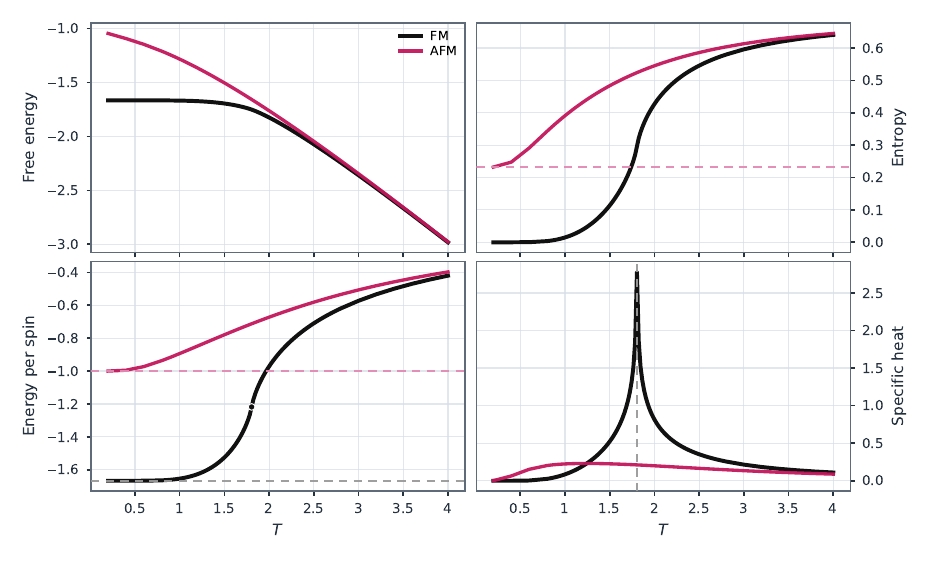}

\vspace{0.6em}
\hbox to \linewidth{%
  \hss
  \masktilemat{{\color{maskOne}w^{(1)}}}{9}{%
    {1,1,1,1,1,0,1,1,1},%
    {1,0,1,1,1,1,1,1,1},%
    {1,1,1,1,1,1,0,1,1},%
    {1,1,0,1,1,1,1,1,1},%
    {1,1,1,1,1,1,1,0,1},%
    {1,1,1,0,1,1,1,1,1},%
    {1,1,1,1,1,1,1,1,0},%
    {1,1,1,1,0,1,1,1,1},%
    {0,1,1,1,1,1,1,1,1}}
  \hss
  \masktilemat{{\color{maskTwo}w^{(2)}}}{9}{%
    {0,0,1,0,1,1,0,0,1},%
    {1,1,0,0,1,0,0,1,0},%
    {1,0,0,1,0,1,1,0,0},%
    {0,1,1,0,0,1,0,0,1},%
    {0,1,0,0,1,0,1,1,0},%
    {1,0,1,1,0,0,1,0,0},%
    {0,0,1,0,0,1,0,1,1},%
    {0,1,0,1,1,0,0,1,0},%
    {1,0,0,1,0,0,1,0,1}}
  \hss
  \masktilemat{{\color{maskThree}w^{(3)}}}{9}{%
    {1,1,0,0,0,0,0,1,0},%
    {0,0,0,1,0,1,1,0,0},%
    {0,1,1,0,0,0,0,0,1},%
    {0,0,0,0,1,0,1,1,0},%
    {1,0,1,1,0,0,0,0,0},%
    {0,0,0,0,0,1,0,1,1},%
    {0,1,0,1,1,0,0,0,0},%
    {1,0,0,0,0,0,1,0,1},%
    {0,0,1,0,1,1,0,0,0}}%
  \hss
}
\end{atlascard}

\begin{atlascard}{\texttt{krj-d}}{table_icons_static/krj-d.png}{\carddatastruct{12}{864}{144}{240}{bond}\cardphysicsgeneric{1.80746}{-1.21581}{0.46640}{none}{none}{0.23241}}
\atlasplotobject{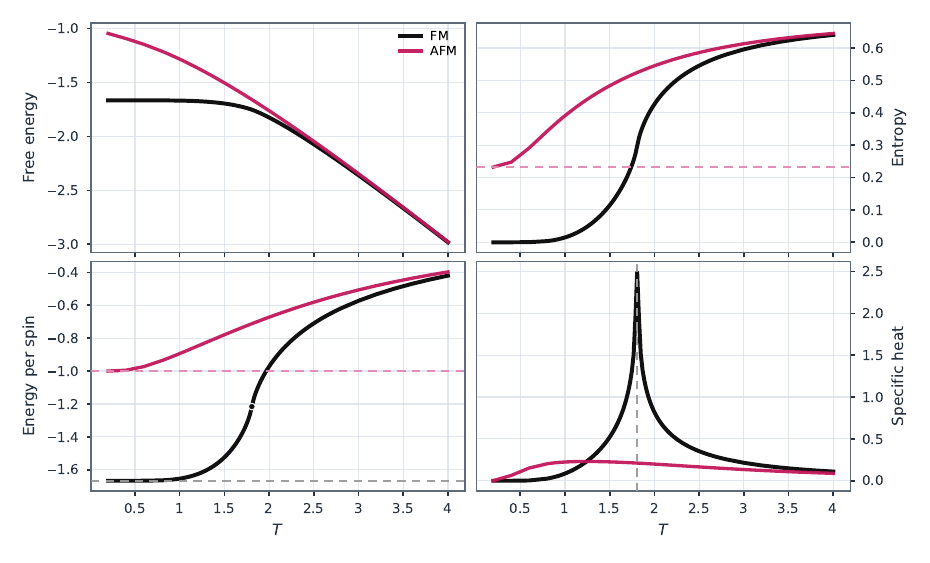}

\vspace{0.6em}
\hbox to \linewidth{%
  \hss
  \masktilemat{{\color{maskOne}w^{(1)}}}{12}{{0,0,0,1,0,1,0,0,0,1,0,1},{1,0,0,0,0,1,1,0,0,0,0,1},{1,0,1,0,0,0,1,0,1,0,0,0},{0,0,1,1,0,0,0,0,1,1,0,0},{0,0,0,1,0,1,0,0,0,1,0,1},{1,0,0,0,0,1,1,0,0,0,0,1},{1,0,1,0,0,0,1,0,1,0,0,0},{0,0,1,1,0,0,0,0,1,1,0,0},{0,0,0,1,0,1,0,0,0,1,0,1},{1,0,0,0,0,1,1,0,0,0,0,1},{1,0,1,0,0,0,1,0,1,0,0,0},{0,0,1,1,0,0,0,0,1,1,0,0}}
  \hss
  \masktilemat{{\color{maskTwo}w^{(2)}}}{12}{{1,0,0,1,1,0,1,0,0,1,1,0},{1,1,0,1,0,0,1,1,0,1,0,0},{1,1,0,1,0,0,1,1,0,1,0,0},{1,0,0,1,1,0,1,0,0,1,1,0},{1,0,0,1,1,0,1,0,0,1,1,0},{1,1,0,1,0,0,1,1,0,1,0,0},{1,1,0,1,0,0,1,1,0,1,0,0},{1,0,0,1,1,0,1,0,0,1,1,0},{1,0,0,1,1,0,1,0,0,1,1,0},{1,1,0,1,0,0,1,1,0,1,0,0},{1,1,0,1,0,0,1,1,0,1,0,0}, {1,0,0,1,1,0,1,0,0,1,1,0}}
  \hss
  \masktilemat{{\color{maskThree}w^{(3)}}}{12}{{1,1,1,0,1,1,1,1,1,0,1,1},{1,1,1,0,1,1,1,1,1,0,1,1},{0,1,1,1,1,1,0,1,1,1,1,1},{0,1,1,1,1,1,0,1,1,1,1,1},{1,1,1,0,1,1,1,1,1,0,1,1},{1,1,1,0,1,1,1,1,1,0,1,1},{0,1,1,1,1,1,0,1,1,1,1,1},{0,1,1,1,1,1,0,1,1,1,1,1},{1,1,1,0,1,1,1,1,1,0,1,1},{1,1,1,0,1,1,1,1,1,0,1,1},{0,1,1,1,1,1,0,1,1,1,1,1},{0,1,1,1,1,1,0,1,1,1,1,1}}%
  \hss
}
\end{atlascard}

\begin{atlascard}{\texttt{krw-d}}{table_icons_static/krw-d.png}{\carddatastruct{15}{1350}{225}{375}{bond}\cardphysicsgeneric{1.80250}{-1.21993}{0.471580}{none}{none}{0.23217}}
\atlasplotobject{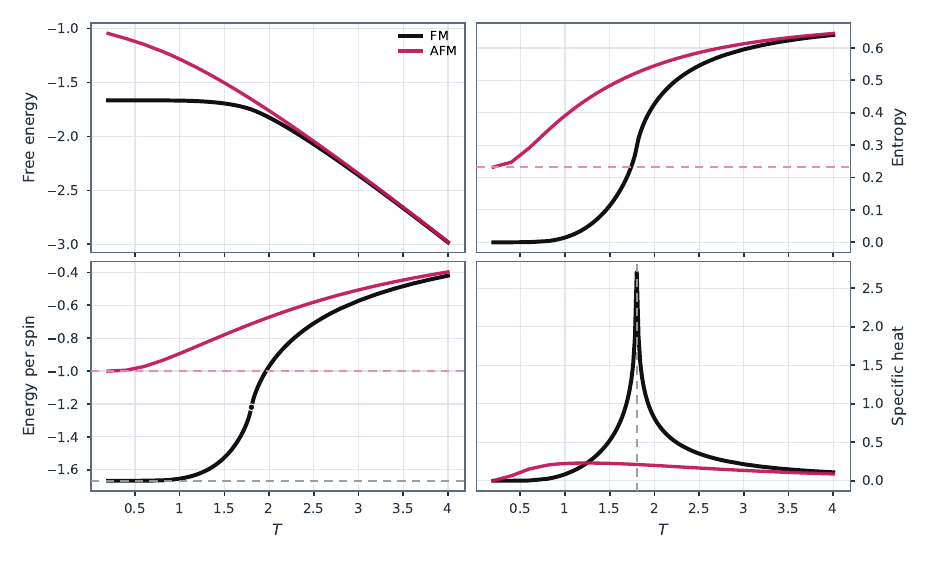}

\vspace{0.6em}
\hbox to \linewidth{%
  \hss
  \masktilemat{{\color{maskOne}w^{(1)}}}{15}{%
    {0,1,1,1,1,1,1,1,1,0,1,1,0,1,1},%
    {1,1,0,1,1,0,1,1,0,1,1,1,1,1,1},%
    {1,0,1,1,1,1,1,1,1,1,0,1,1,0,1},%
    {1,1,1,0,1,1,0,1,1,0,1,1,1,1,1},%
    {1,1,0,1,1,1,1,1,1,1,1,0,1,1,0},%
    {1,1,1,1,0,1,1,0,1,1,0,1,1,1,1},%
    {0,1,1,0,1,1,1,1,1,1,1,1,0,1,1},%
    {1,1,1,1,1,0,1,1,0,1,1,0,1,1,1},%
    {1,0,1,1,0,1,1,1,1,1,1,1,1,0,1},%
    {1,1,1,1,1,1,0,1,1,0,1,1,0,1,1},%
    {1,1,0,1,1,0,1,1,1,1,1,1,1,1,0},%
    {1,1,1,1,1,1,1,0,1,1,0,1,1,0,1},%
    {0,1,1,0,1,1,0,1,1,1,1,1,1,1,1},%
    {1,1,1,1,1,1,1,1,0,1,1,0,1,1,0},%
    {1,0,1,1,0,1,1,0,1,1,1,1,1,1,1}}
  \hss
  \masktilemat{{\color{maskTwo}w^{(2)}}}{15}{%
    {1,1,0,1,1,0,1,1,0,1,0,0,1,0,0},%
    {1,0,1,0,0,1,0,0,1,1,0,1,1,0,1},%
    {0,1,1,0,1,1,0,1,1,0,1,0,0,1,0},%
    {1,1,0,1,0,0,1,0,0,1,1,0,1,1,0},%
    {0,0,1,1,0,1,1,0,1,1,0,1,0,0,1},%
    {0,1,1,0,1,0,0,1,0,0,1,1,0,1,1},%
    {1,0,0,1,1,0,1,1,0,1,1,0,1,0,0},%
    {1,0,1,1,0,1,0,0,1,0,0,1,1,0,1},%
    {0,1,0,0,1,1,0,1,1,0,1,1,0,1,0},%
    {1,1,0,1,1,0,1,0,0,1,0,0,1,1,0},%
    {0,0,1,0,0,1,1,0,1,1,0,1,1,0,1},%
    {0,1,1,0,1,1,0,1,0,0,1,0,0,1,1},%
    {1,0,0,1,0,0,1,1,0,1,1,0,1,1,0},%
    {1,0,1,1,0,1,1,0,1,0,0,1,0,0,1},%
    {0,1,0,0,1,0,0,1,1,0,1,1,0,1,1}}
  \hss
  \masktilemat{{\color{maskThree}w^{(3)}}}{15}{%
    {1,0,1,1,0,1,1,0,0,0,0,0,0,0,0},%
    {0,0,0,0,0,0,0,0,1,0,1,1,0,1,1},%
    {0,1,0,1,1,0,1,1,0,0,0,0,0,0,0},%
    {1,0,0,0,0,0,0,0,0,1,0,1,1,0,1},%
    {0,0,1,0,1,1,0,1,1,0,0,0,0,0,0},%
    {1,1,0,0,0,0,0,0,0,0,1,0,1,1,0},%
    {0,0,0,1,0,1,1,0,1,1,0,0,0,0,0},%
    {0,1,1,0,0,0,0,0,0,0,0,1,0,1,1},%
    {0,0,0,0,1,0,1,1,0,1,1,0,0,0,0},%
    {1,0,1,1,0,0,0,0,0,0,0,0,1,0,1},%
    {0,0,0,0,0,1,0,1,1,0,1,1,0,0,0},%
    {1,1,0,1,1,0,0,0,0,0,0,0,0,1,0},%
    {0,0,0,0,0,0,1,0,1,1,0,1,1,0,0},%
    {0,1,1,0,1,1,0,0,0,0,0,0,0,0,1},%
    {0,0,0,0,0,0,0,1,0,1,1,0,1,1,0}}%
  \hss
}
\end{atlascard}

\begin{atlascard}{\texttt{usm-d}}{table_icons_static/usm-d.png}{\carddatastruct{18}{1944}{324}{540}{bond}\cardphysicsgeneric{1.80439}{-1.21840}{0.469736}{none}{none}{0.23149}}
\atlasplotobject{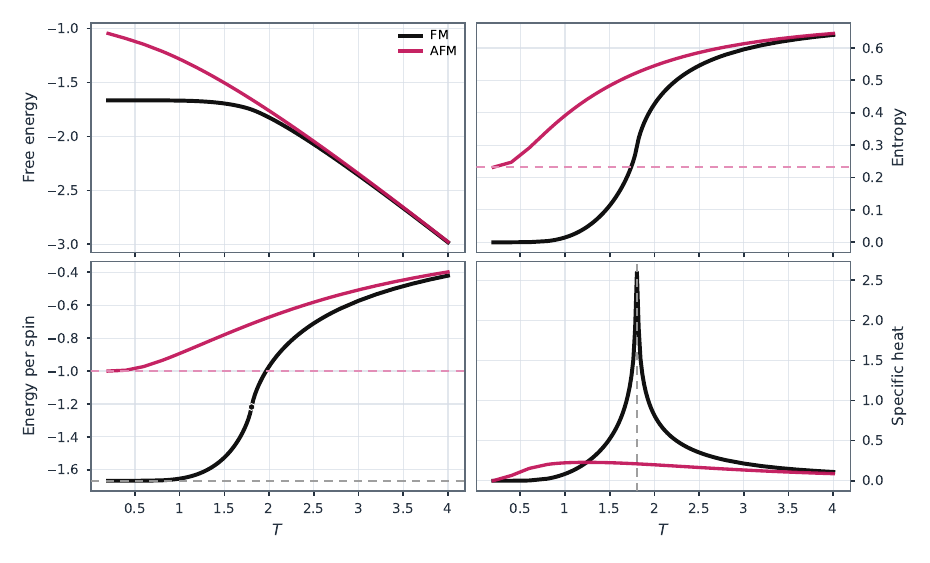}

\vspace{0.6em}
\hbox to \linewidth{%
  \hss
  \masktilemat{{\color{maskOne}w^{(1)}}}{18}{%
    {1,0,1,0,0,0,1,0,0,1,0,1,0,0,0,1,0,0},%
    {0,1,1,1,1,1,0,0,0,0,1,1,1,1,1,0,0,0},%
    {0,1,0,0,1,0,1,0,0,0,1,0,0,1,0,1,0,0},%
    {1,0,0,0,0,1,1,1,1,1,0,0,0,0,1,1,1,1},%
    {0,1,0,0,0,1,0,0,1,0,1,0,0,0,1,0,0,1},%
    {1,1,1,1,1,0,0,0,0,1,1,1,1,1,0,0,0,0},%
    {1,0,0,1,0,1,0,0,0,1,0,0,1,0,1,0,0,0},%
    {0,0,0,0,1,1,1,1,1,0,0,0,0,1,1,1,1,1},%
    {1,0,0,0,1,0,0,1,0,1,0,0,0,1,0,0,1,0},%
    {1,1,1,1,0,0,0,0,1,1,1,1,1,0,0,0,0,1},%
    {0,0,1,0,1,0,0,0,1,0,0,1,0,1,0,0,0,1},%
    {0,0,0,1,1,1,1,1,0,0,0,0,1,1,1,1,1,0},%
    {0,0,0,1,0,0,1,0,1,0,0,0,1,0,0,1,0,1},%
    {1,1,1,0,0,0,0,1,1,1,1,1,0,0,0,0,1,1},%
    {0,1,0,1,0,0,0,1,0,0,1,0,1,0,0,0,1,0},%
    {0,0,1,1,1,1,1,0,0,0,0,1,1,1,1,1,0,0},%
    {0,0,1,0,0,1,0,1,0,0,0,1,0,0,1,0,1,0},%
    {1,1,0,0,0,0,1,1,1,1,1,0,0,0,0,1,1,1}}
  \hss
  \masktilemat{{\color{maskTwo}w^{(2)}}}{18}{%
    {0,1,0,1,1,1,0,0,0,0,1,0,1,1,1,0,0,0},%
    {1,1,0,0,1,1,1,1,0,1,1,0,0,1,1,1,1,0},%
    {1,0,0,0,0,1,0,1,1,1,0,0,0,0,1,0,1,1},%
    {1,1,1,0,1,1,0,0,1,1,1,1,0,1,1,0,0,1},%
    {1,0,1,1,1,0,0,0,0,1,0,1,1,1,0,0,0,0},%
    {1,0,0,1,1,1,1,0,1,1,0,0,1,1,1,1,0,1},%
    {0,0,0,0,1,0,1,1,1,0,0,0,0,1,0,1,1,1},%
    {1,1,0,1,1,0,0,1,1,1,1,0,1,1,0,0,1,1},%
    {0,1,1,1,0,0,0,0,1,0,1,1,1,0,0,0,0,1},%
    {0,0,1,1,1,1,0,1,1,0,0,1,1,1,1,0,1,1},%
    {0,0,0,1,0,1,1,1,0,0,0,0,1,0,1,1,1,0},%
    {1,0,1,1,0,0,1,1,1,1,0,1,1,0,0,1,1,1},%
    {1,1,1,0,0,0,0,1,0,1,1,1,0,0,0,0,1,0},%
    {0,1,1,1,1,0,1,1,0,0,1,1,1,1,0,1,1,0},%
    {0,0,1,0,1,1,1,0,0,0,0,1,0,1,1,1,0,0},%
    {0,1,1,0,0,1,1,1,1,0,1,1,0,0,1,1,1,1},%
    {1,1,0,0,0,0,1,0,1,1,1,0,0,0,0,1,0,1},%
    {1,1,1,1,0,1,1,0,0,1,1,1,1,0,1,1,0,0}}
  \hss
  \masktilemat{{\color{maskThree}w^{(3)}}}{18}{%
    {0,0,1,1,1,0,0,1,1,0,0,1,1,1,0,0,1,1},%
    {1,0,1,1,1,0,1,1,1,1,0,1,1,1,0,1,1,1},%
    {0,0,1,1,0,0,1,1,1,0,0,1,1,0,0,1,1,1},%
    {0,1,1,1,1,0,1,1,1,0,1,1,1,1,0,1,1,1},%
    {0,1,1,1,0,0,1,1,0,0,1,1,1,0,0,1,1,0},%
    {0,1,1,1,0,1,1,1,1,0,1,1,1,0,1,1,1,1},%
    {0,1,1,0,0,1,1,1,0,0,1,1,0,0,1,1,1,0},%
    {1,1,1,1,0,1,1,1,0,1,1,1,1,0,1,1,1,0},%
    {1,1,1,0,0,1,1,0,0,1,1,1,0,0,1,1,0,0},%
    {1,1,1,0,1,1,1,1,0,1,1,1,0,1,1,1,1,0},%
    {1,1,0,0,1,1,1,0,0,1,1,0,0,1,1,1,0,0},%
    {1,1,1,0,1,1,1,0,1,1,1,1,0,1,1,1,0,1},%
    {1,1,0,0,1,1,0,0,1,1,1,0,0,1,1,0,0,1},%
    {1,1,0,1,1,1,1,0,1,1,1,0,1,1,1,1,0,1},%
    {1,0,0,1,1,1,0,0,1,1,0,0,1,1,1,0,0,1},%
    {1,1,0,1,1,1,0,1,1,1,1,0,1,1,1,0,1,1},%
    {1,0,0,1,1,0,0,1,1,1,0,0,1,1,0,0,1,1},%
    {1,0,1,1,1,1,0,1,1,1,0,1,1,1,1,0,1,1}}%
  \hss
}
\end{atlascard}